\documentclass[a4paper,10pt]{report}

\usepackage[inner=3.cm,outer=3.cm,bottom=2cm]{geometry}
\usepackage[utf8]{inputenc}
\usepackage{graphicx}
\usepackage{tikz}
\usepackage{amsmath}
\usepackage{amssymb}
\usepackage{xspace}
\usepackage{slashed}
\usepackage{young}
\usepackage[hidelinks]{hyperref}

\usepackage{bm}
\usepackage{bbm}
\usepackage{amsfonts}

\usepackage[vcentermath]{youngtab}

\usepackage{tensor}

\pdfpageheight\paperheight
\pdfpagewidth\paperwidth

\usepackage{etoolbox}
    
    \patchcmd{\maketitle}{\@fpheader}{}{}{}

\usepackage{amsfonts}
\usepackage{bbm}
\usepackage{amssymb}
\usepackage{amsmath}

\usepackage{graphicx,color}
\usepackage{slashed}
\usepackage{young}
\usepackage[vcentermath]{youngtab}

\makeatletter

\def\a{\alpha}
\def\b{\beta}
\def\g{\gamma}
\def\G{\Gamma}
\def\d{\delta}
\def\D{\Delta}
\def\e{\varepsilon}
\def\ve{\varepsilon}
\def\z{\zeta}
\def\h{\eta}
\def\th{\theta}
\def\Th{\Theta}

\def\l{\lambda}
\def\L{\Lambda}
\def\m{\mu}
\def\n{\nu}
\def\x{\xi}
\def\X{\Xi}

\def\P{\Pi}

\def\r{\rho}

\def\s{\sigma}

\def\vf{\varphi}

\def\o{\omega}
\def\O{\Omega}

\def\cC{{\cal C}}

\def\cF{{\cal F}}
\def\cG{{\cal G}}
\def\cH{{\cal H}}

\def\cJ{{\cal J}}

\def\cL{{\cal L}}

\def\cN{{\cal N}}
\def\cO{{\cal O}}

\def\cQ{{\cal Q}}

\def\cT{{\cal T}}

\def\be{\begin{equation}}
\def\ee{\end{equation}}
\def\bea{\begin{eqnarray}}
\def\eea{\end{eqnarray}}
\def\ba{\begin{array}}
\def\ea{\end{array}}

\def\nn{\nonumber}

\def\nd{\nabla\!\cdot}

\def\pe{\prime}
\def\12{\frac{1}{2}}
\def\pr{\partial}
\def\prd{\partial \cdot}

\newcommand{\bin}[2]{{#1 \choose #2}}
\newcommand{\und}[1]{\underline{#1}}

\newtheorem{theorem}{Theorem}
\newtheorem{lemma}{Lemma}
\newtheorem{corollary}{Corollary}

\usepackage{appendix}
\usepackage{chngcntr}
\usepackage{etoolbox}
\usepackage{lipsum}

\usepackage[nosort]{cite}

 \csname
@addtoreset\endcsname{equation}{section}

\newcommand{\supsuboverbrace}[2][]{%
  {\everymath{\scriptstyle}%
   \underbrace{\scriptstyle#2}_{#1}}%
}

\newcommand{\supsuboverbrac}[2][]{%
  {\everymath{\scriptstyle}%
   \overbrace{\scriptstyle#2}^{#1}}%
}

\def\a{\alpha}
\def\b{\beta}
\def\g{\gamma}
\def\G{\Gamma}
\def\d{\delta}
\def\D{\Delta}
\def\e{\epsilon}
\def\ve{\varepsilon}
\def\z{\zeta}
\def\h{\eta}
\def\th{\theta}
\def\Th{\Theta}

\def\l{\lambda}
\def\L{\Lambda}
\def\m{\mu}
\def\n{\nu}
\def\x{\xi}
\def\X{\Xi}

\def\P{\Pi}

\def\r{\rho}

\def\s{\sigma}

\def\vf{\varphi}

\def\o{\omega}
\def\O{\Omega}

\def\cC{{\cal C}}

\def\cF{{\cal F}}
\def\cG{{\cal G}}
\def\cH{{\cal H}}

\def\cJ{{\cal J}}

\def\cL{{\cal L}}

\def\cN{{\cal N}}
\def\cO{{\cal O}}

\def\cQ{{\cal Q}}

\def\cT{{\cal T}}

\def\be{\begin{equation}}
\def\ee{\end{equation}}
\def\bea{\begin{eqnarray}}
\def\eea{\end{eqnarray}}
\def\ba{\begin{array}}
\def\ea{\end{array}}

\def\nn{\nonumber}

\def\nd{\nabla\!\cdot}

\def\pe{\prime}
\def\12{\frac{1}{2}}
\def\pr{\partial}
\def\prd{\partial \cdot}

\AtBeginEnvironment{subappendices}{%
\chapter*{Appendix}
\addcontentsline{toc}{chapter}{Appendices}
\counterwithin{figure}{section}
\counterwithin{table}{section}
}

\usetikzlibrary{shapes,snakes}
\usetikzlibrary{decorations.pathmorphing}
\usetikzlibrary{decorations.markings}
\graphicspath{ {Figures/} }
\title{Aspects of higher spin Hamiltonian dynamics: \\ conformal geometry, duality and charges}
\author{Amaury Leonard}

\begin{document}


\begin{titlepage}

\begin{tikzpicture}[remember picture,overlay]
  \node[yshift=2cm,xshift=3cm] at (current page.south west)
              {\includegraphics[scale=0.4]{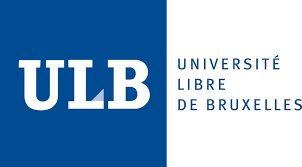}};
\end{tikzpicture}
\begin{center}
\textbf{UNIVERSIT\'E LIBRE DE BRUXELLES}\\
\textbf{Faculté des Sciences}\\
\textbf{Département de Physique}\\
\textbf{Service de Physique Mathématique des Interactions Fondamentales}
\vfill{}\vfill{}

{\Huge Aspects of higher spin Hamiltonian dynamics: Conformal geometry, duality and charges}

{\Huge \par}
\begin{center}{\LARGE Amaury Leonard}\end{center}{\Huge \par}
\vfill{}\vfill{}
\begin{flushright}{\large \textbf{Promoteur:}}\hfill{}{\large Thèse de doctorat présentée en vue de}\\
{\large Marc Henneaux}\hfill{}{\large l'obtention du titre de}\\
{\large \textbf{Jury:}}\hfill{}{\large Docteur en Sciences}\\
{\large Glenn Barnich}\hfill{}$\phantom{a}$\\
{\large Xavier Bekaert}\hfill{}$\phantom{a}$\\
{\large Frank Ferrari}\hfill{}$\phantom{a}$\\
{\large Axel Kleinschmidt}\hfill{}$\phantom{a}$\\
{\large Michel Tytgat}\hfill{}$\phantom{a}$
\end{flushright}{\large\par}
\vfill{}\vfill{}
\textbf{Année académique 2016~-~2017}
\end{center}
\end{titlepage}

\pagenumbering{roman}

\newpage
\thispagestyle{empty} 
\null

\newenvironment{vcenterpage}
{\newpage\thispagestyle{empty} 
\vspace*{\fill}}
{\vspace*{\fill}\par\pagebreak}

\begin{vcenterpage}
\begin{flushright}
    \large\em\null\vskip1in 
    \`A François, Christian, Antigone et Gwladys \vfill
  \end{flushright}
\end{vcenterpage}
\thispagestyle{empty}
\vspace*{5cm}

\thispagestyle{empty} 
\tableofcontents

\newpage
\thispagestyle{empty}

\vspace{10cm} 

This thesis was defended behind closed doors on June 8th and publicly
on July 3rd 2017, in front of the following jury at U.L.B., in Brussels:\\

\begin{itemize}
\item President: Prof. Frank Ferrari (\textit{Université Libre de Bruxelles}), \\
\item Secretary: Prof. Glenn Barnich (\textit{Université Libre de Bruxelles}), \\
\item Advisor: Prof. Marc Henneaux (\textit{Université Libre de Bruxelles}), \\
\item Prof. Michel Tytgat (\textit{Université Libre de Bruxelles}),\\
\item Dr. Xavier Bekaert (\textit{Université François Rabelais}),\\
\item Dr. Axel Kleinschmidt (\textit{Max-Planck-Institut für Gravitationsphysik}).\\
\end{itemize}

\chapter*{Acknowledgements}

\addcontentsline{toc}{chapter}{Acknowledgements}
\thispagestyle{empty} 

If the writing of this thesis sometimes proved to be a long and winding path, it could not have been completed without the help and support of many people. Among them, the decisive role was certainly played by my supervisor Marc Henneaux. For the swift efficiency, outstanding clarity and intellectual depth of his direction, advices, suggestions and explanations, for the kind, patient and generous presence he continuously offered during these years, I am eternally grateful.\\

My deepest thanks also go to Glenn Barnich, Xavier Bekaert, Frank Ferrari, Axel Kleinschmidt and Michel Tytgat, for accepting to constitute my jury and to bring forward the thoughts, commentaries and judgement this modest work inspired them. In addition, my gratitude goes to them regarding their long helping presence as elders and colleagues and, for many of them, as professors.\\

Many more of my colleagues deserve my heartfelt thanks under many regards, not only for the ideas, enlightenment and collaboration they bestowed upon me but also, now and then, for the precious friendship they made me the honor of. Among them, my most appreciative thoughts go to Andrea Campoleoni, Laure-Anne Douxchamps, Victor Lekeu, Gustavo Lucena Gomez, Sergio Hortner, Blagoje Oblak and Rakibur Rahman.\\

Beyond the academic universe, I received crucial assistance from as many origins. My affection is owed without limit to my irreplaceable and unfalteringly loyal friends and brotherly room-mates, Gwladys Lefeuvre, Antigone Germain, François Postic and Christian Denaeyer, for sharing with me for four years various places which shall forever be remembered as home to me. \\

For their brilliant - although too often aloof - company, trusted affection and bond, Antoine Ledent and Ismail Zekhnini rank high in the canopy of my thankful thoughts. For his invaluable intellectual understanding, welcoming friendship and moral support, Jonathan Lindgren remains the object of my boundless debt. As for Antoine Bourget and Antoine Brix, the thought of their role will remain with me as an everlasting symbol of human attachment, closeness of thought and ineffable hospitality.\\

Finally and most essentially, there are those toward whom my debt is primordial and always replenished and my affection limitless and eternal: my family, my sister and my parents, who provided me with the most treasured of all support, which shall always accompany me.

\chapter*{Credits}

\addcontentsline{toc}{chapter}{Credits}

The original results of this thesis were presented in the following papers, on which we heavily rely:

\begin{itemize}

\item   C.~Bunster, M.~Henneaux, S.~H\"ortner and A.~Leonard,
  ``Supersymmetric electric-magnetic duality of hypergravity,''
  Phys.\ Rev.\ D {\bf 90}, no. 4, 045029 (2014)
[\href{http://arxiv.org/abs/arXiv:1406.3952}{arXiv:1406.3952 [hep-th]}].

\item   M.~Henneaux, S.~H\"ortner and A.~Leonard,
  ``Higher Spin Conformal Geometry in Three Dimensions and Prepotentials for Higher Spin Gauge Fields,''
  JHEP {\bf 1601} (2016) 073
[\href{http://arxiv.org/abs/arXiv:1511.07389}{arXiv:1511.07389 [hep-th]}].

\item   M.~Henneaux, S.~H\"ortner and A.~Leonard,
  ``Twisted self-duality for higher spin gauge fields and prepotentials,''
  Phys.\ Rev.\ D {\bf 94}, no. 10, 105027 (2016)
    [\href{http://arxiv.org/abs/arXiv:1609.04461}{arXiv:1609.04461 [hep-th]}].

\item   A.~Campoleoni, M.~Henneaux, S.~H\"ortner and A.~Leonard,
  ``Higher-spin charges in Hamiltonian form. I. Bose fields,''
  JHEP \textbf{1610} (2016) 146
  [\href{http://arxiv.org/abs/arXiv:1608.04663}{arXiv:1608.04663 [hep-th]}].

\item   A.~Campoleoni, M.~Henneaux, S.~H\"ortner and A.~Leonard,
  ``Higher-spin charges in Hamiltonian form. II. Fermi fields,''
  JHEP \textbf{1702} (2017) 058
  [\href{http://arxiv.org/abs/arXiv:1701.05526}{arXiv:1701.05526 [hep-th]}].

\item M. Henneaux, V. Lekeu, A. Leonard, ``Chiral tensors of mixed Young symmetry," Phys.\ Rev.\ D \textbf{95}, no.8, 084040 (2017)
	  [\href{http://arxiv.org/abs/arXiv:1612.02772 }{arXiv:1612.02772  [hep-th]}].

\end{itemize}

\chapter*{Abstract}

\addcontentsline{toc}{chapter}{Abstract}

\thispagestyle{empty} 

We have examined the properties of free higher spin gauge fields through an investigation of various aspects of their Hamiltonian dynamics. Over a flat background space-time, the constraints produced by the Hamiltonian analysis of these gauge systems were identified and solved through the introduction of prepotentials, whose gauge invariance intriguingly contains both linearized generalized diffeomorphisms and linearized generalized Weyl rescalings, which motivated a systematic study of conformal invariants for higher spins. We built these invariants with the Cotton tensor, whose properties (tracelessness, symmetry, divergencelessness; completeness, invariance) we established. With these geometric tools, the Hamiltonian analysis was then brought to completion, and a first order action written down in terms of the prepotentials. This action is observed to exhibit manifest invariance under electric-magnetic duality; this invariance, together with the gauge freedom of the prepotentials, actually completely fixes the action. More generally, this action is also observed to be the same as the one obtained through a rewriting of higher spin equations of motion as (non-manifestly covariant) twisted self-duality conditions.

With an interest in supersymmetric extensions, we began to extend this study to fermions. The spin $5/2$ massless free field was subjected to a similar analysis, and its prepotential found to share the conformal gauge invariance observed in the general bosonic case. The spin $2$-spin $5/2$ supermultiplet was considered, and a rigid symmetry of its action, combining an electric-magnetic duality rotation of the spin $2$ with a chirality rotation of the spin $5/2$, was built to commute with supersymmetry. In another research, we studied the properties of a mixed symmetry chiral tensor field over a flat six-dimensional space-time: the so-called $(2,2)$ form, who appears in the intriguing $(4,0)$ six-dimensional supersymmetric theory. Hamiltonian analysis was performed, prepotentials introduced, and the first order action again turned out to be the same as the one obtained through a rewriting of the equations of motion of the field as (non-manifestly covariant) self-duality conditions.

Finally, we made a study of both fermionic and bosonic higher spin surface charges over a constantly curved background space-time. This was realized through the Hamiltonian analysis of these systems, the constraints being identified as the generators of gauge transformations. Plugging into these generators values of the gauge parameters corresponding to improper gauge transformations (imposing a physical variation of the fields), their finite and non-vanishing on-shell values were computed and recognized as conserved charges of the theory. Their algebra was checked to be abelian.\\

\noindent \textbf{Keywords}: higher spins, duality, chirality, gauge theory, hamiltonian formalism, conformal geometry, surface charges.

\thispagestyle{empty} 

\chapter*{Résumé}

\addcontentsline{toc}{chapter}{Résumé}
\thispagestyle{empty} 

Nous avons investigué les propriétés des champs de jauge de spin élevé libres à travers une étude de divers aspects de leur dynamique hamiltonienne. Pour des champs se propageant sur un espace-temps plat, les contraintes issues de l'analyse hamiltonienne de ces théories de jauge ont été identifiées et résolues par l'introduction de prépotentiels, dont l'invariance de jauge comprend, de façon intrigante, à la fois des difféomorphismes linéarisés généralisés et des transformations d'échelle de Weyl généralisées et linéarisées. Cela a motivé notre étude systématique des invariants conformes pour les spins élevés. Les invariants correspondants ont été construits à l'aide du tenseur de Cotton, dont nous avons établi les propriétés essentielles (symétrie, conservation, trace nulle; invariance, complétude). Avec ces outils géométriques, l'analyse hamiltonienne a pu être complétée et une action du premier ordre écrite en termes des prépotentiels. Nous avons constaté que cette action possédait une invariance manifeste par dualité électromagnétique; cette invariance, combinée à l'invariance de jauge des prépotentiels, fixe d'ailleurs uniquement l'action. En outre, de façon générale, cette action s'est révélée être exactement celle obtenue à travers une réécriture des équations du mouvement des spins élevés comme des conditions d'auto-dualité tordue (non manifestement covariantes).

Avec un intérêt pour les extensions supersymétriques, nous avons amorcé la généralisation de cette étude aux champs fermioniques. Le champ de masse nulle libre de spin $5/2$ a été soumis à la même analyse, et son prépotentiel s'est révélé partager l'invariance de jauge conforme déjà observée dans le cas bosonique général. Le supermultiplet incorporant les spins $2$ et $5/2$ a ensuite été considéré, et une symétrie rigide de son action, combinant une transformation de dualité électromagnétique du spin $2$ avec une transformation de chiralité du spin $5/2$ a été construite pour commuter avec la supersymétrie. Dans une autre direction, nous avons étudié les propriétés d'un champ tensoriel chiral de symétrie mixte dans un espace-temps plat à six dimensions: une $(2,2)$-forme. Son analyse hamiltonienne a été réalisée, des prépotentiels introduits et l'action de premier ordre obtenue s'est encore une fois révélée être la même que celle obtenue à travers une réécriture des équations du mouvement comme des conditions d'auto-chiralité (non manifestement covariante).

Finalement, nous nous sommes penchés sur les charges de surface des champs fermioniques et bosoniques de spin élevé se propageant sur un espace-temps à courbure constante. Cela a été réalisé par une analyse hamiltonienne de ces systèmes, les contraintes étant identifiées aux générateurs des transformations de jauge. Injectant dans ces générateurs des valeurs des paramètres des transformations de jauge correspondant à des transformations impropres de jauge (imposant une réelle variation physique sur les champs) a ensuite permis d'évaluer la valeur de ces générateurs pour des champs résolvant les équations du mouvement: elle s'est bien révélée finie et non-nulle, constituant les charges de surface de ces théories.
\\

\noindent \textbf{Mots-clefs}: spins élevés, dualité, chiralité, théories de jauge, formalisme hamiltonien, géométrie conforme, charges de surface.

\thispagestyle{empty}

\chapter*{General conventions}

\addcontentsline{toc}{chapter}{General conventions}

Our general conventions are as follows.\\

Symmetrization (denoted by parenthesis) and antisymmetrization (denoted by brackets) are carried with weight one: $A_{(\mu\nu)} \equiv \frac{1}{2} \left(A_{\mu\nu} + A_{\nu\mu}\right)$ and $A_{[\mu\nu]} \equiv \frac{1}{2} \left(A_{\mu\nu} - A_{\nu\mu}\right)$.\\

For Minkowski or AdS spacetime, we work with the mostly plus signature $- \ + \cdots \ +$. The AdS radius of curvature is $L$.\\

On Minkowski, in dimension $D$, Greek indices take values from $0$ to $D-1$ while Latin indices run from $1$ to $D-1$ (spatial directions).  The covariant trace is denoted by a prime: $h'\equiv h^{\mu}_{\phantom{\mu}\mu}$. The spatial trace is denoted by a bar: $\bar{h} \equiv h^k_{\phantom{k}k}$. Unless otherwize specified, we are in dimension four, $D=4$.\\

The fully antisymmetric Lorentz tensor in dimension four, $\epsilon_{\mu\nu\rho\sigma}$, satisfies $\epsilon^{0123} = - 1 =  \epsilon_{0123}$.\\

In $D=4$, the Dirac matrices are in a Majorana representation (matrices $\gamma_{\mu}$ real, $\gamma_0^T = - \gamma_0$ and $\gamma_k^T = \gamma_k$ where $T$ denotes the transposition; and so $\gamma^{\mu \dagger} = \gamma^{\mu T} =\gamma^{0} \gamma^{\mu} \gamma^{0}$). 

In addition, $\gamma_5 \equiv  \gamma_0 \gamma_1 \gamma_2 \gamma_3 = - 1/4! \ \epsilon^{\mu \nu \rho \sigma} \gamma_{\mu} \gamma_{\nu} \gamma_{\rho} \gamma_{\sigma}$, which implies $\gamma_5^{\dagger} = -\gamma_5$, $\gamma_5^{T} = - \gamma_5$ et $\left( \gamma_5 \right)^2 =- I$. Finally, $\gamma^{\mu\nu} \equiv \frac{1}{2} \left[ \gamma^{\mu} , \gamma^{\nu} \right] = \gamma^{[\mu}\gamma^{\nu]}$ (and $\delta^{\mu\nu\rho}_{\alpha\beta\gamma} \equiv 6 \ \delta^{[\mu}_{\alpha} \delta^{\nu}_{\beta}\delta^{\rho]}_{\gamma}$), etc. \\

In the internal plane of the prepotentials associated to the electric and magnetic fields, we use indices $a$, $b$ and $c$ (and only them). They are equal to one or two. The $SO(2)$ invariant tensors are $\epsilon_{ab}$ and $\delta_{ab}$. The first one is antisymmetric and satisfies $\epsilon_{12} = 1$. The second is symmetric and diagonal, with eigenvalues both equal to one. $\delta_{ab}$ and its inverse are used to lower and raise this kind of index.

Specific conventions are used in the study of surface charges on AdS, and we explain them in appendices \textbf{\ref{app:conventions}} for bosons and \textbf{\ref{app:conventions_fermi}} for fermions.

\chapter*{Introduction}

\pagenumbering{arabic}

\addcontentsline{toc}{chapter}{Introduction}

Higher spin fields arise as the most natural generalization of the familiar fields involved in the Standard Model of particle physics. The particles appearing therein can be seen as corresponding to irreducible representations of the Poincaré group of spin equal to $0$, $1/2$ or $1$ (scalar, spinor, vector)\footnote{The complete classifications of the unitary representations of the Poincaré group was realized by Wigner in 1939 \cite{Wigner}.}. Linearized general relativity and supergravity are well known to also contain spin $3/2$ and $2$ representations of this group. It seems quite natural to consider particles corresponding to irreducible representations of the Poincaré group of arbitrary spin, especially since string theory is known to incorporate low energy modes that correspond to (increasingly massive) fields of arbitrarily high spin.

We will particularly be interested in massless representations of the Poincaré group, which are the gauge fields. The presence of a variety of no-go theorems, apparently forbidding the construction of non-trivial, interacting theories involving higher spins\footnote{See \cite{Bekaert:2010hw} and its references for a review.} has long reduced the enthusiasm of such investigations - until the discoveries of the last decades rekindle it: a possibly interacting theory of great algebraic richness has been developed, if only at the level of its equations of motion\footnote{See, however, the nonstandard action principle proposed in \cite{actionHS1} and the references therein.}\footnote{This theory was developed by the Lebedev school, see \cite{Fradkin:1986ka,Fradkin:1986qy,Fradkin:1987ks,Vasiliev:1990en,Vasiliev:1999ba}. Many reviews exist, among which \cite{Didenko:2014dwa,Bekaert:2005vh,review-strings,Vasiliev:1995dn,review-Giombi}. \label{note_Vasiliev}}: \textit{Vasiliev theory}. This theory is actually defined on anti-de Sitter space-time (AdS), whose symmetry group, different from Poincaré, also admits massless representations, and it involves an infinite tower of fields of arbitrarily high spin. The construction of interacting theories of massless higher spin fields over a flat space-time presents considerable difficulties, due the abovementionned class of no-go theorems preventing their minimal coupling to gravity or to any form of matter\footnote{See \cite{Weinberg1,Weinberg2,Porrati}.}; nevertheless, attempts exist to circumvent these theorems, notably through non-minimal couplings involving higher derivatives and the consideration of an infinite collection of fields of increasing spin.  In any case, massless higher spins are of interest regarding the tensionless, very high energy limit of string theory, whose symmetries also seem to be extremely rich\footnote{See \cite{Gross:1988ue}.}. 

This work, however, focuses specifically on \textit{free} higher spin gauge fields, whose dynamical and geometrical features we have attempted to clarify for particular aspects, which may hopefully prove to be of some use regarding the ultimately far more interesting and complex interacting theories. Elegant second order covariant equations of motion and Lagrangian action principles have been built describing these massless fields of arbitrary spin, constituting the so-called Fronsdal formalism\footnote{The elaboration of these results has been a long historical process, from Fierz-Pauli program \cite{FierzPauli} to the massive Lagrangians of Singh and Hagen \cite{Singh:1974bosons,Singh:1974rc} and their massless limit worked out by Fang and Fronsdal \cite{Fronsdal:1978rb,Fang:1978wz,fronsdal-AdS,fronsdal-AdS_fermi}, over both flat and constantly curved space-time.}. Fronsdal's formalism naturally relies on gauge invariance to describe with Lorentz invariant tensor fields a system whose physical degrees of freedom form a massless representation of Poincaré group. The gauge variations of free higher spin massless fields, which are necessary in order to remove the overnumerous components of Lorentz tensors, are linearized generalized diffeomorphisms.

The dynamical behaviour of gauge systems is not always fully transparent, and it is why we have been interested in the \textit{Hamiltonian analysis} of higher spin gauge fields. Hamiltonian analysis of gauge systems necessarily presents a certain number of subtleties, which revolve around the presence of a specific type of constraint. The phase space of gauge systems, parametrized by coordinates and their conjugate momenta, is not entirely accessible to the trajectories of the system; in addition, even on the ``constrained" subdomain of the phase space to which authorized motion is restricted, different points actually correspond to the same physical configuration of the system. The points of the constrained surface which are physically indistinguishable are those which can be mapped onto each other by a gauge transformation, whose generators are precisely the constraints, the tangents to the gauge orbits simply being the symplectic gradient of the constraints\footnote{In other words, gauge systems are characterized by constraints whose symplectic gradient ($i.e.$  the vector associated to their gradient by the symplectic two-form of the phase space) is tangent to the constraint surface. See \cite{Dirac:LQM,Henneaux:1994pup}.}.

As far as free higher spin gauge fields are concerned, the constraints unveiled through the Hamiltonian analysis of these theories - taking their Fronsdal second order Lagrangian description as a starting point - take the form of local differential expressions in terms of the Fronsdal fields (or Pauli-Fierz fields)\footnote{The metric-like fields used to describe massless higher spins are sometimes refered to as ``potentials" because of their gauge content: the physical, gauge invariant variables are the curvatures built from these fields. This terminology is the reason why the variables in terms of which these metric-like fields are expressed are called ``prepotentials".} and their momenta. For the well-known spin one case - that is, classical electromagnetism, these constraints are simply Gauss law, setting the divergence of the conjugate momentum of the spatial components of the potential to zero. In a four-dimensional space time, this constraint is easily solved by the introduction of a second, ``dual" potential spatial vector of which the spatial momentum is the curl\footnote{See \cite{Deser:1976iy}.}. In the free spin two case, which is linearized general relativity,  Hamiltonian analysis generates constraints on both the Pauli-Fierz field spatial components and its conjugate momenta, and these constraints have both been solved\footnote{See \cite{Henneaux:2004jw}.}, leading to the introduction of two \textit{prepotentials}: spatial symmetric tensor fields of rank two in terms of which the potential field and its momenta both identically satisfy the constraints. A remarkable feature appeared in this instance\footnote{This feature is absent from the spin one case because the Pauli-Fierz field does not have enough indices in order for a possible non-trivial conformal transformation to exist.} the gauge invariance of these prepotentials not only included linearized diffeomorphisms, but also linearized \textit{Weyl rescalings}. Indeed, the general gauge variation of one of these prepotentials (say $Z_{ij}$) was found to be given by:
\begin{equation}
\delta_{gauge} Z_{ij} = 2 \ \partial_{(i} \xi_{j)} \ + \ \delta_{ij} \lambda .
\label{intro_conf_gauge_2} \nonumber
\end{equation}
The first part of this variation is simply the first order effect of a diffeomorphism on a variation of the metric around flatness, while the second part is the corresponding first order effect of a rescaling of the metric.

The presence of this enlarged, conformal gauge invariance at the level of the prepotentials turned out to be a general feature, although its conceptual origin remains elusive. Nevertheless, it led us to investigate in greater detail the conformal geometry of higher spins - at the linearized level - in dimension three (which is relevant to study a field theory over a four-dimensional space-time). The interest of such an investigation goes beyond the intriguing conformal gauge invariance of the Hamiltonian unconstrained variables of higher spins, since conformal higher spin fields have also been the object of abundant research in the recent past\footnote{See \cite{Fradkin:1985am,Pope:1989vj,Fradkin:1989md,Fradkin:1989xt,Vasiliev:2001zy,Segal:2002gd,Shaynkman:2004vu,Vasiliev:2007yc,Metsaev:2007rw,Marnelius:2008er,Metsaev:2009ym,Vasiliev:2009ck,Florakis:2014kfa,Nutma:2014pua}.}. To be definite, we considered fields propagating over a flat four-dimensional space-time, in which Fronsdal's formalism describes higher spin bosonic massless fields by fully symmetric Lorentz tensors, making the prepotentials symmetric spatial fields, in a three-dimensional Euclidean flat space. The generalization of the spin two prepotential gauge invariance for a spin prepotential ($Z_{i_1 \cdots i_s}$) is:
\begin{equation}
\delta_{gauge} Z_{i_1 \cdots i_s} =
s \ \partial_{(i_1} \xi_{i_2 \cdots i_s )}
\ + \ \frac{s(s-1)}{2} \ \delta_{(i_1 i_2} \lambda_{i_3 \cdots i_s )} .
\label{intro_conf_gauge_s} \nonumber
\end{equation}
The constructions of a complete set of invariants under such transformations were a preliminary task to accomplish before embarking on the Hamiltonian analysis of higher spins. In any dimension, the gauge invariant curvature associated to the first part of this gauge transformation is the generalized Riemann tensor (or Freedman-de Wit tensor\footnote{See \cite{dWF}.}) - or, equivalently in three dimensions, the generalized Einstein tensor. It forms a complete set of gauge invariants in the sense that a necessary and sufficient condition for a field to be pure gauge is for its Riemann tensor to vanish; also, any local gauge invariant function of the field is a function of its Riemann tensor (and of its derivatives). In dimension four or higher, the corresponding \textit{conformal curvature} is simply the traceless part of the Riemann tensor, the so-called Weyl tensor. However, in dimension three, this tensor is identically zero, which leads, in the spin two and three cases, to the construction of a higher order curvature: the Cotton tensor\footnote{The peculiarities of thre-dimensional conformal geometry are well-known; see \cite{Eisenhart}. The Cotton tensor for higher spins was introduced in \cite{Damour:1987vm,Pope:1989vj}.}. We have extended the known properties of the Cotton tensor for spin three\footnote{See \cite{Damour:1987vm}.} to all integer spins: in arbitrary dimension, the Cotton tensor is a tensor whose symmetry type corresponds to a Young tableau of the following form:
\begin{equation}
\overbrace{\yng(10,9)}^{\text{$s$ boxes}} , \nonumber
\end{equation}
whose derivative projected on the symmetry type associated to a rectangular Young tableau (with two rows of length $s$) gives zero.
In three dimensions, it can equivalently be defined as a symmetric, divergenceless and traceless tensor, obtained by taking $2s-1$ derivatives of a spin-$s$ prepotential. It is indeed a complete set of conformal invariants and, in addition, we established that any tensor sharing the algebraic and differential properties of the Cotton tensor \textit{is} the Cotton tensor of some spin-$s$ prepotential\footnote{We gathered these results in our paper \cite{Henneaux:2015cda}.}\footnote{These results have also been obtained in \cite{BBB}, which studies the conformal geometry in arbitrary dimension for an arbitrary integer spin, using the ordinary differential (which squares to zero) in the frame-like formalism. In this approach, the Schouten tensor is shown to be a component of the conformal one-form connection. The authors of \cite{Linander:2016brv} also derived partial results in the same direction.}.

With these tools in hand, we were able to complete the Hamiltonian analysis of all integer spin gauge fields\footnote{In our papers \cite{Henneaux:2015cda,Henneaux:2016zlu}.}: the expansion of Fronsdal's action led us to the spin-$s$ field constraints, and our knowledge of conformal geometry allowed us to solve these constraints in terms of two prepotential fields, which turned out to be symmetric spatial tensors of rank $s$, endowed with exactly the postulated conformal gauge invariance. The emergence of this spin-$s$ Weyl invariance remains as intriguing as it was for the graviton, since we started with Fronsdal's Lagrangian action, in which no sign of higher spin conformal gauge symmetry is present, the resolution of the constraints appearing in the Hamiltonian analysis bringing in prepotentials enjoying somewhat unexpectedly this conformal invariance.\\

Written in terms of the prepotentials, the Hamiltonian action of higher spins is no longer manifestly Lorentz invariant. However, it is manifestly invariant under another type of transformations that could have a deeper significance: $SO(2)$ \textit{electric-magnetic duality} rotations. This symmetry stems from the old observation that Maxwell's equations in the vacuum are invariant under rotations in the internal plane of the electric and magnetic fields:
\begin{eqnarray}
\vec{E} &\rightarrow& \vec{E}\cos \theta \ - \ \vec{B} \sin \theta , \nonumber
\\
\vec{B} &\rightarrow& \vec{B}\cos \theta \ + \ \vec{E} \sin \theta . \nonumber
\end{eqnarray}
This is a symmetry of the equations of motion and of their solutions, and it long seemed dubious whether it could be extended into an off-shell symmetry, leaving invariant the action from which these equations derive, the form these electric-magnetic duality rotations take when acting on the covariant one-form potential from which the electric and magnetic fields derive being far from obvious. However, the remarkable discovery was made that when one goes to the Hamiltonian action and explicitly solves Gauss law by introducing a second potential spatial vector, electric-magnetic duality becomes an apparent symmetry of this action. Indeed, as it turns out, it is represented off-shell by a rotation in the internal plane of the two potential vectors, of which the electric and magnetic fields are simply the curls\footnote{See \cite{Deser:1976iy,Deser:1997mz}.}. This observation was then extended to the free graviton, for which, on-shell, $SO(2)$ electric-magnetic duality transformations take the form of rotations in the internal plane of the Riemann tensor and its dual ($i.e.$ its Hodge dual taken over its first pair of indices); they constitute an on-shell symmetry since the linearized form of Einstein's equations in the vacuum sets the trace of the Riemann tensor to zero, which is equivalent to forcing the dual of the Riemann tensor to satisfy Bianchi algrebraic identity, and reciprocally. Again, the off-shell extension of these transformations is not obviously possible, but it was shown that the Hamiltonian action of the spin two, written in terms of prepotentials, exhibits manifest electric-magnetic invariance\footnote{See \cite{Henneaux:2004jw,Deser:2004xt,Julia:2005ze}. This approach has been applied to other systems, see \cite{Henneaux:1988gg,ScSe,Deser:1997mz,Hillmann:2009zf,Bunster:2011aw,Bunster:2011qp}.}.

A rather noteworthy feature of these invariances is that they completely fix the action. Both for the spin one and two, the gauge invariance of the two prepotentials \eqref{intro_conf_gauge_s} together with the $SO(2)$ electric-magnetic duality invariance under orthogonal rotations in the plane of the two prepotentials makes the action\footnote{Once the number of derivatives is also fixed, as the bilinear dependence in the prepotentials.} determined, up to constant factors that can be absorbed by a redefinition of the spatial and time scales. In particular, spatial conformal invariance and $SO(2)$ electric-magnetic duality invariance seem to impose Lorentz invariance. This is another puzzling fact whose ultimate significance is not yet entirely clear, although some have argued this could be a sign that duality invariance is in some sense deeper than Lorentz invariance\footnote{See \cite{Bunster:2012hm,Gelfond:2015poa}.}.

Although the ultimate reason of this connection remains to be understood, we are left with a first order action whose geometric form is far more simple, transparent and universal than the original Fronsdal Lagrangian action: the kinetic term is simply the contraction of one of the prepotentials with the time derivative of the Cotton tensor of the other, while the Hamiltonian is the sum of the contraction of each prepotential with the curl of its Cotton tensor:
\begin{eqnarray}
S\left[Z^a_{ij} \right] &=&
\frac{1}{2} \ \int \ d^4 x \  Z^{a \ i_1 \cdots i_s} \left\lbrace
\varepsilon_{ab} \ \dot{B}^b_{i_1 \cdots i_s}
\ - \ \delta_{ab} \ \epsilon_{i_1 jk} \partial^j B^{b \ k}_{\phantom{b\ k} i_2 \cdots i_s}
\right\rbrace . \nonumber
\end{eqnarray}

This action finally has one additional interesting aspect: the equations of motions derived from it can be seen as \textit{twisted self-duality conditions}\footnote{See our paper \cite{Henneaux:2016zlu}.}. Twisted self-duality is a general way of rewriting the equations of motion of higher spin bosonic gauge fields, once they have been brought into a higher order form in which they simply set the trace of the (generalized) Riemann tensor of the spin-$s$ field to zero\footnote{See \cite{Damour:1987vm,Bekaert:2003az}. This higher order formalism relies on a larger gauge invariance than Fronsdal's, as the gauge parameter of spin-$s$ diffeomorphism is no longer subject to a tracelessness condition.}. Since the tracelessness of the Riemann tensor is equivalent to its dual\footnote{$i.e.$ Its Hodge dual taken over its first pair of antisymmetric indices.} being of the same symmetry type as itself ($i.e.$ the symmetry type corresponding to a rectangular Young tableau in which each line has $s$ boxes), these equations of motion turn out to be equivalent to requiring the dual of the Riemann tensor of the potential field to be, itself, the Riemann tensor of some other, dual potential. The original field and its dual turn out to have the same symmetry type ($i.e.$ the one associated to a one-row Young tableau) in a four-dimensional space-time, but this rewriting can be carried over in any dimension. In any case, these twisted self-duality conditions equate the dual curvature of one potential to the curvature of the other, and reciprocally, up to a sign (since duality squares to minus one in dimension four, self-duality can not be imposed). This is the covariant form of these conditions, but a non-manifestly covariant subset thereof can in fact be isolated, equating the (generalized) electric field of one potential to the (generalized) magnetic field of the other, and reciprocally (again, up to a sign). These equations were shown to be equivalent to the full covariant set. They are not all dynamical, and those of them not containing time derivatives are again constraints similar to the Hamiltonian ones. Finally, after a manipulation removing the pure gauge components of the potentials and solving the constraints by the introduction of prepotentials, this non-covariant form of the twisted self-duality conditions can be seen to be first order in time and, in reality, exactly the same as the Hamiltonian equations of motions obtained by starting from Fronsdal's action.\\

So far, we have only explored aspects of bosonic fields. Since supersymmetry is generally expected to be present in the most interesting theories, an equivalent treatment of fermions should be made. Although we have not been able to completely realize it yet, it seems to lead to results strikingly similar to the integer spin case. The Hamiltonian analysis of the spin $s+1/2$ massless field leads to constraints solved through the introduction of a single prepotential which is a symmetric spatial spinor-tensor $\Sigma_{i_1 \cdots i_s}$ whose gauge symmetry again combines both the diffeomorphism-like and the Weyl rescaling-like transformations:
\begin{equation}
\delta_{gauge} \Sigma_{i_1 \cdots i_s} \ = \
s \ \partial_{(i_1} \zeta_{i_2 \cdots i_s )}
\ + \ s \ \gamma_{(i_1} \theta_{i_2 \cdots i_s )} . \nonumber
\end{equation}
Here too, this gauge invariance, together with chirality invariance, seems to completely fix the form of the Hamiltonian action in terms of the prepotential.

In this work, we only expound these general results for the simpler case of the spin $5/2$\footnote{See our paper \cite{Bunster:2014fca}.}. Beyond its simplicity, this field presents the interest of being a possible superpartner of the graviton, as an alternative to the well-known spin $3/2$. The theory based on this supermultiplet is known as \textit{hypergravity}\footnote{See \cite{Aragone:1979hx,Berends:1979wu,Berends:1979kg}.}. By itself, in dimension higher than three, it is the object of its own set of no-go theorems\footnote{See \cite{Aragone:1979hx,Aragone:1980rk}.}, forbidding the introduction of interactions. Of course, as we mentionned earlier, these theorems have been circumvented by the work of Lebedev school\footnote{See note \ref{note_Vasiliev}.}. However, the introduction of interaction requires to add an infinite number of fields of arbitrarily high spin. More modestly, we have inquired into the symmetries of the free hypergravity. Our Hamiltonian analysis of the spin $5/2$, together with the well-known first order formalism of linearized gravity , allowed us to observe that the $SO(2)$ electric-magnetic duality rotations of the spin $2$ are actually mirrored through supersymmetry in the chirality rotations of the spin $5/2$: an appropriate combination of each of these two symmetries on its respective field forms a chirality-duality rotation commuting with supersymmetry.\\

In a related work\footnote{See our paper \cite{chiral22form}.}, we have extended our earlier results to another bosonic theory: the so-called $(2,2)$-\textit{Curtright field}\footnote{See \cite{Curtright:1980yk}. Curtright was the first to study gauge fields of mixed symmetry of any type.}, or $(2,2)$-form. It deals with a tensor field of mixed symmetry, having the same symmetry as the Riemann tensor of the graviton:
\begin{equation}
\yng(2,2) , \nonumber
\end{equation}
and propagates over a six-dimensional flat space-time. Its interest comes from its presence as a key element of a promising supersymmetric system, the $(4,0)$ theory\footnote{See \cite{Hull:2000zn,Hull:2000rr,Hull:2001iu}.}. It was argued - by a reasoning based on the equations of motion - that the strong coupling limit of theories having $N=8$ supergravity as their low energy effective theory in five space-time dimensions should be a six-dimensional theory involving, besides chiral $2$-forms, a chiral $(2,2)$-Curtright field in place of the standard graviton. More generally, mixed symmetry tensors are also naturally present in the mode expansion of String theory - and in the dual formulation of gravity in dimension higher than four. 

Our study of this field generalizes the well-established results for chiral $p$-forms\footnote{See \cite{Floreanini:1987as,Henneaux:1988gg}.}: a first order, non-manifestly covariant action principle that gives directly the chirality condition was constructed. This action principle not only automatically yields the chirality condition, which does not need to be separately imposed by hand, but it involves solely the $p$-form gauge potential without auxiliary fields, making the dynamics quite transparent.

Starting with the second order covariant equations of motion of the $(2,2)$-Curtright field, which set to zero the trace of its gauge invariant curvature (its generalized Riemann tensor), we rewrote them as covariant self-duality conditions (since duality squares to one in six dimensions, we can consider non-twisted self-duality). An equivalent non-manifestly covariant subset of these equations was then isolated that was first order in time, equating the electric and magnetic fields of the spatial components of the $(2,2)$-Curtright potential (the other remaining non purely spatial components being removed through some spatial curls). Among these equations, the non-dynamical ones - the constraints - were identified and solved, leading to a prepotential again exhibiting the intriguing conformal gauge invariance. The prepotential turned out to be a spatial tensor of the same symmetry type as the initial field. Written in terms of the (generalized) Cotton tensor of the prepotentials, the non-manifestly covariant self-duality conditions took an elegant and simple form, equating its time derivative to its curl. An action principle from which these equations could be derived, uniquely fixed by chirality and conformal gauge invariance, was written down and found to be exactly the same as the one obtained by the more cumbersome Hamiltonian analysis of the initial covariant Lagrangian action of the $(2,2)$-Curtright field. Again, the simplicity, unicity and the transparency of this action make it a promising start to investigate the supersymmetric extensions of this theory.\\

Finally, we have undertaken an investigation of the \textit{surface charges} of higher spin massless fields over anti-de Sitter, in arbitrary dimension. It is well-known that the conserved charges associated to gauge invariance are surface charges: they correspond to improper gauge transformations, formally analogous to unphysical proper gauge transformations but based on an asymptotic behaviour of the gauge parameter which actually maps a physical configuration of the system onto a different one\footnote{See \cite{review-charges,Regge,benguria-cordero,HT}.}. For instance, the total energy and angular momentum in general relativity, or color charges in Yang-Mills theories,  are given by surface charges\footnote{See \cite{Arnowitt:1962hi,Regge,AD1,AD2}.}.

Conserved charges naturally play an important role in the dynamics of any system, and, with their generalized diffeomorphism gauge invariance, higher spins offer a fascinating application of their analysis. Various studies have already explored this topic\footnote{See \cite{HScharges,unfolded-charges1,unfolded-charges2}.}, but they relied on Fronsdal's formulation of the dynamics and on covariant methods\footnote{See \cite{covariant-charges1,covariant-charges2}. There are also investigations of higher spin surface charges in four space-time dimensions within Vasiliev's unfolded formulation (see \cite{Didenko:2014dwa,Bekaert:2005vh,Vasiliev:1995dn}). In three space-time dimensions higher-spin charges have also been derived from the Chern-Simons formulation of higher-spin models in both constant-curvature \cite{HR,CS3,deSitter} and flat backgrounds \cite{flat1,flat2}.}. We have instead based our computations on the canonical formalism, along the lines developed for general relativity\footnote{See \cite{Regge,AD1,HT,BrownHenneaux,AdS-generic}.}.  

Presenting higher-spin charges in Hamiltonian form may help testing the expected thermodynamical properties of the proposed black hole solutions of  Vasiliev's equations\footnote{See \cite{4D-BH1,4D-BH2}.}, in analogy with what happened for higher-spin black holes in three space-time dimensions\footnote{See \cite{BH1,BH2,BH3,BH4}.}. Quite generally, the canonical formalism provides a solid framework for analysing conserved charges and asymptotic symmetries. One virtue of the Hamiltonian derivation is indeed that the charges are clearly related to the corresponding symmetry. The charges play the dual role of being conserved through Noether's theorem, but also of generating the associated symmetry through the Poisson bracket. This follows from the action principle. For these reasons, our work may also be useful to further develop the understanding of holographic scenarios involving higher-spin fields\footnote{See \cite{review-Giombi}.}. 

In four space-time dimensions or higher, several holographic conjectures indeed anticipated a careful analysis of the Poisson algebra of AdS higher-spin charges, which defines the asymptotic symmetries of the bulk theory to be matched with the global symmetries of the boundary dual. The study of asymptotic symmetries in three dimensions\footnote{See \cite{HR,CS3,GH,Wlambda}.} proved instead crucial to trigger the development of a higher-spin AdS$_3$/CFT$_2$ correspondence\footnote{See \cite{review-3D}.}. Similarly, asymptotic symmetries played an important role in establishing new links between higher-spin theories and string theory, via the embedding of the previous holographic correspondences in stringy scenarios\footnote{See \cite{ABJ-HS,stringy3D}, and also sect.~6.5 of \cite{review-strings} for a review.}.

More precisely, we compute surface charges starting from the rewriting in Hamiltonian form of Fronsdal's action on Anti de Sitter backgrounds of arbitrary dimension\footnote{See \cite{fronsdal-AdS_fermi,fronsdal-AdS-D,Metsaev:2011iz}.}. This is of course the action describing the \emph{free} dynamics of higher-spin particles; nevertheless we expect that the rather compact final expression for the charges\footnote{This expression is displayed in \eqref{Q-final} for the spin-3 example and in \eqref{qfin} for the generic case.} will continue to apply even in the full non-linear theory, at least in some regimes and for a relevant class of solutions. This expectation\footnote{It is further discussed in section \textbf{\ref{sec:conclusions}}.} is supported by several examples of charges linear in the fields appearing in gravitational theories. It also agrees  with a previous analysis of asymptotic symmetries of three-dimensional higher-spin gauge theories\footnote{See \cite{metric3D}.} based on the perturbative reconstruction of the interacting theory within  Fronsdal's formulation\footnote{See \cite{metric-like,metric-like2}.}. Our setup is therefore close to the Lagrangian derivation of higher-spin charges\footnote{See \cite{HScharges}. We compare explicitly the two frameworks in the spin-3 case at the end of section \textbf{\ref{sec:spin3-charges}}.}, while our analysis also proceeds further by showing how imposing boundary conditions greatly simplifies the form of the charges at spatial infinity. This additional step allows a direct comparison between the surface charges of the bulk theory and the global charges of the putative boundary dual theories, which fits within the proposed AdS/CFT correspondences. \\

This work is organized as follows. We begin by reviewing a series of established results on which we built our subsequent developments. The first chapters are devoted to a presentation of the Fronsdal formalism for free higher spins on a flat or constantly curved space-time (chapter \textbf{\ref{Chap:free_hs}}), to an overview of the Hamiltonian analysis of gauge systems (constraints, Dirac brackets, charges), including its application to lower spins (chapter \textbf{\ref{Sec:Ham_form_constr}}) and to an exposition of electric-magnetic duality and twisted self-duality for the free photon and graviton (chapter \textbf{\ref{Chap:dualities}}). 

We then proceed to expound our contributions. A first part gathers those related to bosonic higher spin fields in a four-dimensional space-time: the construction of the spatial conformal invariants (chapter \textbf{\ref{Chap:confgeom}}), their use to compute the Hamiltonian action in terms of prepotentials (chapter \textbf{\ref{Chap:hambos}}) and the equivalent description in terms of twisted self-duality conditions (chapter \textbf{\ref{Chap:tsdc}}). The second part deals with two fields whose study opens on supersymmetric extensions: the spin $5/2$ field and hypergravity in four dimensions (chapter \textbf{\ref{Chap:hygra}}) and the chiral $(2,2)$-form in six dimensions (chapter \textbf{\ref{Chap:chiral_mix}}). Finally, the third part covers our computation of the surface charges of free massless higher spins over AdS in arbitrary dimension, for Bose (chapter \textbf{\ref{Chap:Bose}}) and Fermi fields (chapter \textbf{\ref{Chap:Fermi}}).

\part{Review of fundamentals}

\chapter{Free higher spin gauge fields}

\label{Chap:free_hs}

In this chapter, we will provide an exposition of the free theory of massless higher spin fields in arbitrary dimension, limiting ourselves, for now, to fields whose symmetries correspond to Young tableau of one row, that is, to fully symmetric tensor fields. These are the only possible fields in four space-time dimensions, but that is not the case for higher dimensions, as we shall succinctly explore further on (see chapter \textbf{\ref{Chap:chiral_mix}} for a study of mixed symmetry type tensors in six dimensions). We will not follow a completely constructive approach, but give arguments along the way to motivate the choice of fields, equations of motion and action, and show how these indeed contain the expected degrees of freedom. We will follow this construction for bosons first, and then for fermions. Finally, the corresponding equations of motion and action on AdS will be covered.

\section{Bosonic fields}

\subsection{Flat background space-time}

\label{Sec:Lag_s_bos}

The $D$ dimensional Poincaré group massless irreducible unitary representations correspond to irreducible unitary representations of the rotation group $SO(D-2)$. For $D = 4$ (or even $D = 5$), these can all be described in terms of fully symmetric traceless tensors in the appropriate $D-2$ dimensional space (with, say, $s$ indices): $h_{i_1 \cdots i_s}$. For $D = 4$, such a tensor indeed has two independent components, independently from its rank. For space-time dimension superior or equal to six, representations associated with Young tableau with more than one line can also be considered, but we shall ignore them at first.

The simplest Lorentz group (non unitary) representation in which to embed such a tensor is, certainly, a symmetric tensor: $h_{\mu_1 \cdots \mu_s}$. This will be our covariant field. It obviously contains much more independent components than the two-dimensional representation of the Poincaré little group we want to describe, so we will need to give it a certain gauge invariance. The most natural gauge invariance is the following:
\begin{equation}
\delta h_{\mu_1 \cdots \mu_s} = s \ \partial_{(\mu_1} \xi_{\mu_2 \cdots \mu_s)} , \label{gauge_spin_bos_s_no_trace}
\end{equation}
\noindent which is indeed the natural generalization of the linearized spin $1$ and $2$ gauge invariance.

If we try to write down gauge invariant equations of motion for such a field, we immediately encounter a significant difficulty: the only gauge invariant curvature that can be built from such an object is its generalized \textit{Riemann tensor} (or Freedmann-de Wit tensor), which involves $s$ derivatives:
\begin{equation}
R_{\mu_1 \nu_1 \vert \cdots \vert \mu_s \nu_s} \ \equiv \ 2^s \ \partial_{[\mu_1 \vert} \cdots  \partial_{[\mu_s \vert} h_{\vert \nu_1 ] \cdots \vert \nu_s ]} .
\end{equation}

Equations of motion encoding the relevant degrees of freedom can indeed be expressed in terms of this tensor: requiring its trace (the generalized Ricci tensor) to vanish leads to physical modes ($i.e.$ gauge invariant solutions) that transform in the looked for representation of the Poincaré group. 

However, we are more specifically interested in traditional second order equations of motion, which leads us to define the following object, the \textit{Fronsdal tensor}:
\begin{eqnarray} \label{Fronsdal_bos}
\mathcal{F}_{\mu_1 \cdots \mu_s} &\equiv&
\square h_{\mu_1 \cdots \mu_s} 
\ - \ s \ \partial_{(\mu_1} \partial^{\nu} h_{\mu_2 \cdots \mu_s)\nu}
\ + \ \frac{s\left(s-1\right)}{2} \ \partial_{(\mu_1} \partial_{\mu_2} h'_{\mu_3 \cdots \mu_s)} .
\end{eqnarray}

Note that this tensor is fully symmetric, exactly as the field from which it is built. Under our gauge transformation (\ref{gauge_spin_bos_s_no_trace}), this tensor's variation is given by:
\begin{eqnarray} \label{Fronsdal_bos_var_trace_eps}
\delta \mathcal{F}_{\mu_1 \cdots \mu_s} &=& 
\frac{s\left(s-1\right)\left(s - 2\right)}{2} \ \partial_{(\mu_1} \partial_{\mu_2} \partial_{\mu_3} \xi'_{\mu_4 \cdots \mu_s)} .
\end{eqnarray}

For spin $1$ or $2$, Fronsdal tensor is generally gauge invariant: it is actually equal to the divergence of the fieldstrength tensor of electromagnetism (for $s = 1$) and to the linearized Ricci tensor (for $s = 2$). For spins higher than $2$, on the other hand, it is only invariant under gauge transformations whose parameter is \textit{traceless}\footnote{This is not a fully general statement, since one could limit oneself to requiring the third symmetrized derivative of the trace of the gauge parameter to vanish. These actually only contain a finite number of modes.}. This algebraic condition imposed on the gauge parameter is the source of many subtleties.

We are then led to consider the following gauge transformation:
\begin{eqnarray} \label{gauge_spin_bos_s}
\delta h_{\mu_1 \cdots \mu_s} &=& s \ \partial_{(\mu_1} \xi_{\mu_2 \cdots \mu_s)} ,
\\
0 &=& \xi'_{\mu_4 \cdots \mu_s},
\end{eqnarray}
\noindent under which the following equations of motion are indeed invariant:
\begin{eqnarray}
0 &=& \mathcal{F}_{\mu_1 \cdots \mu_s}. \label{eom_bos_cov}
\end{eqnarray}

Let us check that the physical modes satisfying these equations are the expected ones. We are naturally led to make a partial gauge fixing that brings us into a generalized de Donder gauge, setting:
\begin{eqnarray}
0 &=& \partial^{\nu} h_{\nu \mu_2 \cdots \mu_s} \ - \ \frac{\left(s - 1\right)}{2} \ \partial_{(\mu_2} h'_{\mu_3 \cdots \mu_s)} . \label{deDonder}
\end{eqnarray}

The gauge variation of this quantity is given by $\square \xi_{\mu_2 \cdots \mu_s}$ and can be set to zero. The equations of motion then imply $0 \ = \ \square h_{\mu_1 \cdots \mu_s}$, which brings us on the massless momentum shell. Let us then consider a Fourier mode ($i.e.$ a mode of the form $h_{\mu_1 \cdots \mu_s}\left(x\right) = e^{ip\cdot x} H_{\mu_1 \cdots \mu_s}$, where $x$ is the position vector in some orthonormal coordinates of an inertial frame) and let us go to the light-cone coordinate system\footnote{That is, a coordinate system in which the metric components satisfy $\eta_{++} = \eta_{--} = 0$ and $\eta_{+-} \neq 0$, the other non-vanishing components being diagonal and positive.} in which the only non-vanishing component of the momentum of our mode is $p_+$. The residual gauge freedom allows us to cancel all the components of $H_{\mu_1 \cdots \mu_s}$ with at least one index in the $+$ direction, which completely fixes the gauge. The covariant gauge condition (\ref{deDonder}) then implies that the components of  $H_{\mu_1 \cdots \mu_s}$ with at least one index in the $-$ direction also vanish\footnote{This is the form equation (\ref{deDonder}) takes if one sets none of its free indices equal to $+$: the second term vanishes (since the derivative carries a free lower index, and the derivative is non-zero only if this index is $+$) and the first one has its contracted index set to $-$ by the divergence.} and that $H_{\mu_1 \cdots \mu_s}$ is traceless\footnote{By setting only one free index of (\ref{deDonder}) equal to $+$, which suppresses the first term and reduces the second to the trace of the field.}.

So, all the gauge invariant solutions of (\ref{eom_bos_cov}) can be parametrized without redundancy by traceless Fourier modes $H_{\mu_1 \cdots \mu_s}\left(p\right)$ (with momentum on the massless shell: $p^2 = 0$) whose only non-vanishing components are in the $(D-2)$-dimensional transverse (to $p$) and spatial directions: these indeed form a vector space on which a massless irreducible spin $s$ representation of Poincaré group would act\footnote{To obtain this action, of course, one would have to expand the action in oscillating modes and quantize it, in order to compute the precise action of the isometry generators on the quantum states. This would allow to check that these indeed form the appropriate representation of Poincaré group.}.\\

We can now look for a gauge invariant action principle from which to derive these equations of motion. Similarly to what happens to linearized gravity, where the contraction of the metric with its linearized Ricci tensor (whose vanishing is the equations of motion) is not a suitable Lagrangian, because the non-tracelessness of the Ricci tensor prevents the gauge variation of this Lagrangian from being a divergence, we need to build an \textit{Einstein tensor}, which would be an identically divergenceless curvature. For this, we need a generalized Bianchi identity. 

We can easily see that the following identity holds:
\begin{eqnarray} 
\partial^{\nu} \mathcal{F}_{\nu \mu_2 \cdots \mu_s} 
\ - \ \frac{\left(s-1\right)}{2} \partial_{(\mu_2} \mathcal{F}'_{\mu_2 \cdots \mu_s)}
&=& \ - \ \frac{\left(s-1\right)\left(s-2\right)\left(s-3\right)}{4} \ \partial_{(\mu_2} \partial_{\mu_3} \partial_{\mu_4} h''_{\mu_5 \cdots \mu_s)} . \qquad \qquad \label{Bianchi_bos}
\end{eqnarray}

For $s=2$, this is just Bianchi identity but, starting from $s = 4$, we have a non-vanishing right-side. This naturally leads to a second algebraic constraint, this time imposed on the field itself: we will force its double trace to vanish. This is a necessary condition in order for its second order curvature to satisfy a Bianchi identity:
\begin{eqnarray}
0 &=& h''_{\mu_5 \cdots \mu_s} .
\end{eqnarray}

With this condition\footnote{Which is actually necessary to make the counting of degrees of freedom we performed above, when we made the final gauge fixing, as the careful reader will have noted.} (from which we can infer the vanishing of the double traces of the Fronsdal and Einstein tensor), the identity (\ref{Bianchi_bos}) leads us to define the Einstein tensor:
\begin{eqnarray}
\mathcal{G}_{\mu_1 \cdots \mu_s} &\equiv&
\mathcal{F}_{\mu_1 \cdots \mu_s} 
\ - \ \frac{s\left(s-1\right)}{4} \ \eta_{(\mu_1 \mu_2} \mathcal{F}'_{\mu_3 \cdots \mu_s)} .
\end{eqnarray}

It is of course gauge invariant, ant its divergence is proportional to the metric: $\partial^{\nu} \mathcal{G}_{\nu \mu_2 \cdots \mu_s} = \ - \ \frac{\left(s - 1\right)\left(s - 2\right)}{4} \ \eta_{(\mu_2 \mu_3} \partial^{\nu} \mathcal{F}'_{ \mu_4 \cdots \mu_s)\nu}$. Given the tracelessness we already imposed on the gauge parameter, this leads us to postulate the following lagrangian density:
\begin{eqnarray}
\mathcal{L} &=& \frac{1}{2} \ \mathcal{G}^{\mu_1 \cdots \mu_s} h_{\mu_1 \cdots \mu_s}  ,
\end{eqnarray}
\noindent whose gauge variation is indeed proportional to the divergence of $\xi_{\mu_2 \cdots \mu_s} \mathcal{G}^{\mu_1 \cdots \mu_s}$.

We have arrived to our gauge invariant action, whose variation gives us the vanishing of the Einstein tensor of the field or, equivalently, of its Fronsdal tensor:
\begin{eqnarray} \label{action_bos_compact}
S &=& \frac{1}{2} \ \int d^4 x \ \ \mathcal{G}^{\mu_1 \cdots \mu_s} h_{\mu_1 \cdots \mu_s}, 
\\
0 &=& h''_{\mu_5 \cdots \mu_s} ,
\end{eqnarray}
\noindent which can also be rewritten (through partial integration):
\begin{eqnarray} \label{action_bos_full}
S &=& - \ \frac{1}{2} \ \int \ d^4 x \ \left\lbrace
 \partial_{\nu} h_{\mu_1 \cdots\mu_s}\partial^{\nu} h^{\mu_1 \cdots\mu_s}
\ - \ s \ \partial^{\nu} h_{\nu\mu_2\cdots\mu_s} \partial_{\rho} h^{\rho\mu_2\cdots\mu_s}
\right.
\nonumber \\ && \qquad \qquad \qquad
\left.
\ + \ s\left( s - 1 \right) \ \partial^{\nu} h_{\nu\rho\mu_3\cdots\mu_s} \partial^{\rho} h'^{\mu_3\cdots\mu_s}
\ - \ \frac{s\left(s - 1 \right)}{2} \ \partial_{\rho} h'_{\mu_3 \cdots\mu_s} \partial^{\rho} h'^{\mu_3 \cdots\mu_s}
\right.
\nonumber \\ && \qquad \qquad \qquad
\left.
\ - \ \frac{s\left(s - 1\right)\left(s - 2 \right)}{4} \ \partial^{\nu} h'_{\nu\mu_4\cdots\mu_s} \partial_{\rho} h'^{\rho\mu_4\cdots\mu_s} 
\right\rbrace ,
\\
0 &=& h''_{\mu_5 \cdots \mu_s} .
\end{eqnarray}

It can be useful to rewrite all these equations in the index-free notation in which symmetrization over all free indices is implicit and is carried with a weight equal to the minimal number of terms necessary to write down (this convention is only used here). The definitions take the form:
\begin{eqnarray}
\delta h &=& \partial \xi ,
\\
\mathcal{F} &=& \square h \ - \ \partial \partial \cdot h \ + \ \partial^2 h' ,
\\
\mathcal{G} &=& \mathcal{F} \ - \ \frac{1}{2} \ \eta \mathcal{F}' .
\end{eqnarray}

The traces conditions arise from:
\begin{eqnarray}
\delta \mathcal{F} &=& 3 \ \partial^3 \xi' ,
\\
\partial \cdot \mathcal{F} \ - \ \frac{1}{2} \ \partial \mathcal{F}' &=& \ - \ \frac{3}{2} \ \partial^3 h'' .
\end{eqnarray}

Finally, our gauge-fixing condition was:
\begin{eqnarray}
0 &=& \partial \cdot h \ - \ \frac{1}{2} \ \partial h' .
\end{eqnarray}

To sum things up, we have absorbed the components of a massless irreducible representation of the Poincaré group of spin $s$ into a symmetric covariant tensor field. By giving it the gauge freedom (\ref{gauge_spin_bos_s_no_trace}), it can be subjected to the gauge invariant second order equations of motion (\ref{eom_bos_cov}), which precisely removed the spurious components of the field. These equations of motion can, in turn, be derived from the gauge invariant action (\ref{action_bos_compact}) or, equivalently, (\ref{action_bos_full}).

\subsubsection{Lower spin}

\label{Sec:low_bos}

Let us note that this general formalism for describing free higher spin gauge fields over flat space-time in a Lagrangian formalism reduces to well-known ones for lower spin. Indeed, if we plug into the equations from above the value $s= 1$, we immediately recover the covariant theory of classical electromagnetism in the vacuum, or \textit{photon}: the field (denoted $A_{\mu}$ in agreement with tradition) is a vector, whose gauge variation is given by the gradient of a scalar function $\xi$:
\begin{eqnarray}
\delta A_{\mu} &=& \partial_{\mu} \xi ,
\end{eqnarray}
\noindent and its Fronsdal tensor (which is equivalent to its Einstein tensor, since there are no traces to take) and action are:
\begin{eqnarray}
\mathcal{F}_{\mu} &=& \square A_{\mu} \ - \ \partial_{\mu} \partial^{\nu} A_{\nu} ,
\\
S &=& \frac{1}{2} \ \int \ d^4 x \ A^{\mu} \mathcal{F}_{\mu} .
\end{eqnarray}

These equations are identical to the familiar ones, provided we express them in terms of the strength field tensor $F_{\mu\nu} \equiv \partial_{\mu} A_{\nu} \ - \ \partial_{\nu} A_{\mu}$, giving:
\begin{eqnarray}
\mathcal{F}_{\mu} &=& \partial^{\nu} F_{\nu\mu} ,
\\
S &=&  - \ \frac{1}{4} \ \int \ d^4 x \ F^{\mu\nu} F_{\mu\nu} . 
\end{eqnarray}

Entering the value $s = 2$ similarly generates the free massless spin $2$ field theory, the \textit{graviton}, which is exactly the first order weak field limit of General Relativity. It is described by a symmetric tensor $h_{\mu\nu}$ whose gauge variation is:
\begin{eqnarray}
\delta h_{\mu\nu} &=& \partial_{\mu} \xi_{\nu} \ + \ \partial_{\nu} \xi_{\mu} ,
\end{eqnarray}
\noindent which is indeed the linearized expression of a diffeomorphism performed on a variation of the metric from a flat one (the metric plugged into General Relativity is $g_{\mu\nu} \equiv \eta_{\mu\nu} \ + \ h_{\mu\nu}$, with $\eta_{\mu\nu}$ being flat). The expressions for Fronsdal and Einstein tensors are:
\begin{eqnarray}
\mathcal{F}_{\mu\nu} &=& 
\square h_{\mu\nu} \ - \ 2 \ \partial_{(\mu} \partial^{\rho} h_{\nu )\rho} \ + \ \partial_{\mu} \partial_{\nu} h ,
\\
\mathcal{G}_{\mu\nu} &=& 
\eta_{\mu\nu} \ \left(\partial^{\rho} \partial^{\sigma} h_{\rho\sigma} \ - \ \square h'\right)
\\ &&
 + \ \square h_{\mu\nu} \ - \ 2 \ \partial_{(\mu} \partial^{\rho} h_{\nu )\rho} \ + \ \partial_{\mu} \partial_{\nu} h' ,
\end{eqnarray}
\noindent which are exactly those of their corresponding linearized counterparts in General Relativity (Ricci and Einstein tensors, respectively), as appears once we express them in terms of $R_{\mu\nu} \equiv \square h_{\mu\nu} \ - \ 2 \ \partial_{(\mu} \partial^{\rho} h_{\nu )\rho} \ + \ \partial_{\mu} \partial_{\nu} h$:
\begin{eqnarray}
\mathcal{F}_{\mu\nu} &=&
R_{\mu\nu} ,
\\
\mathcal{G}_{\mu\nu} &=& 
R_{\mu\nu} \ - \ \frac{1}{2} \ \eta_{\mu\nu} R .
\end{eqnarray}

The action we derive from the previous expressions can be obtained by expanding Einstein-Hilbert action quadratically (the first non-vanishing order) in a metric being a variation of a flat one:
\begin{eqnarray}
S &=&
- \ \frac{1}{2} \ \int \ d^4 x \ \left\lbrace
\partial_{\rho} h_{\mu\nu} \partial^{\rho} h^{\mu\nu}
\ - \ 2 \ \partial^{\rho} h_{\rho\mu}\partial_{\nu} h^{\nu\mu}
\ + \ 2 \ \partial^{\mu} h_{\mu\nu} \partial^{\nu} h'
\ - \ \partial_{\mu} h' \partial^{\mu} h'
\right\rbrace . \qquad  \qquad
\end{eqnarray}

\subsection{Constantly curved background space-time}

\label{Sec:bose_lag_AdS}

As we saw, the key element in the Lagrangian description of massless higher spins outlined above is gauge invariance: it is this gauge invariance (present at the level of the action) which allows us to remove spurious components from the physical degrees of freedom of the field, in order for them to fit into a massless representation of the symmetry group of space-time.

The construction introduced above for fields propagating over a flat background space-time - whose symmetry group is simply Poincaré - can actually be very easily extended to a constantly curved background and, in particular, to \textit{anti-de Sitter} space-time, $AdS$.

To be perfectly definite, let us indeed consider this space-time chose curvature is constant and negative, with a radius of curvature $L$. The covariant derivatives $\nabla_{\mu}$ acting on a covariant vector $V_{\mu}$ satisfy the following commutation rule:
\be 
[ \nabla_{\!\m} \,, \nabla_{\!\n}] V_\r = \frac{1}{L^2}\! \left( g_{\n\r} V_\m - g_{\m\r} V_\n \right) .
\ee

This commutation relation shows that the flat space action written above, \eqref{action_bos_full}, can be modified in a very simple way in order to conserve its full gauge invariance: if we replace the partial derivatives in it by covariant derivatives, in order for the action to be well-defined\footnote{The use of partial derivatives over a curved space-time would make the action dependent on the choice of coordinate system in which it would be defined\ldots}, and if we add a ``mass term" appropriately chosen to cancel the additional terms generated by the non-trivial commutation rule under a gauge variation of the field to be defined shortly, we get:
\begin{eqnarray} 
S &=& - \ \frac{1}{2} \ \int \ d^4 x \ \sqrt{-g} \ \left\lbrace
 \nabla_{\nu} h_{\mu_1 \cdots\mu_s}\nabla^{\nu} h^{\mu_1 \cdots\mu_s}
\ - \ s \ \nabla_{\nu} h_{\rho\mu_2\cdots\mu_s} \nabla^{\rho} h^{\nu\mu_2\cdots\mu_s}
\right.
\nonumber \\ && \qquad \qquad \qquad
\left.
\ + \ s\left( s - 1 \right) \ \nabla^{\nu} h_{\nu\rho\mu_3\cdots\mu_s} \nabla^{\rho} h'^{\mu_3\cdots\mu_s}
\ - \ \frac{s\left(s - 1 \right)}{2} \ \nabla_{\rho} h'_{\mu_3 \cdots\mu_s} \nabla^{\rho} h'^{\mu_3 \cdots\mu_s}
\right.
\nonumber \\ && \qquad \qquad \qquad
\left.
\ - \ \frac{s\left(s - 1\right)\left(s - 2 \right)}{4} \ \nabla^{\nu} h'_{\nu\mu_4\cdots\mu_s} \nabla_{\rho} h'^{\rho\mu_4\cdots\mu_s} 
\right.
\nonumber \\ && \qquad \qquad \qquad
\left.
\ - \ \frac{(s-1)(d+s-3)}{L^2} \ \left[
h_{\mu_1 \cdots \mu_s} h^{\mu_1 \cdots \mu_s}
\ - \ \frac{s(s-1)}{4} \ h'_{\mu_3 \cdots \mu_s} h'^{\mu_3 \cdots \mu_s}
\right]
\right\rbrace , \nonumber \\ \label{action_bos_full_AdS}
\\
0 &=& h''_{\mu_5 \cdots \mu_s} ,
\end{eqnarray}
\noindent where $g$ is the determinant of the space-time metric.

This action is easily checked to be invariant under the gauge transformation:
\begin{eqnarray}
\delta h_{\mu_1 \cdots \mu_s} &=&
s \ \nabla_{( \mu_1} \xi_{\mu_2 \cdots \mu_s )} ,
\\
0 &=& \xi'_{\mu_4 \cdots \mu_s} .
\end{eqnarray}

This action, is naturally, valid in any dimension.

\section{Fermionic fields}

\subsection{Flat background space-time}

\label{Sec:Lag_s_fermi}

We follow a completely similar line of development to describe fermionic fields, and we will give the equivalent results. We start with a covariant symmetric tensor-spinor field with $s$ indices, $\psi_{\mu_1 \cdots \mu_s}$ and endow it with a gauge symmetry of the following type (generalizing the spin $3/2$):

\begin{eqnarray}
\delta \psi_{\mu_1 \cdots \mu_s} &=& s \ \partial_{(\mu_1} \zeta_{\mu_2 \cdots \mu_s)} . \label{gauge_spin_fermi_s_no_trace}
\end{eqnarray}

If we try to write down gauge invariant equations of motion for such a field, we again immediately encounter the difficulty that the only gauge invariant curvature that can be built from such an object is its generalized \textit{Riemann tensor} (or Freedmann-de Wit tensor), which involves $s$ derivatives:
\begin{equation}
\mathcal{R}_{\mu_1 \nu_1 \vert \cdots \vert \mu_s \nu_s} \ \equiv \ 2^s \ \partial_{[\mu_1 \vert} \cdots  \partial_{[\mu_s \vert} \psi_{\vert \nu_1 ] \cdots \vert \nu_s ]} .
\end{equation}

We would be interested in a first order curvature, which leads us to define the following object, the \textit{Fronsdal tensor}:
\begin{eqnarray} \label{Fronsdal_fermi}
\mathcal{F}_{\mu_1 \cdots \mu_s} &\equiv&
\slashed{\partial} \psi_{\mu_1 \cdots \mu_s} 
\ - \ s \ \partial_{(\mu_1} \slashed{\psi}_{\mu_2 \cdots \mu_s)}.
\end{eqnarray}

Note that this tensor is fully symmetric, exactly as the field from which it is built. Under our gauge transformation (\ref{gauge_spin_fermi_s_no_trace}), this tensor's variation is given by:
\begin{eqnarray} 
\delta \mathcal{F}_{\mu_1 \cdots \mu_s} &=& 
\ - \ s\left(s-1\right) \ \partial_{(\mu_1} \partial_{\mu_2} \slashed{\zeta}_{\mu_3\cdots \mu_s)} .
\end{eqnarray}

For spin $3/2$, Fronsdal tensor is generally gauge invariant. However, for spins superior or equal to $5/2$, it is only invariant under gauge transformations whose parameter is \textit{gamma-traceless}\footnote{This is not a fully general statement, since one could limit oneself to requiring the second symmetrized derivative of the gamma-trace of the gauge parameter to vanish...}. This algebraic condition imposed on the gauge parameter is, as it was for bosons, the source of many subtleties.

We are then led to consider the following gauge transformation:
\begin{eqnarray} \label{gauge_spin_fermi_s}
\delta \psi_{\mu_1 \cdots \mu_s} &=& s \ \partial_{(\mu_1} \zeta_{\mu_2 \cdots \mu_s)} ,
\\
0 &=& \slashed{\zeta}_{\mu_3 \cdots \mu_s},
\end{eqnarray}
\noindent under which the following equations of motion are indeed invariant:
\begin{eqnarray}
0 &=& \mathcal{F}_{\mu_1 \cdots \mu_s}. \label{eom_fermi_cov}
\end{eqnarray}

Let us check that the physical modes satisfying these equations are the expected ones. We can easily access the gamma-traceless gauge, since $\delta \slashed{\psi}_{\mu_2 \cdots \mu_s} = \slashed{\partial} \zeta_{\mu_2 \cdots \mu_s}$. This partial gauge fixing brings the equations of motion (\ref{eom_fermi_cov}) into the form $0 = \slashed{\partial} \psi_{\mu_1 \cdots \mu_s}$, which shows that we are indeed on the massless shell. The gamma-trace of this equation gives the divergencelessness of the field: $0 = \partial^{\nu} \psi_{\nu \mu_2 \cdots \mu_s}$.

We can consider a Fourier mode ($i.e.$ a mode of the form $\psi_{\mu_1 \cdots \mu_s} = e^{ip\cdot x} \Psi_{\mu_1 \cdots \mu_s}$, where $x$ is the position vector in some orthonormal coordinates of an inertial frame) and go to the light-cone coordinate system\footnote{That is, a coordinate system in which the metric components satisfy $\eta_{++} = \eta_{--} = 0$.} in which the only non-vanishing component of the momentum of our mode is $p_+$. The divergenceless condition implies that any component of $\Psi_{\mu_1 \cdots \mu_s}$ with at least one index in the $-$ direction vanishes. The residual gauge freedom also allows us to cancel all the components of $\Psi_{\mu_1 \cdots \mu_s}$ with at least one index in the $+$ direction, which completely fixes the gauge.

So, all the gauge invariant solutions of (\ref{eom_fermi_cov}) can be parametrized without redundancy by gamma-traceless Fourier modes $\Psi_{\mu_1 \cdots \mu_s}\left(p\right)$ (with momentum on the massless shell: $p^2 = 0$) whose only non-vanishing (tensorial) components are in the $(D-2)$-dimensional transverse (to $p$) and spatial directions, satisfying $0 = \slashed{p}\Psi_{\mu_1 \cdots \mu_s}\left(p\right)$: these indeed form a vector space on which a massless irreducible spin $s + 1/2$ representation of Poincaré group could act.\\

We can now look for a gauge invariant action principle from which to derive these equations of motion. Similarly to what happens to linearized gravity, where the contraction of the metric with its linearized Ricci tensor (whose vanishing is the equations of motion) is not a suitable Lagrangian, because the non-tracelessness of the Ricci tensor prevents the gauge variation of this Lagrangian from being a divergence, we need to build an \textit{Einstein tensor} (which is already necessary for spin $3/2$). For this, we need a generalized Bianchi identity. 

We can easily see that the following identity holds:
\begin{eqnarray} 
\partial^{\nu} \mathcal{F}_{\nu \mu_2 \cdots \mu_s} 
\ - \ \frac{1}{2} \ \slashed{\partial} \slashed{\mathcal{F}}_{\mu_2 \cdots \mu_s}
\ - \ \frac{\left(s - 1\right)}{2} \ \partial_{(\mu_2} \mathcal{F}'_{\mu_3 \cdots \mu_s)} 
&=&  \frac{\left(s-1\right)\left(s-2\right)}{2} \ \partial_{(\mu_2} \partial_{\mu_3} \slashed{\psi}'_{\mu_4 \cdots \mu_s)} . \qquad \qquad \label{Bianchi_fermi}
\end{eqnarray}

Starting from $s = 3$ (spin $7/2$), we have a non-vanishing right-side. This naturally leads to a second algebraic constraint, this time imposed on the field itself: we will force its triple gamma-trace to vanish. This is a necessary condition in order for its first order curvature to satisfy a Bianchi identity:
\begin{eqnarray}
0 &=& \slashed{\psi}'_{\mu_4 \cdots \mu_s} .
\end{eqnarray}

With this condition\footnote{Also necessary for the counting of physical modes performed above to be consistent.} (from which we can infer the vanishing of the triple gamma-traces of the Fronsdal and Einstein tensor), the identity (\ref{Bianchi_fermi}) leads us to define the Einstein tensor:
\begin{eqnarray}
\mathcal{G}_{\mu_1 \cdots \mu_s} &\equiv&
\mathcal{F}_{\mu_1 \cdots \mu_s} 
\ - \ \frac{s}{2} \ \gamma_{(\mu_1} \slashed{\mathcal{F}}_{\mu_2 \cdots \mu_s)}
\ - \ \frac{s\left(s-1\right)}{4} \ \eta_{(\mu_1 \mu_2} \mathcal{F}'_{\mu_3 \cdots \mu_s)} .
\end{eqnarray}

It is of course gauge invariant, ant its divergence is proportional to gamma-matrices: 
\begin{eqnarray}
\partial^{\nu} \mathcal{G}_{\nu \mu_2 \cdots \mu_s} &=& \ - \ \frac{\left(s-1\right)}{2} \ \gamma_{(\mu_2} \partial^{\nu} \slashed{\mathcal{F}}_{\mu_3 \cdots \mu_s)\nu}\ - \ \frac{\left(s - 1\right)\left(s - 2\right)}{4} \ \eta_{(\mu_2 \mu_3} \partial^{\nu} \mathcal{F}'_{ \mu_4 \cdots \mu_s)\nu}. \qquad \qquad
\end{eqnarray} 

Given the gamma-tracelessness we already imposed on the gauge parameter, this leads us to postulate the following Lagrangian density:
\begin{eqnarray}
\mathcal{L} &=& \ - \ i \ \bar{\psi}^{\mu_1 \cdots \mu_s} \mathcal{G}_{\mu_1 \cdots \mu_s}  ,
\end{eqnarray}
\noindent whose gauge variation is indeed proportional to the divergence of $\bar{\zeta}_{\mu_2 \cdots \mu_s} \mathcal{G}^{\mu_1 \cdots \mu_s}$.

We have arrived to our gauge invariant action, whose variation gives us the vanishing of the Einstein tensor of the field or, equivalently, of its Fronsdal tensor:
\begin{eqnarray} \label{action_fermi_compact}
S &=& \ - \ i \ \int d^4 x \ \ \bar{\psi}^{\mu_1 \cdots \mu_s} \mathcal{G}_{\mu_1 \cdots \mu_s}, 
\\
0 &=& \slashed{\psi}'_{\mu_4 \cdots \mu_s} ,
\end{eqnarray}
\noindent which can also be rewritten (through partial integration):
\begin{eqnarray} \label{action_fermi_full}
S &=& - \ i \ \int \ d^4 x \ \left\lbrace
\bar{\psi}^{\mu_1 \cdots \mu_s} \slashed{\partial} \psi_{\mu_1 \cdots \mu_s}
\ + \ s \ \bar{\slashed{\psi}}^{\mu_2 \cdots \mu_s} \slashed{\partial} \slashed{\psi}_{\mu_2 \cdots \mu_s}
\ - \ \frac{s\left(s-1\right)}{4} \ \bar{\psi}'^{\mu_3 \cdots \mu_s} \slashed{\partial} \psi'_{\mu_3 \cdots \mu_s}
\right.
\nonumber \\ && \qquad \qquad
\left.
\ - \ s \ \left(\bar{\psi}^{\nu \mu_2 \cdots \mu_s} \partial_{\nu} \slashed{\psi}_{\mu_2 \cdots \mu_s}
\ + \ \bar{\slashed{\psi}}^{\mu_2 \cdots \mu_s} \partial^{\nu} \psi_{\nu \mu_2 \cdots \mu_s} \right)
\right.
\nonumber \\ && \qquad \qquad
\left.
\ + \ \frac{s\left(s - 1 \right)}{2} \ \left(\bar{\slashed{\psi}}^{\nu \mu_3 \cdots \mu_s} \partial_{\nu} \psi'_{\mu_3 \cdots \mu_s} \ + \ \bar{\psi}'^{\mu_3 \cdots \mu_s} \partial^{\nu} \slashed{\psi}_{\nu \mu_3 \cdots \mu_s} \right)
\right\rbrace ,
\\
0 &=& \slashed{\psi}'_{\mu_4 \cdots \mu_s} .
\end{eqnarray}

It can be useful to rewrite all these equations in the index-free notation in which symmetrization over all free indices is implicit and is carried with a weight equal to the minimal number of terms necessary to write down (this convention is only used here). The definitions take the form:
\begin{eqnarray}
\delta \psi &=& \partial \zeta ,
\\
\mathcal{F} &=& \slashed{\partial} \psi \ - \ \partial \slashed{\psi} ,
\\
\mathcal{G} &=& \mathcal{F} \ - \ \frac{1}{2} \ \gamma \slashed{\psi} \ - \ \frac{1}{2} \ \eta \mathcal{F}' .
\end{eqnarray}

The traces conditions arise from:
\begin{eqnarray}
\delta \mathcal{F} &=& - \ 2 \ \partial^2 \slashed{\zeta} ,
\\
\partial \cdot \mathcal{F} \ - \ \frac{1}{2} \ \slashed{\partial} \slashed{\psi} \ - \ \frac{1}{2} \ \partial \mathcal{F}' &=& \ \partial^2 \slashed{\psi}' .
\end{eqnarray}

To sum things up, we have absorbed the components of a massless irreducible representation of the Poincaré group of spin $s + 1/2$ into a symmetric covariant tensor field. By giving it the gauge freedom (\ref{gauge_spin_fermi_s_no_trace}), it can be subjected to the gauge invariant first order equations of motion (\ref{eom_fermi_cov}), which precisely removed the spurious components of the field. These equations of motion can, in turn, be derived from the gauge invariant action (\ref{action_fermi_compact}) or, equivalently, (\ref{action_fermi_full}).

\subsubsection{Lower spin}

\label{Sec:low_fermi}

Plugging $s = 0$ into the previous expression yields Dirac field of spin $1/2$, of course without gauge invariance. The field is a spinor $\psi$ and its Lagrangian is $\mathcal{L} = - \ i \ \bar{\psi} \slashed{\partial} \psi$.

With $s = 1$, we get the massless spin $3/2$, the free $gravitino$ propagating over a flat space-time. The field is a spinor-vector $\psi_{\mu}$ whose gauge variation is the gradient of a spinor: 
\begin{eqnarray}
\delta \psi_{\mu} &=& \partial_{\mu} \zeta .
\end{eqnarray}

Its Fronsdal and Einstein tensors are:
\begin{eqnarray}
\mathcal{F}_{\mu} &=& 
\slashed{\partial} \psi_{\mu} \ - \ \partial_{\mu} \slashed{\psi} ,
\\
\mathcal{G}_{\mu} &=&
\slashed{\partial} \psi_{\mu} \ - \ \partial_{\mu} \slashed{\psi}
\ - \ \gamma_{\mu} \left(\partial^{\nu} \psi_{\nu} \ - \ \slashed{\partial} \slashed{\psi} \right)
\\ &=& \gamma_{\mu\nu\rho} \partial^{\nu} \psi^{\rho} .
\end{eqnarray}

The last form allows us to see that the general action we wrote above indeed reduces to:
\begin{eqnarray}
S &=& - \ i \ \int \ d^4 x \ \bar{\psi}_{\mu} \gamma^{\mu\nu\rho} \partial_{\nu} \psi_{\rho} .  
\end{eqnarray}

It is under this form that this action is more well-known.

\subsection{Constantly curved background space-time}

\label{Sec:fermi_lag_AdS}

As we saw, the key element in the Lagrangian description of massless higher spins outlined above is gauge invariance: it is this gauge invariance (present at the level of the action) which allows us to remove spurious components from the physical degrees of freedom of the field, in order for them to fit into a massless representation of the symmetry group of space-time.

The construction introduced above for fields propagating over a flat background space-time - whose symmetry group is simply Poincaré - can actually be very easily extended to a constantly curved background and, in particular, to \textit{anti-de Sitter} space-time, $AdS$.

To be perfectly definite, let us indeed consider this space-time chose curvature is constant and negative, with a radius of curvature $L$. The covariant derivatives $D_{\mu}$\footnote{Which includes the spin connection. See Appendix \textbf{\ref{app:conventions_fermi}} for more details.} acting on a covariant vector $V_{\mu}$ satisfy the following commutation rule:
\be 
[ D_{\!\m} \,, D_{\!\n}] V_\r = \frac{1}{L^2}\! \left( g_{\n\r} V_\m - g_{\m\r} V_\n \right) .
\ee

This commutation relation shows that the flat space action written above, \eqref{action_fermi_full}, can be modified in a very simple way in order to conserve its full gauge invariance: if we replace the partial derivatives in it by covariant derivatives, in order for the action to be well-defined\footnote{The use of partial derivatives over a curved space-time would make the action dependent on the choice of coordinate system in which it would be defined\ldots}, and if we add a ``mass term" appropriately chosen to cancel the additional terms generated by the non-trivial commutation rule under a gauge variation of the field to be defined shortly, we get:
\begin{eqnarray}
S &=& - \ i \ \int \ \sqrt{-g} \ d^4 x \ \left\lbrace
\bar{\psi}^{\mu_1 \cdots \mu_s} \slashed{D} \psi_{\mu_1 \cdots \mu_s}
\ + \ s \ \bar{\slashed{\psi}}^{\mu_2 \cdots \mu_s} \slashed{D} \slashed{\psi}_{\mu_2 \cdots \mu_s}
\ - \ \frac{s\left(s-1\right)}{4} \ \bar{\psi}'^{\mu_3 \cdots \mu_s} \slashed{D} \psi'_{\mu_3 \cdots \mu_s}
\right.
\nonumber \\ && \qquad \qquad
\left.
\ - \ s \ \left(\bar{\psi}^{\nu \mu_2 \cdots \mu_s} D_{\nu} \slashed{\psi}_{\mu_2 \cdots \mu_s}
\ + \ \bar{\slashed{\psi}}^{\mu_2 \cdots \mu_s} D^{\nu} \psi_{\nu \mu_2 \cdots \mu_s} \right)
\right.
\nonumber \\ && \qquad \qquad
\left.
\ + \ \frac{s\left(s - 1 \right)}{2} \ \left(\bar{\slashed{\psi}}^{\nu \mu_3 \cdots \mu_s} D_{\nu} \psi'_{\mu_3 \cdots \mu_s} 
\ + \ \bar{\psi}'^{\mu_3 \cdots \mu_s} D^{\nu} \slashed{\psi}_{\nu \mu_3 \cdots \mu_s} \right)
\right.
\nonumber \\ && \qquad \qquad
\left.
\ + \ \frac{d+2(s-2)}{2L} \ \left[
\bar{\psi}_{\mu_1 \cdots \mu_s}\psi^{\mu_1 \cdots \mu_s}
\ - \ s \ \bar{\slashed{\psi}}_{\mu_2 \cdots \mu_s}\slashed{\psi}^{\mu_2 \cdots \mu_s}
\ - \ \frac{s(s-1)}{4} \ \bar{\psi}'_{\mu_3 \cdots \mu_s}\psi'^{\mu_3 \cdots \mu_s}
\right]
\right\rbrace , \nonumber \\ \label{action_fermi_full_AdS}
\\
0 &=& \slashed{\psi}'_{\mu_4 \cdots \mu_s} .
\end{eqnarray}
\noindent where $g$ is the determinant of the space-time metric.

This action is easily checked to be invariant under the gauge transformation:
\begin{eqnarray}
\delta h_{\mu_1 \cdots \mu_s} &=&
s \ \left[ D_{( \mu_1} \epsilon_{\mu_2 \cdots \mu_s )}
\ + \ \frac{1}{2L} \ \gamma_{(\mu_1} \epsilon_{\mu_2 \cdots \mu_s )} \right] ,
\\
0 &=& \xi'_{\mu_4 \cdots \mu_s} .
\end{eqnarray}

This action, is naturally, valid in any dimension.

\chapter{Hamiltonian formalism for constrained systems}

\label{Sec:Ham_form_constr}

We are now going to consider the Hamiltonian formalism of constrained systems\footnote{The classic expositions of this topic are \cite{Dirac:LQM} and \cite{Henneaux:1994pup}.}, whose trajectory is limited to a certain subdomain   of their phase space and whose equations of motion are not determinist: the number of physical degrees of freedom of these systems is in fact smaller than the number of variables used to to describe them. The description is redundant and different values of the variables used represent the same physical state: this is gauge invariance; the range of value of some variables may also be limited. This is precisely the type of system to which belong massless fields of spin higher or equal to one (photon, gravitino, graviton, etc.). Dirac developed a formalism allowing us to obtain the Hamiltonian formalism of this kind of system.

We will begin by expounding Dirac formalism in full generality, showing how constraints appear when one starts with a Lagrangian. We will then consider the two categories of constraints that come into play: first class constraints, related to \textit{gauge invariance}, and second class constraints, which we will explain how to get rid of. In order to keep the development clear, we will present Dirac formalism in the case of system with a finite dimensional phase space.

We will then illustrate this formalism on the simplest gauge theories: the free massless field of spin one and two. This will allow us to appreciate the efficiency of the procedure just introduced and give us results which we will use in the following chapter.

We will then consider what happens when one studies gauge systems whose action is already first order, which is typically the case of massless fermionic fields of spin superior or equal to $3/2$. We will also introduce anticommuting Grassman variables and fix their conventions.\\

In order to be perfectly definite, let us fix a point of nomenclature: in this chapter and the following, what is meant by the number of \textit{degrees of freedom} is the dimension of the physically relevant phase space. Thus, the number of degrees of freedom of a system will be equal to the number of initial conditions that need to be fixed in order to determine its time evolution. This will prevent us from having to deal with ``half degrees of freedom" in the case of fermionic systems.

\section{Dirac formalism}

\subsection{From Lagrangian to Hamiltonian}

\label{Sec:Ham_form_constr_gen}

Generally, when one is given the Lagrangian formulation of a dynamical system (\textit{i.e.} a function $L\left(q^i , \dot{q}^j \right)$ of the coordinates $q^i$ and the velocities $\dot{q}^i$, where $i = 1, \ldots , N$), one gets to the Hamiltonian formulation in two steps: first, one defines the momenta by $p_i\left( q^j , \dot{q}^j \right) \equiv \frac{\partial L}{\partial \dot{q}^i}$; then, one builds the Hamiltonian function as $H \equiv p_i \dot{q}^i - L$. \textit{A priori}, the Hamiltonian is a function of coordinates and velocities, but, as a computation of its differential easily shows, it actually depends on the velocities only through the functions $\left( p_i , q^i \right)$. We will call the space parametrized by the coordinates and velocities the \textit{configuration space} and the one parametrized by the coordinates and momenta the \textit{phase space}.

Up to this point, we have considered a fully general Lagrangian. However, once we try to obtain the explicit form of the dependence of $H$ in terms of the coordinates and momenta, two situations can arise: either the relation defining the momenta is invertible, allowing us to express the velocities as functions of the momenta and coordinates ($\dot{q}^i = \dot{q}^i \left( q^j , p_j \right)$), in which case we can directly obtain the looked for Hamiltonian function $H\left( q^j , p_j \right)$ as a uniquely defined expression; or this relation is not invertible. This is the case we are interested in.

The non-invertibility of this relation typically corresponds to the case where the matrix $\frac{\partial^2 L}{\partial \dot{q}^i \partial \dot{q}^j}$ is also non invertible, from which we can infer that the equations of motion ($i.e.$ Euler-Lagrange equations), $\frac{\partial L}{\partial q^i} = \frac{\partial^2 L}{\partial \dot{q}^i \partial \dot{q}^j} \ddot{q}^j + \frac{\partial^2 L}{\partial \dot{q}^i \partial q^j} \dot{q}^j$, do not uniquely determine the accelerations $\ddot{q}^j$ in terms of the coordinates and velocities. In other words, all the coordinates $q^i$ can not correspond to degrees of freedom, since their evolution is not uniquely determined: the description of the system we are using must contain superfluous variables.

Since the application $\left( q^j , \dot{q}^j \right) \rightarrow \left( q^j , p_j \right)$ is not invertible, the dimension of its image space must be inferior to that of the initial space ($2N$), which is the configuration space: let us note this dimension $2N-K$. This means the image space can be defined by $K$ independent constraints $\phi_k \left( q^j , p_j \right) = 0$, $k = 1, \ldots , K$ (in other words, if one expresses the momenta as functions of the coordinates and velocities, these constraints will be identically satisfied). These constraints are said to be $primary$, because they appear immediately once one has defined the momenta. Limited to the surface defined by the constraints (the \textit{constraint surface}), the relation between the velocities and momenta is invertible. The independence of the functions $\phi_k$ requires that any function cancelling on the constraint surface be a linear combination of the constraints\footnote{The coefficients involved in the linear combination will generally be functions over the full phase space.}.
We will say of an identity satisfied on the constraint surface that it is \textit{weakly} satisfied. If it is true over the whole phase space, we will say that it is \textit{strongly} satisfied. Weakly satisfied equations will be written with the symbol $\approx$ (\textit{e.g.} two functions coinciding over the constraint surface will satisfy $f \approx g$).

As for the Hamiltonian, let us observe that, on the constraint surface, it is uniquely defined. However, if one tries to extend its definition over the whole phase space, we will get a continuous family of possible Hamiltonian functions, differing by a linear combination of the constraints. Let us assume, for now, to have picked one of these authorized Hamiltonians, $H$, keeping in mind that other possible choices remain, at this point, equivalent.

Let us then turn our attention to the Hamiltonian form of the equations of motion. The Hamiltonian was initially defined uniquely as a function of coordinates and velocities and, in full generality, its differential is given by $\delta H = \dot{q}^i \delta p_i - \left( \frac{\partial L}{\partial q^i}\right) \delta q^i$. When one writes the Hamiltonian as a function of the coordinates and momenta, which is always feasible but not automatically unique, this formula for the differential of $H$ is only valid for variations of the coordinates and momenta tangent to the surface constraint.
\footnote{Indeed, the configuration space being sent to the constraint surface, all the variations of the coordinates and velocities must correspond to variations of the coordinates and momenta tangent to the constraint surface.}.
Therefore, the coefficients of $\delta p_i$ and $\delta q^i$ in the differential of $H$ written above will differ from the partial derivatives of $H$ by a linear combination of the corresponding partial derivatives of the constraints: $\dot{q}^i = \frac{\partial H}{\partial p_i} + u^k \frac{\partial \phi_k}{\partial p_i}$ and $- \frac{\partial L}{\partial q^i} = \frac{\partial H}{\partial q^i} + u^k \frac{\partial \phi_k}{\partial q^i}$, where the $u^k$ are Lagrange multipliers. By the Lagrangian equations of motion, $\frac{\partial L}{\partial q^i} = \frac{d}{dt}\left(\frac{\partial L}{\partial \dot{q}^i}\right)$, we see that, in Hamiltonian form the equations of motion become:
\begin{eqnarray}
\dot{p}_i &=& 
- \ \frac{\partial H}{\partial q^i} \ - \ u^k \frac{\partial \phi_k}{\partial q^i} , \label{eqmv1}
\noindent \\
\dot{q}^i &=& 
\frac{\partial H}{\partial p_i} \ + \ u^k \frac{\partial \phi_k}{\partial p_i} , \label{eqmv2}
\end{eqnarray}
\noindent where $u^k$ are arbitrary functions of time: their presence confirms that not all phase space coordinates represent physical degrees of freedom, since their time evolution contains arbitrary functions. 

As usual, we would like to express these equations of motion in the form of Poisson brackets. If we define these for any pair of phase space functions as $\left\lbrace f , g \right\rbrace = \frac{\partial f}{\partial q^i}\frac{\partial g}{\partial p_i} - \frac{\partial g}{\partial q^i}\frac{\partial f}{\partial p_i}$, and if we also choose a total Hamiltonian $H_T$ defined as $H_T \equiv H + u^k \phi_k$, the equations of motion then take the general form, for any phase space function $f$, $\dot{f} = \left\lbrace f , H_T \right\rbrace$
\footnote{As long as we restric ourselves to considering this expression on the constraint surface, where the Poisson brackets containing $u^k$ do  not need to be considered, since they are multiplied by the constraints.}, 
the $u^k$ being manipulated by following the formal rules satisfied by the Poisson brackets: additivity, antisymmetry and product law 
\footnote{$\left\lbrace fg , h \right\rbrace = f \left\lbrace g , h \right\rbrace + g\left\lbrace f , h \right\rbrace$}.

Let us now consider the consistency question: if we are given as initial conditions a point of phase space on the constraint surface, will the evolution controlled by (\ref{eqmv1}) and (\ref{eqmv2}) maintain it there ? In other words, will the constraints be conserved ? Do we have $\dot{\phi}_k = 0$ on the constraint surface, or, equivalently, $\left\lbrace \phi_k , H_T \right\rbrace \approx 0$ ? We see that the consistency condition is: 
\begin{eqnarray}
0 &\approx& \left\lbrace \phi_k , H \right\rbrace \ + \ u^l \left\lbrace \phi_k , \phi_l \right\rbrace .
\label{constcond}
\end{eqnarray}

These conditions may take four possible forms. They can be identically satisfied ($i.e.$ be equivalent to an identity such as $0 \approx 0$), which will not lead to anything new; they can lead to a contradiction, ($e.g. \ 0 \approx 1$), in which case we must conclude that the initial Lagrangian did not describe a consistent system. They can also take a form from which the $u^k$ are absent, of the type $\varphi \left( q^i , p_j \right) = 0$ for a new function $\varphi$ over phase space. The requirement $\dot{\varphi} \approx 0$ will then also have to be imposed, eventually generating new constraints, calles $secondary$ (because they only appear after the equations of motion have been used), which will be noted $\varphi_m = 0$ ($m = 1, \ldots , M$). Finally, we can obtain a condition on the $u^k$, which we shall discuss later.

Once all these consistency conditions have been examined, we are left with a series of primary and secondary constraints, which will henceforward be treated on an equal footing. They can be gathered into a single series of (independent) constraints:$\phi_r$, $r = 1, \ldots, J$, defining a new constraint surface. We can also define an extended Hamiltonian, $H_E$, equal to $H + u^r \phi_r$ (the sum running now over both primary and secondary constraints).

Let us introduce an additional element of terminology: a phase space function will be said to be \textit{first class} if its Poisson bracket with all the constraints weakly cancels: $\left\lbrace f , \phi_r \right\rbrace \approx 0$. The constraints can themselves be first or second class (when their Poisson brackets with the other constraints do or do not cancel $modulo$ the constraints).

Let us prove a theorem: the Poisson bracket of two first class functions ($R$ and $S$) is also first class. Indeed, since $\left\lbrace R , \phi_k \right\rbrace \approx 0$ (resp. $\left\lbrace S , \phi_k \right\rbrace \approx 0$), we have $\left\lbrace R , \phi_k \right\rbrace = r^l_k \phi_l$ (resp. $\left\lbrace S , \phi_k \right\rbrace = s^l_k \phi_l$), from which it follows that $\left\lbrace \left\lbrace R , S \right\rbrace , \phi_k \right\rbrace $ $= - \left\lbrace \left\lbrace S , \phi_k \right\rbrace , R \right\rbrace - \left\lbrace \left\lbrace \phi_k , R \right\rbrace , S \right\rbrace = - \left\lbrace s^l_k \phi_l , R \right\rbrace + \left\lbrace r^l_k \phi_l , S \right\rbrace = - s^l_k  \left\lbrace \phi_l , R \right\rbrace - \phi_l \left\lbrace s^l_k , R \right\rbrace + r^l_k \left\lbrace \phi_l , S \right\rbrace + \phi_l \left\lbrace r^l_k , S \right\rbrace \approx 0$ (by using Jacobi identity\footnote{$\left\lbrace f , \left\lbrace g , h \right\rbrace \right\rbrace + \left\lbrace g , \left\lbrace h , f \right\rbrace \right\rbrace + \left\lbrace h , \left\lbrace f , g \right\rbrace \right\rbrace = 0$}, the product law and the fact that $R$ et $S$ are first class).

\subsection{Gauge systems}

\label{Sec:Ham_form_constr_gauge}

First class constraints are related to gauge invariance: as soon as a constraint is first class, the system considered has a gauge invariance. That is the point we are now going to establish.

We are indeed going (in this section) to take a closer look at a particular instance of constrained system: one in which all constraints are first class. In that case, we have $\left\lbrace \phi_r , \phi_s \right\rbrace = c^t_{rs} \phi_t \approx 0$ for any $r$, $s$. The secondary constraints are of the form $\left\lbrace \phi_k , H \right\rbrace \approx 0$ and $\left\lbrace \phi_k , \phi_l \right\rbrace \approx 0$ and the Hamiltonian is automatically first class.

In order to see why we say that such a system has a gauge invariance, let us observe that if we take as initial conditions in time $t$ a point of phase space (on the constraint surface) and evolve it over an infinitesimal time interval $\delta t$, the variation of a phase space function $g$ between the initial and final points of this infinitesimal trajectory will be given by (from (\ref{eqmv1}) and (\ref{eqmv2})): $g\left(t + \delta t \right) = g\left( t \right) + \delta t \ \dot{g} = g\left( t \right) + \delta t \left\lbrace g , H_T \right\rbrace = g\left( t \right) + \delta t \left\lbrace g , H \right\rbrace + \delta t \ u^k \left\lbrace g , \phi_k \right\rbrace$, where the $u^k$ are arbitrary. If we compare two different time evolutions obtained by picking different values of the $u^k$, we will see that the value of the function $g$ in $t + \delta t$ will shift by $\Delta g \left( t + \delta t \right) = \delta t \ \Delta u^k \left\lbrace g , \phi_k \right\rbrace = \epsilon^k \left\lbrace g , \phi_k \right\rbrace = \left\lbrace g , \epsilon^k \phi_k \right\rbrace$ (where we defined $\epsilon^k \equiv \delta t \Delta u^k$). These two values of $g$, having evolved from the same phase space point - and so from the same physical state, must obviously be physically indistinguishable. As a corollary, we see that any transformation of the form $\Delta g = \left\lbrace g , \epsilon^k \phi_k \right\rbrace$ (the parameters $\epsilon^k$ being arbitrary) must leave the physically relevant information unchanged: this is precisely what is called a gauge transformation, which we conclude to be generated by primary first class functions $\epsilon^k \phi_k$. Primary first class constraints are the generators of gauge transformations.

A natural question arises: are the transformations generated by secondary constraints also gauge transformations ($i.e.$ contact transformations which do not alter the physical state of the system) ? It seems plausible enough, given the fact that if we look at the commutator of two gauge transformations (generated by the functions $\epsilon^k_1 \phi_k$ and $\epsilon^k_2 \phi_k$), we observe it to be a transformation generated by the Poisson bracket of the generating functions of the initial transformations: $\left[ \delta_1 , \delta_2 \right] g = \left\lbrace \left\lbrace g , \epsilon^k_1 \phi_k \right\rbrace , \epsilon^l_2 \phi_l \right\rbrace - \left\lbrace \left\lbrace g , \epsilon^k_2 \phi_k \right\rbrace , \epsilon^l_1 \phi_l \right\rbrace = - \left\lbrace \left\lbrace \epsilon^k_1 \phi_k , \epsilon^l_2 \phi_l \right\rbrace , g \right\rbrace - \left\lbrace \left\lbrace \epsilon^l_2 \phi_l , g \right\rbrace , \epsilon^k_1 \phi_k \right\rbrace - \left\lbrace \left\lbrace g , \epsilon^k_2 \phi_k \right\rbrace , \epsilon^l_1 \phi_l \right\rbrace = \left\lbrace g , \left\lbrace \epsilon^k_1\phi_k , \epsilon^l_2 \phi_l \right\rbrace \right\rbrace$ (where we used Jacobi identity, antisymmetry and linearity of Poisson brackets). 

It is obvious that the commutator of two gauge transformations is a gauge transformation (the commutator of two transformations without physical modification must be a transformation without physical modification): the Poisson bracket of primary constraints must generate a gauge transformation. By our theorem on the Poisson bracket of first class functions, we know that the Poisson bracket of two primary constraints is a first class function. Moreover, the constraints being first class, $\left\lbrace \phi_k , \phi_l \right\rbrace \approx 0$. So, $\left\lbrace \phi_k , \phi_l \right\rbrace$ is a first class function that cancels on the constraint surface. In other words, it is a first class constraint, and so a linear combination of the constraints, combination possibly including secondary ones. A similar reasoning will show that secondary first class constraints of the form $\left\lbrace \phi_k , H \right\rbrace$ also generate gauge transformations, since they generate transformations obtained by commuting time evolution and a gauge transformation generated by the first class primary constraint $\phi_k$, which must obviously be a gauge transformation, on the same physical basis.

If the secondary (first class) constraints indeed generate gauge transformations, we can include their action in the dynamical evolution (since their presence will only make explicit the physical equivalence of the states they relate), and use the extended Hamiltonian $H_E$.

Let us still observe that each first class constraint removes two degrees of freedom: one degree of freedom is removed directly by the constraint, which decreases by one unit the dimension of the subdomain of the phase space in which the trajectory must be located; another degree of freedom is showed to be unphysical by the presence of an arbitrary function in the time evolution (on the constraint surface).

\subsection{Second class constraints}

\label{Sec:Ham_form_constr_2nd_cl}

Let us now consider the more general situation where some of the constraints are second class. In this case, the consistency conditions contain some constraints on the $u^r$, of the form (\ref{constcond}): it is a system of $J$ linear inhomogeneous equations (with $K$ unknown: the $u^k$\footnote{We remind the reader that $J$ and $K$ are respectively the total number of constraints and the number of primary constraints.}), whose coefficients are functions over phase space. We will begin by solving this system (if there is no solution, the system is not consistent): let us consider a particular solution ($U^r\left(p_j , q^i \right)$) and a complete set of independent solutions of the equations without their inhomogeneous part ($V^r_a \left(p_j , q^i \right)$, $a = 1, \ldots, A$): the general solution will then be $u^k = U^k + v_a V^k_a$, where the $v_a$ remain arbitrary. If we replace $u^k$ by this solution in $H_T$, we find a Hamiltonian of the form $H_T = H + U^k \phi_k + v_a V^k_a \phi_k \equiv H' + v_a \phi_a$, where $H' \equiv H + U^k \phi_k$ et $\phi_a \equiv V^k_a \phi_k$
\footnote{The sum over $k$ only ranges over primary constraints.}).
The new Hamiltonian $H'$ (which of course satisfies $H' \approx H$) is now first class (indeed, $\left\lbrace H' , \phi_r \right\rbrace = \left\lbrace H , \phi_r \right\rbrace + \left\lbrace U^k \phi_k , \phi_r \right\rbrace = \left\lbrace H , \phi_r \right\rbrace + U^k \left\lbrace \phi_k , \phi_r \right\rbrace + \phi_k \left\lbrace U^k , \phi_r \right\rbrace \approx 0$ by the definition of $U^k$), and so are the $\phi_a$ (again, $\left\lbrace \phi_a , \phi_r \right\rbrace \approx V^k_a \left\lbrace \phi_k , \phi_r \right\rbrace \approx 0$, by the definition of the $V^k_a$) and $H_T$. The $\phi_a$ form a maximal set of first class constraints (since they represent all the independent solutions of the system $V^k_a \left\lbrace \phi_k , \phi_r \right\rbrace \approx 0$): we can reorganize the constraints in order to have a list $\phi_a$, $\chi_r$, the $\chi_r$ being the (independent) remaining constraints, or the ``irreducible" second class constraints ($i.e.$ of which no first class linear combination exists).

The total Hamiltonian $H_T$ and the dynamical evolution of the system only contain $A$ arbitrary functions $v_a$ (completely unconstrained, since all consistency conditions have already been satisfied): there is no arbitrary function associated to second class constraints. Indeed, we will see that each second class constraint is related to the presence of a non physical degree of freedom, and we will now see how to remove it
\footnote{Contrary to first class constraints, the second class constraints do not imply the presence of an arbitrary function in the time evolution of the variables, but they still force some of these variables to be expressed as functions of the others (see below).}.

We show in Appendix \textbf{\ref{App:2nd_class_constr_invert}} that the matrix of the Poisson brackets of the irreducible second class constraints, $\left\lbrace \chi_r , \chi_s \right\rbrace$, is invertible (this matrix being antisymmetric, this implies that it must be of even dimension, and that the number of second class constraints is even): let us note its inverse $C^{rs}$. This inverse can be used to define \textit{Dirac brackets}: 
\begin{eqnarray}
\left\lbrace f , g \right\rbrace_D &\equiv&
\left\lbrace f , g \right\rbrace \ - \
\left\lbrace f , \chi_r \right\rbrace C^{rs} \left\lbrace \chi_s , g \right\rbrace .
\end{eqnarray}

The antisymmetry and linearity of this expression are automatic, but Jacobi identity needs to be checked. Moreover, $H_T$ being first class, $\left\lbrace g , H_T \right\rbrace_D \approx \left\lbrace g , H_T \right\rbrace$, and replacing the Poisson backets by the Dirac brackets does not alter the dynamics. The distinctive property of Dirac bracket is that it strongly cancels whenever it is applied to a second class constraint: $\left\lbrace f , \chi_r \right\rbrace_D = \left\lbrace f , \chi_r \right\rbrace - \left\lbrace f , \chi_s \right\rbrace C^{st} \left\lbrace \chi_t , \chi_r \right\rbrace = \left\lbrace f , \chi_r \right\rbrace - \left\lbrace f , \chi_s \right\rbrace \delta^s_r = 0$.

That means we can actually work with these brackets on the second class constraint surface $\chi_r = 0$ (which we shall note $\mathcal{S}$ and on which one may define an intrinsic coordinate system), completely ignoring the remaining phase space, which can not be done with first class constraints, whose Poisson bracket must be computed in the full phase space before being evaluated on the constraint surface. Once one restricts oneself to $\mathcal{S}$, one can completely forget the second class constraints. In conclusion, we see that each second class constraint indeed removes only one degree of freedom, as opposed to first class constraints, which remove two of them. Since we have seen that there is always an even number of second class constraints, an even number of physical degrees of freedom will be subtracted, and the parity of the initial formal phase space dimension will be preserved in the number of dynamical variables physically relevant and non redundant\footnote{This parity is necessary, since a set of initial conditions must contain an even number of independent data for a second order dynamical system.}.

The reasoning followed in this section can naturally be generalized to the case where we would use $H_E$ rather than $H_T$ as Hamiltonian (which, as we argued at the end of the previous section, is the natural thing to do).

\section{Illustrations of Dirac formalism}

\subsection{Classical electromagnetism}

\label{Sec:EM_ham}

In order to illustrate the general formalism expounded above, we shall now consider its most classical aplication: Maxwell theory of the free electromagnetic field in four-dimensional space-time. Obviously, this is a field theory, and, in the formal developments made above, we only considered dynamical theories with a finite number of degrees of freedom. However, the transition is completely straightforward.

The (real-valued) fields are the spatial potential-vector $\vec{A}$ and the spatial scalar potential $V$ (gathered into a space-time four-vector $A^{\mu} \equiv \left( V , \vec{A} \right)$). The action is given by $S = \int dt \ L =  \int d^4 x \ \mathcal{L}$, where $\mathcal{L} = - \frac{1}{4} F_{\mu \nu} F^{\mu \nu}$ (the strength field being $F_{\mu \nu} \equiv \partial_{\mu} A_{\nu} - \partial_{\nu} A_{\mu}$). In a non manifestly covariant way, this gives:
\begin{eqnarray}
L &=& \frac{1}{2} \ \int \ d^3 x \
\left\lbrace \Vert \dot{\vec{A}} + \vec{\nabla} V \Vert^2 \ - \ \Vert \vec{\nabla} \times \vec{A} \Vert^2 \right\rbrace, \label{lagrangian_maxwell}
\end{eqnarray}

\noindent whence one extracts the conjugate momenta:
\begin{eqnarray}
\Pi_V &\equiv& \frac{\delta L}{\delta \dot{V}} \ = \ 0 , \label{primary_constraint_maxwell}
\\
\vec{\Pi} &\equiv& \frac{\delta L}{\delta \vec{\dot{A}}} \ = \ \dot{\vec{A}} + \vec{\nabla} V .
\end{eqnarray}

We immediately notice that (\ref{primary_constraint_maxwell}) gives us a primary constraint. Let us now build the Hamiltonian:
\begin{eqnarray}
H &=& \int \ d^3 x \ \left\lbrace \Pi_V V \ + \ \vec{\Pi} \cdot \dot{\vec{A}}\right\rbrace \ - \ L
\nonumber \\&=&
\int \ d^3 x \ \left\lbrace \frac{\Vert \vec{\Pi} \Vert^2}{2} + \frac{\Vert \vec{\nabla} \times \vec{A}\Vert^2}{2} + V \left( \vec{\nabla} \cdot \vec{\Pi} \right)\right\rbrace . \label{hamiltonian_maxwell}
\end{eqnarray}
\noindent where an integration by part was performed on the last term.

Consequently, we have $\left\lbrace \Pi_V , H \right\rbrace = - \vec{\nabla}.\vec{\Pi} = - \Pi^k_{,k}$, and so the consistency condition imposes the secondary constraint $\vec{\nabla} \cdot \vec{\Pi} = 0$. Since these constraints have a vanishing Poisson bracket among themselves ($\left\lbrace \Pi_V , \vec{\nabla} \cdot \vec{\Pi} \right\rbrace = 0$) and with the Hamiltonian ($\left\lbrace \Pi_V , H \right\rbrace = \vec{\nabla} \cdot \vec{\Pi} \approx 0$ et $\left\lbrace \vec{\nabla} \cdot \vec{\Pi} , H \right\rbrace = 0$), there are no additional constraints. Moreover, they are all first class, and the Hamiltonian given by  (\ref{hamiltonian_maxwell}) also is: this is indeed a gauge system.

The counting of degrees of freedom is trivial: we started with eight variable, the $A_k$, $V$, $\Pi_k$ et $\Pi_V$. Since we found two first class constraints, we are left with four ($= 8 - 2 \times 2$) physical phase space dimensions, corresponding to two physically distinguishable configurations (whose amplitudes and phases - or velocities - are parametrized by the four coordinates of the reduced phase space): the states of helicity $\pm 1$.

We then proceed to write the total and extended Hamiltonian:
\begin{eqnarray}
H_T &=& H \ + \ \int \ d^3 x \ u \ \Pi_V,
\\
H_E &=& H_T \ + \ \int \ d^3 x \ v \left( \vec{\nabla} \cdot \vec{\Pi} \right),
\end{eqnarray}

\noindent where $u$ and $v$ are arbitrary functions. Let us observe that one could absorb $V$ into $v$ by an appropriate redefinition, when one uses $H_E$, which is very natural in this case, the secondary constraint manifestly generating a gauge transformation (since it generates transformations obtained by commuting time evolution with a gauge transformation generated by the primary first class constraint). The corresponding equations of motion, obtained by taking Poisson brackets with $H_E$, are:
\begin{eqnarray}
\dot{V} &=& u,
\\
\dot{\Pi}_V &=& - \ \vec{\nabla} \cdot \vec{\Pi} ,
\\
\dot{\vec{A}} &=& \vec{\Pi} \ - \ \vec{\nabla} \left( v + V \right) ,
\\
\dot{\vec{\Pi}} &=& \vec{\nabla} \times \left( \vec{\nabla} \times \vec{A} \right) .
\end{eqnarray}

In other words, $\Pi_V = 0$ and the time evolution of $V$ is given by an arbitrary function: the non-physical nature of the degrees of freedom associated to this pair of variables is manifest (especially considering the vanishing of their Poisson brackets with all the other variables). Similarly, in a non local way, by decomposing the fields $\vec{\Pi}$ and $\vec{A}$ into transverse and longitudinal components, we see that the longitudinal component of $\vec{\Pi}$ cancels and that the time evolution of its conjugate variable, the longitudinal component of $\vec{A}$, is given by an arbitrary function: the associated degrees of freedom are also clearly non-physical, and we could explicitly remove them, only keeping as variables the transverse parts of $\vec{\Pi}$ and $\vec{A}$ and the Hamiltonian $H' = \frac{1}{2} \ \int \ d^3 x \ \left[ \Vert \vec{\Pi}_{\perp} \Vert^2 + \Vert \vec{\nabla} \times \vec{A}_{\perp} \Vert ^2 \right]$. This description is free of any gauge invariance, but it is non-local.\\

To sum things up, we have two first class constraints ($\Pi_V = 0$ and $\vec{\nabla} \cdot \vec{\Pi} = 0$), to which correspond the presence of two arbitrary functions in the time evolution of the remaining, unconstrained variables. The combinations of these whose time evolution is undetermined are $V$ and $\vec{\nabla} \cdot \vec{A}$.\\

We will be much interested, in generalizing these results to higher spins, in the following approach: explicitly removing the variables $V$ and $\Pi_V$, and solving the constraint $\vec{\nabla} \cdot \vec{\Pi} = 0$, through the introduction of a second potential-vector, $\vec{A}^2$ (the initial potential vector being relabelled $\vec{A}^1$), defined by $\vec{\Pi} = \vec{\nabla} \times \vec{A}^2$. Working with $\vec{A}^2$ is equivalent to using a proper coordinate system on the constraint surface: the new variables, $\vec{A}^1$ and $\vec{A}^2$, are unconstrained, but non uniquely defined (two vectors $\vec{A}^2$ differing through a gradient are clearly equivalent). In this formalism, the Hamiltonian given by (\ref{hamiltonian_maxwell}) takes the particularly simple and symmetric form (in which the terms proportional to $u$ and $v$ in $H_E$ have disappeared since, in terms of the new variables, the constraints are identically satisfied):
\begin{eqnarray}
H &=& \int \ d^3 x \ \left[ \frac{\left\Vert \vec{\nabla} \times \vec{A}^1 \right\Vert ^2}{2}
\ + \ \frac{\left\Vert \vec{\nabla} \times \vec{A}^2 \right\Vert ^2}{2} 
\right] .
\end{eqnarray}

The Hamiltonian action then writes:
\begin{eqnarray}
S_H &=& \int \ d^4 x \ \vec{\Pi} \cdot \dot{\vec{A}}^1 \ - \ \int \ dt \ H
\nonumber \\ &=&
\int \ d^4 x \ \left\lbrace \left(\vec{\nabla} \times \vec{A}^2 \right) \cdot \dot{\vec{A}}^1
\ - \  \frac{\left\Vert \vec{\nabla} \times \vec{A}^1 \right\Vert ^2}{2}
\ - \ \frac{\left\Vert \vec{\nabla} \times \vec{A}^2 \right\Vert ^2}{2} 
\right\rbrace . \label{hamiltonian_action_maxwell}
\end{eqnarray}

If we define the ``curvature" (the magnetic field) of each potential as $\vec{B}^a \equiv \vec{\nabla} \times \vec{A}^a$ ($a = 1,2$) and introduce the antisymmetric two-dimensional tensor $\epsilon_{ab}$ (with $\epsilon_{12} = 1$) in the plane of the two potentials, this action can again be written:
\begin{eqnarray}
S_H &=& \frac{1}{2} \ \int \ d^4 x \ \left\lbrace \epsilon_{ab} \dot{\vec{A}}^a  \cdot \vec{B}^b
\ - \  \delta_{ab} \vec{B}^a \cdot \vec{B}^b
\right\rbrace .
\end{eqnarray}

Except for the relative factor between the kinetic and Hamiltonian parts, this action is completely fixed by its gauge invariance $\delta \vec{A}^a = \vec{\nabla} f^a$ and its invariance under the rigid duality rotation in the plane of the two potentials.

\subsection{Linearized gravity}

\label{Sec:Grav_ham}

A completely similar investigation can be made for linearized gravity ($i.e.$ for the free massless spin $2$ theory)\footnote{The original study of the Hamiltonian formulation of gravity is \cite{Arnowitt:1962hi}.}. Its lagrangian action is:
\begin{eqnarray}
S &=&
\int \ d^4 x \ \mathcal{L}
\\ &=&
- \ \frac{1}{2} \ \int \ d^4 x \ \left\lbrace
\partial_{\rho} h_{\mu\nu} \partial^{\rho} h^{\mu\nu}
\ - \ 2 \ \partial^{\rho} h_{\rho\mu}\partial_{\nu} h^{\nu\mu}
\ + \ 2 \ \partial^{\mu} h_{\mu\nu} \partial^{\nu} h'
\ - \ \partial_{\mu} h' \partial^{\mu} h'
\right\rbrace , \qquad  \qquad
\end{eqnarray}
\noindent where $h' \equiv h^{\mu}_{\phantom{\mu}\mu}$.

The Hamiltonian analysis of this action requires us to break explicit covariance, giving:
\begin{eqnarray}
S &=&
\frac{1}{2} \ \int \ d^4 x \ \left\lbrace
\dot{h}_{kl} \dot{h}^{kl}
\ - \ \dot{h}^2
\ - \ 4 \ \dot{h}^{kl} \partial_k h_{0l}
\ + \ 4 \ \dot{h} \partial^k h_{0k}
\ + \ \ldots
\right\rbrace , \qquad  \qquad
\end{eqnarray}
\noindent where the unwritten terms only contain spatial derivatives and $\bar{h}\equiv h^k_{\phantom{k}k}$. This gives the only non-vanishing momenta as:
\begin{eqnarray}
\Pi_{kl} &=& \dot{h}_{kl} \ - \ \delta_{kl} \dot{\bar{h}}
\ - \ 2 \ \partial_{(k} h_{l)0} \ + \ 2 \ \delta_{kl} \partial^m h_{0m} ,
\end{eqnarray}
\noindent which can easily be inverted to yield:
\begin{eqnarray}
\dot{h}_{kl} &=&
\Pi_{kl} \ - \ \frac{1}{2} \ \delta_{kl} \bar{\Pi} \ + \ 2 \ \partial_{(k} h_{l)0} .
\end{eqnarray}

Substituting these and redefining $N_k \equiv h_{0k}$ and $N \equiv h_{00}$, we get:
\begin{eqnarray}
\int \ d^3 x \ \left[ \Pi^{kl} \dot{h}_{kl} \ - \ \mathcal{L} \right]
&=&
\int \ d^3 x \ \left[\mathcal{H} \ + \ N^k \mathcal{C}_k \ + \ N \mathcal{C} \right] ,
\end{eqnarray}
\noindent where:
\begin{eqnarray}
\mathcal{H} &=& \frac{\Pi^{kl} \Pi_{kl} }{2} \ - \ \frac{\bar{\Pi}^2}{4}
\nonumber \\ && 
\ + \ \frac{1}{2} \ \partial_{m} h_{kl} \partial^m h^{kl}
\ - \ \partial^m h_{km} \partial_l h^{kl}
\ + \ \partial^k \bar{h} \partial^l h_{kl}
\ - \ \frac{1}{2} \ \partial_k \bar{h} \partial^k \bar{h},
\\
\mathcal{C}_k &=& - \ 2 \ \partial^l \Pi_{kl} ,
\\
\mathcal{C} &=& \partial^k \partial^l h_{kl} \ - \ \Delta \bar{h} .
\end{eqnarray}

The consistency checks (imposing the conservation in time of the primary constraints, which requires the vanishing of the Poisson brackets of $\Pi_{0\mu}$ with the Hamiltonian written above) then give us the secondary constraints, which are just $0 = \mathcal{C}_k$ and $0 = \mathcal{C}$. The constraints and the Hamiltonian are then easily seen to be first class. The gauge transformations generated by the primary constraints simply show that the variables $N$ and $N_k$ (canonically conjugate to $\Pi_{0\mu}$) are pure gauge, their value being arbitrary. As for the secondary constraints, they generate the following independent gauge transformations. If we define:
\begin{eqnarray}
\mathcal{G}_1 &=& \int \ d^3 x \  \xi_k \mathcal{C}^k
\nonumber \\ &=& 2 \ \int \ d^3 x \ \Pi^{kl} \partial_{(k} \xi_{l)} ,
\\
\mathcal{G}_2 &=& \int \ d^3 x \  \theta \mathcal{C} 
\nonumber \\ &=& \int \ d^3 x \ h^{kl} \left( \partial_{k} \partial_l \theta \ - \ \delta_{kl} \Delta \theta\right) ,
\end{eqnarray}
\noindent we easily see that they generate the following gauge transformations (with $\mathcal{G} \equiv \mathcal{G}_1 \ + \ \mathcal{G}_2$):
\begin{eqnarray}
\delta h_{kl} &=&
\left\lbrace h_{kl} , \mathcal{G} \right\rbrace
\nonumber \\ &=& 2 \ \partial_{(k} \xi_{l)} , \label{spin2_non_cov_gauge_freed_h}
\\
\delta \Pi_{kl} &=&
\left\lbrace \Pi_{kl} , \mathcal{G} \right\rbrace
\nonumber \\ &=&
\delta_{kl} \Delta \theta \ - \ \partial_k \partial_l \theta . \label{spin2_non_cov_gauge_freed_pi}
\end{eqnarray}

Each of these gauge transformation clearly preserves the constraints - which are indeed first class.

The counting of degrees of freedom is satisfactory: we started with $20$ variables ($h_{\mu\nu}$ and $\Pi^{\mu\nu}$) and we end up with eight first class constraints ($0 = \Pi^{0\mu} = \mathcal{C}_k = \mathcal{C}$), removing $16$ degrees of freedom and leaving us with a four-dimensional physical phase space, which corresponds with the expected first order description of the two helicities of a massless spin $2$ field.

In a way completely similar to what was done in the case of electromagnetism, we then solve these secondary constraints\footnote{It was done in \cite{Henneaux:2004jw}.} and simply remove the $N_k$, $N$ and their conjugate momenta. The \textit{momentum constraint} $0 = \mathcal{C}_k$ is easily solved, since it naturally implies that $\Pi_{kl}$ is the Einstein tensor of some symmetric tensor $P_{kl}$ (any symmetric divergenceless tensor is the Einstein tensor of some field; see the Appendix \textbf{\ref{graviton_momentum_constraint_solv}} for an explicit derivation):
\begin{eqnarray}
\Pi_{kl} &=&
G_{kl} \left[P_{mn}\right]
\nonumber \\ &=&
\epsilon_{kmp} \epsilon_{lnq} \partial^m \partial^n P^{pq}
\nonumber \\ &=&
\delta_{kl} \left(\Delta \bar{P} \ - \ \partial^m \partial^n P_{mn}\right)
\ + \ 2 \ \partial_{(k} \partial^m P_{l)m}
\ - \ \Delta P_{kl}
\ - \ \partial_k \partial_l \bar{P} .
\end{eqnarray}

The gauge invariance of $P_{kl}$ is easily identified by observing that the gauge variation of $\Pi_{kl}$ is precisely the Einstein tensor of a field of the form $\delta_{kl} \theta$:
\begin{eqnarray}
\delta \Pi_{kl} &=& G_{kl} \left[\delta_{mn} \theta \right] ,
\end{eqnarray}
\noindent from which it follows that (the vanishing of its (linearized) Einstein tensor - which, in three dimensions, is equivalent to the vanishing of its (linearized) Riemann tensor - being a necessary and sufficient condition for a tensor field to be equal to a (linearized) diffeomorphism - that is, the symmetrized derivative of a vector field) the gauge invariance of $P_{kl}$ is:
\begin{eqnarray}
\delta P_{kl} &=& 2 \ \partial_{(k} \lambda_{l)} \ + \ \delta_{kl} \theta .
\end{eqnarray}

The \textit{Hamiltonian constraint} $0 = \mathcal{C}$ is slightly more involved to solve. We can start by going into a gauge where $h$, the trace of $h_{kl}$, is equal to zero. Up to a gauge transformation, the constraint then solves as\footnote{This is quite straightforward: a first integration of the vanishing of the double divergence yields $\partial^l h_{kl} = \epsilon_{klm} \partial^l v^m$, which is itself a divergence whose integration gives $h_{kl} = \epsilon_{klm} v^m + 1/2 \ \epsilon_{lmn} \partial^m \phi^n_{\phantom{n}k}$. The symmetrization of this expression over $kl$ finally gives the looked for expression (and the vanishing of the trace of $h_{kl}$ imposes the symmetry of $\phi_{kl}$). The $1/2$ factor is introduced for later convenience.}:
\begin{eqnarray}
h_{kl} &=& \frac{1}{2} \ \epsilon_{(k\vert mn} \partial^m \phi^{n}_{\phantom{n} \vert l)} \ + \ 2 \ \partial_{(k} \xi_{l)} . \label{2_h_Z}
\end{eqnarray}

The gauge invariance of $\phi_{kl}$ can be computed (see Appendix \textbf{\ref{graviton_prepotential_gauge_inv}}) and is:
\begin{eqnarray}
\delta \phi_{kl} &=& 2 \ \partial_{(k} \eta_{l)} \ + \ \delta_{kl} \sigma .
\end{eqnarray} 

The fields $P_{kl}$ and $\phi_{kl}$, which are symmetric spatial tensors in terms of which the constraints are identically satisfied, are called \textit{prepotentials}. Let us note that they have the same gauge invariance (with independent parameters), which is precisely the gauge invariance of the metric field (the ``potential") in linearized conformal gravity.

If we substitute these (partially gauge fixed) solutions into our Hamiltonian density $\mathcal{H}$, it becomes (up to total derivatives):
\begin{eqnarray}
\mathcal{H} 
&=& 
\Delta \phi_{kl} \Delta \phi^{kl}
\ - \ 2 \ \partial_n \partial^l \phi_{kl} \partial^n \partial_m \phi^{km}
\nonumber \\ &&
 - \ \frac{1}{2} \ \left(\Delta \bar{\phi} \right)^2
\ + \ \partial^k \partial^l \phi_{kl} \Delta \bar{\phi}
\ + \ \frac{1}{2} \ \left(\partial^k \partial^l \phi_{kl}\right)^2
\nonumber \\ &&
+ \ \Delta P_{kl} \Delta P^{kl}
\ - \ 2 \ \partial_n \partial^l P_{kl} \partial^n \partial_m P^{km}
\\ &&
 - \ \frac{1}{2} \ \left(\Delta \bar{P} \right)^2
\ + \ \partial^k \partial^l P_{kl} \Delta \bar{P}
\ + \ \frac{1}{2} \ \left(\partial^k \partial^l P_{kl}\right)^2 .
\end{eqnarray}

We can now look at the kinetic part of the Hamiltonian action, which now takes the form (again, up to boundary terms):
\begin{eqnarray}
\int \ d^4 x \ \Pi^{kl}\left[ P_{ij}\right] \dot{h}_{kl} \left[\phi_{mn}\right] &=&
 \int \ d^4 x \ \phi_{kl} B^{kl} \left[\dot{P}_{mn}\right] ,
\end{eqnarray}
\noindent where we have introduced the Cotton tensor:
\begin{eqnarray}
B_{kl} \left[P_{mn}\right] &\equiv&
\frac{1}{2} \ \epsilon_{(k\vert mn} \partial^m \left(\Delta P^n_{\phantom{n}l)} 
\ - \ \partial_{l)} \partial^r P^{n}_{\phantom{n}r} \right) .
\end{eqnarray}

This tensor is identically traceless and divergenceless, and it vanishes when the field that is plugged into it is a (linearly) conformally flat metric, up to a (linearized) diffeomorphism. It follows from these properties that this part of the Hamiltonian action has the same gauge invariance (in terms of the prepotentials) as the Hamiltonian.

This action can be rewritten as:
\begin{eqnarray}
S_H &=&
\frac{1}{2} \ \int \ d^4 x \ \left\lbrace
\epsilon_{ab} \dot{Z}^{a kl} B^b_{kl}
\ - \  \delta_{ab} \left[G^{a kl} G^b_{kl} \ - \ \frac{1}{2} \ \bar{G}^a \bar{G}^b \right]
\right\rbrace , \label{action_2_ham_prepot}
\end{eqnarray}
\noindent where $G^a_{kl} \equiv G_{kl} \left[Z^a_{ij}\right]$ is the Einstein tensor and $B^a_{kl} \equiv B_{kl} \left[Z^a_{ij}\right]$ the Cotton tensor of one of the prepotentials that we have gathered into a two-components vector:
\begin{eqnarray}
Z^a_{kl} &\equiv& \left( \begin{array}{c} P_{kl} \\ \phi_{kl} \end{array} \right) .
\end{eqnarray}

\section{First order action}

\subsection{General formalism}

\label{Sec:1st_order_gen}

For free fermionic field theories, the Lagrangian action is already linear in the time derivatives and, in this case, there is a way to shortcut Dirac formalism, as we are now going to show. We will again consider the case of a system with a finite number of degrees of freedom, in order not to overcomplicate notations.

Let us consider a system whose state is described by the (bosonic) variables $\lbrace z^A\rbrace$ and whose action is given by:
\begin{eqnarray}
S\left[ z \right] &=& \int \ dt \ \left\lbrace \theta_A \left( z \right) \dot{z}^A \ - \ H \left( z \right) \right\rbrace
\label{general_action_first_order} .
\end{eqnarray}

Such an action can be obtained from a Hamiltonian action in canonical form $S\left[ p , q \right]  = \int dt \ \left\lbrace p_a \dot{q}^a - H\left( p , q \right) \right\rbrace $ by performing a change of variables of the form $z = z \left( p , q \right)$. In particular, one should keep in mind that the number of variables $z^A$ gives the total dimension of the phase space, that is, twice the number of canonical coordinates. Therefore, we shall note $A = 1, ..., 2N$ (or again $a = 1, ..., N$). Each $z^A$ should accordingly be seen as representing half a physical configuration.

The action (\ref{general_action_first_order}) gives, through its variation, the following equations of motion:
\begin{eqnarray}
\omega_{AB} \ \dot{z}^B &=& \frac{\partial H}{\partial z^A},
\end{eqnarray}
\noindent where $\omega_{AB} \equiv \frac{\partial \theta_B}{\partial z^A} - \frac{\partial \theta_A}{\partial z^B}$ is the  \textit{symplectic 2-form}; its inverse (assuming one exists
\footnote{Which indeed implies the evenness of the number of $z^A$.}: if such is not the case, it is due to the presence of first class constraints, some of the $z$ variables actually being arbitrary functions of type $u^k$; see below)
will be written $\omega^{AB}$ ($\omega^{AB}\omega_{BC} = \delta^A_C$). The equations of motion then take the form:
\begin{eqnarray}
\dot{z}^A &=& \omega^{AB} \frac{\partial H}{\partial z^B}.
\end{eqnarray}

(Let us observe that in the case in which the symplectic 2-form takes the canonical form $\omega = \left( \begin{array}{cc} 0 & - I \\ I & 0 \end{array}\right)$, if we relabel the phase space coordinates $z \equiv \left( \begin{array}{c} q \\ p \end{array} \right) $, these equations reproduce Hamilton equations: $\dot{p} = - \frac{\partial H}{\partial q}$ et $\dot{q} = \frac{\partial H}{\partial p}$: we are indeed dealing with a general Hamiltonian formalism.)

The time derivative of any phase space function $f(z)$ will be: $\dot{f} = \frac{\partial f}{\partial z^A} \dot{z}^A = \omega^{AB} \frac{\partial f}{\partial z^A} \frac{\partial H}{\partial z^B}$, which leads us to define the Dirac bracket of two phase space functions $f$ and $g$ as: 
\begin{equation}
\left\lbrace f , g \right\rbrace_D \equiv \omega^{AB} \frac{\partial f}{\partial z^A} \frac{\partial g}{\partial z^B}
\label{dirac_bracket_definition_first_order},
\end{equation}
\noindent in order to have $\dot{f} = \left\lbrace f , H \right\rbrace_D$. This way, we also have: $\left\lbrace z^A , z^B \right\rbrace_D = \omega^{AB}$.
\\

We did not rely on the general formalism developped above, since Dirac bracket could be defined through a more direct path. However, Dirac formalism would have led to the same brackets. Indeed, computing the conjugate momenta from the action (\ref{general_action_first_order}) gives $\Pi_A = \frac{\partial L}{\partial \dot{z}^A} = \theta_A \left( z \right)$, and we immediately obtain $2N$ primary constraints: $\phi_A \left( \Pi , z \right) = \Pi_A - \theta_A \left( z \right) = 0$. Moreover, the Hamiltonian is given by $H(z)$. To study the consistency conditions, let us observe that (reminding ourselves that the Poisson bracket is defined as $\left\lbrace f , g \right\rbrace _P \equiv \frac{\partial f}{\partial z^A} \frac{\partial g}{\partial \Pi_A} - \frac{\partial g}{\partial z^A} \frac{\partial f}{\partial \Pi_A}$):
\begin{eqnarray}
\left\lbrace \phi_A , H \right\rbrace &=& - \ \frac{\partial H}{\partial z^A} ,
\\ 
\left\lbrace \phi_A , \phi_B \right\rbrace &=& \omega_{AB}.
\end{eqnarray}

Consequently, when the symplectic 2-form is invertible, all the constraints are second class (since the first class constraints are associated to the null eigenvectors of the matrix of the Poisson brackets of the constraints), and there are no secondary constraints (the consistency conditions $\left\lbrace \phi_A , H \right\rbrace_P + u^B \left\lbrace \phi_A , \phi_B \right\rbrace_P$ $\approx 0$ imposing no new relation on the $\Pi$ and $z$). We must then impose $strongly$ ($i.e.$ before computing Dirac brackets) the constraints (which comes down to substituting the $\theta_A (z)$ for the $\Pi_A$), and use Dirac brackets, which are (observing that $\left\lbrace f , \phi_A \right\rbrace_P = \frac{\partial f}{\partial z^A} + \frac{\partial \theta_A}{\partial z^B} \frac{\partial f}{\partial\Pi_B}$):
\begin{eqnarray}
\left\lbrace f , g \right\rbrace_D &=& 
\left\lbrace f , g \right\rbrace_P 
\ - \ \omega^{AB} \left\lbrace f , \phi_A \right\rbrace_P  \left\lbrace \phi_B , g \right\rbrace_P
\nonumber \\ &=&
\frac{\partial f}{\partial z^A} \frac{\partial g}{\partial \Pi_A} - \frac{\partial g}{\partial z^A} \frac{\partial f}{\partial \Pi_A}
\ + \ \omega^{AB} 
\left( \frac{\partial f}{\partial z^A} + \frac{\partial \theta_A}{\partial z^C} \frac{\partial f}{\partial\Pi_C}\right) 
\left( \frac{\partial g}{\partial z^B} + \frac{\partial \theta_B}{\partial z^D} \frac{\partial g}{\partial\Pi_D}\right) 
\nonumber \\ &=&
\omega^{AB} \frac{\partial f}{\partial z^A} \frac{\partial g}{\partial z^B}.
\end{eqnarray}

We indeed recover our previous definition of the Dirac brackets. \\

A case more relevant for the study of fermionic higher spin gauge fields is, of course, one in which some first class constraints are also present, which is embodied by an action of the following form:
\begin{eqnarray}
S\left[ z , u \right] &=& \int \ dt \ \left\lbrace \theta_A \left( z \right) \dot{z}^A \ - \ u^k f_k\left( z \right) \ - \ H \left( z \right) \right\rbrace .
\label{gauge_action_first_order_constraint}
\end{eqnarray}

It is the form the Lagrangian action will take for massless fields of half-integer spin superior to $\frac{1}{2}$, even though we are currently looking at ``bosonic" ($i.e.$ commuting) variables. The $u^k$ will manifestly be unphysical variables: they are Lagrange multipliers imposing the constraints $f_k = 0$. The equations of motions derived from the variation of the action (\ref{gauge_action_first_order_constraint}) are (with the same Dirac brackets as those defined by (\ref{dirac_bracket_definition_first_order})):
\begin{eqnarray}
\dot{z}^A &=& \left\lbrace z^A , H \right\rbrace_D
\ + \ u^k \left\lbrace z^A , f_k \right\rbrace_D ,
\\
f_k &=& 0 .
\end{eqnarray}

The $u^k$ are free, unconstrained, arbitrary, and their presence in the equations of motion signals gauge freedom. In a general way, the evolution of a function of the $z$ will be given by $\dot{g} = \left\lbrace g , H \right\rbrace_D + u^k \left\lbrace g , f_k \right\rbrace_D$: the time evolution is determined by the Hamiltonian, up to a gauge transformation generated by the constraints $f_k$ and parametrized by the $u^k$. The case we are interested in is the one in which the constraints $f_k$ and the Hamiltonian $H$ are first class, that is, gauge invariant on the constraint surface. The consistency conditions are then identically satisfied.

Let us still observe that the gauge transformations could be identified through a direct examination of the action, since it is gauge invariant. Indeed, two trajectories differing by a gauge transformation will clearly have the same action, since the extremization of the action can not fix which one of these trajectories will be followed. Explicitly, we can check the invariance of the action under the transformation $\delta z^A = \left\lbrace z^A , v^k f_k \right\rbrace_D = \omega^{AB} \frac{\partial f_k}{\partial z^B} v^k$: the constraints and the Hamiltonian being gauge invariant, their gauge variation cancels on the constraint surface and must be a linear combination of the constraints ($\delta H = h_k^{\ l} v^k f_l$ and $\delta f_k = c^{m}_{\ \ kl} v^l f_m$). The variation of the kinetic term is:
\begin{eqnarray}
\delta \left( \int \ dt \ \theta_A \dot{z}^A \right) &=&
\int \ dt \ \omega_{AB} \dot{z}^B \delta z^A 
\nonumber \\ &=&   
- \ \int \ dt \ \dot{z}^A v^k \frac{\partial f_k}{\partial z^A}
\nonumber \\ &=& 
- \ \int \ dt \ v^k \dot{f}_k .
\end{eqnarray}

In order for the action to be invariant, it is enough for the variables $u^k$ to transform under a gauge transformation of parameters $v^k$ as $\delta u^k = \dot{v}^k \ - \ h^k_{\ l} v^l \ - \ c^{k}_{\ lm} u^l v^m$, which will give $\delta S = - \int \ dt \ \left(v^k \dot{f}_k + \dot{v}^k f_k \right) = - \left[v^k f_k \right] = 0$ (up to ``boundary terms").

In particular, once the gauge symmetries of a first order action have been found, one can directly identify which variables are Lagrange multipliers by isolating those whose gauge variation contains a time derivative of the transformation parameters
\footnote{It is actually a completely logical result, since, the action being first order and gauge invariant, it can not contain time derivatives of the variables (which must then be Lagrange multipliers) whose gauge variation contains a time derivative, because, if it did, the variation of these terms would generate second order derivatives in time that could not cancel with anything.}.

Finally, let us still remark that each Lagrange multiplier $u^k$ removes three degrees of freedom from the system: two through the first class constraint it implements and one because it is not itself a dynamical variable (its conjugate momentum vanishes).\\

In the case which will interest us, we will need to apply this formalism to Grassmann complex variables, which does not present any significant subtlety, but which requires a careful handling of notations. Let us now fix those.

Consider an action of the form:
\begin{equation} \label{grassmann_action_first_order}
S \left[\Psi , \bar{\Psi}\right] = \int dt \left\lbrace
\theta^A\!\left( \bar{\Psi}  \right) \dot{\Psi}_A + \dot{\bar{\Psi}}_A\, \bar{\theta}^A\!\left(\Psi\right) - H\left(\Psi , \bar{\Psi}\right)
\right\rbrace ,
\end{equation}
\noindent where the $\Psi_A$ are complex Grassmann variables, their complex conjugate being denoted by a bar. $H$ is a real-valued function.

Its variation is given by:
\begin{equation}
\delta S = \int dt \left\lbrace
\delta \bar{\Psi}_A \left[ \omega^{AB} \dot{\Psi}_B \, - \, \frac{\partial^L H}{\partial \bar{\Psi}_A} \right]
\, + \, \left[ - \ \dot{\bar{\Psi}}_A \omega^{AB} \, - \, \frac{\partial^R H}{\partial \Psi_B}\right] \delta \Psi_B 
\right\rbrace ,
\end{equation}
\noindent where
\begin{equation}
\omega^{AB} =
\frac{\partial^L \theta^B}{\partial \bar{\Psi}_A}
\, - \, \frac{\partial^R \bar{\theta}^A}{\partial \Psi_B} \ .
\end{equation}

Note that we have $\omega^{AB} = - \ \bar{\omega}^{BA}$. Our symplectic 2-form can now be seen as an anti-hermitian matrix. A noteworthy consequence of this is that it no longer needs to be even-dimensional in order to be invertible (typically, it could be a one by one matrix of imaginary value). The phase space (complex) dimension can indeed be odd.

If we again define the inverse of this 2-form as $\omega^{AB} \omega_{BC} = \delta^A_C$, the equations of motion become:
\begin{eqnarray}
\dot{\Psi}_A &=&
\omega_{AB}\, \frac{\partial^L H}{\partial \bar{\Psi}_B}, \label{eom1o}
\\
- \ \dot{\bar{\Psi}}_A &=&
\frac{\partial^R H}{\partial \Psi_B}\ \omega_{BA} .
\end{eqnarray}

This suggests to define a Dirac bracket such that $\dot{F} = \left\lbrace F , H \right\rbrace_D$. It is:
\begin{equation}
\left\lbrace F , G \right\rbrace_D =
\frac{\partial^R F}{\partial \Psi_A} \,\omega_{AB}\, \frac{\partial^L G}{\partial \bar{\Psi}_B}
\ - \ \frac{\partial^R G}{\partial \Psi_A} \ \omega_{AB}\, \frac{\partial^L F}{\partial \bar{\Psi}_B} .
\end{equation}

It is antisymmetric and satisfies $\left\lbrace \Psi_A , \bar{\Psi}_B \right\rbrace_D = \omega_{AB}$, the other brackets vanishing identically.

If we add to the action (\ref{grassmann_action_first_order}) a term of the form $\ - \ \int dt \left\lbrace \bar{u}_a\, f^a\!\left[ \Psi \right] + \bar{f}^a \!\left[ \bar{\Psi} \right] u_a \right\rbrace$, with Lagrange multipliers $u_a$, it will generate gauge transformations according to:
\begin{equation}
\delta \Psi_A = \left\lbrace \Psi_A , \bar{f}^a\! \left[ \bar{\Psi} \right] u_a \right\rbrace_D
=  \omega_{AB}\, \frac{\partial^L \bar{f}^a}{\partial \bar{\Psi}_B} u_a \, .
\label{gauge1o}
\end{equation}

Let us still observe the useful fact that, if the kinetic term of the action is quadratic (as will be the case for free theories), the ``momenta" $\theta^A$ are linear in the positions $\Psi_A$ and we have:
\begin{eqnarray}
\theta_A &=&
\frac{1}{2} \ \bar{\Psi}_B \omega^{BA} \, ,
\\
\bar{\theta}_A &=& \frac{1}{2} \ \bar{\omega}^{AB} \Psi_B 
\ = \ - \ \frac{1}{2} \ \omega^{AB} \Psi_B \, .
\end{eqnarray}

In this case, the symplectic 2-form is a (matrix of) $c$-number. The full initial action can then be rewritten:
\be \label{free_grassmann_action_first_order}
S = \!\int\! dt \left\{ \frac{1}{2}\! \left( \bar{\Psi}_{A}\, \o^{AB} \dot{\Psi}_B - \dot{\bar{\Psi}}_{A}\, \o^{AB} \Psi_B \right) - \bar{u}_a\, f^a\!\left[ \Psi \right] - \bar{f}^a \!\left[ \bar{\Psi} \right] u_a  - H[\Psi,\bar{\Psi}] \right\} .
\ee

\subsection{Illustration of constrained first order formalism: spin $3/2$ gauge field on flat space}

\label{Sec:spin_3_2_ham}

Let us consider the following action, where the field $\psi_{\mu}$ is a spinor-vector:
\begin{eqnarray}
S &=& 
- \ i \ \int \ d^4 x \ 
\bar{\psi}_{\mu} \gamma^{\mu \nu \rho} \partial_{\nu} \psi_{\rho} . 
\end{eqnarray}

As we saw, this action is invariant under the gauge transformation $\delta \psi_{\mu} = \partial_{\mu} \zeta$, where $\zeta$ is an arbitrary spinor. Since only $\delta \psi_0 = \dot{\zeta}$ contains a time derivative, we are led to assume that, the action being first order, $\psi_0$ must play the role of a Lagrange multiplier, in agreement with the considerations of the previous section.

This is indeed what the explicit decomposition of this action yields:
\begin{eqnarray}
S &=&
\int d^4 x \ \left[ 
\frac{1}{2} \ \left(\psi^{\dagger}_k \omega^{kl} \dot{\psi}_l \ - \  \dot{\psi}^{\dagger}_{k} \omega^{kl} \psi_l \right)
\ - \ \psi_0^{\dagger} \mathcal{F} \ - \ \mathcal{F}^{\dagger} \psi_0 
\ - \ \mathcal{H} 
\right],
\nonumber \\ 
\end{eqnarray}

\noindent where an integration by parts was performed and the following notations are introduced for the symplectic 2-form, Hamiltonian and constraints, respectively given by:
\begin{eqnarray}
\omega^{kl} &=& - \ i \ \gamma^{kl} ,
\\
\mathcal{H} &=& i \ \bar{\psi}_{k} \gamma^{klm} \partial_{l} \psi_{m} , 
\\
\mathcal{F} &=& - \ i \ \gamma^{kl} \partial_k \psi_l . 
\end{eqnarray}

The symplectic 2-form is indeed anti-hermitian ($\left(\omega^{kl}\right)^{\dagger} = - \ \omega^{lk}$).

As far as the counting of the degrees of freedom is concerned, we see that, starting with a spinor-vector $\psi_{\mu}$ containing four spinors, we find that one of these is a Lagrange multiplier ($\psi_0$), which removes three spinors in total (with one spinorial constraint and one spinorial gauge transformation acting on the remaining three spinors $\psi_k$), leaving us with the appropriate  $4 - 3 \times 1 = 1$ spinor encoding the physical degrees of freedom, which indeed corresponds to the first order formalism of a system with two physical states (the states of helicity $\pm 3/2$), since a Dirac spinor has four complex components in four dimensions.

Let us now identify the inverse $\omega_{kl}$ of the symplectic 2-form, which is easily seen to be equal to:
\begin{eqnarray}
\omega_{kl} &=& - \ i \ \delta_{kl} \ + \ \frac{i}{2}\gamma_{k} \gamma_l
\\ &=&
- \ \frac{i}{2} \ \delta_{kl} \ + \ \frac{i}{2} \ \gamma_{kl}
\\&=& - \ \frac{i}{2} \ \gamma_l \gamma_k .
\end{eqnarray}

We conclude that the Dirac bracket must be:
\begin{eqnarray}
\left\lbrace \psi_k\left(\vec{x}\right) , \psi^{\dagger}_l \left(\vec{x}'\right)\right\rbrace_D &=& 
- \ \frac{i}{2} \ \gamma_l \gamma_k \ \delta\left( \vec{x} - \vec{x}' \right) .
\end{eqnarray}

We can now check that the constraints do generate the gauge transformations through these Dirac brackets, noting that, if we rewrite the constraint part of the action containing $\psi^{\dagger}_k$ (renaming $\psi_0$ as $\epsilon$, removing the time integral and performing some partial integrations), we get:
\begin{eqnarray}
\int \ d^3 x \  \mathcal{F}^{\dagger} \epsilon 
&=& - \ i \ \int \ d^3 x \ \psi_k^{\dagger} \gamma^{kl} \partial_l \epsilon .
\end{eqnarray}

From that expression, we reach:
\begin{eqnarray}
\delta \psi_k \left(\vec{x}\right) &=&
\left\lbrace \psi_k \left(\vec{x}\right) , \int \ d^3 x' \  \left[
\mathcal{F}^{\dagger} \left(\vec{x'}\right) \epsilon \left(\vec{x'}\right)  
\ + \ \epsilon^{\dagger} \left(\vec{x'}\right)  \mathcal{F} \left(\vec{x'}\right) \right] \right\rbrace_D
\nonumber \\ &=&
\omega_{kl} \frac{\partial^L }{\partial \psi^{\dagger l}\left(\vec{x}\right)} \int \ d^3 x' \ \mathcal{F}^{\dagger}\left(\vec{x'}\right)\epsilon \left(\vec{x'}\right)
\nonumber \\ &=& 
- \ \frac{i}{2} \ \gamma_l \gamma_k \frac{\partial^L }{\partial \psi^{\dagger l}\left(\vec{x}\right)} \int \ d^3 x' 
\left( - \ i \  \psi_n^{\dagger}\left(\vec{x'}\right) \gamma^{nm} \partial'_m \epsilon \left(\vec{x'}\right) \right)
\nonumber \\ &=& 
- \ \frac{i}{2} \ \gamma_l \gamma_k \left(- \ i \  \gamma^{lm} \partial_m \epsilon \left(\vec{x}\right)\right) 
\nonumber \\ &=& \partial_k \epsilon \left(\vec{x}\right) .
\end{eqnarray}

This is exactly the transformation under which our action is manifestly invariant (together with an appropriate variation of the Lagrange multiplier), as is immediately seen from its covariant form.

We can also check that this transformation leaves the Hamiltonian and the constraints invariant (up to total derivatives), which shows that they are indeed first class, completing our consistency check. \\

Let us observe that the constraint is easily solved: $\gamma^{kl} \partial_k \psi_l = 0$ is equivalent to $\partial_k \left( \gamma^{kl} \psi_l \right) = 0$, from which we extract $\gamma^{kl} \psi_l = \epsilon^{klm} \partial_l \chi_m$, or $\psi_n = - \frac{1}{2}\epsilon^{klm} \gamma_k \gamma_n \partial_l \chi_m$.

\section{Surface charges}

\label{Sec:Ham_surf_charges}

As we saw in section \textbf{\ref{Sec:Ham_form_constr_gauge}}, first class constraints generate gauge transformations through their Poisson brackets. The most general general generator of gauge transformations is:
\begin{equation}
G\left[u\right] \equiv u^k \phi_k \left(p,q\right) .
\end{equation}
\noindent where $\phi_k \approx 0$ are the constraints.

The gauge variation of a phase space function $f\left(p,q\right)$ is:
\begin{equation}
\delta f = \left\lbrace f , G\left[u\right] \right\rbrace ,
\end{equation}
\noindent the arbitrary variables $u^k$ being the parameters of the gauge transformation.

As we saw, a gauge transformation maps onto each other two phase space points corresponding to the \textit{same} physical state. Gauge transformations are unphysical, being simply an effect of the use of a redundant parametrization of the system. Gauge invariance is not, strictly speaking, a \textit{symmetry} of the system, since a symmetry is the invariance of the dynamics of a system under a non-trivial physical transformation, which maps onto each other physically distinguishable phase space points.

This appears, in particular, from the fact that no non-trivial conserved quantity is associated to gauge invariance: its charges are the gauge generators, which, being proportional to the constraints, vanish on-shell.\\

However, a more interesting situation arises for systems with a infinite number of degrees of freedom and of gauge invariances\footnote{For further details and illustrations, see \cite{Regge,benguria-cordero,HT}. We follow the exposition of the section 3 of \cite{benguria-cordero}.}. Indeed, in this case, certain transformations formally similar to gauge transformations can turn out to actually be true symmetries of the system, to which non-trivial, non-vanishing conserved charges are associated.

To be definite, let us consider a field theory defined over a four-dimensional space-time, for which the phase space is parametrized by coordinates $Q(\vec{x})$ and conjugate momenta $P\left(\vec{x}\right)$ which are fields over a three-dimensional manifold - the spatial section of space-time. The constraints and the gauge parameters will generally be local functions of these fields, and we can note them $\phi\left(\vec{x}\right)$ and $u\left(\vec{x}\right)$. A gauge generating functional will generically be:
\begin{equation}
G\left[ u \right] \equiv \int \ d^3 x \ u\left(\vec{x}\right) \phi\left(\vec{x}\right) . \label{Ham_sc_G}
\end{equation}
The variation of the canonical variables shall be given as before by Poisson brackets:
\begin{eqnarray}
\delta Q \left(\vec{x}\right) &=&
\left\lbrace Q \left(\vec{x}\right) , G \left[u\right] \right\rbrace
\nonumber \\ &=& 
\frac{\delta G\left[u\right\rbrace}{\delta P \left(\vec{x}\right)} ,
\\
\delta P \left(\vec{x}\right) &=&
\left\lbrace P \left(\vec{x}\right) , G \left[u\right] \right\rbrace
\nonumber \\ &=& 
- \ \frac{\delta G\left[u\right\rbrace}{\delta Q \left(\vec{x}\right)} .
\end{eqnarray}

In order for these functional derivatives appearing in the Poisson brackets to be well-defined, it will generally turn out to be necessary to add a surface term to the gauge generator introduced above, since the constraints will generally consist of derivatives of the canonical variables (see section \textbf{\ref{Sec:EM_ham}} for the electric-magnetic example, in which $\vec{\nabla} \cdot \vec{\Pi} \left(\vec{x} \right) \approx 0$ is seen to be a constraint). In order to compute the Poisson brackets of $G$, integrations by parts - generating surface terms - will have to be performed to bring it into the following form:
\begin{equation}
G = \int \ d^3 x \ \left\lbrace A\left(\vec{x}\right) Q \left(\vec{x}\right) \ + \ B \left(\vec{x}\right) P \left(\vec{x}\right) \right) 
\ + \ \int \ d^2 x \ \ldots 
\end{equation}
The surface term prevents the Poisson bracket to be taken, since it does not have a well-defined functional derivative in the bulk. So, a gauge generator of the form \eqref{Ham_sc_G} can not be used: the surface term that partial integrations will generate when computing the brackets must be removed from it. The correct generator shall be:
\begin{equation}
G\left[ u \right] \equiv \int \ d^3 x \ u\left(\vec{x}\right) \phi\left(\vec{x}\right) \ - \ \int \ d^2 x \ \ldots .
\end{equation}
Remarkably, this quantity no longer automaticall vanishes on-shell, since the vanishing of the constraints (in the bulk) does not force the surface term to be zero. We do not necessarily have $\bar{G}\left[u\right] \approx 0$.\\

Now, of course, the question arises of the function space to which the $u\left(\vec{x}\right)$ must be confined: it will naturally be identified with the dual space of the space of the constraints - which is itself determined by the boundary conditions to which the canonical variables are subjected. If the parameters $u\left(\vec{x}\right)$ are chosen in this set, $\bar{G}\left[u\right]$ will generate \textit{proper} gauge transformations - that is, physically irrelevant phase space transformations. In this case, $\bar{G}\left[u\right]$ shall still vanish on-shell.

On the other hand, if we pick our gauge parameters $u\left(\vec{x}\right)$ outside of this dual space - but still subject to the requirement that the transformation they parametrize map a physical state onto another physical state ($i.e.$ such that it preserves the boundary conditions of the canonical variables) - we will obtain a generator of \textit{improper} gauge transformations, which shall be true symmetries of the system, sending a physical state to a different one. In this case, the on-shell value of $\bar{G}\left[u\right]$ shall be non-zero and finite: it will be a conserved surface charge, being the on-shell value of the generator of a physical symmetry of the system.

\begin{subappendices}

\section{Invertibility of the second class constraints Poisson brackets matrix}

\label{App:2nd_class_constr_invert}

We are now going to show that the determinant of the second class constraints Poisson brackets ($\chi_r , \chi_s$), with $r$ and $s$ ranging from $1$ to $S$, does not cancel (even weakly). Let us note this determinant:
\begin{eqnarray}
D &\equiv&
\left\vert \begin{array}{cccc}
0 & \left\lbrace \chi_1 , \chi_2 \right\rbrace & \ldots & \left\lbrace \chi_1 , \chi_S \right\rbrace \\ 
\left\lbrace \chi_2 , \chi_1 \right\rbrace &  0 & \ldots & \left\lbrace \chi_2 , \chi_S \right\rbrace \\ 
\vdots & \vdots & \ddots & \vdots \\
\left\lbrace \chi_S , \chi_1 \right\rbrace & \left\lbrace \chi_S , \chi_2 \right\rbrace & \ldots & 0
\end{array} \right\vert .
\end{eqnarray} 

Indeed, let us assume that $D \approx 0$: the Poisson brackets matrix would then be of rank $T < S$ (on the constraint surface). Then, it would be possible to find a minor of $D$ of rank $T$ which does not cancel. Let us assume, to keep notations simple, that this minor be in the upper left block $\left( T + 1 \right) \times T$ of $D$, which allows us to build the following non trivial linear combination of the second class constraints:
\begin{eqnarray}
A &\equiv&
\left\vert \begin{array}{ccccc}
\chi_1 & 0 & \left\lbrace \chi_1 , \chi_2 \right\rbrace & \ldots & \left\lbrace \chi_1 , \chi_T \right\rbrace \\
\chi_2 & \left\lbrace \chi_2 , \chi_1 \right\rbrace & 0 & \ldots & \left\lbrace \chi_2 , \chi_T \right\rbrace \\
\vdots & \vdots & \vdots & \ddots & \vdots \\
\chi_{T+1} & \left\lbrace \chi_{T+1} , \chi_1 \right\rbrace & \left\lbrace \chi_{T+1} , \chi_2 \right\rbrace & \ldots & \left\lbrace \chi_{T+1} , \chi_T \right\rbrace 
\end{array} \right\vert ,
\end{eqnarray}
\noindent the coefficient of each $\chi_s$ being, once we develop this determinant in terms of its first column, one of the minors of rank $T$ of $D$, of which we know that they do not all cancel.

We can also show that $A$ is first class. Indeed, if we take its Poisson bracket with a constraint $\phi$, by applying the product rule, we will obtain a series of terms of two forms, depending on whether the Poisson bracket acts on the first column or one of the others. In the first case, either $\phi$ is first class, and its Poisson bracket with the $\chi_s$ vanishes (weakly), or $\phi$ is second class, and is then one of the $\chi_s$, and this term must cancel since it is a minor of rank $T + 1$ of $D$ (which is of rank $T$). In the second case, we have a determinant whose first column is made of $\chi_s$, which automatically cancels (weakly). So, we have $\left\lbrace A , \phi \right\rbrace \approx 0$, and $A$ is first class, which is impossible, since we assumed the $\chi_s$ to form a minimal set of second class constraints and $A$ is a non trivial linear combination of them.

\newpage
\section{Resolution of the momentum constraint \label{graviton_momentum_constraint_solv}}

The momentum constraint is $\partial^k \Pi_{kl} = 0$. It can be solved through Poincaré lemma, considering the index $l$ as a spectator index: 
\begin{eqnarray}
\Pi_{kl} &=& \partial^m U_{mk\vert l},
\end{eqnarray}
\noindent where the tensor $U_{mk\vert l}$ is antisymmetric in its first pair of indices. The requirement that $\Pi_{kl}$ be symmetric then imposes $0 = \partial^m U_{m[k\vert l]}$, which is solved by:
\begin{eqnarray}
U_{m[k \vert l]} = \partial^n V_{nm\vert kl},
\end{eqnarray}
\noindent where $V_{nm\vert kl} = V_{[nm] \vert [kl]}$. Now, any tensor satisfying $U_{mk\vert l} = U_{[mk]\vert l}$ will also satisfy $U_{mk\vert l} = 3 U_{[mk\vert l]} - 2 U_{m[k\vert l]}$ which leads to:
\begin{eqnarray}
\Pi_{kl} &=& \partial^m \partial^n W_{nk \vert ml} ,
\end{eqnarray}
\noindent where:
\begin{eqnarray}
W_{nk\vert ml} &=& - \ \left(V_{nk\vert ml} \ + \ V_{ml\vert nk} \right) .
\end{eqnarray}

Finally, by double dualization, we define $W_{nk\vert lm} \equiv \epsilon_{nkp}\epsilon_{lmq} P^{pq}$, where $P^{pq}$ is obviously symmetric, obtaining:
\begin{eqnarray}
\Pi_{kl} &=& \epsilon_{nkp}\epsilon_{lmq} \partial^k \partial^l P^{pq},
\end{eqnarray}
\noindent which is precisely the Einstein tensor of $P^{pq}$.

\newpage
\section{Gauge invariance of the prepotential $\phi_{kl}$ \label{graviton_prepotential_gauge_inv}}

For $h_{kl}$ to be pure gauge, we must have:
\begin{eqnarray}
0 &=& 2 \ \epsilon_{(k\vert mn} \partial^m \phi^n_{\phantom{n}\vert l)}
\ + \ 2 \ \partial_{(k} u_{l)} .
\end{eqnarray}

The trace of this expression gives $\partial^k u_k = 0$, which is solved as $u_k = - \ \epsilon_{klm} \partial^l v^m$. If we redefine $\psi_{kl} \equiv \phi_{kl} \ - \ 2 \ \partial_{(k} v_{l)}$, we have:
\begin{eqnarray}
0 &=& 2 \ \epsilon_{(k\vert mn} \partial^m \psi^n_{\phantom{n}\vert l)} . \label{graviton_gauge_inv_prepot_interm}
\end{eqnarray}

The divergence of this equation implies that $\epsilon_{kmn} \partial^m \left(\partial^l \psi^n_{\phantom{n} l} \right) = 0$, which leads to $\partial^l \psi^n_{\phantom{n} l} = \partial^n C$. Using this, we can contract equation (\ref{graviton_gauge_inv_prepot_interm}) with $\epsilon^{kij}$, yielding:
\begin{eqnarray}
0 &=& 4 \ \partial^{[i} \left( \psi^{j]}_{\phantom{j]} l} \ + \ \delta^{j]}_l B\right) ,
\end{eqnarray}
\noindent where $B \equiv 1/2 \left(\psi^k_{- \ \phantom{k}k} \ + \ C \right)$. Integrating this equation (with $l$ as a spectator index) gives:
\begin{eqnarray}
\psi_{kl} &=& \partial_k w_l \ - \ \delta_{kl} B , 
\end{eqnarray}
\noindent whose antisymmetrization shows that $w_l = \partial_l t$. So:
\begin{eqnarray}
\phi_{kl} &=& 2 \ \partial_{(k} v_{l)} \ + \ \partial_k \partial_l t \ - \ \delta_{kl} B ,
\end{eqnarray}
\noindent which, up to some redefinition of the fields (the double derivative term $\partial_k \partial_l t$ can be eliminated by a redefinition of the $v_k$), is exactly the pure gauge state of linearized conformal gravity:
\begin{eqnarray}
\phi_{kl} &=& 2 \ \partial_{(k} \eta_{l)} \ + \ \delta_{kl} \sigma .
\end{eqnarray}

\end{subappendices}

\chapter{Dualities}

\label{Chap:dualities}

\section{Electric-magnetic duality}

\subsection{Spin $1$}

\label{Sec:EMduality_1}

Maxwell's equations in the vacuum are well-known for their invariance under rotations in the plane of the electric and magnetic fields, which is called \textit{electric-magnetic duality invariance}. Indeed, in a non (manifestly) covariant form, these equations are:
\begin{eqnarray}
\vec{\nabla} \times \vec{E} &=& - \ \frac{\partial \vec{B}}{\partial t} , \label{Faraday}
\\
\vec{\nabla} \times \vec{B} &=&  \frac{\partial \vec{E}}{\partial t} , \label{Ampere}
\\
\vec{\nabla} \cdot \vec{E} &=& 0 , \label{Gauss}
\\
\vec{\nabla} \cdot \vec{B} &=& 0. \label{Gauss_B}
\end{eqnarray}

On the other hand, it is not obvious that this symmetry of the equations of motion can be carried to the level of the action and turned into an off-shell symmetry, with an associated conserved Noether charge. The relevant variables in terms of which to write down a second order action principle for electromagnetism are the potentials obtained by solving equations (\ref{Faraday}) and (\ref{Gauss_B}), which are not modified by the presence of charges and currents. Their general solution is:
\begin{eqnarray}
\vec{E} &=& - \ \vec{\nabla} V \ - \ \frac{\partial \vec{A}}{\partial t} ,
\\
\vec{B} &=& \vec{\nabla} \times \vec{A} .
\end{eqnarray}

The appropriate Lagrangian action whose extremization yields the remaining two equations is:
\begin{eqnarray}
S_L\left[ \vec{A} , V \right] &=& \frac{1}{2} \ \int \ d^4 x \ \left\lbrace \Vert \vec{E} \Vert^2 \ - \ \ \Vert \vec{B} \Vert^2 \right\rbrace . \label{non_cov_lag_action_EM}
\end{eqnarray}

The first thing that strikes us is that this action is invariant under hyperbolic rotations in the electric-magnetic plane, not under Euclidean rotations. However, the variables on which this action depends are the potential scalar and vector, in terms of which we do not yet know how to express an electric-magnetic duality rotation. This expression could be built, but it would be non-local. 

The most natural way to show that electric-magnetic duality is also a symmetry at the level of the action is to go to Hamiltonian formalism, in which duality rotations take a simple and local form. The price of this transition is to lose manifest Lorentz invariance, since the action (\ref{non_cov_lag_action_EM}) can be rewritten in a manifestly covariant form:
\begin{eqnarray}
S_L\left[ A^{\mu} \right] &=& \ - \ \frac{1}{4} \ \int \ d^4 x \ F^{\mu\nu} F_{\mu\nu} , \label{cov_lag_action_EM}
\end{eqnarray}
\noindent where $F_{\mu\nu} \equiv \partial_{\mu} A_{\nu} \ - \ \partial_{\nu} A_{\mu}$ and the prepotential is $A^{\mu} \equiv \left( V , \vec{A} \right)$.

Indeed, this manifestly covariant form is simply a rewriting of the older form from which we started, which treats spatial and time-like directions on a different footing. In terms of $F_{\mu\nu}$, Maxwell's equations take the following simple form:
\begin{eqnarray}
0 &=& \partial^{\mu} F_{\mu\nu} ,
\\
0 &=& \partial^{\mu} \tilde{F}_{\mu\nu} ,
\end{eqnarray}
\noindent where we have introduced the Hodge dual of the strenght field, $\tilde{F}_{\mu\nu} \ \equiv \ \frac{1}{2} \ \epsilon_{\mu\nu\rho\sigma} F^{\rho\sigma}$. The first equation is just a rewriting of Ampère's law and of Gauss law for the electric field - that is, (\ref{Ampere}) and (\ref{Gauss}) ; the second is a rewriting of Gauss law for the magnetic field and Faraday's law, or (\ref{Faraday}) and (\ref{Gauss_B}). The second equation is equivalent to $0 = \partial_{[\mu} F_{\nu\rho ]}$, which is also equivalent to $F_{\mu\nu} = 2 \ \partial_{[\mu} A_{\nu ]}$.

In any case, we already saw that Dirac procedure led to a rewriting of this action in the first order form in terms of two spatial vector potentials (the second one being obtained through the solving of Gauss law, equation (\ref{Gauss}), which appears as a secondary first class constraint on the conjugate momentum of the initial potential vector):
\begin{eqnarray}
S_H \left[\vec{A}^a\right] &=& \frac{1}{2} \ \int \ d^4 x \ \left\lbrace \epsilon_{ab} \dot{\vec{A}}^a  \cdot \vec{B}^b
\ - \  \delta_{ab} \vec{B}^a \cdot \vec{B}^b
\right\rbrace ,
\end{eqnarray}
\noindent where the index $a = 1, 2$, $\vec{B}^a \equiv \vec{\nabla} \times \vec{A}^a$ and $\epsilon_{ab}$ is the antisymmetric tensor defined with $\epsilon_{12} = 1$; the dot denotes a time derivative.

This first order action has a manifest gauge invariance under transformations of the form:
\begin{eqnarray}
\delta \vec{A}^a &=& \vec{\nabla} \lambda^a,
\end{eqnarray}
\noindent and an invariance under the rigid transformation:
\begin{eqnarray}
\delta \vec{A}^a &=& \xi \ \epsilon^{a}_{\phantom{a}b} \vec{A}^b ,
\end{eqnarray}
\noindent which is precisely electric-magnetic duality, since the $\vec{B}^a$ are precisely the electric and magnetic fields.

Bringing the lagrangian second order action of the massless free spin $1$ into a first order Hamiltonian form, in which the constraints have been explicitly solved, makes the off-shell duality invariance of the system manifest. This will actually be a general feature, and generalizing this result to higher spin gauge fields has been one of our results.

\subsection{Spin $2$}

\label{Sec:EMduality_2}

A completely similar investigation can be performed for linearized gravity - $i.e.$ for the free massless spin $2$ field over flat space-time - in four dimensions. Here, we can directly consider the covariant formalism. The field - the ``potential" - is a symmetric tensor $h_{\mu\nu}$, and its equations of motion are the vanishing of its Ricci tensor $R_{\mu\nu}$, which is the trace of its Riemann tensor $R_{\mu\nu\rho\sigma} \equiv 4 \ \partial_{[\mu \vert} \partial_{[\rho} h_{\sigma ] \vert \nu ]}$: $R_{\mu\nu} \equiv R^{\rho}_{\phantom{\rho} \mu \rho\nu}$.

As in the spin $1$ case, where the expression $F_{\mu\nu} = 2 \ \partial_{[\mu} A_{\nu ]}$ (for some vector $A_{\mu}$) was equivalent to $\partial_{[\mu} F_{\nu\rho ]} = 0$ (provided that $F_{\mu\nu}$ be antisymmetric), we see that $R_{\mu\nu\rho\sigma} = 4 \ \partial_{[\mu \vert} \partial_{[\rho} h_{\sigma ] \vert \nu ]}$ is equivalent to $\partial_{[\mu} R_{\nu\rho ] \sigma \lambda} = 0$, provided that $R_{\mu\nu\rho\sigma}$ have the right algebraic symmetries (antisymmetric in the first and second pairs of indices; gives zero when antisymmetrized over three indices). So, if $R_{\mu\nu\rho\sigma}$ is a tensor whose only required property is to be antisymmetric in the first and second pair of its indices, the equations of motion the spin $2$ are equivalent to:
\begin{eqnarray}
0 &=& R_{\mu [\nu\rho\sigma ]} , \label{spin_2_Bianchi_alg_R}
\\
0 &=& R_{\mu\nu [\rho\sigma , \lambda ]}, \label{spin_2_Bianchi_dif_R}
\\
0 &=& R_{\mu\nu} , \label{spin_2_trace_0_R}
\end{eqnarray}
\noindent the first two equations implying the existence of a ``potential" $h_{\mu\nu}$ whose equations of motion are given by the third equation.

Now, if we define the dual of $R_{\mu\nu\rho\sigma}$ as $\tilde{R}_{\mu\nu\rho\sigma} \equiv \frac{1}{2} \epsilon_{\mu\nu\lambda\tau} R^{\lambda\tau}_{\phantom{\lambda\tau}\rho\sigma}$, these equations can be seen to be equivalent to those:
\begin{eqnarray}
0 &=& \tilde{R}_{\mu [\nu\rho\sigma ]} , \label{spin_2_Bianchi_alg_S}
\\
0 &=& \tilde{R}_{\mu\nu [\rho\sigma , \lambda ]}, \label{spin_2_Bianchi_dif_S}
\\
0 &=& \tilde{R}_{\mu\nu} . \label{spin_2_trace_0_S}
\end{eqnarray}

Indeed, (\ref{spin_2_Bianchi_alg_S}) is equivalent to (\ref{spin_2_trace_0_R}), (\ref{spin_2_trace_0_S}) to (\ref{spin_2_Bianchi_alg_R}) and (\ref{spin_2_Bianchi_dif_S}) to (\ref{spin_2_Bianchi_dif_R}). Let us note that, together, equations (\ref{spin_2_Bianchi_alg_S}) and (\ref{spin_2_Bianchi_dif_S}) imply the existence of a dual ``potential" (or metric variation), which would be a second symmetric tensor $f_{\mu\nu}$ satisfying $\tilde{R}_{\mu\nu\rho\sigma} = 4 \ \partial_{[\mu \vert} \partial_{[\rho} f_{\sigma ] \vert \nu ]}$.

So, we again have equations of motion which are invariant under rotations in the plane of the $R_{\mu\nu\rho\sigma}$ and $\tilde{R}_{\mu\nu\rho\sigma}$ fields (or, equivalently, in the plane of the $h_{\mu\nu}$ and $f_{\mu\nu}$ fields): this is the \textit{electric-magnetic duality invariance of linearized gravity}.

This duality invariance is not a manifest symmetry of the spin $2$ Lagrangian action:
\begin{eqnarray}
S_L \left[h_{\mu\nu}\right] &=&
- \ \frac{1}{2} \ \int \ d^4 x \ \left\lbrace
\partial_{\rho} h_{\mu\nu} \partial^{\rho} h^{\mu\nu}
\ - \ 2 \ \partial^{\rho} h_{\rho\mu}\partial_{\nu} h^{\nu\mu}
\right. \nonumber \\ && \qquad \qquad \qquad \left.
\ + \ 2 \ \partial^{\mu} h_{\mu\nu} \partial^{\nu} h'
\ - \ \partial_{\mu} h' \partial^{\mu} h'
\right\rbrace , \qquad  \qquad
\end{eqnarray}
\noindent since it is actually not clear how this transformation, which was defined on the curvatures, should act on the field $h_{\mu\nu}$.

However, as we saw, we can perform the Hamiltonian analysis of this action, obtaining its momenta and constraints. The explicit solving of these leads to the following first order action: 
\begin{eqnarray}
S_H \left[Z^a_{kl}\right] &=&
\frac{1}{2} \ \int \ d^4 x \ \left\lbrace
\epsilon_{ab} \dot{Z}^{a kl} B^b_{kl}
\ - \  \delta_{ab} \left[G^{a kl} G^b_{kl} \ - \ \frac{1}{2} \ G^a G^b \right]
\right\rbrace ,
\end{eqnarray}
\noindent where $G^a_{kl} \equiv G_{kl} \left[Z^a_{ij}\right]$ is the Einstein tensor and $B^a_{kl} \equiv B_{kl} \left[Z^a_{ij}\right]$ the Cotton tensor of one of the two prepotentials $Z^a_{kl}$, which are symmetric spatial tensors. This action is invariant (up to surface terms) under the gauge transformations:
\begin{eqnarray}
\delta Z^a_{kl} &=&
2 \ \partial_{(k} \lambda^a_{l)} \ + \ \delta_{kl} \ \mu^a .
\end{eqnarray}

It also has the looked for invariance under (Euclidean) rotations in the plane of the two prepotentials:
\begin{eqnarray}
\delta Z^a_{kl} &=&
\xi \ \epsilon^a_{\phantom{a}b} Z^b_{kl} .
\end{eqnarray}

One can check that, on-shell, this rotation indeed reproduces a rotation in the plane of the Riemann tensor (of the initial field $h_{\mu\nu}$, which can itself be expressed in terms of the prepotentials) and its dual.

\section{Twisted self-duality conditions}

\subsection{Spin $1$}

\label{Sec:twistselfdual_1}

The equations of motion of the free massless spin $1$ field $A_{\mu}$ in four dimensions are $\partial^{\mu} F_{\mu\nu}\left[A\right] = 0$, where $F_{\mu\nu}\left[A\right] \equiv \partial_{\mu} A_{\nu} \ - \ \partial_{\nu} A_{\mu}$. In terms of $\tilde{F}_{\mu\nu}\left[ A \right]$, the Hodge dual of $F_{\mu\nu}\left[ A \right]$, this equation is equivalent to $\partial_{[\mu} \tilde{F}_{\nu\rho ]} = 0$, which is itself equivalent to $\tilde{F}_{\mu\nu} \left[ A \right] = \partial_{\mu} A'_{\nu} \ - \ \partial_{\nu} A'_{\mu}$ for some vector field $A'_{\mu}$. So, the equations of motion of the field $A_{\mu}$ are equivalent to the requirement that the Hodge dual of its strength field be itself the strength field of some other (``dual") vector field:
\begin{eqnarray}
0 &=& \partial^{\mu} F_{\mu\nu} \left[A\right] 
\Leftrightarrow
\tilde{F}_{\mu\nu} \left[A\right] = F_{\mu\nu} \left[A'\right] .
\end{eqnarray}

If we relabel our two potential vectors $A_{\mu}^a$, with $a = 1, \ 2$, with $A^1_{\mu} \equiv A_{\mu}$ and $A^2_{\mu} \equiv  A'_{\mu}$, this equation is easily seen to be identical to (with the convention $F_{\mu\nu}^a \equiv F_{\mu\nu}\left[A^a\right]$):
\begin{eqnarray}
\tilde{F}^a_{\mu\nu} \equiv \epsilon^a_{\phantom{a}b} F^b_{\mu\nu} ,
\end{eqnarray}
\noindent ($\epsilon^{ab}$ is an antisymmetric tensor satisfying $\epsilon^{12} = 1$ and its indices are lowered with a two-dimensional Euclidean orthogonal metric $\delta_{ab}$: $\delta_{11} = \delta_{22} = 1$).

This is of course a set of two equations, but they are actually identical to each other, given the fact that taking twice the Hodge dual of an antisymmetric tensor changes its sign ($\tilde{\tilde{F}}_{\mu\nu} = - \ F_{\mu\nu}$, which comes from $\epsilon_{\mu\nu\rho\sigma}\epsilon^{\rho\sigma\lambda\tau} = - \ 4 \ \delta^{\lambda\tau}_{\mu\nu}$). The presence of the antisymmetric matrix $\epsilon_{ab}$ (which can not be diagonalized using real numbers) in this equation is the reason why we talk of \textit{twisted} self-duality conditions.

Finally, let us observe that these equations have a manifest gauge invariance under independent variations of the two potential vector fields in the form of a gradient:
\begin{eqnarray}
\delta A^a_{\mu} &=& \partial_{\mu} \lambda^a .
\end{eqnarray}

In any case, we see that the equations of motion of the spin $1$ field can be rewritten as such \textit{twisted self-duality conditions} (applied on the strength fields - or curvatures - of the potential field and its dual). Let us note that, under this new form, the equations of motion are of first order.\\

We would like to write down an action principle for the spin $1$ from which the extracted equations of motion would immediately be of the form obtained above. They are written in a covariant formalism, but we can also break explicit Lorentz invariance and express them in terms of the spatial electric and magnetic vector fields. Indeed, these are just $E^i \left[A\right] \equiv F^{0i}\left[ A \right]$ and $B^i \left[A\right] \equiv \frac{1}{2} \epsilon^{ijk} F_{jk} \left[A\right]$, and taking the Hodge dual of the strength field of a potential vector is easily seen to be the same as switching (up to a sign) its electric and magnetic fields. In particular, our twisted self-duality conditions become (with $\vec{E}^a \equiv \vec{E} \left[A^a\right] = - \ \dot{\vec{A}}^a \ - \ \vec{\nabla} A^{a0}$, etc.):
\begin{eqnarray}
\vec{E}^a &=& \epsilon^a_{\phantom{a}b} \vec{B}^b .
\end{eqnarray}

We could be tempted to derive these equations from a Lagrangian density of the form $\mathcal{L} = \vec{A}_a \cdot \left( \vec{E}^a -\epsilon^a_{\phantom{a}b} \vec{B}^b\right)$. However, such an action is not gauge invariant, since the electric fields $\vec{E}^a$ are not identically divergenceless off-shell, as the magnetic field is\footnote{The magnetic field is divergenceless off-shell since it is the curl of the spatial part of $A^a_{\mu}$. The electric field is only divergenceless on-shell, since the equations of motion equate it (up to a sign) to the magnetic field of the dual potential.}. This difficulty could be overcome by taking the curl of our equations of motion, which would make them identically divergenceless, but we have to check that this would not entail a loss of information. In order to see it, let us note that, if we separate the parts of the electric field depending on $\vec{A}_a$ and $V_a \equiv A^0_a$, the equations of motion become:
\begin{eqnarray}
\dot{\vec{A}}^a \ + \ \epsilon^a_{\phantom{a}b} \vec{B}^b &=& - \ \vec{\nabla} V^a .
\end{eqnarray}

The curl of these equations is:
\begin{eqnarray}
0 &=& \dot{\vec{B}}^a \ + \ \epsilon^a_{\phantom{a}b} \vec{\nabla} \times \vec{B}^b .
\end{eqnarray}

We can remove the curl from these equations and go back to the previous ones simply by introducing $V^a$ as the two scalar functions whose gradients fill the information loss due to the curl\footnote{Under this form, this equation is actually Ampère's law, (\ref{Ampere}), once Gauss law (\ref{Gauss}) has been solved by introducing a dual potential vector $\vec{A}^2$ whose magnetic field ($i.e.$ its curl) is equal to the electric field of the initial potential.}. 

Under this new form, the equations of motion are identically divergenceless, ensuring that their contraction with $\vec{A}_a$ will be gauge invariant (up to a total derivative), and we can derive them from the action:
\begin{eqnarray}
S\left[\vec{A}^a\right] &=& 
\frac{1}{2} \ \int \ d^4 x \ \epsilon_{ab} \vec{A}^a \cdot \left(\dot{\vec{B}}^b \ + \ \epsilon^b_{\phantom{b}c} \vec{\nabla} \times \vec{B}^c \right) .
\end{eqnarray}

The one-half factor was introduced for normalization and we made an antisymmetric contraction of the equations of motion with the fields (with the $\epsilon_{ab}$ tensor) in order for the action to be non-zero (up to a surface term).

Performing some contractions and partial integrations (and redefining $A^2_{\mu}$ to switch its sign), we see that this action, obtained through a careful examination of the equations of motion, is exactly the action obtained earlier through the \textit{Hamiltonian analysis} of the spin $1$!

\subsection{Spin $2$}

\label{Sec:twistselfdual_2}

We saw that the equations of motion of the massless free spin $2$ field $h_{\mu\nu}$ over a flat four-dimensional space are the vanishing of the trace of its Riemann tensor $R_{\mu\nu\rho\sigma}\left[h\right]$. As we know, this is equivalent to the requirement that the Hodge dual $\tilde{R}_{\mu\nu\rho\sigma}\left[h\right]$ of its Riemann tensor satisfy the algebraic Bianchi identity, $0 = \tilde{R}_{\mu [\nu\rho\sigma ]}\left[h\right]$. However, this is also equivalent to the requirement that $\tilde{R}_{\mu\nu\rho\sigma}\left[h\right]$ be itself the Riemann tensor of some other, \textit{dual} field $h'_{\mu\nu}$ (since, by its definition, $\tilde{R}_{\mu\nu\rho\sigma}\left[h\right]$ is already antisymmetric in its first and second pairs of indices and satisfies the differential Bianchi identity over its second pair of indices). So, we have the two equivalent forms of the equations of motion:
\begin{eqnarray}
0 &=& R^{\rho}_{\phantom{\rho}\mu\rho\nu} \left[h\right] 
\Leftrightarrow
\tilde{R}_{\mu\nu\rho\sigma} \left[h\right] = R_{\mu\nu\rho\sigma} \left[h'\right] .
\end{eqnarray}

If we relabel our two potential fields $h_{\mu\nu}^a$, with $a = 1, \ 2$, with $h^1_{\mu\nu} \equiv h_{\mu\nu}$ and $h^2_{\mu\nu} \equiv  h'_{\mu\nu}$, this equation is easily seen to be identical to (with the convention $R_{\mu\nu\rho\sigma}^a \equiv R_{\mu\nu\rho\sigma}\left[h^a\right]$):
\begin{eqnarray}
\tilde{R}^a_{\mu\nu\rho\sigma} \equiv \epsilon^a_{\phantom{a}b} R^b_{\mu\nu\rho\sigma} , \label{twist_self_dual_spin2_cov}
\end{eqnarray}
\noindent ($\epsilon^{ab}$ is an antisymmetric tensor satisfying $\epsilon^{12} = 1$ and its indices are lowered with a two-dimensional euclidean orthogonal metric $\delta_{ab}$: $\delta_{11} = \delta_{22} = 1$).

This is of course a set of two equations, but they are actually identical to each other, given the fact that taking twice the Hodge dual of an antisymmetric tensor changes its sign ($\tilde{\tilde{R}}_{\mu\nu\rho\sigma} = - \ R_{\mu\nu\rho\sigma}$, which comes from $\epsilon_{\mu\nu\rho\sigma}\epsilon^{\rho\sigma\lambda\tau} = - \ 4 \ \delta^{\lambda\tau}_{\mu\nu}$). The presence of the antisymmetric matrix $\epsilon_{ab}$ (which can not be diagonalized using real numbers) in this equation is the reason why we talk of \textit{twisted} self-duality conditions.

Finally, let us observe that these equations have a manifest gauge invariance under independent variations of the two potential vector fields in the form of a symmetrized gradient (or linearized diffeomorphism):
\begin{eqnarray}
\delta h^a_{\mu\nu} &=& 2 \ \partial_{(\mu} \xi^a_{\nu )} .
\end{eqnarray}

In any case, we see that the equations of motion of the spin $2$ field can be rewritten as such \textit{twisted self-duality conditions} (applied on the strength fields - or curvatures - of the potential field and its dual). Let us note that, under this new form, the equations of motion are of first order.\\

In order to explore the structure of this new form of the equations of motion - and to build a satisfactory action principle from which to derive them, we shall now lose explicit Lorentz invariance and introduce the two generalized ``electric" and ``magnetic fields":
\begin{eqnarray}
\mathcal{E}_{ij}\left[h\right] &\equiv& \frac{1}{4} \ \epsilon_{ikl} \epsilon_{jmn} R^{klmn}\left[h\right]
\nonumber \\ &=&  \epsilon_{ikl} \epsilon_{jmn} \partial^k \partial^m h^{ln} 
\nonumber \\ &=& \delta_{ij}\left(\Delta h \ - \ \partial^k \partial^l h_{kl}\right)
\ + \ 2 \ \partial_{(i} \partial^k h_{j)k}
\ - \ \partial_i \partial_j h
\ - \ \Delta h_{ij} ,
\\
\mathcal{B}_{ij}\left[h\right] &\equiv& \frac{1}{2} \ \epsilon_{jkl} R_{0i}^{\phantom{0i}kl}\left[h\right] 
\nonumber \\ &=&  \epsilon_{jkl} \partial^k \dot{h}^l_{\phantom{l}i}
\ - \ \epsilon_{jkl} \partial_i \partial^k h^l_{\phantom{l}0} ,
\end{eqnarray}
\noindent where $h\equiv h^k_{\phantom{k}k}$.

Such as they are defined - that is, off-shell, these fields have the following properties: the electric tensor is symmetric and divergenceless; the magnetic field is traceless and transverse on its second index. As we see, the electric field is actually the Einstein tensor of the spatial components of the potential field $h_{\mu\nu}$:
\begin{eqnarray}
\mathcal{E}_{ij} \left[h_{\mu\nu}\right] &=& G_{ij} \left[h_{kl}\right] .
\end{eqnarray}

The twisted self-duality conditions (\ref{twist_self_dual_spin2_cov}) separate into equations with $\mu\nu\rho\sigma$ equal to $klmn$ or $kl0m$ (the equation where they are equal to $0k0l$ is the same as the one where they are equal to $kl0m$...). In the first case, we can see that they are equivalent to:
\begin{eqnarray}
\mathcal{E}^a_{ij} &=& \epsilon^a_{\phantom{a}b} \mathcal{B}^b_{ij} . \label{twist_self_dual_spin2_E_B}
\end{eqnarray}

The other independent component of the equations (\ref{twist_self_dual_spin2_cov}) can be rewritten as:
\begin{eqnarray}
\mathcal{B}^a_{ij} &=& \epsilon^a_{\phantom{a}b} R^b_{0i0j}
\nonumber \\ &=& 
\epsilon^a_{\phantom{a}b} 
\left(\ddot{h}^b_{ij} \ - \ 2 \ \partial_{(i} \dot{h}^b_{j)0} \ + \ \partial_i \partial_j h^b_{00}\right) .  
\label{twist_self_dual_spin2_00}
\end{eqnarray}

So, the covariant twisted self-duality conditions (\ref{twist_self_dual_spin2_cov}) are equivalent to the two sets of conditions (\ref{twist_self_dual_spin2_E_B}) and (\ref{twist_self_dual_spin2_00}). Actually, as we are going to see, (\ref{twist_self_dual_spin2_00}) can be obtained from (\ref{twist_self_dual_spin2_E_B}).

In order to reach this result, let us  note (\ref{twist_self_dual_spin2_00}) as $0 = \mathcal{K}^a_{ij}$, where: $$\mathcal{K}^a_{ij} = \mathcal{B}^a_{ij} \ - \ \epsilon^a_{\phantom{a}b} \left(\ddot{h}^b_{ij} \ - \ 2 \ \partial_{(i} \dot{h}^b_{j)0} \ + \ \partial_i \partial_j h^b_{00}\right)$$.

Let us first observe that taking the time derivative of (\ref{twist_self_dual_spin2_E_B}) gives us precisely the curl of (\ref{twist_self_dual_spin2_00}) over its first index:
\begin{eqnarray}
0 &=& \epsilon^{kli} \partial_l \mathcal{K}_{ij}^a .
\end{eqnarray}

Next, let us notice that (\ref{twist_self_dual_spin2_E_B}) implies that the magnetic fields $\mathcal{B}^a_{ij}$ are symmetric, since the electric fields to which they are equated are identically symmetric. 

So, (\ref{twist_self_dual_spin2_E_B}) implies the symmetry of $\mathcal{K}^a_{ij}$ in its $ij$ indices, and it also implies the vanishing of its curl. From this, we can conclude that $\mathcal{K}^a_{ij}$ vanishes up to the double derivative of a (pair of) scalar(s):
\begin{eqnarray}
\mathcal{K}_{ij}^a &=& \partial_i \partial_j \rho^a .
\end{eqnarray}

But this is precisely the form of the contribution of $h^a_{00}$ to $\mathcal{K}_{ij}^a$. Since $h^a_{00}$ only appears in the right-hand side of (\ref{twist_self_dual_spin2_00}) and is absent from the electric and magnetic fields, and since we know it to be pure gauge, we can always redefine it to absorb $\rho^a$.

In summary, the equations (\ref{twist_self_dual_spin2_E_B}) do not contain $h^a_{00}$, but if it is chosen appropriately, (\ref{twist_self_dual_spin2_E_B}) implies (\ref{twist_self_dual_spin2_00}). So, we can say that (\ref{twist_self_dual_spin2_E_B}) captures the full twisted self-duality conditions.\\

The equations (\ref{twist_self_dual_spin2_E_B}) are not all dynamical: some of them are constraints. Indeed, by taking the trace of (\ref{twist_self_dual_spin2_E_B}), we see that, since the magnetic field is identically traceless, they imply the tracelessness of the electric field. However, the electric field is the Einstein tensor of $h^a_{ij}$, so that this constraint is exactly the Hamiltonian constraint we encountered in making the Hamiltonian analysis of the Pauli-Fierz spin two action (see section \textbf{\ref{Sec:Grav_ham}}). We already saw that its solution is given by:
\begin{eqnarray}
h^a_{kl} &=& \frac{1}{2} \ \epsilon_{(k\vert mn} \partial^m Z^{a \ n}_{\phantom{a \ n} \vert l)} \ + \ 2 \ \partial_{(k} \xi^a_{\phantom{a} l)} , \label{tsdc_pp}
\end{eqnarray}
in terms of two prepotentials, $Z^a_{ij}$, which are symmetric spatial tensors. Their gauge invariance is:
\begin{eqnarray}
\delta Z^a_{kl} &=& 2 \ \partial_{(k} \eta^a_{\phantom{a}l)} \ + \ \delta_{kl} \sigma^a .
\end{eqnarray}

Finally we can find an action principle from which to extract (\ref{twist_self_dual_spin2_E_B}). The Lagrangian density $h_a^{ij} \left(\mathcal{E}_{ij}^{a} \ - \ \epsilon^a_{\phantom{a}b} \mathcal{B}^b_{ij}\right)$ is not satisfactory because it is not gauge invariant, since the magnetic field is not divergenceless on both its indices off-shell. So, the equations of motion that can be extracted from an action principle are not (\ref{twist_self_dual_spin2_E_B}) but their curl on their first index (which is identically divergenceless on all its indices). Together with the requirement that $\mathcal{B}_{ij}$ be symmetric, the curl of (\ref{twist_self_dual_spin2_E_B}) implies (\ref{twist_self_dual_spin2_E_B}), up to terms that are of the form of the contribution of $h^a_{0i}$ to (\ref{twist_self_dual_spin2_E_B}), and which can be absorbed into them. Once the curl of (\ref{twist_self_dual_spin2_E_B}) has been taken, only the spatial components $h_{ij}$ appear. So, the action will be:
\begin{eqnarray}
S &=& \frac{1}{2} \ \int \ d^4x \
 h_{aij} \epsilon^{ikl} \partial_{k} \left(\mathcal{E}_l^{aj} \ - \ \epsilon^a_{\phantom{a}b} \mathcal{B}_l^{bj}\right) .
\end{eqnarray}

Written in terms of the prepotentials through equation \eqref{tsdc_pp}, this exactly the Hamiltonian action of the spin computed earlier, \eqref{action_2_ham_prepot}.

\part{Bosonic higher spin on flat space}

\chapter{Conformal curvature in $D = 3$}

\label{Chap:confgeom}

As we saw, the spin $3/2$ and $2$ unconstrained Hamiltonian formulation in four dimensions involves fields - the prepotentials - whose gauge invariance contains both a linearized diffeomorphism and a linearized Weyl rescaling. For the spin $2$, for example, we have two prepotentials $Z^a_{ij}$ whose gauge transformation is:
\begin{eqnarray}
\delta Z^a_{ij} &=&
2 \ \partial_{(i} \lambda^a_{j)} \ + \ \delta_{ij} \mu^a .
\end{eqnarray}

The spin $3/2$ prepotential $\chi_i$ is a spatial spinor-vector whose gauge invariance is:
\begin{eqnarray}
\delta \chi_i &=& \partial_i \theta \ + \ \gamma_i \rho .
\end{eqnarray}

These features will actually turn out to be very general, characterizing the first order formulation of all free higher spin in any dimension. In order to be able to study the first order formalism of these theories in a straightforward way, it will then be of interest to have a systematic knowledge of this Weyl invariance.

We will consider conformal invariance in three dimension\footnote{The peculiarity of dimension three for conformal invariance is well known - see $e.g.$ \cite{Eisenhart}.}, which is of interest for higher spins propagating over a four-dimensional space-time (whose Hamiltonian variables - the prepotentials - are tensor fields living in a three-dimensional Euclidean space). 

We are going to build a complete set of gauge invariant curvature associated to (linearized) Weyl rescaling and diffeomorphism for an arbitrary spin: the \textit{Cotton tensor}\footnote{The Cotton tensor was introduced for general spins in \cite{Damour:1987vm,Pope:1989vj} and \cite{Henneaux:2015cda} (where it was denoted $B$), and used extensively in three-dimensional higher spin models in \cite{Bergshoeff:2009tb,Bergshoeff:2011pm,Nilsson:2013tva,Nilsson:2015pua,Linander:2016brv,Kuzenko:2016bnv,Kuzenko:2016qdw}.  }. This tensor will be fully symmetric, traceless and divergenceless. Its vanishing will be a necessary and sufficient condition for the field from which it was built to be pure gauge. Any gauge invariant local function of the field will be a function of its Cotton tensor. Finally, any symmetric, divergenceless and traceless tensor field will be the Cotton tensor of some existing field.

The proof of all these properties will require using a generalized Poincaré lemma for a generalized differential, and we give the necessary mathematical results in Appendix \textbf{\ref{App:generalized_differential}}.

\section{Spin-$s$ diffeomorphism invariance}

Let us begin with generalized (linearized) diffeomorphism gauge invariance\footnote{The results expounded in this section were established in \cite{dWF}.}, which, for a spin $s$ field $h_{i_1\cdots i_s}$, takes the form of the symmetrized derivative of a symmetric tensor $\xi_{i_2 \cdots i_s}$:
\begin{eqnarray}
\delta h_{i_1 \cdots i_s} &=& s \ \partial_{(i_1} \xi_{i_2 \cdots i_s)} 
. \label{diffeomorphism_s_latin}
\end{eqnarray}

From now on, we will refer to such a gauge transformation as a spin $s$ diffeomorphism. Please note that in this section the gauge parameters are not constrained by an tracelessness requirement.

We are going to build a complete set of curvatures invariant under these gauge transformations. We will fist define these in arbitrary dimension, where they are the components of the Riemann tensor, and then consider the three dimensional case, where they reduce to the Einstein tensor.

\subsection{Riemann tensor}

\label{Sec:Riemann}

In terms of the differential complex defined in Appendix \textbf{\ref{App:generalized_differential}}, the field and the gauge parameter belong to the complex - noted $\Omega_s$ - of tensor fields whose symmetry type is given by Young diagrams with at most $s$ columns (and whose rows are all of length $s$, with the possible exception of the last): in the language of this Appendix, the field $h$ is of type $Y^s_s$ and the gauge parameter $\xi$ of type $Y^{s-1}_s$\footnote{The Young diagram $Y^p_s$ has $k$ rows of length $s$ and one (the last) of length $l$, where $p = sk + l$, $k$ and $l$ being integers, with $l$ strictly smaller than $s$.}. Diagrammatically, we have: 
\begin{eqnarray}
\xi &\backsim& \overbrace{\yng(9)}^{\text{$s-1$ boxes}} \equiv Y^{s-1}_s ,
\\
h &\backsim& \overbrace{\yng(10)}^{\text{$s$ boxes}} \equiv Y^{s}_s .
\end{eqnarray}

We saw that a nilpotent differential $d_{(s)}$: $Y^p_s \rightarrow Y^{p+1}_s$ could be defined over such a family of fields, satisfying:
\begin{eqnarray}
0 &=& d_{(s)}^{s+1} .
\end{eqnarray}

In terms of this generalized exterior derivative, the gauge transformation of the field can be written in an index-free form as:
\begin{eqnarray}
\delta h &=& s \ d_{(s)} \xi .
\end{eqnarray}

Now, the cohomology of $d_{(s)}$ is empty when $d_{(s)}$ acts on well-filled tensors\footnote{More precisely, there is no closed but not exact well-filled tensor - that is, a well-filled tensor annihilated by $d_{(s)}^k$, for some positive power $k$ strictly smaller than $s+1$, which would be equal to $d_{(s)}^{s+1-k}$ acting on some tensor of appropriate symmetry.}, whose Young diagram is rectangular (all its rows being of length $s$). The field $h$ is precisely such a tensor\footnote{The complex $\Omega_N$ with $N=s$ was, of course, chosen so that $h$ be well-filled\ldots}. So, we see that a necessary and sufficient condition for the spin-$s$ field to be pure gauge is for its $s$th (generalized) exterior derivative to vanish (and, of course, the $s$th derivative of $h$ is gauge invariant). This leads us to define the \textit{Riemann tensor} of the field as:
\begin{eqnarray}
R &\equiv& 2^s \ d^s_{(s)} h . \label{Riemann_s_no_index}
\end{eqnarray}

In components, this gives:
\begin{eqnarray}
R_{i_1 j_1 i_2 j_2 \cdots i_s j_s} \equiv 
2^s \ 
\partial_{[i_1\vert} \partial_{[i_2 \vert} \cdots \partial_{[ i_s \vert}
h_{\vert j_1 ] \vert j_2 ] \vert \cdots \vert j_s ]}  
, \label{Riemann_s_components}
\end{eqnarray}
\noindent where the antisymmetrizations are to be carried on each pair of indices $\left(i_k , j_k\right) (k = 1 , \ldots , s)$.

Remarkably, we see that the Riemann tensor (or Freedmann-de Wit tensor) of the spin $s$ involves $s$ derivatives of the field.

The Riemann tensor has the following Young symmetry:
\begin{eqnarray}
R &\thicksim& \overbrace{\yng(10,10)}^{\text{$s$ boxes}} . \label{Riemann_s_Young}
\end{eqnarray}

The nilpotency of degree $s$ the differential ($i.e.$ the fact that is satisfies $d_{(s)}^{s+1} = 0$) implies that the Riemann tensor fulfills the Bianchi identity:
\begin{eqnarray}
0 &=& d_{(s)} R , \label{Riemann_s_Bianchi_no_index}
\end{eqnarray} 
\noindent or, more explicitly:
\begin{eqnarray}
0 &=& \partial_{[k_1} R_{i_1 j_1 ] i_2 j_2 \cdots i_s j_s} . \label{Riemann_s_Bianchi_components}
\end{eqnarray}

On the other hand, we see that $R$ is also well-filled (its symmetry type is $Y^{2s}_s$), so that the cohomology of $d_{(s)}$ acting on it is also empty. From this, we can immediately conclude that any tensor with the algebraic symmetries of the Riemann tensor fulfilling the Bianchi identity will be the Riemann tensor of some field:
\begin{equation}
d_{(s)} R = 0 \Rightarrow \exists h, \ R = d_{(s)}^s h .
\end{equation}

In conclusion, we have the following lemmas:

\begin{lemma}
The Riemann tensor $R\left[h\right]$ of a spin $s$ field, defined by (\ref{Riemann_s_components}), has the Young symmetry (\ref{Riemann_s_Young}) and satisfies the Bianchi identity (\ref{Riemann_s_Bianchi_components}). It is invariant under spin-$s$ diffeomorphism of $h$ (given by (\ref{diffeomorphism_s_latin})). Moreover, a necessary and sufficient condition for $h$ to be pure gauge is that its Riemann tensor vanish.
\end{lemma}

\begin{lemma}
For any tensor $R$ of Young symmetry (\ref{Riemann_s_Young}) satisfying the Bianchi identity (\ref{Riemann_s_Bianchi_components}), there exists a symmetric tensor $h$ of which it is the Riemann tensor.
\end{lemma}

\subsection{Einstein tensor}

\label{Sec:Einstein_s}

In dimension three, the Riemann tensor can be dualized (through the Levi-Civita volume form) over each of its antisymmetric pairs of indices, giving the Einstein tensor:
\begin{eqnarray}
G_{k_1 \cdots k_s} &\equiv& 2^{-s} \  \varepsilon_{k_1 i_1 j_1} \varepsilon_{k_2 i_2 j_2} \cdots
\varepsilon_{k_s i_s j_s} R^{i_1 j_1 i_2 j_2 \cdots i_s j_s}
\nonumber \\ &=&
\varepsilon_{k_1 i_1 j_1} \varepsilon_{k_2 i_2 j_2} \cdots
\varepsilon_{k_s i_s j_s} \partial^{i_1} \partial^{i_2 } \cdots \partial^{ i_s }
h^{ j_1  j_2  \cdots  j_s }  . \label{Einstein_s_components}
\end{eqnarray}

In an index free notation, one can also write:
\begin{eqnarray}
G \equiv \ast R ,
\end{eqnarray}
\noindent where $\ast$ means the dual has been taken on each pair of indices.

The expression of the Einstein tensor in terms of the Riemann is invertible, giving:
\begin{eqnarray}
R_{i_1 j_1 i_2 j_2 \cdots i_s j_s} = \varepsilon_{k_1 i_1 j_1} \varepsilon_{k_2 i_2 j_2} \cdots
\varepsilon_{k_s i_s j_s} G^{k_1 k_2\cdots k_s} ,
\end{eqnarray}
\noindent (we can also write this equation in the compact form $R = \ast G$).

The Einstein tensor is a fully symmetric tensor, its associate Young diagram being:
\begin{eqnarray}
G &\thicksim& \overbrace{\yng(10)}^{\text{$s$ boxes}} . \label{Einstein_s_Young}
\end{eqnarray} 

The Bianchi identity satisfied by the Riemann tensor, (\ref{Riemann_s_Bianchi_components}), is easily seen to be equivalent to the ``contracted Bianchi identity" of the Einstein tensor, which demands its divergencelessness:
\begin{eqnarray}
0 &=& \partial^{i_1} G_{i_1 i_2 \cdot i_s} . \label{Einstein_s_Bianchi}
\end{eqnarray}

Since the Einstein tensor is equivalent to the Riemann tensor and Bianchi identity for the Riemann tensor is equivalent to the divergencelessness of the Einstein tensor, the previous theorems for the Riemann tensor immediately transpose themselves to the Einstein tensor.

\begin{corollary}
The Einstein tensor $G\left[h\right]$ of a spin $s$ field, defined by (\ref{Einstein_s_components}), is fully symmetric in its $s$ indices and divergenceless. It is invariant under spin-$s$ diffeomorphism of $h$ (given by (\ref{diffeomorphism_s_latin})). Moreover, a necessary and sufficient condition for $h$ to be pure gauge is that its Einstein tensor vanish. \label{corollary_Einstein1}
\end{corollary}

\begin{corollary}
For any tensor $G$ symmetric and divergenceless, there exists a symmetric tensor $h$ (of same rank) of which it is the Einstein tensor. \label{corollary_Einstein2}
\end{corollary}

\subsubsection{Consequences of the tracelessness of the Einstein tensor}

\label{Sec:Einstein_traceless}

Finally, let us make the following observation, which will be useful later. If we take $n$ traces of the Einstein tensor of $h$, we obtain a symmetric divergenceless tensor with $s-2n$ indices. The dual of this tensor has the following algebraic symmetry:
\begin{eqnarray}
\ast G^{[n]} &\backsim& \overbrace{\yng(6,6)}^{\text{$\left(s-2n\right)$ boxes}} ,
\end{eqnarray}
\noindent where the dual is explicitly given by:
\begin{eqnarray}
\left(\ast G^{[n]}\right)_{i_{2n+1} j_{2n+1} i_{2n+2} j_{2n+2} \cdots i_s j_s} = \varepsilon_{k_{2n+1} i_{2n+1} j_{2n+1}} \varepsilon_{k_{2n+2} i_{2n+2} j_{2n+2}} \cdots
\varepsilon_{k_s i_s j_s} G^{[n] k_{2n+1} k_{2n+2}\cdots k_s} .
\end{eqnarray}

All the traces of the Einstein tensor are divergenceless, so that their dual satisfies Bianchi identity:
\begin{eqnarray}
0 &=& \partial_{[k_{2n+1}} \left(\ast G^{[n]}\right)_{i_{2n+1} j_{2n+1}] i_{2n+2} j_{2n+2} \cdots i_s j_s} .
\end{eqnarray}

However, this identity applies to a tensor whose Young diagram has $s-2n$ columns. This means that $\ast G^{[n]}$ is annihilated by a differential of nilpotency degree $s - 2n +1$:
\begin{eqnarray}
0 &=& d_{(s-2n)} \ast G^{[n]} .
\end{eqnarray}

For this differential, $\ast G^{[n]}$ is still well-filled, so that we can conclude, by the absence of cohomology, that:
\begin{eqnarray}
\ast G^{[n]} &=& d_{(s-2n)}^{s-2n} \Psi ,
\end{eqnarray}
\noindent where $\Psi$ is a symmetric tensor with $s-2n$ indices.

In particular, this means that the vanishing of the $n$th trace of the Einstein tensor is equivalent to $\Psi$ being closed ($0 = d_{(s-2n)}^{s-2n} \Psi$). But, as far as $d_{(s-2n)}$ is concerned, $\Psi$ is well-filled, so that this leads to $\Psi = d_{(s-2n)} \Pi$, where $\Pi$ is a fully symmetric tensor of rank $s-3n$.

\section{Spin-$s$ Weyl invariance}

\label{Sec:Weyl_s}

Let us now consider a spin $s$ field endowed with a larger set of gauge transformations, including generalized linearized Weyl rescaling - which we will call spin-$s$ Weyl transformations. The gauge symmetry will then be:
\begin{eqnarray}
\delta h_{i_1 \cdot i_s} &=& 
s \ \partial_{(i_1} \xi_{i_2 \cdot i_s )}
\ + \ \frac{s\left(s-1\right)}{2} \ \delta_{(i_1 i_2} \lambda_{i_3 \cdots i_s )} , \label{gauge_Weyl_diff_s}
\end{eqnarray}
\noindent where, once more, we stress that there are no conditions on the traces of the parameters.

Since we already saw that the most general invariant under the first part of these transformations was the Riemann tensor, we know that an invariant under the full variation (\ref{gauge_Weyl_diff_s}) will be a function of the Riemann tensor. Let therefore take a look at the transformation of the Riemann tensor under a spin-$s$ Weyl transformation:

\begin{equation}
\delta R _{i_1 j_1 i_2 j_2 \cdots i_s j_s}= 2^s \frac{s(s-1)}{2} \Pi \left( \partial_{i_1 \cdots i_s} \delta_{(j_1 j_2} \lambda_{j_3  \cdots j_s)} \right),
\end{equation}
\noindent where the projection operator $\Pi$ carries the antisymmetrizations within each pair of indices $(i_k, j_k)$.  The variation  of the Riemann tensor is clearly pure trace since all its terms contain a $\delta_{mn}$ factor with a pair of indices in $(i_1, j_1, i_2, j_2, \cdots, i_s, j_s)$. Therefore, the \textit{Weyl tensor}, which is the trace-free part of the Riemann tensor, is Weyl invariant.  As the Riemann tensor, it contains  $s$ derivatives of the spin-$s$ field. In dimension $D \geq 4$, the Weyl tensor vanishes if and only if the spin-$s$ field is  pure gauge taking into account all the gauge symmetries of conformal spin $s$, i.e. those of (\ref{gauge_Weyl_diff_s})\footnote{This is well known for $s=2$ and was established in \cite{Damour:1987vm} for $s=3$.   The demonstration proceeds along the same lines for $s>3$.}.

However, we are more specifically interested in the three dimensional case, which is less direct because the Weyl tensor vanishes identically\footnote{
The vanishing of the Weyl tensor in three dimensions is a well-known fact.  It is a direct consequence of the identities valid in three dimensions
\begin{eqnarray}
&& R_{i_1 j_1 i_2 j_2 \cdots i_s j_s} \nonumber \\
&& = \frac14\epsilon_{i_1 j_1 k_1} \epsilon^{k_1 m_1 n_1} \epsilon_{i_2 j_2 k_2} \epsilon^{k_2 m_2 n_2} R_{m_1 n_1 m_2 n_2 \cdots m_s n_s} \nonumber \\
&& = \frac14\epsilon_{i_1 j_1 k_1}  \epsilon_{i_2 j_2 k_2} \epsilon^{k_1 m_1 n_1} \epsilon^{k_2 m_2 n_2} R_{m_1 n_1 m_2 n_2 \cdots m_s n_s} \nonumber \\
&& = \frac12 \epsilon_{i_1 j_1 k_1}  \epsilon_{i_2 j_2 k_2} \delta^{k_1 k_2} R_{[2] m_3 n_3  \cdots m_s n_s}  \nonumber \\
&& \hspace{.4cm} - \epsilon_{i_1 j_1 k_1}  \epsilon_{i_2 j_2 k_2} R^{\;\; \; k_1 k_2}_{ [1]\; \; \; \; \; \; \;   m_3 n_3  \cdots m_s n_s}  \nonumber
\end{eqnarray}
showing that the Riemann tensor is entirely expressible in terms of its trace $R_{[1] k_1 k_2 \vert k_3 j_3  \cdots k_s j_s} =   \delta^{j_1 j_2} R_{k_1 j_1 k_2 j_2 \cdots k_s j_s}$ and thus that its traceless part is zero.  Here, $R_{ [2] m_3 n_3  \cdots m_s n_s}$ is the double trace $ \delta^{k_1 k_2} R_{[1] k_1 k_2 \vert m_3 n_3  \cdots m_s n_s}$.}. In order to encode the Weyl invariants, we will need to build a new, higher order tensor.

\subsection{Schouten tensor}

In three dimensions, the Einstein tensor completely captures the spin-$s$ diffeomorphism invariance, so that we can look for our Weyl invariant as a function of the Einstein tensor. Under a spin-$s$ Weyl transformation,  the Einstein tensor transforms as:

\begin{equation} 
\delta G_{i_1 \cdots i_s} = \frac{s(s-1)}{2}\left(- \partial_{(i_1} \partial_{i_2} \mu_{i_3 \cdots i_s)} +  \delta_{(i_1 i_2} \Delta \mu_{i_3 \cdots i_s)}\right) \label{Einstein_s_var_Weyl}
\end{equation}
\noindent where:
 
\begin{equation}
\mu_{i_3 \cdots i_s} =   \varepsilon_{i_3 j_3 k_3} \cdots
\varepsilon_{i_s j_s k_s}  \partial^{j_3}  \cdots \partial^{ j_s} \lambda^{k_3  \cdots k_s} . \label{FormOfMu}
\end{equation}

The tensor $\mu_{i_3 \cdots i_s}$ fulfills $\partial^{i_3} \mu_{i_3 \cdots i_s} = 0$ and conversely, any tensor that fulfills that equation can be written as in (\ref{FormOfMu}) with $\lambda_{k_3  \cdots k_s}$ completely symmetric\footnote{This follows again from the cohomological theorem \textbf{\ref{Theorem_cohomlogy}} of Appendix \textbf{\ref{App:generalized_differential}} (see \cite{DuboisViolette:1992ye,DuboisViolette:1999rd}) applied now to the differential operator $d_{(s-2)}$ defined in the space of tensors with $s-2$ columns, which fulfills $d_{(s-2)}^{s-1} = 0$. $\mu$ is actually the Einstein tensor of $\lambda$.}.

The Schouten tensor $S_{i_1 \cdots i_s}$ is then defined through:
\begin{eqnarray}
S_{i_1 \cdots i_s} &=& G_{i_1 \cdots i_s} 
\nonumber \\ && 
\ + \ \sum_{n=1}^{[\frac{n}{2}]} a_n \ \delta_{(i_1 i_2}  \cdots \delta_{i_{2n-1} i_{2n}} G^{[n]}_{ i_{2n+1}  \cdots i_s)} , \hspace{.5cm}\label{Schouten_s_def}
\end{eqnarray}
\noindent where the terms added to  $ G_{i_1 \cdots i_s}$ to define $S_{i_1 \cdots i_s}$ involve the successive higher traces of $ G_{i_1 \cdots i_s}$ and are recursively adjusted in such a way that the Schouten tensor fulfill the crucial  property of  transforming as:

\begin{equation}
\delta S_{i_1 i_2 \cdots i_s} = -\frac{s(s-1)}{2}\partial_{(i_1} \partial_{i_2} \nu_{i_3 \cdots i_s)} ,  \label{Schouten_s_var}
\end{equation}
\noindent under spin-$s$ Weyl transformations,  where $\nu$ is related to $\mu$ as follows:
\begin{eqnarray}
\nu_{i_3 \cdots i_s} &=&  \mu_{i_3 \cdots i_s}  \nonumber \\  
&& \ + \  \sum_{n=1}^{[\frac{n}{2}]-1} b_n  \delta_{(i_3 i_4}  \cdots \delta_{i_{2n+1} i_{2n+2}} \mu^{[n]}_{i_{2n+3}  \cdots i_s)} . \hspace{.5cm}
\end{eqnarray} 

One finds that the coefficients $a_n$ are explicitly given by:

\begin{equation}
a_n = \frac{(-1)^n}{4^n} \frac{s}{n!} \frac{(s-n-1)!}{(s-2n)!}, \; \; \; (n\geq 1) 
\end{equation}
\noindent and that the coefficients $b_n$ are then:

\begin{equation}
b_n = a_n \frac{(s-2n) (s-2n-1)}{s(s-1)}, \; \; \;  (n \geq 1)
\end{equation}

The recursive procedure amounts to successively eliminating the terms $\Delta \mu_{i_3 \cdots i_s}$, $\Delta \bar{\mu}^{i_5 \cdots i_s}$ involving the Laplacian by adding symmetrized products of $\delta^{ij}$'s with multiple traces of the Einstein tensors, with suitable coefficients that are determined uniquely. 

In terms of the variables $S^{i_1 i_2 \cdots i_s}$ and $\nu^{i_3 i_4 \cdots i_s}$, the Bianchi identity and the condition $\partial_{i_3} \mu^{i_3 i_4 \cdots i_s}$ $ = 0$ read respectively:

\be
\partial_{i_1} S^{i_1 i_2 \cdots i_s} - (s-1) \partial^{(i_2}\bar{S}^{i_3 \cdots i_s)} =0 \label{Schouten_s_Bianchi}
\ee
and
\be 
\partial_{i_1} \nu^{i_1 i_2 \cdots i_{s-2}} + \frac{s-3}{3} \partial^{(i_2} \bar{\nu}^{i_3 \cdots i_{s-2})} =0 . \label{Nu_s_Bianchi}
\ee

The easiest way to prove these important relations is to observe that they follow uniquely from the requirement of invariance under (\ref{Schouten_s_var}), in much the same way as the Bianchi identity $\partial_{i_1} G^{i_1 i_2 \cdots i_s} = 0$ and the condition $\partial_{i_3} \mu^{i_3 i_4 \cdots i_s} = 0$ are the unique identity and condition compatible with the transformation (\ref{Einstein_s_var_Weyl}) within the class
$$ \partial_{i_1} G^{i_1 i_2 \cdots i_s} +a \partial^{(i_2}\bar{G}^{i_3 \cdots i_s)}  + b \delta^{(i_2 i_3} \partial^{i_4} \bar{\bar{G}}^{i_5 \cdots i_s)} + \cdots =0,$$ 
$$ \partial_{i_3} \mu^{i_3 i_4 \cdots i_s} +k \partial^{(i_4}\bar{\mu}^{i_5 \cdots i_s)}  + \ell \delta^{(i_4 i_5} \partial^{i_6} \bar{\bar{\mu}}^{i_7 \cdots i_s)} + \cdots =0$$ 
(invariance under (\ref{Einstein_s_var_Weyl}) forces $a=b=\cdots = k = \ell = \cdots = 0$).

\subsection{Cotton tensor}

In index-free notations, the transformation of the Schouten tensor reads:

\begin{equation}
\delta S = - \ \frac{s\left(s-1\right)}{2} \ d_{(s)}^2 \nu .
\end{equation}

Indeed, $\nu$ is a symmetric tensor with $s - 2$ indices and $S$ is a symmetric tensor with $s$ indices.

In its original formulation, the Cotton tensor $C$ is defined as $d^{s-1}_{(s)} S$.  It is a tensor of mixed symmetry type $\left(s , s-1\right)$:

\begin{equation}
C \backsim \underbrace{\overbrace{\yng(10,9)}^{\text{$s$ boxes}}}_{\text{$s-1$ boxes}} .
\end{equation}

It is invariant under spin-$s$ Weyl transformations as it follows from $d_{(s)}^{s+1} = 0$:

\be
\delta C = d_{(s)}^{s-1} \delta S = -  \ \frac{s\left(s-1\right)}{2} \ d_{(s)}^{s+1} \nu = 0.
\ee

It contains $2s-1$ derivatives of the spin-$s$-field $h_{i_1 \cdots i_s}$.  As a consequence of the Bianchi identity (\ref{Schouten_s_Bianchi}),  it can be verified to be  traceless on the last index of the first row with any other index (i.e., one gets zero when the last index of the first row is contracted with any other index). 

In the dual representation on the first $s-1$ indices of $C$ which we shall adopt, the Cotton tensor $B^{i_1 \cdots i_s}$   is explicitly given by:

\begin{eqnarray}
B^{i_1 i_2  \cdots i_s} &=& \varepsilon^{i_1 j_1 k_1}\varepsilon^{i_2 j_2 k_2} \cdots \varepsilon^{i_{s-1} j_{s-1} k_{s-1}} \nonumber \\
&& \hspace{.5cm} \partial_{j_1} \partial_{j_2} \cdots \partial_{j_{s-1}} S^{\hspace{1.38cm} i_s}_{k_1 k_2 \cdots k_{s-1}} 
. \label{Cotton_s_def}
\end{eqnarray}

This tensor is manifestly symmetric in its first $s-1$ indices.  Symmetry in $i_{s-1}$, $i_s$ is a direct consequence of the Bianchi identity (\ref{Schouten_s_Bianchi}) (this is equivalent to the tracelessness property of $C$ just mentioned).  Hence, the tensor $B^{i_1 \cdots i_s}$ is fully symmetric  i.e.,   is of symmetry type:

$$
B \backsim \overbrace{\yng(10)}^{\text{$s$ boxes}} 
$$
Furthermore, it is easily proved to be conserved on the first index (i.e., its divergence on the first index is zero).  It is also traceless on the last two indices because of the Young symmetries of $C$.  Since $B$ is fully symmetric, one thus gets:

\be
\delta_{i_k i_m} B^{i_1 i_2  \cdots i_s} = 0, \; \; \; \partial_{i_p } B^{i_1 i_2  \cdots i_s} = 0, 
\ee
\noindent  with $ 1 \leq  k<m\leq s $ and $1 \leq p \leq s$.
 
We stress that, as we have shown,  the Cotton tensor $B$ is completely symmetric as a consequence of the Bianchi identity.  Hence, it is not necessary to enforce symmetrization in its definition since it is automatic. Enforcing complete symmetrization, as done in \cite{Bergshoeff:2009tb,Bergshoeff:2011pm}, is of course permissible, but is not needed. \\

By construction, the Cotton tensor is invariant under the gauge transformation (\ref{gauge_Weyl_diff_s}), and the results obtained up to this point can be summarized as follows:

\begin{lemma}
The Cotton tensor $B$ of a spin $s$ field is given by the expression (\ref{Cotton_s_def}). It is divergenceless, fully symmetric in its $s$ indices and traceless. It is invariant under spin $s$ diffeomorphism and Weyl transformations.
\end{lemma}

We still have to prove that the Cotton tensor constitutes a complete set of invariants, and that any symmetric, traceless and divergenceless tensor is the Cotton tensor of some spin $s$ field. Before we establish these crucial properties, we will illustrate the general development just followed on the cases of spin $2$, $3$ and $4$.

\subsection{Spin-$2$}

\label{Sec:Spin2}

The above construction reproduces the familiar  spin-$2$ formulas.  One finds for the Schouten tensor

$$ 
S^{ij} = G^{ij} - \frac12 \delta^{ij} \bar{G}, \; \; \;  G^{ij} = S^{ij} - \delta^{ij} \bar{S} , 
$$
\noindent and $\delta S_{ij} = - \partial_i \partial_j \lambda$, $\partial_i S^{ij} - \partial^j \bar{S} = 0$.

The Cotton tensor $C_{ijk} $ is $C_{ijk} = \partial_i S_{jk} - \partial_j S_{ik} $ and is a Weyl-invariant tensor of type $\yng(2,1)$, which is traceless on $(j,k)$ (or $(i,k)$) because of the Bianchi identity, $\bar{C}_i = 0$.  It involves three derivatives of $h_{ij}$. In the dual representation, the Cotton tensor $B^{ij}$ is:

$$
B^{ij} =  \varepsilon^{i}_{\; \, mn} \partial^{m} S^{nj} 
$$
\noindent and is easily checked to be indeed symmetric, traceless and divergenceless. Note that this can be rewritten in the more explicit form:
\begin{eqnarray}
B^{ij} &=& 
\epsilon^{(i \vert kl} \partial_k \left( 
\partial^{\vert j)} \partial^m h_{ml} \ - \ \Delta h^{\vert j)}_{\phantom{\vert j )}l} \right) .
\end{eqnarray}

\subsection{Spin-$3$}

We now move to the spin 3 case,
$$
h \thicksim \yng(3) ,
$$
where the above derivation reproduces  the results of \cite{Damour:1987vm}. We derive below the form the general formulas take for $s=3$.

The Schouten tensor for a spin-$3$ field reads explicitly
\be
S^{i_1 i_2 i_3} = G^{i_1 i_2 i_3} - \frac{3}{4} \delta^{(i_1 i_2} \bar{G}^{i_3)}  \label{Schouten3}
\ee
Its trace is equal to
$
\bar{S}^i = - \ \frac14 \ \bar{G}^i
$ so that the inverse formula to (\ref{Schouten3}) is
$
G^{i_1 i_2 i_3} = S^{i_1 i_2 i_3} \ - \ 3 \ \delta^{(i_1 i_2} \bar{S}^{i_3)} 
$.
The Schouten tensor transforms as
\be
\delta S_{i_1 i_2 i_3} = - \ 3 \ \partial_{(i_1} \partial_{i_2} \mu_{i_3)}
\ee
under Weyl transformations, where $\mu_i$ is given by
\be
\mu^i =\varepsilon^{ijk} \partial_j \lambda_k
\ee
and fulfills $\partial_k \mu^k = 0$.

The Bianchi identity implies $\partial_i \bar{S}^i = 0$ and can  equivalently be written in terms of $S^{i_1 i_2 i_3}$ as
\be
\partial_{i_1} S^{i_1 i_2 i_3} - \partial^{i_2} \bar{S}^{i_3} - \partial^{i_3} \bar{S}^{i_2} = 0 \label{BianchiSSpin3}
\ee

According to the above general definition, the Cotton tensor $C= d_{(3)}^2 S$ is explicitly given by 
\begin{eqnarray}
&& C_{i_1 j_1 \vert i_2 j_2 \vert i_3} = \partial_{i_1} \partial_{i_2} S_{j_1 j_2 i_3} - \partial_{j_1} \partial_{i_2} S_{i_1 j_2 i_3} \nonumber \\ 
&& \hspace{1cm} - \partial_{i_1} \partial_{j_2} S_{j_1 i_2 i_3} + \partial_{j_1} \partial_{j_2} S_{i_1 i_2 i_3} \nonumber
\end{eqnarray}
and has Young symmetry type:

$$
C \backsim \yng(3,2) .
$$

In the dual representation, the Cotton tensor $B^{k_1 k_2 k_3}$ reads
\be 
B^{k_1 k_2 k_3} = \varepsilon^{k_1 i_1 j_1} \varepsilon^{k_2 i_2 j_2}  \partial_{i_1} \partial_{i_2} S_{j_1 j_2}^{\; \; \; \; \; \; k_3} \label{DualCottonSpin3}
\ee
and, using the Bianchi identity (\ref{BianchiSSpin3}) and its consequence $\partial_j \bar{S}^j = 0$, is easily seen to be equal to 
\be
B^{k_1 k_2 k_3} = 3 \partial^{(i_1} \partial^{i_2} \bar{S}^{i_3)} - \Delta S^{i_1 i_2 i_3}
\ee
an expression that is manifestly symmetric.
Transverseness and tracelessness follow again from the Bianchi identity (\ref{BianchiSSpin3}).

\subsection{Spin-$4$}

We now write the formulas in the spin-$4$ case. The
Schouten tensor is
\be
S^{i_1 i_2 i_3 i_4} = G^{i_1 i_2 i_3 i_4} \ - \  \delta^{(i_1 i_2} \bar{G}^{i_3  i_4)} \ + \ \frac{1}{8} \  \delta^{(i_1 i_2} \delta^{i_3 i_4)} \bar{\bar G} 
\ee
and transforms as
\be
\delta S_{i_1 i_2 i_3 i_4} = - \ 6 \ \partial_{(i_1} \partial_{i_2} \nu_{i_3 i_4)}
\ee
under Weyl transformations, with $\nu_{ij} = \mu_{ij} - \frac16 \ \delta_{ij} \bar{\mu}$.

In terms of the Schouten tensor, the Bianchi identity reads
\be
\partial_{i_1} S^{i_1 i_2 i_3 i_4} - \partial^{i_2} \bar{S}^{i_3 i_4} - \partial^{i_3} \bar{S}^{i_2 i_4} -  \partial^{i_4} \bar{S}^{i_2 i_3} = 0 \label{BianchiSpin4a}
\ee
and one has $\partial_{i} \nu^{ij }+ \frac13 \partial^j \bar{\nu} = 0$.

The spin-$4$ Weyl invariant Cotton tensor is $d^3_{(4)} S$. Writing the explicit formulas directly in the dual representation, one finds
\be 
B^{k_1 k_2 k_3 k_4} = \varepsilon^{k_1 i_1 j_1} \varepsilon^{k_2 i_2 j_2}  \varepsilon^{k_3 i_3 j_3} \partial_{i_1} \partial_{i_2} \partial_{i_3} S_{j_1 j_2 j_3}^{\; \; \; \; \; \; \; \; k_4} \, . \label{DualCottonSpin4}
\ee
 Again, the symmetry in $(k_1, k_2, k_3)$ is manifest, while the symmetry in the last index $k_4$ with any other index is a consequence of the Bianchi identity (\ref{BianchiSpin4a}).  The Cotton tensor is transverse and traceless,
\be
\partial_i B^{ijkl} = 0, \; \; \; \bar{B}^{ij} = 0.
\ee

\section{Cotton tensor as a complete set of Weyl invariants}

\label{Sec:4}

The Cotton tensor is quite important because it is not only gauge invariant, but it also completely captures higher spin Weyl invariance.  By this, we mean that any function of the higher spin field and its derivatives that is invariant under higher spin diffeomorphisms and Weyl transformations is necessarily a function of the Cotton tensor and its derivatives,

\be
\delta_{\xi, \lambda}  f([h]) = 0 \; \; \Rightarrow f= f([B]) .
\ee

Equivalently, a necessary and sufficient condition for a spin-$s$ field to be pure gauge (equal to zero up to spin-$s$ diffeomorphisms and Weyl transformations) is that its Cotton tensor vanishes.

The first version of this property is demonstrated in Appendix {\bf \ref{App:Completeness_s_Weyl}}.  We show here how to prove the second version\footnote{We were kindly informed by Xavier Bekaert that the property ``Cotton tensor = 0 $\Leftrightarrow$ spin-$s$ field is diffeomorphism and Weyl pure gauge" can also be viewed as a consequence as the cohomological theorems of \cite{Shaynkman:2004vu} on the representations of the conformal group, see \cite{Beccaria:2014jxa,BekaertChengdu}.  We are grateful to him for this information.}:

\begin{lemma}
A necessary and sufficient condition of a spin-$s$ field to be pure gauge (with the gauge symmetry (\ref{gauge_Weyl_diff_s})) is that its Cotton tensor vanish.
\end{lemma}

Assume, then,  that the Cotton tensor $B$ (or equivalently, $C$) is equal to zero.  Using the cohomological theorem \textbf{\ref{Theorem_cohomlogy}} of Appendix \textbf{\ref{App:generalized_differential}}\footnote{See \cite{DuboisViolette:1992ye,DuboisViolette:1999rd}.}, one gets:
 
\be
C = d_{(s)}^{s-1} S = 0  \Rightarrow S = - d_{(s)}^2 \nu
\ee for some $\nu$. The Bianchi identity implies that one can choose $\nu$ in such a way that the corresponding tensor $\mu$ fulfills $\partial_{i_1} \mu^{i_3 \cdots i_s} = 0$ and so,  can be written as in (\ref{FormOfMu}) for some completely symmetric $\lambda_{k_3  \cdots k_s}$.
This implies in turn that $R[h-  \lambda \star \delta] = 0$, or equivalently $d_{(s)}^s (h - \lambda \star \delta) = 0$.  Here, $ \lambda \star \delta$ stands for the Weyl transformation term $\delta_{(i_1 i_2} \lambda_{i_3 \cdots i_s)}$.  Using again the cohomological theorem \textbf{\ref{Theorem_cohomlogy}} of Appendix \textbf{\ref{App:generalized_differential}}\footnote{See \cite{DuboisViolette:1992ye,DuboisViolette:1999rd}.}, one finally obtains:

\be
h = d_{(s)}  \xi +  \lambda \star \delta,
\ee
i.e., $h$ is pure gauge.  Conversely, if $h$ is pure gauge, the Cotton tensor vanishes.
We can thus conclude that a necessary and sufficient condition for the spin-$s$ field to be pure gauge is that its Cotton tensor vanishes.

We illustrate explicitly the derivations in the spin-3 and spin-4 cases.

\subsection{Spin $3$}

Consider a spin-$3$ field $h_{i_1 i_2 i_3}$ with vanishing  Cotton tensor.  According to the theorem \textbf{\ref{Theorem_cohomlogy}} of Appendix \textbf{\ref{App:generalized_differential}}\footnote{See \cite{DuboisViolette:1992ye,DuboisViolette:1999rd}.}, the Schouten tensor reads
\be
 S_{i_1 i_2 i_3} = - \partial_{(i_1} \partial_{i_2} \mu_{i_3)} \label{EqForS}
\ee
for some $\mu_i$.  We want to prove that $\mu_i$ can be chosen so that $\partial_i \mu^i = 0$.  From the Bianchi identity, one gets
 
$$
\partial_i \partial_j (\partial_k \mu^k) = 0
$$
It follows that $\partial_k \mu^k$ is at most linear in the coordinates,

$$
\partial_k \mu^k = a + b_k x^k
$$
Define $\tilde{\mu}^k = \frac13 a x^k + \frac 12 c^k_{\; \; ij} x^i x^j$ where $ c^k_{\; \; ij} =  c^k_{\; \; ji}$, $ c_{(kij)} = 0$ and $ c^k_{\; \; kj} = b_j$. [Such a $ c^k_{\; \; ij}$ exists.  It has the Young symmetry $\yng(2,1)$ in the dual conventions where symmetry is manifest while antisymmetry is not.  The trace of such a tensor is unconstrained and so can be taken to be equal to $b_j$.]  By construction, $\partial_k \tilde{\mu}^k = \partial_k \mu^k$ and $\partial_{(i_1} \partial_{i_2} \tilde{\mu}_{i_3)}=0$, so that $S_{i_1 i_2 i_3} = - \partial_{(i_1} \partial_{i_2} (\mu_{i_3)} - \tilde{\mu}_{i_3)})$, implying that we can assume that $\mu^k$ in (\ref{EqForS}) fulfills $\partial_k \mu^k = 0$, which will be done from now on.  We then have $\mu^k = \varepsilon^{kij} \partial_i \lambda_j$ for some $\lambda_j$, and so the Einstein tensor of $h_{ijk}$ is equal to the Einstein tensor of $3 \ \delta_{(ij} \lambda_{k)}$, implying $h_{ijk} = 3 \ \partial_{(i} \xi_{jk)} + 3 \ \delta_{(ij} \lambda_{k)}$ for some $\xi_i$, as announced in the general discussion above.

\subsection{Spin $4$}

We now illustrate the produre for the spin-$4$ field.  The vanishing of the Cotton tensor implies again (theorem \textbf{\ref{Theorem_cohomlogy}} of Appendix \textbf{\ref{App:generalized_differential}}\footnote{See \cite{DuboisViolette:1992ye,DuboisViolette:1999rd}.}), 

\be
 S_{i_1 i_2 i_3 i_4} = - \partial_{(i_1} \partial_{i_2} \nu_{i_3 i_4)} \label{EqForS4}
\ee
for some $ \nu_{i_3 i_4}$.  The Bianchi identity yields then

\be
\partial_{(i}\partial_j N_{k)} = 0, \; \; \; N_{k} \equiv \partial^m \nu_{km} + \frac13 \partial_k \bar{\nu}.
\ee
This does not imply that $N_k =0$ since, as for spin-$3$, there are non trivial solutions of the equation $\partial_{(i}\partial_j N_{k)} = 0$.   These solutions have been analyzed in section {\bf 6} of \cite{DuboisViolette:2001jk}.   The space of solutions is finite-dimensional; one easily gets from the equation that the third derivatives of $N_k$ vanish, so that $N_k$ is at most quadratic in the $x^i$'s,
$$ N_k = a_k + b_{k\vert m} x^m + c_{k \vert mn}x^m x^n,$$
for some constants $a_k$,  $b_{k\vert m}$ and $ c_{k \vert mn} = c_{k \vert nm}$ which have the respective symmetry $\yng(1)$, $\yng(1) \otimes \yng(1)$, and $\yng(1) \otimes \yng (2)$.  Now, let 
$$
\tilde{\nu}_{km} = \rho (a_k x_m + a_m x_k) + \sigma_{km \vert rs} x^r x^s  + \theta_{km \vert rsp} x^r x^s x^p
$$
where the constants $\rho$, $\sigma_{km \vert rs}$ (with symmetry $\yng(2) \otimes \yng(2)$) and $\theta_{km \vert rsp}$ (with symmetry $\yng(2) \otimes \yng(3)$) are chosen such that (i) $\sigma_{(km \vert ij)}=0$, $\theta_{(km\vert ij)p} = 0$ so that $\partial_{(i}\partial_j \tilde{\nu}_{km)} = 0$; and (ii) $\tilde{N}_k = N_k$.  This is always possible since this second condition restricts only the traces, which are left free by the first condition.  Then, by substracting $\tilde{\nu}_{km}$ from $\nu_{km}$, one sees that one can assume $N_k=0$.  This implies that the corresponding $\mu^{km}$ can be assumed to fulfill $\partial_m \mu^{km} = 0$ and thus is equal to $\mu^{km} = \varepsilon^{krs} \varepsilon^{mpq} \partial_r \partial_p \lambda_{sq}$ for some $\lambda_{sq} = \lambda_{qs}$.  Therefore, the Einstein tensor  of $h_{ijkm}$ is equal to the Einstein tensor of $\delta_{(ij} \lambda_{km)}$, implying that
$$h_{ijkm} = 4\ \partial_{(i} \xi_{jkm)} + 6 \ \delta_{(ij} \lambda_{km)},$$
which is the result that we wanted to prove.

\section{Cotton tensors as a complete set of transverse-traceless symmetric tensors}

\label{Sec:Cotton_for_any_traceless}

We have proven so far the vanishing of its Cotton tensor was a necessary and sufficient condition for a spin-$s$ field to be conformally flat. Let us now establish that a tensor that has the properties of the Cotton tensor $is$ the Cotton tensor of an existing field:

\begin{lemma}
For any symmetric, divergenceless and traceless tensor $B$ with $s$ indices, there exists a symmetric tensor of same rank of which $B$ is the Cotton tensor. \label{lemma_Cotton_for_any_traceless}
\end{lemma}

\subsection{The problem}

We have recalled with corollary \ref{corollary_Einstein2} of section \textbf{\ref{Sec:Einstein_s}} that if a completely symmetric tensor $G^{i_1 \cdots i_s}$ fulfills the equation
\be
\partial_{i_1} G^{i_1 i_2 \cdots i_s} = 0,
\ee
then there exists $h_{i_1 \cdots i_s}$ such that $G = G[h]$.

We want to address the question: let $B^{i_1 i_2 \cdots i_s}$ be a completely symmetric tensor that is both transverse 
\be
\partial_{i_1} B^{i_1 i_2 \cdots i_s} = 0,  \label{EQ1}
\ee
and traceless,
\be
\delta_{i_1 i_2} B^{i_1 i_2 i_3\cdots i_s} = 0, \label{EQ2}
\ee
Does there exist a totally symmetric tensor $Z_{i_1 \cdots i_s}$ such that $B^{i_1 i_2 \cdots i_s}$ is the Cotton tensor of $Z_{i_1 \cdots i_s}$?

We prove here that  the answer is affirmative, starting with the spin-$2$ case.

\subsection{Spin $2$}

The Cotton tensor $B^{ij}$ is dual to $C_{ijk} = - C_{jik}$ on the first index.  The tracelessness condition on $B^{ij}$ implies that $C_{ijk}$ has Young symmetry type $\yng(2,1)$, while the symmetry in $(i,j)$ of $B^{ij}$ implies that $C_{ijk}$ is traceless. The divergenceless condition on $B^{ij}$ implies then, by Poincar\'e lemma ($k$ being a ``spectator" index) that
\be
C_{ijk} = \partial_i S_{jk} - \partial_j S_{ik}
\ee
where $S_{ik}$ is not a priori symmetric.  However, the ambiguity in $S_{ik}$ is $S_{ik} \rightarrow S_{ik} + \partial_i T_k$,  and using this ambiguity, the condition $C_{[ijk]} = 0$ and Poincar\'e lemma, one easily sees that $S_{ik}$ can be assumed to be symmetric.  Then, the tracelessness condition implies the Bianchi identity $\partial_i S^{ij} - \partial^j \bar{S} = 0$ for $S^{ij}$ (or $\partial_i G^{ij} = 0$ for $G^{ij}$), from which follows the existence of $Z_{ij}$ such that $S = S[Z]$ and thus $B = B[Z]$.  This establishes the result.

\subsection{Higher spin}
The same steps work for higher spins.  For instance, for spin $3$, the reasoning proceeds as follows:
\begin{itemize}
\item Define $C_{i_1 j_1 \vert i_2 j_2 \vert k}$ from $B^{ijk}$ by dualizing on the first two indices, with $k$ ``spectator",
\be
C_{i_1 j_1 \vert i_2 j_2 \vert k}= \epsilon_{i_1 i_2 i} \epsilon_{j_1 j_2 j} \delta_{km} B^{ijm}
\ee
\item Because $B^{ijm}$ is completely symmetric and traceless, $C_{i_1 i_2 \vert j_1 j_2 \vert k}$ has Young symmetry type $\yng(3,2)$ and is traceless on $k$ and any other of its indices.
\item The transverse condition on $B^{ijk}$ is equivalent to  $d_{(2)} C = 0$ where $d_{(2)}$ is acting on $C$ as if it was a collection of tensors of type $\yng(2,2)$ parametrized by $k$. The Poincar\'e lemma implies then the existence of a tensor $R_{ijk}$ such that 
\be C_{i_1 j_1 \vert i_2 j_2 \vert k}=  \partial_{[i_1} \partial_{[j_1} R_{i_2] j_2] k} \label{FormOfC}
\ee
where the antisymmetrizations 	are on the pairs of indices $(i_1, i_2)$ and $(j_1,j_2)$. At this stage, the tensor $R$ has a component $S$ that is completely symmetric $S \thicksim \yng(3)$,
\begin{eqnarray}
S_{i_2 j_2 k} &=& \frac13(R_{i_2 j_2 k}  + R_{ j_2 k i_2}  + R_{k i_2 j_2} ) \\
&=& S_{(i_2 j_2 k)} 
\end{eqnarray}
and a component $T$ that has Young symmetry type $\yng(2,1)$,
\be
T_{ijk} = \tilde{T}_{kji} - \tilde{T}_{kij}
\ee
\be
3\tilde{T}_{kji} = T_{ijk} +T_{ikj}
\ee
with
\be
\tilde{T}_{i_2 j_2 k} = 2 R_{i_2 j_2 k} - R_{k j_2 i_2} - R_{k i_2 j_2}
\ee
Explicitly,
\be
R_{i_2 j_2 k} = S_{i_2 j_2 k} + \frac13 \tilde{T}_{i_2 j_2 k}.
\ee
The change from $T$ to $\tilde{T}$ corresponds to the change of conventions in which either antisymmetry or symmetry is manifest.  We shall call $T$ the ``Curtright tensor".
\item The tensor $R_{i_2 j_2 k}$ is not completely determined by $C_{i_1 j_1 \vert i_2 j_2 \vert k}$ since one may add to it 
\be
R_{i_2 j_2 k} \rightarrow R_{i_2 j_2 k} + \partial_{i_2} \mu_{j_2 k} + \partial_{j_2} \mu_{i_2 k} \label{GaugeR}
\ee
without violating (\ref{FormOfC}). This is the only ambiguity.
\item Furthermore, the condition $C_{i_1 j_1 \vert [i_2 j_2 \vert k]} = 0$ is easily verified to imply that the strength field of the Curtright tensor is equal to zero, so that $T$ is pure gauge and can be set equal to zero by a gauge transformation of the type (\ref{GaugeR}).  
Therefore, one can assume
\be C_{i_1 j_1 \vert i_2 j_2 \vert k}=  \partial_{[i_1} \partial_{[j_1} S_{i_2] j_2] k} \label{FormOfCbis}
\ee
with $S$ completely symmetric.
\item The residual gauge symmetry after $T$ has been set equal to zero is given by
\be
S_{ijk} \rightarrow S_{ijk} + \partial_{(i} \partial_j \mu_{k)}  \label{Residual}
\ee (which is still present because the gauge symmetries of the Curtright tensor are reducible).
\item Finally, the tracelessness condition of $C$ on $k$ and any other index yields
\be
\partial_{[i_1} U_{i_2]j} = 0 \label{BianchiU}
\ee
where 
\be
U_{ij} = \partial^k S_{ijk} - \partial_i \bar{S}_j - \partial_j \bar{S}_i
\ee
is such that $U_{ij} =0$ is the Bianchi identity for the Schouten tensor (see (\ref{BianchiSSpin3})).   Now, (\ref{BianchiU}) implies
\be
U_{ij} = \partial_{i} \partial_{j} \rho
\ee 
for some $\rho$ ( cohomology of $d_{(2)}$ for $d_{(2)}$ with $d_{(2)}^3 = 0$).   On the other hand, $U_{ij}$ transforms as
\be
U_{ij} \rightarrow U_{ij} - 3 \partial_i \partial_{j} (\partial_k \mu^k)
\ee
under (\ref{Residual}).  This enables one to chose $S$ to obey  the Bianchi identity of the Schouten tensor (take $\mu^k$ such that $3\partial_k \mu^k = \rho$), implying the existence of $Z_{ijk}$ such that $S = S[Z]$ and hence $B = B[Z]$. [Note that $\delta U =0$ when $\partial_k \mu^k =0$, as it should.]  This ends the demonstration of  the property that we wanted to prove.
\end{itemize}

\begin{subappendices}

\section{Generalized differential}

\label{App:generalized_differential}

\subsection{$p$-forms}

If we consider tensor fields over a flat manifold of dimension $D$ - on which ordinary partial derivatives provide us with a satisfactory connection, we know that, if we limit ourselves to fully antisymmetric (covariant) tensors (also called $p$-forms, $p$ being their number of indices), not only can we define an internal product over this space (that associates to a $p$-form and a $q$-form a $\left(p+q\right)$-form in an associative way), turning it into a graded Grassmann algebra, but we can also define a unique exterior derivative $d$ that send $p$-forms to $\left(p+1\right)$-forms and which satisfies a series of properties (its action on a $0$-form, which is a scalar function, gives the gradient of this function; it obeys Leibnitz rule when acting on a product of $p$-forms; \ldots). It can be explicitly built in the following way: if $\boldsymbol{\omega}$ is a $p$-form with components $\omega_{i_1 \ldots i_s}$\footnote{Which means that in the corresponding base of $1$-forms $\boldsymbol{e}^{i}$, we have $\omega = \omega_{i_1 \ldots i_s} \boldsymbol{e}^{i_1} \bigotimes \ldots \bigotimes \boldsymbol{e}^{i_p}$.}, its exterior derivative will be the $\left(p+1\right)$-form $d\boldsymbol{\omega}$ with components given by $\left(d\boldsymbol{\omega}\right)_{i_1 \ldots i_{p+1}} = \left(p+1\right)\partial_{[ i_1} \omega_{i_2 \ldots i_{p+1}]}$.

The most remarkable feature of this differential operator is, of course, that it squares to zero: $d^2 = 0$.  This naturally leads us to look at the cohomology of $d$, that is, to ask whether there exist $p$-forms annihilated by $d$ which are not themselves the exterior derivative of a $\left(p-1\right)$-form. In other words, are there closed $p$-forms that are not exact ? As is well-known, as long as we consider antisymmetric tensor fields over a manifold which has the topology of $\mathbb{R}^D$, the answer turns out to be negative for forms of non-zero degree (for $p = 0$, the kernel of $d$ - which is the set of constant functions, $\mathbb{R}$ - is also the cohomology of $d$, since $0$-forms can not be obtained by taking the exterior derivative of anything).

The precise notation of the cohomology is the following: $H^p\left(d\right)$ is the set of closed $p$-forms $modulo$ exact ones. We then have the following result over $\mathbb{R}^D$:

\begin{eqnarray}
0 &=& H^p\left(d\right) \ \mathrm{for} \ p\geq 1 ,
\\
\mathbb{R} &=& H^0\left(d\right) .
\end{eqnarray}

These results are of particular interest when one considers the spin $1$ massless field, the photon, whose potential is a one-form $\mathbf{A}$ (that is, a vector field). Its curvature is the strength field two-form $\mathbf{F}$ which is the exterior derivative of the potential: $\mathbf{F} \equiv d\mathbf{A}$. Since the physically observable object is the strength field, one easily identifies the gauge invariance of the potential as the kernel of $d$ of form degree one, which, thanks to the absence of cohomology, is the set of exact one-forms, or the set of external derivative of zero-form. In other words, a gauge variation of the potential is given by $\delta \mathbf{A} = d f$, $f$ being a scalar function.

In addition, a necessary and sufficient condition for a two-form $\mathbf{F}$ to be the strength field tensor of a potential is that its exterior derivative should vanish, $d\mathbf{F}= 0 \Leftrightarrow \exists \mathbf{A}$ such that $\mathbf{F} = d \mathbf{A}$.

\subsection{Beyond fully antisymmetric tensors}

The construction outlined above applies to fully antisymmetric tensors, whose Young diagram consists of a single column. We are now going to give the results of interest concerning the generalization of the concept of exterior derivative to tensors of more complex symmetry\footnote{These tools have been introduced in \cite{DuboisViolette:1999rd,DuboisViolette:2001jk}. See also \cite{XBNB,Coho} for a general discussion of cohomological techniques adapted to tensor fields with mixed Young symmetry of general type.}.

In full generality, one should consider an arbitrary sequence of Young diagrams $Y^p$, $p$ being a positive integer and the number of boxes of $Y^p$. The exterior derivative would send tensor fields of symmetry given by $Y^p$ to tensor fields of symmetry given by $Y^{p+1}$. However, we do not need such an elaborate construction, and we will limit ourselves to the following type of sequence: we will only consider Young diagrams with a maximal number of columns, say $N$. In addition, we will require these Young diagrams to have rows with exactly $N$ boxes, except for the last one. In other words, we are interested in Young diagrams of the form $\left(N , \ldots , N , l\right) \equiv \left(N^k , l\right)$, with $0\leq l \leq N-1$. $d$ will simply be the composition of the projector on the appropriate Young diagram with the partial derivative.

So, we have in mind tensor fields over $\mathbb{R}^D$ whose symmetry is given by Young diagrams of the following structure:

\begin{equation}
\overbrace{\yng(5,5,5,5,3)}^{\text{$N$ boxes}}  . 
\end{equation}

(In the diagram shown above, $N = 5$, $k = 4$ and $l = 3$).

We can then arrange these diagrams in a sequence $Y_N^p$, $p$ being the number of boxes (equivalently, $Y_N^p \equiv \left(N^k , l\right)$, where $p = Nk + l$, with $0\leq k$ and $0\leq l \leq N-1$). If we denote by $\mathbf{Y}_N^p$ the projector of tensors with $p$ indices on the space of tensors whose symmetry is given by the Young diagram $Y_N^p$, our external derivative is given by:

\begin{eqnarray}
d_{(N)} &\equiv& \mathbf{Y}_N^{p+1} \ \circ \ \partial \ \ \mathrm{:} \ \ Y_N^p \ \rightarrow \ Y_N^{p+1} .
\end{eqnarray}

The set of tensor fields over $\mathbb{R}^D$ of symmetry type $Y^p_N$ for any $p$ (for a given $N$) will be denoted by $\Omega_N \left(\mathbb{R}^D\right)$. We defined an internal derivative acting on it, and we could easily define an internal product over it, turning it into a graded (by $p$) Grassmann algebra. It is also called a differential $N$-complex, for reasons that we are going to explain.\\

Since partial derivatives commute and indices are antisymmetrized in each column, and we have no more than $N$ columns, taking $N +1$ times the generalized exterior derivative $d$ must result in the antisymmetrization of two partial derivatives, giving zero. So, we have:

\begin{eqnarray}
0 &=& d_{(N)}^{N+1} .
\end{eqnarray}

As before, this suggests to consider cohomologies. We would like to find $H^p_{(k)} \left(\Omega_N \left(\mathbb{R}^D\right) , d_{(N)}\right)$, which is the set of tensor fields of symmetry type $Y^p_N$ annihilated by $d_{(N)}^k$ \textit{modulo} those which are the image of tensor fields of symmetry type $Y^{p+k-N-1}_n$ through the action of $d_{(N)}^{N+1-k}$.

Contrarily to the simple $N = 1$ case of the $p$-forms, which was developed above, this cohomology is not generally empty for non zero $p$. However, it is empty as long as we consider \textit{well-filled} tensors, whose Young diagram is rectangular. Equivalently,  $H^p_{(k)} \left(\Omega_N \left(\mathbb{R}^D\right) , d_{(N)}\right)$ is empty if $N$ divides $p$ without rest. Thus, we have the following theorem (whose proof can be carried on through the explicit construction of the relevant homotopies):

\begin{theorem}
$H^{mN}_{(k)} \left(\Omega_N \left(\mathbb{R}^D\right) , d_{(N)}\right)=0$, $\forall m \geq 1$. \label{Theorem_cohomlogy}
\end{theorem}

As for $H^0_{(k)}\left(\Omega_N \left(\mathbb{R}^D\right) , d_{(N)}\right)$, one can easily convince oneself that it consists solely of the real polynomials function on $\mathbb{R}^D$ of degree strictly less than $k$.

\newpage

\section{Complete set of gauge invariant functions}
\label{App:Completeness_s_Weyl}

\subsection{Generalities}

Let $\varphi^A$ be some fields invariant under some gauge symmetries,
\be \delta_\xi \varphi^A = k^A_\alpha \xi^\alpha + k^{A i}_\alpha \partial_i \xi^\alpha + k^{Aij}_\alpha \partial_i \partial_j \xi^\alpha \label{AppGauge}
\ee
where for definiteness, we have assumed that the gauge parameters and their derivatives up to second order appear in the gauge transformations.  The discussion would proceed in the same way if there were higher derivatives present in (\ref{AppGauge}).  We also assume that the coefficients $k^A_\alpha$, $k^{A i}_\alpha$ and $k^{Aij}_\alpha$ do not involve the fields, so that the gauge transformations are of zeroth order in the fields (and of course linear in the gauge parameters).

We consider local functions, i.e., functions $f(\varphi^A, \partial_i \varphi^A, \cdots , \partial_{i_1} \partial_{i_2} \cdots \partial_{i_k} \varphi^A)$ of the fields and their derivatives up to some finite but unspecified order.  That unspecified order can depend on $f$.  We denote such local funtions as $f([\varphi^A)])$.  Among the local functions, the gauge invariant ones are particularly important. In our linear theories, non trivial (i.e., not identically constant) local functions that are gauge invariant exist.  For instance, the components of the (linearized) Riemann tensor are local gauge invariant functions under the (linearized) diffeomorphisms.    [Gauge symmetries that involve the fields might not allow for non trivial  local gauge invariant functions.  This occurs for example in the case of full diffeomorphism invariance where even the scalar curvature (say) transforms under change of coordinates, $\delta_\xi  R = \xi^i \partial_i R$ (transport term).]

The local functions are functions on the ``jet spaces'' $J^k$, which can be viewed, in the free theories investigated here, as the vector spaces with coordinates given by the field components $\varphi^A$ and their successive derivatives up to order $k$. The gauge orbits obtained by integrating the gauge transformations are $m$-dimensional planes in those vector spaces $J^k$, where $m$ is the number of independent gauge parameter components and their derivatives (effectively) appearing in the gauge transformations of the fields and their derivatives up to order $k$.  

For instance, for a free spin $3$-field in $3$ dimensions, $J^0$ has dimension $10$ because there are $10$ independent undifferentiated field components $h_{ijk}$.  The gauge orbits have also dimension $10$ since there are $18$ independent derivatives $\partial_{k}\xi_{ij}$ of the gauge parameters  but only $10$ of them, the symmetrized ones $\partial_{(k}\xi_{ij)}$ effectively appear in the gauge transformations. Accordingly, $J^0$ is a single gauge orbit.  Similarly, $J^1$ has dimension $10 + 30 = 40$, the new coordinates being the $30$ derivatives $\partial_m h_{ijk}$ of the fields.  There are $36$ independent second derivatives of the gauge parameters but only $30$ of them effectively act in the gauge transformations of the $\partial_m h_{ijk}$. The jet space $J^1$ reduces again to a single gauge orbit. This is also true for $J^2$. It is only in the jet spaces $J^k$ with $k \geq 3$ that the gauge orbits have a dimension strictly smaller than the dimension of the corresponding jet spaces.  For $J^3$, which has dimension $10$ (number of undifferentiated field components $h_{ijk}$)  $+ 30$ (number of $\partial_m h_{ijk}$) $+ 60$ (number of $\partial_{m}\partial_{n} h_{ijk}$) $ + 100$ (number of $\partial_{m}\partial_{n} \partial_q h_{ijk}$) $ = 200$, the gauge orbits have dimension $10$ (number of effective $\partial_{k}\xi_{ij}$)  $+ 30$ (number of effective $\partial_{k} \partial_m \xi_{ij}$)  $ + 60 $ (number of $\partial_{k} \partial_m \partial_q \xi_{ij}$, which are all effective) $+ 90 $ (number of $\partial_{k} \partial_m \partial_q \partial_r \xi_{ij}$, which are all effective) $= 190$.  Accordingly, the quotient space of $J^3$ by the $190$-dimensional planes generated by the gauge transformations has dimension $10$, which is -- as it should -- the number of independent components of the Riemann tensor, which has Young symmetry
$$\yng(3,3).$$

Without loss of generality, we can assume that the gauge invariant functions vanish when the fields $\varphi^A$ and their derivatives vanish (just substract from $f$ the gauge invariant constant $f(0, 0, \cdots, 0)$). 

A set of gauge invariant functions $\{f_\Delta \}$ is said to form a complete set if any gauge invariant function $f$ can be expressed as a function of the $f_\Delta$, $\delta_\xi f = 0 \Rightarrow f= f (f_\Delta)$.  There might be relations among the $f_\Delta$'s (redundancy) but this will not be of concern to us.   In the linear theories considered here, we can construct complete sets of gauge invariant functions that are linear in the fields and their derivatives. 

Consider a definite jet space $J^k$, with $k$ fixed but arbitrary.  Let $\{f_\Delta \}$ be a complete set of gauge invariant functions. The functions $f^{(k)}_\Delta$ in this complete set that involve derivatives of the fields up to order $k$ are defined in $J^k$.  They provide a coordinate system of the linear quotient space of $J^k$ by the gauge orbits $O_k$ (in case of redundancy, one must take a subset of independent $f^{(k)}_\Delta$).   If this were not the case, one could find a gauge invariant function in $J^k$ not expressible in terms of the functions in the complete set.  The trivial orbit of the pure gauge field configurations is the orbit of $0$, on which the gauge invariant functions have been adjusted to vanish.  It follows from these observations that {\em a set  $\{f_\Delta \}$ of gauge invariant functions is a complete set if and only if the condition $f_\Delta = 0$ implies that the fields are pure gauge.}

\subsection{Spin-$s$ Weyl invariance}
We now turn to the proof that a complete set of invariants for higher spin conformal fields in three dimensions is given by the Cotton tensor and its successive derivatives. As we just shown, this is equivalent to the statement that the vanishing of the Cotton tensor implies that the spin-$s$ field is pure gauge.

To determine a complete set of invariants,  we reformulate  the problem as a problem of cohomogy in the successive jet spaces augmented by new fermionic variables, ``the ghosts", and decompose the successive derivatives in irreducible representations of $GL(3)$.  This approach is standard and has been developed successively in the case of spin-$s$ diffeormorphism invariance for  spin $1$ \cite{Dixon:1991wi,Bandelloni:1986wz,Bandelloni2,Brandt:1989gy,Brandt:1990,Brandt:1991,DuboisViolette:1992ye,Henneaux:1993jn}, spin $2$ \cite{Boulanger:2000rq} and spin $s$ \cite{Bekaert:2005ka}.

Weyl invariance for spin-$2$ was treated in \cite{Boulanger:2001he}.  By the same techniques as those developed in that reference, one first takes care of spin-$s$ diffeomorphism invariance and concludes  that diffeomorphism invariance forces the local functions to be functions of the Riemann tensor and its derivatives, or, what is the same in $D=3$, of the Schouten tensor and its derivatives, $f = f([S])$.  Spin-$s$ Weyl invariance becomes then the condition $\delta_\nu f = 0$ for $\delta_\nu S_{i_1 \cdots i_s} = -\partial_{(i_1 i_2} \nu_{i_3 \cdots i_s)}$ (for convenience, we absorb the factor $\frac{s(s-1)}{2}$  in a redefinition of $\nu$).  Furthermore, neither the Schouten tensor nor the gauge parameter $\nu$ are independent since their divergences are constrained by (\ref{Schouten_s_Bianchi}) and (\ref{Nu_s_Bianchi}).

To investigate the problem of Weyl invariance, we shall first consider the problem $\delta_\nu f = 0$ for $\delta_\nu S_{i_1 \cdots i_s} = -\partial_{(i_1 i_2} \nu_{i_3 \cdots i_s)}$ for unconstrained $S$ and $\nu$. We shall then analyse the implications of the constraints (\ref{Schouten_s_Bianchi}) and (\ref{Nu_s_Bianchi}) on the divergences.

We thus consider the problem of computing the cohomology at ``ghost number" zero of the differential $\gamma$ defined by 
\be
\gamma S_{i_1 \cdots i_s} = \partial_{(i_1 i_2} C_{i_3 \cdots i_s)}, \, \; \; \gamma C_{i_1 \cdots i_{s-2}} = 0 \label{AppGammaForS} \ee
We introduce a derivative degree that gives weight zero to the ghosts and weight two to the Schouten tensor.

At derivative degree $0$, we have only the ghosts in the cohomology, but these are at ghost number one, so there is no cohomology at ghost number zero.  At derivative degree $1$, there is again no cohomology at ghost number zero for a similar reason.  The ghost-number-zero variables (Schouten tensor) appear only in derivative degree $2$ and higher.

At derivative degree $2$,  the second derivatives of the ghosts transform in the representation
\begin{eqnarray}
&&  \overbrace{\yng(8)}^{\text{$s-2$ boxes}} \otimes \yng(2) \nonumber \\
&& = \overbrace{\yng(10)}^{\text{$s$ boxes}} \nonumber \\
&& \oplus \overbrace{\yng(9,1)}^{\text{$s-1$ boxes}} \nonumber \\
&& \oplus  \overbrace{\yng(8,2)}^{\text{$s-2$ boxes}} \nonumber
\end{eqnarray}
while the undifferentiated Schouten tensor components transform in the representation
$$ \overbrace{\yng(10)}^{\text{$s$ boxes}}.$$
It follows that the undifferentiated Schouten tensor components form contractible pairs with the derivatives of the ghosts transforming in the same representation and disappear from the cohomology.   There is no cohomology at ghost number zero.  The same story proceeds in the same way, with the derivatives of the Schouten tensor being all ``eaten" through contractible pairs with the corresponding derivatives of the ghosts and no cohomology at ghost number zero, with non trivial generators at ghost number one left over, up to derivative degree $s$.  There one finds for the ghosts:
\begin{eqnarray}
&&  \overbrace{\yng(8)}^{\text{$s-2$ boxes}} \otimes \overbrace{\yng(10)}^{\text{$s$ boxes}} \nonumber \\
&& = \overbrace{\yng(18)}^{\text{$2s -2$ boxes}} \nonumber \\
&& \oplus \overbrace{\yng(17,1)}^{\text{$2s-3$ boxes}} \nonumber \\
&& \oplus \cdots \nonumber \\
&& \oplus  \overbrace{\yng(11,7)}^{\text{$s+1$ boxes}} \nonumber \\
&& \oplus  \overbrace{\yng(10,8)}^{\text{$s$ boxes}} \label{AppWeylGhost}
\end{eqnarray}
and exactly the same decomposition for the representation in which the derivatives of order $s-2$ of the Schouten tensor transform,
$$ \overbrace{\yng(10)}^{\text{$s$ boxes}} \otimes \overbrace{\yng(8)}^{\text{$s-2$ boxes}}.$$
There is exact matching and the generators of derivatives order $s$ form contractible pairs and do not contribute to the cohomology.

At higher derivative order, it is now some of the derivatives of the Schouten tensor that are unmatched, namely those which contain the Cotton tensor
$$\underbrace{ \overbrace{\yng(10,9)}}^{\text{$s$ boxes}}_{\text{$s-1$ boxes}}$$
since these representations (and only those) cannot arise in the decomposition of the derivatives of the ghosts of order $t>s$ 
$$\overbrace{\yng(8)}^{\text{$s-2$ boxes}} \otimes \overbrace{\yng(12)}^{\text{$t>s$ boxes}}$$
(the lower line can have at most length $s-2$ as shown by (\ref{AppWeylGhost})).

Accordingly, we can conclude that the $\gamma$-cohomology of the differential defined by (\ref{AppGammaForS}), with unconstrained variables, is given at ghost number zero by the functions $f([C])$ of the Cotton tensor and its derivatives.

We did not take into account so far the constraints (\ref{Schouten_s_Bianchi}) and (\ref{Nu_s_Bianchi}) that the Schouten tensor should obey the Bianchi identity and that the divergence of  the ghost is also determined by its trace.   One must verify that the derivatives of the Schouten tensor that were trivial in the $\gamma$-cohomology because they had an independent ghost partner equal to their $\gamma$-variation, either vanish on account of the constraints or, if they do not vanish, that their ghost partner in the trivial pair also remains different from zero so that both elements in the trivial pair continue being trivial.   

It is easy to convince oneself that this is the case.  The derivatives of the Schouten tensor that remain non-zero after the Bianchi identity has been taken into account may be assumed not to involve a contraction of one derivative index $\partial_i$ with an index of the Schouten tensor, since such terms can be eliminated using the Bianchi identity.  In fact, once we have eliminated such contractions, the remaining derivatives are unconstrained.  A similar situation holds on the ghost side.  If the $\gamma$-variation of a derivative of the Schouten tensor without such contractions involves the ghosts and so is not $\gamma$-closed before the constraints are taken into account, it will remain so after the constraints are taken into account because its $\gamma$-variation necessarily produce independent derivatives of the ghosts without such contractions (in addition to possible terms with such contractions coming from possible traces).

\end{subappendices}

\chapter{Hamiltonian analysis}

\label{Chap:hambos}

Now equipped with the knowledge we have acquired on higher spin conformal geometry, we are ready to make a systematic study of the Hamiltonian dynamics of free massless higher spin fields over a flat four dimensional space-time. 

Indeed, similarly to what happened with the well-known lower spin cases (which we studied in section \textbf{\ref{Sec:EM_ham}} and \textbf{\ref{Sec:Grav_ham}}), and as can be implied from the general theory of gauge systems (as we saw in section \textbf{\ref{Sec:Ham_form_constr_gauge}}), we will find out that massless higher spin have a constrained first order formalism, whose constrains will be solved in a very natural way through the use of conformal curvatures.

In order to keep the logic visible, we will begin with an extensive study of the spin-$3$ case, before moving on to the arbitrary spin. The development will be along the same lines in each case: expand Fronsdal action (\ref{action_bos_compact}) in order to extract the momenta and compute the Hamiltonian and the constraints; then, check that these are first class (that the constraints generate gauge transformations, which preserve both the constraints and the Hamiltonian); eventually, solve the constraints through the introduction of prepotentials in terms of which the Hamiltonian action will be rewritten. 

The more technical features observed in the case of spin one or two (sections \textbf{\ref{Sec:EM_ham}} and \textbf{\ref{Sec:Grav_ham}}) will also reappear: the prepotentials will be two spatial symmetric tensors, of the same rank as the initial field, and they will enjoy a gauge invariance including both diffeomorphism and Weyl rescaling.

\section{Spin-$3$}

Our starting point is the Lagrangian formulation introduced for arbitrary spin in section \textbf{\ref{Sec:Lag_s_bos}}. The spin-$3$ field is described by a symmetric covariant tensor field $h_{\mu\nu\rho}$ with the gauge symmetry:
\begin{eqnarray} \label{gauge_lag_3}
\delta h_{\mu\nu\rho} &=& 3 \ \partial_{(\mu} \xi_{\nu\rho )},
\end{eqnarray}
\noindent where $\xi_{\mu\nu}$ is a symmetric traceless tensor: $\xi'=0$. The field itself is unconstrained.

With such a gauge invariance, we can build the second order gauge invariant curvature (which is the Fronsdal tensor, defined for any spin in (\ref{Fronsdal_bos})):
\begin{eqnarray}
\mathcal{F}_{\mu\nu\rho} &=&
\square h_{\mu\nu\rho} 
\ - \ 3 \ \partial^{\sigma} \partial_{(\mu} h_{\nu \rho ) \sigma}
\ + \ 3 \ \partial_{(\mu} \partial_{\nu} h'_{\rho )} .
\end{eqnarray}

The equations of motion of the spin-$3$ are the vanishing of this tensor, and they can be seen to follow from the following gauge invariant action (introduced for an arbitrary spin in (\ref{action_bos_full})):
\begin{eqnarray} \label{action_lag_3}
S &=& - \ \frac{1}{2} \ \int \ d^4 x \ \left\lbrace
 \partial_{\sigma} h_{\mu\nu\rho}\partial^{\sigma} h^{\mu\nu\rho}
\ - \ 3 \ \partial^{\sigma} h_{\sigma\mu\nu} \partial_{\rho} h^{\rho\mu\nu}
\ + \ 6 \ \partial^{\rho} h_{\rho\mu\nu} \partial^{\mu} h'^{\nu}
\right.
\nonumber \\ && \qquad \qquad \qquad
\left.
\ - \ 3 \ \partial_{\mu} h'_{\nu} \partial^{\mu} h'^{\nu}
\ - \ \frac{3}{2} \ \partial^{\nu} h'_{\nu} \partial_{\mu} h'^{\mu} 
\right\rbrace  .
\end{eqnarray}

\subsection{Hamiltonian and constraints}

In order to make the dynamic of this system manifest, we have to break manifest Lorentz covariant and expand this action in order to identify how many time derivatives each term incorporates. We will quickly observe that the only terms of the action (\ref{action_lag_3}) containing two time derivatives are:
\begin{eqnarray}
S &=&  \int \ d^4 x \ \left\lbrace
\frac{1}{2} \ \dot{h}^{ijk} \left(\dot{h}_{ijk} \ - \ 3 \ \delta_{ij} \dot{\bar{h}}_k \right)
\ + \ \frac{1}{4} \ \left(\dot{h}_{000} \ - \ 3 \ \dot{\bar{h}}_0\right)^2 
\ + \ \ldots
\right\rbrace .
\end{eqnarray}

This suggests to define the dynamical variable:
$$\alpha \equiv h_{000} \ - \ 3 \ \bar{h}_0 ,$$
\noindent in order to have the following non vanishing momenta (from the Lagrangian $S \equiv \int \ dt \ L$):
\begin{eqnarray} \label{momenta_3}
\Pi^{ijk} &\equiv& \frac{\delta L}{\delta \dot{h}_{ijk}}
\nonumber \\ &=&
\dot{h}^{ijk} \ - \ 3 \ \delta^{(ij} \dot{\bar{h}}^{k)}
\ + \ \frac{3}{2} \ \delta^{(ij} \partial^{k)} \alpha ,
\\
\tilde{\Pi} &\equiv& \frac{\delta L}{\delta \dot{\alpha}}
= \frac{\dot{\alpha}}{2} .
\end{eqnarray}

The Legendre transformation then brings us to the following Hamiltonian action (if we also rename the Lagrange multipliers $N_{ij} \equiv h_{0ij}$ and $\mathcal{N}_i \equiv h_{00i}$):
\begin{eqnarray}
S\left[h_{ijk}, \alpha, \Pi^{ijk}, \tilde{\Pi}, N^{ij} , \mathcal{N}^i\right] &=&
\int \ d^4 x \ \left\lbrace\Pi^{ijk} \dot{h}_{ijk} \ + \ \tilde{\Pi} \dot{\alpha}
\ - \ \mathcal{H} \ - \ \mathcal{N}^i \mathcal{C}_i \ - \ N^{ij} C_{ij} \right\rbrace 
, \label{action_3_ham}
\end{eqnarray}
\noindent where the Hamiltonian and the constraints are given by:
\begin{eqnarray}
\mathcal{H} &=&
\frac{1}{2} \ \Pi_{ijk} \Pi^{ijk}
 - \frac{3}{8} \ \bar{\Pi}_k \bar{\Pi}^k
 + \frac{3}{8} \ \bar{\Pi}^k \partial_k \alpha
 + \frac{17}{32} \ \partial_k \alpha \partial^k \alpha + \tilde{\Pi}^2  
 + \frac{1}{2} \ \partial_{k} h_{lmn}\partial^{k} h^{lmn}
 \nonumber\\ \qquad &&
  - \frac{3}{2} \  \partial^{l} h_{lmn} \partial_{k} h^{kmn}
 + 3 \ \partial^{l} h_{klm} \partial^{k} \bar{h}^{m}
 - \frac{3}{2} \ \partial_{k} \bar{h}_{l} \partial^{k} \bar{h}^{l}
 - \frac{3}{4} \ \partial_{k} \bar{h}_{l} \partial^{l} \bar{h}^{k} 
 , \label{ham_3_h}
\\
\mathcal{C}_i &=& 3 \ \left\lbrace \partial_i \tilde{\Pi} - \Delta \bar{h}_i  + \partial^j \partial^k h_{ijk}  - \frac12 \partial_i \partial^j \bar{h}_j  \right\rbrace , \label{constr_mom_3}
\\
C_{ij} &=& - \ 3 \ \left\lbrace \partial^k \pi _{ijk} + \frac{1}{2} \ \delta_{ij} \Delta \alpha \right\rbrace 
. \label{constr_ham_3}
\end{eqnarray}

The constraint $0 \approx \mathcal{C}_i$ shall be referred to as the \textit{Hamiltonian constraint} and the constraint $0 \approx C_{ij}$ as the \textit{Momentum constraint}.

As we shall see, this Hamiltonian and these constraints are all first class\footnote{Strictly speaking, in the formalism of section \textbf{\ref{Sec:Ham_form_constr_gen}}, the primary constraints are $0 \approx \Pi^{0ij}$ and $0 \approx \Pi^{00i}$. The consistency checks (enforcing the conservation in time of these constraints) then leads us to the secondary constraints $0 \approx \mathcal{C}_i$ and $0 \approx C_{ij}$. Together, all these constraints are first class (and also each set separately). We will not mention the primary constraints any more, since the gauge transformations they generate only serve to show that the Lagrange multipliers $\mathcal{N}_i$ and $N_{ij}$ are pure gauge, which we take into account in the following.}. This will be made clear by examining the transformations generated by the constraints.

\subsection{Gauge transformations}

Let us define the following generator:
\begin{eqnarray}
\mathcal{G} &=& 
\int \ d^3 x \ \left(\xi^{ij} C_{ij}
\ + \ \theta^i \mathcal{C}_i \right).
\end{eqnarray}

The transformations it generates are obtained by taking its Poisson bracket (neglecting, for now, boundary terms) with the variables:
\begin{eqnarray} \label{gauge_transf_ham_3}
\delta h_{ijk} &=& \left\lbrace h_{ijk} , \mathcal{G} \right\rbrace
= 3 \ \partial_{(i} \xi_{jk)} ,
\\
\delta \alpha &=& \left\lbrace \alpha , \mathcal{G} \right\rbrace
= - \ 3 \ \partial^k \theta_k ,
\\ 
\delta \Pi^{ijk} &=& \left\lbrace \Pi^{ijk} , \mathcal{G} \right\rbrace
= 
- \ 3 \ \partial^{(i} \partial^j \theta^{k)}
\ + \ 3 \ \delta^{(ij} \left( \Delta \theta^{k)} \ + \ \frac{1}{2} \ \partial^{k)} \partial^l \theta_l \right) ,
\\
\delta \tilde{\Pi} & =& \left\lbrace \tilde{\Pi} , \mathcal{G} \right\rbrace = \frac{3}{2} \ \Delta \bar{\xi} .
\end{eqnarray}

The first two of these transformations, generated by the momentum constraint (\ref{constr_mom_3}), are easily seen to precisely reproduce the spin-$3$ spatial diffeomorphism, while the last two, generated by the Hamiltonian constraint (\ref{constr_ham_3}), correspond to the ``temporal" spin-$3$ diffeomorphism (associated to the Lagrangian gauge parameter $\xi_{0i} \thicksim \theta_i$).

These transformations are straightforwardly checked to preserve the constraints and the Hamiltonian, which are confirmed to be first class, these transformations being identified as a change of gauge.

Incidentally, the transformations (\ref{gauge_transf_ham_3}) are precisely those one would have obtained from the covariant, Lagrangian transformation (\ref{gauge_lag_3}) (with the redefinition $\theta_i \equiv \xi_{0i}$, $\xi_{00}$ being removed through the solving of the tracelessness constraint), using the expression of the momenta in terms of the fields (\ref{momenta_3}).

\subsection{Momentum constraint}

\label{Sec:3_solv_mom}

We first solve the momentum constraint. Using the $\theta$-gauge transformations, one can set $\alpha = 0$. In that gauge, the constraint reduces to $ \partial_i \pi^{ijk} = 0$, which implies $\pi^{ijk} = G^{ijk}[P]$ for some prepotential $P_{ijk}$, which is at this stage determined up to a spin-$3$ diffeomorphism (as follows immediately from the corollary \textbf{\ref{corollary_Einstein2}} of section \textbf{\ref{Sec:Einstein_s}}).

In a general gauge, one has therefore
\begin{eqnarray}
&&\pi^{ijk} = G^{ijk}[P]  -  \partial^{(i} \partial^j \Xi^{k)} \nonumber \\
&& \hspace{1cm} +  \delta^{(ij}  \left(  \Delta \Xi^{k)} + \frac12 \partial^{k)} \partial_m \Xi^m \right)  \label{PrepoForpi}\\
&&\alpha = -  \partial_m \Xi^m 
\end{eqnarray}
where $\Xi_k$ is a second prepotential that describes the gauge freedom of $\pi^{ijk}$ and $\alpha$.

Now, the vector $\Xi^k$ can be decomposed into a transverse and a longitudinal piece,
$$\Xi^k = \varepsilon^{kij} \partial_i \lambda_j + \partial^k \theta.$$  The $\lambda^k$-terms in (\ref{PrepoForpi}) are easily checked to be of the form $G^{ijk}[\varphi]$, where $\varphi_{ijk}=\delta_{(ij} \lambda_{k)}$ has just the form of a spin-$3$ Weyl transformation.  This shows that the prepotential $P_{ijk}$ is determined up to a spin-$3$ Weyl transformation -- in addition to the spin-$3$ diffeomorphism invariance pointed out above.  Therefore, the gauge freedom of the prepotential is
\be
\delta P_{ijk} = 3 \ \partial_{(i} \xi_{jk)} + 3 \ \delta_{(ij} \lambda_{k)},
\ee
i.e., the gauge symmetries of a conformal spin-$3$ field.

The fact that the spin-$3$ diffeomorphisms of the prepotential $P_{ijk}$ have no action on the canonical variables, while its conformal transformations generate (some of) the gauge transformations associated with the Hamiltonian constraint, parallels the situation found in the case of spin 2, as we recalled in section \textbf{\ref{Sec:Grav_ham}}  \footnote{This is of course well known. See \cite{Henneaux:2004jw,Bunster:2012km,Bunster:2013oaa}.}.  There, however, the Weyl transformations of the prepotential accounted for {\em all} the gauge symmetries generated by the Hamiltonian constraint. 

\subsection{Hamiltonian constraint}

\label{Sec:solv_contr_ham_3}

We now turn to solving the Hamiltonian constraint. Its curl $\epsilon^{ijk} \partial_j {\mathcal C}_k $ does not involve $\tilde{\Pi}$ and turns out to be equal to $\bar{G}^i$, the trace of the Einstein tensor of $h_{ijk}$, so that the Hamiltonian constraint implies:
\be 
\bar{G}^i[h] = 0.
\ee

In fact, one may rewrite the Hamiltonian constraint as:
\be
\partial_i \tilde{\Pi} - \Psi_i = 0
\ee 
in terms of the $\Psi$ introduced in section {\bf \ref{Sec:Einstein_traceless}}, such that $^* \bar{G} = d \Psi$.  One has explicitly:
\be
\Psi_i = \Delta \bar{h}_i - \partial^j \partial^k h_{ijk} + \frac12 \partial_i \partial^j \bar{h}_j .
\ee
Therefore, the equations $\bar{G}^i =0 \Leftrightarrow d \Psi = 0$ and $d \Pi - \tilde{\Psi}=0$ are two equivalent versions of the Hamiltonian constraint\footnote{Let us rephrase the argument, in order to be perfectly clear: the vanishing of the trace of the Einstein tensor of $h$ is equivalent to the vanishing of $d\Psi$. Because of the absence of cohomology, this is equivalent to $\Psi$ being exact, that is, equal to $d\tilde{\Pi}$ for some scalar $\tilde{\Pi}$ - which is precisely what the Hamiltonian constraint requires.}.  

The form $\bar{G}^i = 0$  is more amenable to solution because it falls precisely under the lemma {\bf \ref{lemma_Cotton_for_any_traceless}} of section \textbf{\ref{Sec:Cotton_for_any_traceless}}.  According to what we proved in section \textbf{\ref{Sec:Cotton_for_any_traceless}}, it implies the existence of a (second) prepotential $\Phi_{ijk}$ such that the Einstein tensor of $h$ is the Cotton tensor of that prepotential,
\be G^{ijk}[h] = B^{ijk}[\Phi] \label{3_G=B} .
\ee
A particular solution of (\ref{3_G=B}) is given by
\begin{eqnarray}
h_{ijk} &=& - \Delta \Phi_{ijk} + \frac34 \delta_{(ij} \triangle \bar{\Phi}_{k)} \nonumber \\
&& - \frac34 \delta_{(ij} \partial^r \partial^s \Phi_{k)rs} + \frac{3}{10} \delta_{(ij} \partial_{k)} \partial^r \bar{\Phi}_r . \label{3_SolG=B}
\end{eqnarray}

The last term in (\ref{3_SolG=B}) is not necessary but included so that $\delta h_{ijk} = 0$ under Weyl transformation of $\Phi$. 

Now, what are the ambiguities? It is clear that the spin-$3$ field $h_{ijk}$ is determined by (\ref{3_G=B})  up to a spin-$3$ diffeomorphism, so that the general solution of (\ref{3_G=B}) is given by (\ref{3_SolG=B}) plus $\partial_{(i} u_{jk)}$ where $u_{jk}$ may be thought of as another prepotential that drops out because of gauge invariance.   Conversely, the prepotential $\Phi_{ijk}$  itself is determined by (\ref{3_G=B}), i.e., by its Cotton tensor, up to a diffeomorphism and a Weyl rescaling,
\be
\delta \Phi_{ijk} = 3 \ \partial_{(i} \xi'_{jk)} + 3 \ \delta_{(ij} \lambda'_{k)} \label{GaugePhi}
\ee
with independent gauge parameters $\xi'_{ij}$ and $\lambda'_j$. Thus, we see that the resolution of the Hamiltonian constraints also introduces a prepotential possessing the gauge symmetries of a conformal spin-$3$ field.  Note that we have adjusted the ambiguity in the dependence of $h_{ijk}$ on $\Phi_{ijk}$ in such a way that the conformal spin-$3$ transformations of the prepotential leave $h_{ijk}$ invariant, while the spin-$3$ diffeomorphisms of the prepotential induce particular spin-$3$ diffeomorphisms of $h_{ijk}$, as is the case for spin 2 \cite{Henneaux:2004jw,Bunster:2012km,Bunster:2013oaa}.

Once $h_{ijk}$ is determined, one may work one's way up to the constraint and solve for $\Pi$ in terms of the prepotential.  One finds \be
\tilde{\Pi} = - \ \frac18 \ \partial^{i}\partial^j\partial^{k} \Phi_{ijk} + \frac{3}{40} \ \triangle \partial^i \bar{\Phi}_i \ee

The use of conformal techniques to solve the Hamiltonian constraint is somewhat reminiscent of the approach to the initial value problem for full general relativity developed in  \cite{Arnowitt:1962hi,Deser:1967zzb,York:1971hw}.

\subsection{Hamiltonian action in terms of prepotentials}

When their expressions in terms of the prepotentials is substituted for the fields and the momenta, the Hamiltonian action (\ref{action_3_ham}) takes the following form:
\begin{eqnarray}
S &=&
\frac{1}{2} \ \int \ d^4 x \ \left\lbrace
 \varepsilon_{ab} \dot{Z}^a_{ijk} B^{b ijk} 
\ - \ \delta_{ab} \ \left(G^{a ijk} G^b_{ijk} \ - \ \frac{3}{4} \ \bar{G}^{ai} \bar{G}^b_i\right)
\right\rbrace ,
\end{eqnarray}
\noindent where the prepotentials were gathered into a two-vector $(Z^a_{ijk}) \equiv (P_{ijk}, \Phi_{ijk})$ ($a= 1,2)$ and $\varepsilon_{ab}$ and $\delta_{ab}$ are respectively the Levi-Civita tensor and the Euclidean metric in the internal plane of the two prepotentials, while $G^a_{ijk} \equiv G_{ijk}[Z^a]$ and $B^a_{ijk} \equiv B_{ijk}[Z^a]$.  In terms of the prepotentials, the action possesses exactly the same structure as the action for spin 2 (\ref{action_2_ham_prepot}) \footnote{This expression was, as we said, originally obtained in \cite{Bunster:2012km}}.

The kinetic term in the action is manifestly invariant under the gauge symmetries of the prepotentials. The Hamiltonian is  manifestly invariant under the spin-$3$ diffeomorphisms, since it involves the Einstein tensors of the prepotentials.  It is also invariant under spin-$3$ Weyl transformations up to a surface term, as it can easily be verified.

In terms of the prepotentials, the equations of motion read
\be
\dot{B}^a_{ijk} = \epsilon^{ab} \epsilon_{ilm} \partial^l B_{b\, jk}^{\ \ \ m}  \label{EOM_3_prepot}
\ee
and equate the time derivative of the Cotton tensor of one prepotential to the ``curl" of the Cotton tensor of the other (defined as the right-hand side of (\ref{EOM_3_prepot})).

\section{Arbitrary spin}\label{sec:spins}

Again, our starting point is the Lagrangian formulation introduced for arbitrary spin in section \textbf{\ref{Sec:Lag_s_bos}}. The spin-$s$ field is described by a symmetric covariant tensor field $h_{\mu_1 \cdots \mu_s}$ with the gauge symmetry:
\begin{eqnarray} \label{gauge_lag_s}
\delta h_{\mu_1 \cdots \mu_s} &=& s \ \partial_{(\mu_1} \xi_{\mu_2 \cdots \mu_s )},
\end{eqnarray}
\noindent where $\xi_{\mu1 \cdots \mu_s}$ is a symmetric traceless tensor: $\xi_{\mu_3 \cdots \mu_s}'=0$. The field itself is subjected to the constraint that its double trace vanish: $h''_{\mu_5 \cdots \mu_s} = 0$.

The equations of motion of the spin-$s$ are the vanishing of its Fronsdal tensor (\ref{Fronsdal_bos}), and they can be seen to follow from the following gauge invariant action (this is exactly (\ref{action_bos_full})):
\begin{eqnarray} \label{action_lag_s}
S &=& - \ \frac{1}{2} \ \int \ d^4 x \ \left\lbrace
 \partial_{\nu} h_{\mu_1 \cdots\mu_s}\partial^{\nu} h^{\mu_1 \cdots\mu_s}
\ - \ s \ \partial^{\nu} h_{\nu\mu_2\cdots\mu_s} \partial_{\rho} h^{\rho\mu_2\cdots\mu_s}
\right.
\nonumber \\ && \qquad \qquad \qquad
\left.
\ + \ s\left( s - 1 \right) \ \partial^{\nu} h_{\nu\rho\mu_3\cdots\mu_s} \partial^{\rho} h'^{\mu_3\cdots\mu_s}
\ - \ \frac{s\left(s - 1 \right)}{2} \ \partial_{\rho} h'_{\mu_3 \cdots\mu_s} \partial^{\rho} h'^{\mu_3 \cdots\mu_s}
\right.
\nonumber \\ && \qquad \qquad \qquad
\left.
\ - \ \frac{s\left(s - 1\right)\left(s - 2 \right)}{4} \ \partial^{\nu} h'_{\nu\mu_4\cdots\mu_s} \partial_{\rho} h'^{\rho\mu_4\cdots\mu_s} 
\right\rbrace   .
\end{eqnarray}

\subsection{Change of variables and momenta}

We have to break explicit Lorentz invariance and expand this action, to obtain its Hamiltonian formalism. In order to do that, one has to solve the double-trace constraint on the field $h$; one can choose \textit{e.g.} to encode the independent components of the spin-$s$ field in the traceful spatial tensors $h_{i_1 \cdots i_s}$, $h_{0i_2 \cdots i_{s}}$, $h_{00i_3 \cdots i_{s}}$ and $h_{000i_4 \cdots i_s}$. The remaining components of the field are dependent on those and are recursively checked to be equal:
\begin{eqnarray}
h_{0\ldots 0 k_{2n + 1} \ldots k_s} &=&
n \ h^{[n-1]}_{00 k_{2n + 1} \ldots k_s}
\ - \ \left(n - 1 \right) \ h^{[n]}_{k_{2n + 1} \ldots k_s} ,
\nonumber \\ 
h_{0\ldots 0 k_{2n + 2} \ldots k_s} &=&
n \ h^{[n-1]}_{000 k_{2n + 2} \ldots k_s}
\ - \ \left(n - 1 \right) \ h^{[n]}_{0 k_{2n + 2} \ldots k_s} .
\end{eqnarray}

Within this set one has to distinguish between canonical variables - having conjugate momenta such that the Legendre transformation is invertible - and Lagrange multipliers. Comparison with what happened for the spin $3$ and the observation that combinations whose gauge variation \eqref{action_lag_s} does not contain time derivatives are canonical variables leads to the conclusion that these are the spatial components $h_{i_1 \cdots i_s}$ of the field and:
\begin{equation}\label{alpha_s_def}
\a_{i_4 \cdots i_{s}} \equiv  h_{000i_4 \cdots i_{s}} - 3 \ \bar{h}_{0i_4 \cdots i_{s}} .
\end{equation}

The remaining independent components of the covariant field are Lagrange multipliers (see their gauge transformations in \eqref{gauge_ham_s_N1} and \eqref{gauge_ham_s_N2}). We denote them as
\begin{eqnarray}
N_{i_2 \cdots i_s} &\equiv& h_{0 i_2 \cdots i_s},
\\
\mathcal{N}_{i_3 \cdots i_s} &\equiv& h_{00i_3 \cdots i_s} .
\end{eqnarray}

The expansion of the action (\ref{action_lag_s}) confirms our choice of variables, the time derivatives of the Lagrange multipliers indeed being absent. The non vanishing momenta are (with the Lagrangian $S \equiv \int \ dt \ L$):
\begin{eqnarray}
\Pi^{i_1 \cdots i_s} &\equiv& 
\frac{\delta L}{\delta \dot{h}_{i_1\cdots i_s}}
\nonumber \\ &=&
\sum_{n = 0}^{\left[s/2\right]}
\left(\begin{array}{c} s \\ 2n \end{array}\right) \delta^{(i_1 i_2} \cdots \delta^{i_{2n-1} i_{2n}\vert} \left\lbrace
\left(1-2n\right) \ \dot{h}^{[n] \vert i_{2n+1}\cdots i_s )}
\right.
\nonumber \\ && 
\left.
\ + \ \left(2n-1\right)\left(s-2n\right) \ \partial^{\vert i_{2n+1}} N^{[n] i_{2n+2} \cdots i_s )}
\ + \ 2n\left(2n-1\right) \ \partial_k N^{[n_1]k \vert i_{2n+1} \cdots i_s )}
\right.
\nonumber \\ && 
\left.
\ + \ \frac{n\left(s-2n\right)}{2} \ \partial^{\vert i_{2n+1}} \alpha^{[n-1]i_{2n+2} \cdots i_s )}
\ + \ n\left(n-1\right) \ \partial_k \alpha^{[n-2]k \vert i_{2n+1} \cdots i_s )}
\right\rbrace , \label{momenta_s_index_1}
\\
\tilde{\Pi}^{i_4 \cdots i_s} &\equiv& 
\frac{\delta L}{\delta \dot{\alpha}_{i_4\cdots i_s}}
\nonumber \\ &=&
\sum_{n = 1}^{\left[(s-1)/2\right]}
\frac{n}{2}
\left(\begin{array}{c} s \\ 2n + 1 \end{array}\right) \delta^{(i_4 i_5} \cdots \delta^{i_{2n} i_{2n+1}\vert} \left\lbrace
\dot{\alpha}^{[n-1] \vert i_{2n+2} \cdots i_s )}
\right.
\nonumber \\ && 
\left.
\ + \ \left(s-2n-1\right) \ \partial^{\vert i_{2n+2}} \mathcal{N}^{[n] i_{2n+3} \cdots i_{s} )}
\ + \ \left(2n + 1\right) \partial_k \mathcal{N}^{[n-1] k \vert i_{2n+2} \cdots i_s )}
\right\rbrace . \label{momenta_s_index_2}
\end{eqnarray}

It is quite apparent from such expressions that an index-free notation would be much preferable, and we will use it to write sums of this type. In order to be explicit, here is a rewriting of these formulas in our convention (where symmetrization is carried with weight one):
\begin{eqnarray} 
\Pi &=& \sum_{n=0}^{\left[\frac{s}{2}\right]} \!\binom{s}{2n} \delta^n 
\left\lbrace 
(1-2n) \ \dot{h}^{[n]} 
\ + \ (2n-1)(s-2n) \ \partial N^{[n]} 
\ + \ 2n(2n-1) \ \partial\cdot\! N^{[n-1]}
\right.
\nonumber \\ && 
\left. \qquad \qquad \qquad
\ + \ \frac{n(s-2n)}{2}\   \partial \a^{[n-1]} 
\ + \ n(n-1) \ \partial\cdot \a^{[n-2]}  \right\rbrace 
, \label{momenta_s_no_index_1}
\\
\tilde{\Pi} &=& 
\sum_{n=1}^{\left[\frac{s-1}{2}\right]} 
\frac{n}{2} \binom{s}{2n+1} \delta^{n-1} 
\left\lbrace 
\dot{\alpha}^{[n-1]} 
\ + \ (s-2n-1) \ \partial \cN^{[n]} 
\ + \ (2n+1) \ \partial \cdot \cN^{[n-1]} \right\rbrace 
. \qquad \label{momenta_s_no_index_2}
\end{eqnarray}

\subsection{Gauge transformations}

The change of variables (\ref{alpha_s_def}) was confirmed by the computations of the momenta, but we also argued that it could be guessed from an examination of the gauge transformations, the canonical variables being precisely those whose gauge variation does not contain a time derivative of the gauge parameters. We are now going to write down exactly these gauge transformations, extracted from the covariant expression (\ref{gauge_lag_s}). We will then use these expressions to compute the constraints of the Hamiltonian formalism (in the previous section, when studying the spin $3$, we instead computed the constraints directly from the Hamiltonian procedure, before checking that they generate the same transformations as the ones obtained from their Lagrangian form).

Starting from (\ref{gauge_lag_s}) and redefining $\theta_{i_3 \cdots i_s} \equiv \xi_{0i_3 \cdots i_s}$, we easily obtain (using, again, our index-free notation - all the indices being spatial):
\begin{eqnarray}
\delta h &=& 
s \ \partial \xi , \label{gauge_ham_s_h}
\\
\delta \alpha &=& 
- \ 3 \ \partial \cdot \theta \ - \ \left(s-3\right) \ \partial \bar{\theta}, \label{gauge_ham_s_alpha}
\\
\delta N &=& 
\dot{\xi} \ + \ \frac{\left(s-1\right)}{2} \ \partial \theta , \ \label{gauge_ham_s_N1}
\\
\delta \mathcal{N} &=& 
\dot{\theta} \ + \ \left(s-2\right) \ \partial \bar{\xi} , \label{gauge_ham_s_N2}
\end{eqnarray}
\noindent and, using the expression of the momenta in terms of the fields (\ref{momenta_s_index_1}) and (\ref{momenta_s_index_2}):
\begin{eqnarray}
\delta \Pi &=&
\sum_{n = 0}^{\left[s/2\right]} \ 
\left(\begin{array}{c} s \\ 2n \end{array}\right) \ 
\delta^n \ \left\lbrace
\frac{\left(n-1\right)\left(s-2n\right)\left(s-2n-1\right)}{2} \ \partial^2 \theta^{[n]}
\right.
\nonumber \\ && \qquad \qquad \qquad
\left. 
\ + \ n \ \left[
\left(n-1\right)\left(2n-1\right) \ \partial \cdot \partial \cdot \theta^{[n-2]}
\ + \ n \bigtriangleup \theta^{[n-1]}
\right. \right.
\nonumber \\ && \qquad \qquad \qquad
\left. \left.
\ + \ \frac{\left(4n-3\right)\left(s-2n\right)}{2} \ \partial \partial \cdot \theta^{[n-1]}
\right]
\right\rbrace , \label{gauge_ham_s_momenta_1}
\\
\delta \tilde{\Pi} &=&
\sum_{n=1}^{\left[\frac{s-1}{2}\right]} 
\frac{n}{2} \binom{s}{2n+1} \delta^{n-1} 
\left\lbrace 
2\left(n-1\right)\left(2n+1\right) \ \partial \cdot \partial \cdot \xi^{[n-1]}
\ + \ \left(2n+1\right) \bigtriangleup \xi^{[n]}
\right.
\nonumber \\ && \qquad \qquad \qquad
\left.
\ + \ \left(s-2n-1\right)\left[
\left(4n+1\right)\partial \partial \cdot \xi^{[n]}
\ + \ \left(s-2n-2\right) \ \partial^2 \xi^{[n+1]}
\right]
\right\rbrace . \qquad \label{gauge_ham_s_momenta_2}
\end{eqnarray} 

Note that the parameters $\xi$ and $\theta$ appearing in these equations are \textit{traceful} spatial symmetric tensors with respectively $s-1$ and $s-2$ indices.

These variations indeed confirm, again, the identification of $N$ and $\mathcal{N}$ as the Lagrange multipliers.

\subsection{Hamiltonian and constraints}

The Legendre transformation made with the momenta given by \eqref{momenta_s_no_index_1} and \eqref{momenta_s_no_index_2} on the Lagrangian action \eqref{action_lag_s} will bring it into the Hamiltonian (constrained) form:
\begin{equation}
S  \left[h, \alpha ,\Pi , \tilde{\Pi} , N ,\mathcal{N}\right]
= \int \ d^4 x \ \left\lbrace \Pi \dot{h} \ + \ \tilde{\Pi} \dot{\alpha}
\ - \ \mathcal{H} \ - \ \mathcal{N} \mathcal{C} \ - \ NC 
\right\rbrace ,
\end{equation}
\noindent where the Hamiltonian density $\mathcal{H}$ and the constraints $C$ and $\mathcal{C}$ only depend on the canonical variables.

The variations obtained above from the expression of the momenta (\ref{momenta_s_no_index_1})-(\ref{momenta_s_no_index_2}) and the Lagrangian gauge transformations (\ref{gauge_lag_s}) allow us to directly identify the constraints, without having to compute them explicitly from the Legendre transformation. We will follow this way (and not compute the Hamiltonian either, even though we will later obtain its precise form in terms of the prepotentials through invariance arguments).

The gauge transformations must be realized by taking the Poisson bracket of the variables with the gauge transformations generator, which is precisely the constraints contracted with the gauge parameters (see section \textbf{\ref{Sec:Ham_form_constr_gauge}}). If we define:
\begin{equation}
\mathcal{G} \equiv
\int \ d^3 x \ \left(\xi C \ + \ \theta \mathcal{C}
\right) ,
\end{equation}
\noindent we expect the gauge transformations of the variables to be given by:
\begin{eqnarray}
\delta h &=& \left\lbrace h , \mathcal{G} \right\rbrace
= \frac{\delta \mathcal{G}}{\delta \Pi} ,
\\
\delta \alpha &=& \left\lbrace \alpha , \mathcal{G} \right\rbrace
= \frac{\delta \mathcal{G}}{\delta \tilde{\Pi}} ,
\\
\delta \Pi &=& \left\lbrace \Pi , \mathcal{G} \right\rbrace
= - \ \frac{\delta \mathcal{G}}{\delta h} ,
\\
\delta \tilde{\Pi} &=& \left\lbrace \tilde{\Pi} , \mathcal{G} \right\rbrace
= - \ \frac{\delta \mathcal{G}}{\delta \alpha} .
\end{eqnarray}

The identification of these variations with those computed in the previous section naturally leads to the following expression for the constraints:
\begin{eqnarray}
\mathcal{C} &=&
3 \ \partial \tilde{\Pi} \ + \ \left(s-3\right) \ \delta \ \partial \cdot \tilde{\Pi}
\nonumber \\ &&
- \ \sum_{n = 1}^{\left[s/2\right]} \ 
n \left(\begin{array}{c} s \\ 2n \end{array}\right) \ 
\delta^{n-1} \ \left\lbrace
n \ h^{[n]}
\ + \ \left(n-2\right)\left(2n-1\right) \ \partial \cdot \partial \cdot h^{[n-1]}
\right.
\nonumber \\ && \qquad \qquad \
\left.
\ + \ \frac{\left(4n-3\right)\left(s-2n\right)}{2} \ \partial \partial \cdot h^{[n]}
\ + \ \frac{\left(s-2n\right)\left(s-2n-1\right)}{2} \ \partial^2 h^{[n+1]}
\right\rbrace , \label{constr_s_ham}
\\
C &=&
- \ s \ \partial \cdot \Pi
\nonumber \\ &&
 - \ \sum_{n=1}^{\left[\frac{s-1}{2}\right]} 
n \ \binom{s}{2n+1} \delta^{n} 
\left\lbrace
\frac{\left(2n+1\right)}{2} \ \bigtriangleup \alpha^{[n-1]}
\ + \ \left(n-1\right)\left(2n+1\right) \ \partial \cdot \partial \cdot \alpha^{[n-2]}
\right.
\nonumber \\ && \qquad \qquad \
\left.
\ + \ \frac{\left(s-2n-1\right)\left(4n+1\right)}{2} \ \partial \partial \cdot \alpha^{[n-1]}
\ + \ \frac{\left(s-2n-1\right)\left(s-2n-2\right)}{2} \ \partial^2 \alpha^{[n]}
\right\rbrace . \nonumber \\ \label{constr_s_mom}
\end{eqnarray}
Let us again draw attention to the fact that each of these expressions is actually a symmetric tensor with $s-2$ (for \eqref{constr_s_ham}) or $s-1$ indices (for \eqref{constr_s_mom}).

In analogy with the spin $2$ case, we will call the constraint \eqref{constr_s_ham} - which is the generator of ``temporal" diffeomorphism, given by \eqref{gauge_ham_s_momenta_1} - the \textit{Hamiltonian constraint}, while the constraint \eqref{constr_s_mom} - which is the generator of ``spatial" diffeomorphism \eqref{gauge_ham_s_h} - shall be referred to as the \textit{momentum constraint}.

\subsection{Solving the momentum constraint}
We first solve the momentum constraint \eqref{constr_s_mom}.  This constraint reads (switching back, for convenience, to an explicit index notation) 
:
\be \partial_k \pi^{k i_2 \cdots i_s} + \hbox{``more"}= 0
\ee
where ``more" stands for terms that are linear in the second order derivatives of $\alpha_{i_4 \cdots i_s}$, which one can set to zero by a suitable gauge transformation.  In the gauges where ``more" vanishes, the constraint reduces to 
\be
\partial_k \pi^{ki_2 \cdots i_s} = 0 ,
\ee
the general solution of which is given by $\pi^{i_1 \cdots i_{s}} = G^{i_1 \cdots i_{s}}[P]$, according to our corollary \textbf{\ref{corollary_Einstein2}} of section \textbf{\ref{Sec:Einstein_s}}.  Here $G^{i_1 \cdots i_{s}}[P]$ is the Einstein tensor of some prepotential  $P_{i_1 \cdots i_s}$ which is totally symmetric. Its explicit expression is given by \eqref{Einstein_s_components}.

For fixed momentum $\pi^{i_1 \cdots i_s}$, the prepotential $P_{i_1 \cdots i_s}$ is determined up to a spin-$s$ diffeomorphism.  However, there is a residual gauge freedom in the above gauges, so that $\pi^{i_1 \cdots i_s}$ is not completely fixed.  It is straightforward but somewhat tedious to check that the residual gauge symmetry is accounted for by a spin-$s$ Weyl transformation of the prepotential $P_{i_1 \cdots i_s}$, which therefore enjoys all the gauge symmetries of a conformal spin-$s$ field.

These results extend what was  found earlier  for spins $2$\footnote{In \cite{Henneaux:2004jw}} (see Section \textbf{\ref{Sec:Grav_ham}}) and $3$ \footnote{In \cite{Henneaux:2015cda}} (see Section \textbf{\ref{Sec:3_solv_mom}}).  It is instructive to exhibit explicitly the formulas in the case of spin $4$,  which illustrates all the points and still yields readable expressions.

The momentum constraint reads in this case
\begin{eqnarray} 
0 &=& 
4 \ \partial^n \pi_{klmn}
\ + \ 6 \ \delta_{(kl} \Delta \alpha_{m)} \nonumber \\
&& - \ 10 \ \delta_{(kl} \partial_{m)} \partial^n \alpha_n . \label{constr_2}
\end{eqnarray}
and the  gauge freedom is:
\begin{eqnarray}
\delta \pi_{klmn} &=&
- \ 12 \ \partial_{(k} \partial_l \Xi_{mn)} \nonumber \\
&& + \ 12 \ \delta_{(kl} \left( \Delta \Xi_{mn)}\ + \ \partial_m \partial^p \Xi_{n)p} \right)
\nonumber \\ 
& + & 4 \ \delta_{(kl} \delta_{mn)} \left(2 \ \Delta \bar{\Xi} \ + \ 3 \ \partial^p \partial^q \Xi_{pq} \right) ,
\\
\delta \alpha_k &=& 
- \ 6 \ \partial^l \Xi_{kl}
\ - \ 2 \ \partial_k \bar{\Xi} .
\end{eqnarray}

The residual gauge transformations in the gauge $3 \ \Delta \alpha_k \ - \ 5 \ \partial_k \partial^l \alpha_l = 0$, which eliminates the $\alpha$-terms from the constraint,  must fulfill
$$
0 = - \ 18 \ \partial^l \Delta \Xi_{kl}
\ + \ 4 \ \partial_k \Delta \bar{\Xi}
\ + \ 30 \ \partial_k \partial^l \partial^m \Xi_{lm} .
$$
The divergence of this equation gives (after acting with $\Delta^{-1}$) $
3 \ \partial^k \partial^l \Xi_{kl}
\ + \ \Delta \bar{\Xi} = 0$.
Substituting this finding in the previous equation yields, after acting again with $\Delta^{-1}$, 
$3 \ \partial^l \left(\Xi_{kl} \ + \ \frac{1}{3} \ \delta_{kl} \bar{\Xi}\right) = 0$.
This is the divergence of a symmetric tensor, so the solution is the double divergence of a tensor with the symmetry of the Riemann tensor:
\begin{eqnarray}
\Xi_{kl} \ + \ \frac{1}{3} \ \delta_{kl} \bar{\Xi} &=&
\partial^m \partial^n \Theta_{mknl} .
\end{eqnarray}
Therefore, one has
\begin{eqnarray}
\Xi_{kl} &=&
\partial^m \partial^n \Theta_{mknl}
\ - \ \frac{1}{6} \ \delta_{kl} \partial^m \partial_n \Theta_{mp}^{\phantom{mp}np}.
\end{eqnarray}
This class of gauge transformations can be checked  to give a zero variation not only to the contribution of $\alpha_k$ to the constraint but in fact also to $\alpha_k$ itself.

We can dualize $\Theta_{klmn} = \epsilon_{klp}\epsilon_{mnq} \theta^{pq}$, with a symmetric $\theta_{kl}$, to obtain:
\begin{eqnarray}
\Xi_{kl} &=&
\frac{5}{6} \ \delta_{kl} \left(\Delta \bar{\theta} 
\ - \ \partial^m \partial^n \theta_{mn}\right) \nonumber \\
& + &\ 2 \ \partial_{(k} \partial^m \theta_{l)m}
\ - \ \partial_k \partial_l \bar{\theta}
\ - \ \Delta \theta_{kl} . \qquad
\end{eqnarray}
The gauge transformation of $\pi_{klmn}$ with this parameter is found then to be exactly the Einstein tensor of a Weyl diffeomorphism
\begin{eqnarray}
\delta \pi_{klmn} &=&
G_{klmn} \left[12 \ \delta_{(pq} \theta_{rs)}\right] .
\end{eqnarray}

Once the spin-$4$ momentum constraint has been brought in the standard form $\partial^k \pi_{klmn} = 0$ by partial gauge fixing, it can be solved by the familiar techniques recalled at the beginning of this section for general $s$. From the corollary \textbf{\ref{corollary_Einstein2}} of section \textbf{\ref{Sec:Einstein_s}}, the general solution of the equation $\partial^k \pi_{klmn} = 0$ is indeed the Einstein tensor of a symmetric tensor $P_{klmn}$, which is the prepotential for the momenta: 
$$\pi_{klmn} = G_{klmn} \left[ P \right].$$ 
Since the gauge freedom of $\pi_{klmn}$ is given by the Einstein tensor of a Weyl transformation, the gauge freedom of the prepotential $P_{klmn}$ will be given by a spin-$4$ Weyl transformation and a spin-$4$ diffeomorphism:
\begin{eqnarray}
\delta P_{klmn} &=&
4 \ \partial_{(k} \xi_{lmn)}
\ + \ 6 \ \delta_{(kl} \lambda_{mn)} ,
\end{eqnarray}
\noindent the first term corresponding to all the transformations of the prepotential leaving $\pi_{klmn}$ invariant and the second to those realizing on it a residual gauge transformation.

The solution in a general gauge will be given by these expressions to which are added general gauge transformation terms, parametrized by further prepotentials that drop from the action by gauge invariance and which are usually not considered for that reason.

\subsection{Solving the Hamiltonian constraint}

\label{Sec:solv_ham_constr_s}
We now solve the Hamiltonian constraint.  The functions ${\mathcal C}_{i_2 \cdots i_s}$ are linear in the second order derivatives of the spin-$s$ field $h_{i_1 \cdots i_s}$ and linear in the first order derivatives of the conjugate momenta $\tilde{\Pi}^{i_4 \cdots i_s}$.   One may rewrite these constraints in the equivalent form (suppressing indices) 
\be
\Psi - d_{(s-2)} \tilde{\Pi} = 0 \label{HamBis}
\ee
where $\Psi$ is the function of the second order derivatives of the spin-$s$ field with Young symmetry
$$\Psi \thicksim \overbrace{\yng(8)}^{(s-2) \text{ boxes}},$$
introduced in \textbf{\ref{Sec:Einstein_traceless}}, such that $\! \ast \bar{G}  \equiv  \! \ast  G^{[1]}= d_{(s-2)}^{s-2} \Psi $, where $\ast$ denotes here the dual on {\em all} indices (on which no trace was taken).  

The properties of $d_{(s-2)}$ were recalled in \textbf{\ref{App:generalized_differential}}: this differential is nilpotent, $d_{(s-2)}^{s-1}=0$, and its cohomology is empty when it acts on a symmetric tensor with $s-2$ indices, such as $\Psi$. This implies that the equation (\ref{HamBis}) is completely equivalent to $ d_{(s-2)}^{s-2} \Psi = 0$, i.e., $G^{[1]}[h] = 0$.  As we have recalled, this equation implies in turn  the existence of a prepotential $\Phi$ for $h$ (continuing to omit indices) such that the Einstein tensor of $h$ is equal to the Cotton tensor of $\Phi$: any symmetric, traceless and divergenceless tensor (a category to which belongs a traceless Einstein tensor) is the Cotton tensor of some field - this is our lemma \textbf{\ref{lemma_Cotton_for_any_traceless}} of section \textbf{\ref{Sec:Cotton_for_any_traceless}}. Once $h$ is expressed in terms of $\Phi$, the expression $ d_{(s-2)}^{s-2} \Psi[\Phi]$ identically vanishes, implying according to  the generalized algebraic Poincar\'e lemma of theorem \textbf{\ref{Theorem_cohomlogy}} of Appendix \textbf{\ref{App:generalized_differential}}\footnote{See \cite{DuboisViolette:1992ye,DuboisViolette:1999rd}.} that one can write $\Psi[\Phi] = d_{(s-2)} \tilde{\Pi}[\Phi]$, for some $\tilde{\Pi}[\Phi]$.  

This completely solves the Hamiltonian constraint in terms of the prepotential $\Phi_{i_1 \cdots i_s}$.   By construction, this prepotential has the gauge symmetry of a conformal spin-$s$ field.

The procedure is direct for spins $1$, where there is no Hamiltonian constraint,  and $2$, where one gets directly $G^{[1]}[h] = 0$ without having to differentiate. It was detailed for spin $3$ in section \textbf{\ref{Sec:solv_contr_ham_3}}. It is again instructive to illustrate it explicitly for the spin $4$ field, where the formulas remain readable. 

The hamiltonian constraint for the spin $4$ is:
\begin{eqnarray}
\mathcal{C}_{kl} &\equiv& 
3 \ \partial_{(k} \tilde{\Pi}_{l)}
\ + \ \delta_{kl} \partial^m \tilde{\Pi}_m \nonumber \\
&& - \ 6 \ \left(
\Delta \bar{h}_{kl}
\ - \ \partial^m \partial^n h_{klmn}
\ + \ \partial_{(k} \partial^m \bar{h}_{l)m}
\ + \ \partial_k \partial_l \bar{\bar{h}} \right) \nonumber \\
&& - \ 4 \ \delta_{kl} \Delta \bar{\bar{h}}  = 0 , \qquad \label{4_constr_ham}
\end{eqnarray}
The gauge freedom of $ h_{klmn}$ and $\tilde{\Pi}_k$  is:
\begin{eqnarray}
\delta \tilde{\Pi}_k &=& 
6 \ \Delta \bar{\xi}_k \ + \ 10 \ \partial_k \partial^l \bar{\xi}_l ,
\\
\delta h_{klmn} &=& 
4 \ \partial_{(k} \xi_{lmn)} .
\end{eqnarray}
One can equivalently rewrite the constraint as
\be
\mathcal{C}_{kl} \ - \ \frac{1}{6} \ \delta_{kl} \bar{\mathcal{C}} \ \equiv \ 3 \ \partial_{(k} \tilde{\Pi}_{l)} \ - \ 6 \ \Psi_{kl} = 0,
\ee
where $\bar{\mathcal{C}}$ is the trace of $\mathcal{C}_{kl}$ and
\begin{eqnarray}
\Psi_{kl} &\equiv&
\Delta \bar{h}_{kl}
\ - \ \partial^m \partial^n h_{klmn} \nonumber \\
&+ & \partial_{(k} \partial^m \bar{h}_{l)m}
\ + \ \partial_k \partial_l \bar{\bar{h}} .
\end{eqnarray}
One has 
\begin{eqnarray}
\epsilon_{k mp} \epsilon_{l nq} \partial^m \partial^n \Psi^{pq} &=&
\bar{G}_{kl} \left[ h \right]. \qquad \qquad
\end{eqnarray}
where $\bar{G}_{kl}$ is the trace of the Einstein tensor $G_{klmn} \left[ h \right]$ of $h_{klmn}$. So, one gets:
\begin{eqnarray}
\bar{G}_{kl} \left[h\right] &=&
\ - \ \frac{1}{6} \ \epsilon_{k mp} \epsilon_{l nq} \partial^m \partial^n \tilde{\mathcal{C}}^{pq}  .
\end{eqnarray}
with $\tilde {\mathcal C}^{pq} \equiv \mathcal{C}^{pq} \ - \ \frac{1}{6} \ \delta^{pq} \bar{\mathcal{C}}$.

The Hamiltonian constraint implies $\bar{G}_{kl} \left[ h \right] = 0$.  Therefore, the lemma \textbf{\ref{lemma_Cotton_for_any_traceless}} of section \textbf{\ref{Sec:Cotton_for_any_traceless}} yields $G_{ijkl} \left[h\right] = B_{ijkl} \left[\Phi\right]$  for some prepotential $\Phi_{klmn}$, where $B_{ijkl}$ is the Cotton tensor.  Explicitly,
\begin{eqnarray}
h_{ijkl}\left[\Phi\right] &=&
\epsilon_{(i\vert mn} \partial^m \left[
- \ \Delta \Phi^n_{\phantom{n} \vert jkl)}
\ + \  \frac{1}{2} \ \delta_{\vert jk} \Delta \bar{\Phi}^n_{\phantom{n}l)} \right. \nonumber \\
 & - &\left.  \frac{1}{2} \ \delta_{\vert jk} \partial^p \partial^q \Phi^n_{\phantom{n} l)pq} 
\right] + \ 4 \ \partial_{(i} \omega_{jkl)}, \qquad
\end{eqnarray}
where we have added a spin-$4$ diffeomorphism term parametrized by $\omega_{jkl}$.

Direct substitution gives then
\begin{eqnarray}
\Psi_{ij}  &=&
\partial_{(i\vert} \left[
3 \  \Delta \bar{\omega}_{j)}
\ + \ 5 \ \partial_{\vert j)} \partial^k \bar{\omega}_k \right. \nonumber \\
& - & \left. \frac{1}{8} \ \epsilon_{\vert j ) mn} \partial^m \left(
 \partial^k \Delta \bar{\Phi}^n_{\phantom{n}k}
\ + \ \partial^p \partial^q \partial^k \Phi^n_{\phantom{n} kpq} \right)
\right] . \qquad \qquad
\end{eqnarray}
from which one derives, using the constraint, the following expression for $\tilde{\Pi}_i$,
\begin{eqnarray}
\tilde{\Pi}_i &=&
6 \  \Delta \bar{\omega}_{i}
\ + \ 10 \ \partial_{i} \partial^k \bar{\omega}_k \nonumber \\
& - & \frac{1}{4} \ \epsilon_{i mn} \partial^m \left(
  \partial^k \Delta \bar{\Phi}^n_{\phantom{n}k}
\ + \ \partial^p \partial^q \partial^k \Phi^n_{\phantom{n} kpq} \right) 
, \qquad \qquad
\end{eqnarray}
One could in fact add  a solution $\kappa_i$ of the equation $\partial_{(i} \kappa_{j)} = 0$ (Killing equation) to $\tilde{\Pi}_i $ but we do not consider this possibility here by assuming for instance appropriate boundary conditions (vanishing of all fields at infinity eliminates $\kappa_i \sim C_i + \mu_{ij} x^j$, where $C_i$ and $\mu_{ij} = - \mu_{ji}$ are constants). \\

The expression for the spin-$s$ field $h_{i_1 \cdots i_s}$ in terms of the prepotential $\Phi_{i_1 \cdots i_s}$ contains $s-1$ derivatives in order to match the number of derivatives ($s$) of the Einstein tensor $G[h]$  with the number of derivatives ($2s-1$) of the Cotton tensor $B[\Phi]$. This number is odd (even) when $s$ is even (odd) and therefore, in order to match the indices of $h_{i_1 \cdots i_s}$ with those of $\partial_{k_1} \cdots \partial_{k_{s-1}} \Phi_{j_1 \cdots j_s}$, one needs one $\epsilon^{ijk}$ when $s$ is even and  no $\epsilon^{ijk}$ when it is odd, together with products of $\delta_{ij}$'s. 

\subsubsection{Even Spins}

We first turn to the even $s$ case.  We recall that in the spin-$2$ case, a particular solution is given by \eqref{2_h_Z}:
\begin{eqnarray}
h_{ij} &=& \epsilon_{(i\vert kl} \partial^k \Phi^l_{\ \vert j)} .
\end{eqnarray}

The gauge freedom of the prepotential is given by 
\begin{eqnarray}
\delta \Phi_{ij} &=&
\delta_{ij} \sigma 
\ + \ 2 \ \partial_{(i} \rho_{j)} ,
\end{eqnarray}
\noindent which generates the particular diffeomorphism $\delta h_{ij} = \partial_{(i} \theta_{j)}$ of the field, where $\theta_i = \epsilon_{ikl} \partial^k \rho^l$ (it is a diffeomorphism whose parameter is divergenceless).  The generalization of this formula to general even spin $s=2n$ is given in Appendix \ref{EvenSpin}.

Let us also reexamine the expression of the spin $4$ field $h_{ijkl}$ in terms of its prepotential $\phi_{ijkl}$.  One has
\begin{eqnarray}
h_{ijkl} &=&
\epsilon_{(i\vert mn} \partial^m \left[
- \ \Delta \Phi^n_{\phantom{n} \vert jkl)}
\ + \  \frac{1}{2} \ \delta_{\vert jk} \Delta \bar{\Phi}^n_{\phantom{n}l)}
\right. \nonumber \\
&& \left. - \ \frac{1}{2} \ \delta_{\vert jk} \partial^p \partial^q \Phi^n_{\phantom{n} l)pq} 
\right] . \label{hZ4}
\end{eqnarray}
The gauge freedom of the prepotential is given by:
\begin{eqnarray}
\delta \Phi_{ijkl} &=& 
4 \ \partial_{(i} \rho_{jkl)}
\ + \ 6 \ \delta_{(ij} \sigma_{kl)} ,
\end{eqnarray}
which implies:
\be
\delta h_{ijkl} =
\partial_{(i} \theta_{jkl)} ,
\ee
where 
\begin{eqnarray}
\theta_{ijk} &=&
\epsilon_{(i\vert mn} \partial^m \mu^n_{\phantom{n} \vert jk)} ,
\\
\mu_{ijk} &=&
- \ 3 \  \Delta \rho_{ijk}
\   + \ \frac{1}{2}\delta_{(ij} \left[  
\Delta \bar{\rho}_{k)}
\ - \ \partial^p \partial^q \rho_{k)pq} 
\ \right. \nonumber \\
&& \left. \hspace{2cm} - \ 4 \  \partial^p  \sigma_{k) p} \right] .
\end{eqnarray}
In fact, as discussed in Appendix \ref{EvenSpin}, the expression (\ref{hZ4}) is, up to a multiplicative factor,  the only one (with the requested number of derivatives) that implies that a gauge variation of $\Phi$ gives a gauge variation of $h$.

\subsubsection{Odd Spins}
In the odd spin case, the number of derivatives on the prepotential is even, so that the expression relating $h$ to $\Phi$ does not involve the Levi-Civita tensor.  The expression for the spin-$3$ field in terms of its prepotential is explicitly given by (this is the formula \eqref{3_SolG=B}):
\begin{eqnarray}
h_{ijk} &=& - \Delta \Phi_{ijk} + \frac34 \delta_{(ij} \Delta \bar{\Phi}_{k)} \nonumber \\
&& - \frac34 \delta_{(ij} \partial^r \partial^s \Phi_{k)rs} + \frac{3}{10} \delta_{(ij} \partial_{k)} \partial^r \bar{\Phi}_r . \label{SolG=B}
\end{eqnarray}
We repeat that the last term in (\ref{SolG=B}) is actually not necessary but included so that $\delta h_{ijk} = 0$ under Weyl transformations of $\Phi$.   One easily verifies that a gauge transformation of the prepotential induces a gauge transformation of the spin-$3$ field.

The expression of $h_{i_1 \cdots i_s}$ in terms of $\Phi_{i_1 \cdots i_s}$ is given in  Appendix \ref{OddSpin} for general odd spin.

\subsection{Hamiltonian Action in terms of prepotentials}

\label{Sec:action_prepot_bos_s}

The resolution of the constraints led us to the introduction of two prepotentials, which are symmetric spatial tensors, free of any trace constraint. We shall naturally gather them into a two-vector $(Z^a_{i_1 \cdots i_s}) \equiv (P_{i_1 \cdots i_s}, \Phi_{i_1 \cdots i_s})$ ($a= 1,2)$. Their gauge transformations are:
\begin{equation}
\delta Z^a_{i_1 \cdots i_s} =
s \ \partial_{(i_1} \rho^a_{i_2 \cdots i_s )}
\ + \ \frac{s\left(s-1\right)}{2} \ \delta_{(i_1 i_2} \sigma^a_{i_3 \cdots i_s )} . \label{gauge_s_prepot}
\end{equation}

The elimination of the canonical variables in terms of the prepotentials in the action is a rather burdensome task.  However, the derivation can be considerably short-cut  by invariance arguments.

The action will actually be:
\be
S[Z] = \int \ d^4 x \ \left[  \frac{1}{2}\varepsilon_{ab} \dot{Z}^a_{i_1 \cdots i_s} B^{b\, i_1 \cdots i_s}  \ - \ \delta_{ab} \left(\sum_{k=0}^{[\frac{s}{2}]} a_k  G^{[k] a \, i_1 \cdots i_{s-2k}}G^{[k] b}_{\; \; \; \; \; \; \; i_1 \cdots i_{s-2k}}  \right) \right] \label{action_s_prepot}
\ee
where $G^{[k] a \, i_1 \cdots i_{s-2n}}$ stands for the $k$-th trace of the Einstein tensor  $G^{a \, i_1 \cdots i_s}[Z]$ of the prepotential $Z^a_{i_1 \cdots i_s}$ and $B^{a\, i_1 \cdots i_s}$ for its Cotton tensor. The $a_k$ are explicitly given by
$$
 a_k =  \left( - \right)^k \frac{n!}{\left(n-k\right)!k!}
\frac{\left(2n-k-1\right)! \left(2n-1\right)!! }{2^k \left(2n-1\right)!\left(2n-2k-1\right)!!} \frac{1}{2}
$$
for even spin $s=2n$, and 
$$
a_k =
\left( - \right)^k \frac{n!}{\left(n-k\right)!k!}
\frac{\left(2n-k\right)! \left(2n+1\right)!! }{2^k \left(2n\right)!\left(2n-2k+1\right)!!} \frac{1}{2}
$$ for odd spin $s = 2n +1$. We derive these expressions in Appendix \textbf{\ref{App:conf_ham}}.
\\

Let us now see why the Hamiltonian action of the spin $s$ in terms of the prepotentials must be equal to this expression.

The kinetic term in the action is quadratic in the prepotentials $\Phi$ and $P$ and involves $2s -1$ spatial derivatives, and one time derivative.  Furthermore, it must be invariant under spin-$s$ diffeomorphisms and spin-$s$ Weyl transformations of both prepotentials.  This implies, making integrations by parts if necessary, that the kinetic term has necessarily the form of the kinetic term of the action (\ref{action_s_prepot}): only the overall factor is free, and it can easily be fixed by comparison of one term computed in each case.

Similarly, the Hamiltonian is the integral of a quadratic expression in the prepotentials $\Phi$ and $P$ involving $2s$ spatial derivatives.  By spin-$s$ diffeomorphism invariance,  it can be written as the integral of a quadratic expression in their Einstein tensors and their successive traces  -- or equivalently, the Schouten tensors and their successive traces. A lengthy but conceptually direct computation then shows that the coefficients  $a_k$ must be given by the expression written above:
they are in fact uniquely determined up to an overall factor, so that the Hamiltonian takes necessarily the form of the second term of (\ref{action_s_prepot}), but with $\delta_{ab}$ that might be replaced by a diagonal $\mu_{ab}$ with eigenvalues different from $1$. Conformal invariance can not lift this indetermination.

However, we will see in the next chapter that the equations of motion derived from the action \eqref{action_s_prepot} - twisted self-duality conditions - are actually consequences of Fronsdal equations \eqref{eom_bos_cov} (see section \textbf{\ref{Sec:action_tsc}}). This requires that the Hamiltonian analysis of Fronsdal action \eqref{action_lag_s} can not be in contradiction with the action \eqref{action_s_prepot}, and this can only be the case if $\mu_{ab} = \delta_{ab}$.\\

The equations of motion following from the action \eqref{action_s_prepot} are:
\begin{equation}
\dot{B}^{a \ i_1 \cdots i_s} =
\varepsilon^{a}_{\phantom{a}b} \ \varepsilon^{i_1}_{\phantom{i_1}j_1 k_1} \partial^{j_1} B^{b \ k_1 i_2 \cdots i_s} .
\end{equation}

Indeed, the action \eqref{action_s_prepot} can actually rewritten directly as the contraction of the prepotentials with these equations of motion. A careful examination of the $a_k$ coefficients and a comparison of their value with those appearing in the definition of the Schouten tensor will show that the action can actually be rewritten:
\begin{eqnarray}
S\left[Z^a_{ij} \right] &=&
\frac{1}{2} \ \int \ d^4 x \  Z^{a \ i_1 \cdots i_s} \left\lbrace
\varepsilon_{ab} \ \dot{B}^b_{i_1 \cdots i_s}
\ - \ \delta_{ab} \ \epsilon_{i_1 jk} \partial^j B^{b \ k}_{\phantom{b\ k} i_2 \cdots i_s}
\right\rbrace . 
\end{eqnarray}

\begin{subappendices}

\section{Prepotentials for even spins}

\label{App:even_spin_prepot}
\label{EvenSpin}

We give in this appendix the form of the spin-$s$ field $h_{i_1 \cdots i_s}$ in terms of the corresponding prepotential $\Phi_{i_1 \cdots i_s}$ when $s$ is even.  The case of an odd $s$ is treated in Appendix \ref{OddSpin}. 

Because of the gauge symmetries, the expression $h[\Phi]$ is not unique.  To any solution, one may add a gauge transformation term.  Our particular solution corresponds to a definite choice.

Our strategy is as follows: (i) First, one writes the most general form for $h$ in terms of $\Phi$ compatible with the index structure and the fact that it contains $s-1$ derivatives. (ii) Second, one fixes the coefficients of the various terms such that a gauge transformation of $\Phi$ induces a gauge transformation of $h$.  This turns out to completely fix $h[\Phi]$ up to an overall multiplicative constant.  (iii) Third, one fixes that multiplicative constant through the condition $G[h[\Phi]] = B[\Phi]$, which we  impose and verify in a convenient gauge for $\Phi$.

\subsection{First Step}

A generic term in the expression for $h_{i_1 \cdots i_s}$  in terms of $\Phi_{i_1 \cdots i_s}$ involves one Levi-Civita tensor when $s$ is even, as well as  $s-1$ derivatives of $\Phi$.  It can also contain a product of $p$  $\delta_{i_j i_k}$'s with free indices among $i_1, i_2, \cdots, i_s$. Hence a generic term takes the form
\be
 \delta_{i_1 i_2} \cdots \delta_{i_{2p-1} i_{2p}}\, \epsilon_{k_1 k_2 k_3} \, \partial_{m_1} \cdots \partial_{m_{s-1}} \Phi_{j_1 \cdots j_s}  \label{GenericTermEven}
\ee
for some $p$ such that $0 \leq p \leq n-1$ where $s=2n$ ($p$ cannot be equal  to $n$ since the Levi-Civita symbol must necessarily carry a free index, see below, so that there must be at least one free index left). Among the indices $k_1, k_2, k_3,  m_1,  \cdots, m_{s-1}, j_1, j_2, \cdots, j_s$, there are $s-2p$ indices equal to the remaining $i_a$'s, and the other indices are contracted with $\delta^{ab}$'s.   There is also an implicit symmetrization over the free indices $i_a$, taken as before to be of weight one.

The structure of the indices of the Levi-Civita symbol is very clear: because of the symmetries, one index is a free index $i_b$, one index is contracted with a derivative operator, and one index is contracted with an index of $\Phi$. Furthermore, if an index $m_b$ on the derivatives is equal to one of the free indices $i_b$, then, the term can be removed by a gauge transformation.  This means that apart from one index contracted with an index of the Levi-Civita tensor, the remaining indices on the derivative operators are necessarily contracted either among themselves to produce Laplacians or with indices of $\Phi$. In other words, the remaining free indices, in number $s - 2p -1$ are carried by $\Phi$.  One index on $\Phi$ is contracted with one index of the Levi-Civita tensor as we have seen, and  the other indices on $\Phi$, in number $2p$, are contracted either among themselves to produce traces or with the indices carried by the derivative operators. Thus, if we know the number of traces that occur in $\Phi$, say $q$, the structure of the term (\ref{GenericTermEven}) is completely determined,
 \begin{eqnarray}
&& \hspace{-1cm} \delta_{i_1 i_2} \cdots \delta_{i_{2p-1} i_{2p}}\, \epsilon_{i_{2p+1} k} ^{\; \; \; \; \; \;  \; \; \; \; t} \, \partial^k \partial^{j_1} \cdots \partial^{j_{2p-2q}} \Delta^{n-1-p+q}  \Phi^{[q]}_{i_{2p+2}\cdots i_s t j_1 \cdots j_{2p-2q}}, \label{GenericTermEven2}
\end{eqnarray}
or, in symbolic form,
\be
\delta^p  \left(\epsilon \cdot \partial \cdot\right) \left(\partial \cdot \partial \cdot\right)^{p-q} \Delta^{n-1-p+q} \Phi^{[q]} 
\ee
One has $0 \leq q \leq p$ and complete symmetrisation on the free indices $i_b$ is understood.

Accordingly, the expression for $h_{i_1 \cdots i_s}$ in terms of $\Phi_{i_1 \cdots i_s}$ reads
\begin{eqnarray}
&& h =  \sum_{p = 0}^{n-1}
\sum_{ q = 0}^p
a_{p,q} \ \delta^p  \left(\epsilon \cdot \partial \cdot\right) \left(\partial \cdot \partial \cdot\right)^{p-q} \Delta^{n-1-p+q} \Phi^{[q]} . \hspace{1cm} 
\end{eqnarray}
where the coefficients $a_{p,q}$ are determined next.

\subsection{Second Step}
By requesting that a gauge transformation of $\Phi_{i_1 i_2 \cdots i_s}$ induces a gauge transformation of $h_{i_1 \cdots i_s}$, the coefficients $a_{p,q}$ are found to be given by
\begin{eqnarray}
&& a_{p,q} = 2^{-p}  \left(-\right)^q  
\frac{\left(n - 1\right)!\left(2n - p - 1\right)!\left(2n- 1\right)!!}
{q!\left(p - q\right)!\left(n - p -1\right)!\left(2n - 1\right)! \left(2n - 2p -1\right)!!} a  \hspace{1cm}
\end{eqnarray}
where the multiplicative constant $a$ is undetermined at this stage. The double factorial of an odd number $2k+1$ is equal to the product of all the odd numbers up to $2k+1$,
$$
(2k+1)!! = 1 \cdot 3 \cdot 5 \cdots (2k-1) \dot (2k+1)
$$
The computation is fastidious but conceptually straightforward and left to the reader.

\subsection{Third Step}
Finally, we fix the remaining coefficient $a$ by imposing that $G[h[\Phi]] = B[\Phi]$.  This is most conveniently done in the gauge 
\be
\partial^{i_1} \Phi_{i_1 \cdots i_s} = 0, \; \; \; \Phi^{i_1}_{\; \; \; i_1 i_3 \cdots i_s} = 0
\ee
(transverse, traceless gauge). This gauge is permissible given that the gauge transformations of the prepotential involves both spin-$s$ diffeomorphisms and spin-$s$ Weyl transformations.  In that gauge, the Cotton tensor reduces to 
\begin{eqnarray}
B\left[ \Phi \right] _{i_1 i_2 \cdots i_s}&=&
- \ \epsilon_{(i_1 \vert jk} \, \partial^j \,  \Delta^{2n-1} \Phi^{k}_{\; \; \; \vert i_2 \cdots i_s)}
\end{eqnarray}
or in symbolic form,
\begin{eqnarray}
B\left[ \Phi \right] &=&
- \ \left(\epsilon \cdot \partial\cdot\right) \Delta^{2n-1} \Phi ,
\end{eqnarray}
while $h_{i_1 \cdots i_s}$ is also divergenceless and traceless (which shows, incidentally, that on the $\bar{G}[h]=0$ surface, one may impose both conditions also on $h$) and its Einstein tensor, expressed in terms of $\Phi$, becomes
\begin{eqnarray}
G\left[ h\left[ \Phi \right] \right]_{i_1 \cdots i_s} &=&
\left(-\right)^n \Delta^n h\left[ \Phi \right]_{i_1 \cdots i_s}
\nonumber \\ &=&
\left(- \right)^n a \  \epsilon_{(i_1 \vert jk} \, \partial^j \,  \Delta^{2n-1} \Phi^{k}_{\; \; \; \vert i_2 \cdots i_s)} \nonumber
\end{eqnarray}
i.e.,
\begin{eqnarray}
G\left[ h\left[ \Phi \right] \right] &=&
\left(- \right)^n a \ \epsilon \cdot \partial\cdot \Delta^{2n-1} \Phi .
\end{eqnarray}
This shows that 
\be
a = - (-)^n 
\ee
and completes the determination of $h$ in terms of its prepotential $\Phi$.

\subsection{Summary}

The field $h$ expression in terms of its prepotential is given by:
\begin{eqnarray}
&& h =  \sum_{p = 0}^{n-1}
\sum_{ q = 0}^p
a_{p,q} \ \delta^p  \left(\epsilon \cdot \partial \cdot\right) \left(\partial \cdot \partial \cdot\right)^{p-q} \Delta^{n-1-p+q} \Phi^{[q]} , \hspace{1cm} 
\end{eqnarray}
\noindent where the coefficients are equal to:
\begin{eqnarray}
&& a_{p,q} = 2^{-p}  \left(-\right)^{q+n+1}  
\frac{\left(n - 1\right)!\left(2n - p - 1\right)!\left(2n- 1\right)!!}
{q!\left(p - q\right)!\left(n - p -1\right)!\left(2n - 1\right)! \left(2n - 2p -1\right)!!} .  \hspace{1cm}
\end{eqnarray}

\newpage

\section{Prepotentials for odd spins}

\label{App:odd_spin_prepot}
\label{OddSpin}

The procedure for odd spins follows the same steps, but now there is no Levi-civita tensor involved in the expression $h[\Phi]$ since there is an even number of derivatives.

\subsection{First Step} 

A generic term in the expression for $h_{i_1 \cdots i_s}$  in terms of $\Phi_{i_1 \cdots i_s}$ involves  $s-1$ derivatives of $\Phi$.  It can also contain a product of $p$  $\delta_{i_j i_k}$'s with free indices among $i_1, i_2, \cdots, i_s$. Hence a generic term takes the form
\be
 \delta_{i_1 i_2} \cdots \delta_{i_{2p-1} i_{2p}}\,  \partial_{m_1} \cdots \partial_{m_{s-1}} \Phi_{j_1 \cdots j_s}  \label{GenericTermOdd}
\ee
for some $p$ such that $0 \leq p \leq n$ where $s=2n+1$. Among the indices $m_1,  \cdots, m_{s-1}, j_1, j_2, \cdots, j_s$, there are $s-2p$ indices equal to the remaining $i_a$'s, and the other indices are contracted with $\delta^{ab}$'s.   There is also an implicit symmetrization over the free indices $i_a$.

Again, if an index $m_b$ on the derivatives is equal to one of the free indices $i_b$, then, the term can be removed by a gauge transformation.  This means that the indices on the derivative operators are necessarily contracted either among themselves to produce Laplacians or with indices of $\Phi$. In other words, the remaining free indices, in number $s - 2p$ are carried by $\Phi$.  The other indices on $\Phi$, in number $2p $, are contracted either among themselves to produce traces or with the indices carried by the derivative operators. Thus, if we know the number of traces that occur in $\Phi$, say $q$, the structure of the term (\ref{GenericTermOdd}) is completely determined, as in the even spin case
 \begin{eqnarray}
&& \hspace{-1cm} \delta_{i_1 i_2} \cdots \delta_{i_{2p-1} i_{2p}}\, \partial^{j_1} \cdots \partial^{j_{2p-2q}} \Delta^{n-p+q}  \Phi^{[q]}_{i_{2p+1}\cdots i_s  j_1 \cdots j_{2p-2q}}, \label{GenericTermOdd2}
\end{eqnarray}
or, in a more compact way:
\begin{eqnarray}
 \delta^p   \left(\partial \cdot \partial \cdot\right)^{p-q} \Delta^{n-p+q} \Phi^{[q]} .
\end{eqnarray}
One has $0 \leq q \leq p$ and complete symmetrisation on the free indices $i_b$ is understood.

Accordingly, the expression for $h_{i_1 \cdots i_s}$ in terms of $\Phi_{i_1 \cdots i_s}$ reads
\begin{eqnarray}
&& h =  \sum_{p = 0}^{n}
\sum_{ q = 0}^p
a_{p,q} \ \delta^p   \left(\partial \cdot \partial \cdot\right)^{p-q} \Delta^{n-p+q} \Phi^{[q]}  . \hspace{1cm} 
\end{eqnarray}
where the coefficients $a_{p,q}$ are determined in the second step.

\subsection{Second Step}
By requesting that a gauge transformation of $\Phi_{i_1 i_2 \cdots i_s}$ induces a gauge transformation of $h_{i_1 \cdots i_s}$, the coefficients $a_{p,q}$ are found to be given up to an overall multiplicative  constant $a$ by
\begin{eqnarray}
&& a_{p,q} =
\left(-\right)^q
2^{-p} \frac{n!\left(2n - p\right)!\left(2n+ 1\right)!!}{q!\left(p - q\right)!\left(n - p\right)!\left(2n\right)! \left(2n - 2p +1\right)!!} a .
\end{eqnarray}
The computation is again somewhat fastidious but conceptually straightforward and left to the reader.  

\subsection{Third Step}
Finally, we fix the remaining coefficient $a$ by imposing that $G[h[\Phi]] = B[\Phi]$.  This is most conveniently done in the transverse, traceless gauge for $\Phi$
\be
\partial^{i_1} \Phi_{i_1 \cdots i_s} = 0, \; \; \; \Phi^{i_1}_{\; \; \; i_1 i_3 \cdots i_s} = 0
\ee
which is again permissible.  In that gauge, the Cotton tensor reduces to 
\begin{eqnarray}
D\left[ \Phi \right] &=&
\left(\epsilon \cdot \partial\cdot\right) \Delta^{2n} \Phi.
\end{eqnarray}
while the Einstein tensor of $h[\Phi]$ becomes
\begin{eqnarray}
G\left[ h\left[ \Phi \right] \right] &=&
\left(- \right)^n a \ \epsilon \cdot \partial\cdot \Delta^{2n} \Phi.
\end{eqnarray}This leads to
\begin{eqnarray}
a &=& \left(-\right)^n.
\end{eqnarray}and completes the determination of $h$ in terms of its prepotential $\Phi$.

\newpage

\section{Conformally invariant Hamiltonian}

\label{App:conf_ham}

We will use an index-free notation in this Appendix. Also, in order to avoid confusion, we shall use the symbol $\Gamma$ to note gauge variation.

We want to determine the form of the Hamiltonian density $\mathcal{H}$ - quadratic in the field and containing $2s-1$ derivatives - of a prepotential $Z$ - which is a symmetric tensor with $s$ indices - invariant (up to total derivatives) under a gauge variation:
\begin{equation}
\Gamma Z = s \ \partial \rho \ + \ \frac{s\left(s-1\right)}{2} \ \delta \sigma .
\end{equation}

The first part of this gauge invariance, diffeomorphism invariance, forces the Hamiltonian density to be of the form:
\begin{equation}
\mathcal{H} = \frac{1}{2} \ \sum_{k=0}^{[s/2]} a_k G^{[k]} \cdots G^{[k]} .
\end{equation}

We can then use the second part of the gauge invariance, Weyl invariance, to fix the coefficients $a_k$ (up to a global factor).

The variation of the Einstein tensor of $Z$ under a Weyl transformation is:
\begin{equation}
\Gamma G = \frac{s\left(s-1\right)}{2} \ \left[\delta \Delta \mu \ - \ \partial^2 \mu \right] ,
\end{equation}
\noindent where $\mu$ is the Einstein tensor of $\sigma$, so that it is traceless.

\subsection{Even spin: $s = 2n$}

The variation of the trace of the Einstein tensor is:
\begin{eqnarray}
\Gamma G^{[k]} &=&
\frac{s\left(s-1\right)}{2} \ \left[ \delta \Delta \mu \ - \ \partial^2  \mu  \right]^{[k]} 
\nonumber \\ &=&
\left(3k \ + \ 2n\left(n-1\right) \ - \ 2 \left(n-k\right)\left(n-k-1\right)  \right) \Delta \mu^{[k-1]}
\ + \ \left(n-k\right)\left(2n - 2k -1\right) \delta \Delta \mu^{[k]}
\nonumber \\ &&
\ - \ k \ \Delta  \mu^{[k-1]}
\ - \ \left(n - k \right)\left(2n - 2k -1\right) \ \partial^2  \mu^{[k]}
\end{eqnarray}
\noindent where we have used the divergencelessness of $\mu$.

The variation of the Hamiltonian density under a Weyl transformation will be, up to a total divergence:
\begin{eqnarray}
\Gamma \mathcal{H} &=&
\sum_{k = 0}^{n} \ a_k \ G^{[k]} \cdot \Gamma G^{[k]} 
\nonumber \\ &=&
\sum_{k = 0}^{n} \ \left(3k \ - \ k \ + \ 2n\left(n-1\right) \ - \ 2 \left(n-k\right)\left(n-k-1\right)  \right)  \ a_k \ G^{[k]} \cdot \Delta \mu^{[k-1]}
\nonumber \\ &&
\ + \ \sum_{k = 0}^{n} \ \left(n-k\right)\left(2n - 2k -1\right)  \ a_k \ G^{[k+1]} \cdot \Delta \mu^{[k]}
\ + \ \ldots
\nonumber \\ &=&
\sum_{k = 0}^{n} \ \left(3k \ - \ k \ + \ 2n\left(n-1\right) \ - \ 2 \left(n-k\right)\left(n-k-1\right)  \right) \ a_k \ G^{[k]} \cdot \Delta \mu^{[k-1]}
\nonumber \\ &&
\ + \ \sum_{k = 1}^{n} \ \left(n-k +1\right)\left(2n - 2k +1\right)  a_{k-1} \ G^{[k]} \cdot \Delta \mu^{[k-1]}
\ + \ \ldots
\end{eqnarray}

In order for it to vanish, we need:
\begin{eqnarray}
0 &=& \left(3k \ - \ k \ + \ 2n\left(n-1\right) \ - \ 2 \left(n-k\right)\left(n-k-1\right)  \right) \ a_k 
\nonumber \\ &&
\ + \  \left(n-k +1\right)\left(2n - 2k +1\right)  a_{k-1}  ,
\end{eqnarray}
\noindent from $k = 1$ to $n$. This implies:
\begin{eqnarray}
a_k &=&
\ - \ \frac{\left(n-k +1\right)\left(2n - 2k +1\right)}{2k\left(2n - k\right) } \ a_{k-1} .
\end{eqnarray}

The solution of this recursion relation is:
\begin{eqnarray}
a_k &=&
\left( - \right)^k \frac{n!}{\left(n-k\right)!k!}
\frac{\left(2n-k-1\right)! \left(2n-1\right)!! }{2^k \left(2n-1\right)!\left(2n-2k-1\right)!!} .
\end{eqnarray}

\subsection{Odd spin: $s = 2n+1$}

The variation of the trace of the Einstein tensor is:
\begin{eqnarray}
\Gamma G^{[k]} &=&
\frac{s\left(s-1\right)}{2} \ \left[ \delta \Delta \mu \ - \ \partial^2  \mu  \right]^{[k]}
\nonumber \\ &=&
\left(3k \ + \ 2k \ + \ 2n\left(n-1\right) \ - \ 2 \left(n-k\right)\left(n-k-1\right)  \right) \Delta \mu^{[k-1]}
\ + \ \left(n-k\right)\left(2n - 2k +1\right) \delta \Delta \mu^{[k]}
\nonumber \\ &&
\ - \ k \ \Delta  \mu^{[k-1]}
\ - \ \left(n - k \right)\left(2n - 2k -1\right) \ \partial^2  \mu^{[k]}
\end{eqnarray}
\noindent where we have used the divergencelessness of $\mu$.

The variation of the hamiltonian density under a Weyl transformation will be, up to a total divergence:
\begin{eqnarray}
\Gamma \mathcal{H} &=&
\sum_{k = 0}^{n} \ a_k \ G^{[k]} \cdot \Gamma G^{[k]} 
\nonumber \\ &=&
\frac{2n\left(2n-1\right)}{2} \ \left[ \delta \Delta \mu \ - \ \partial^2  \mu \right]^{[k]} 
\nonumber \\ &=&
\sum_{k = 0}^{n} \ \left(3k \ + \ 2k  \ - \ k \ + \ 2n\left(n-1\right) \ - \ 2 \left(n-k\right)\left(n-k-1\right)  \right)  \ a_k \ G^{[k]} \cdot \Delta \mu^{[k-1]}
\nonumber \\ &&
\ + \ \sum_{k = 0}^{n} \ \left(n-k\right)\left(2n - 2k +1\right)  \ a_k \ G^{[k+1]} \cdot \Delta \mu^{[k]}
\ + \ \ldots
\nonumber \\ &=&
\sum_{k = 0}^{n} \ \left(3k \ + \ 2k \ - \ k \ + \ 2n\left(n-1\right) \ - \ 2 \left(n-k\right)\left(n-k-1\right)  \right) \ a_k \ G^{[k]} \cdot \Delta \mu^{[k-1]}
\nonumber \\ &&
\ + \ \sum_{k = 1}^{n} \ \left(n-k +1\right)\left(2n - 2k +3\right)  a_{k-1} \ G^{[k]} \cdot \Delta \mu^{[k-1]}
\ + \ \ldots
\end{eqnarray}

In order for it to vanish, we need:
\begin{eqnarray}
0 &=& \left(3k \ + \ 2k \ - \ k \ + \ 2n\left(n-1\right) \ - \ 2 \left(n-k\right)\left(n-k-1\right) \right) \ a_k 
\nonumber \\ &&
\ + \  \left(n-k +1\right)\left(2n - 2k +3\right)  a_{k-1}  ,
\end{eqnarray}
\noindent from $k = 1$ to $n$. This implies:
\begin{eqnarray}
a_k &=&
\ - \ \frac{\left(n-k +1\right)\left(2n - 2k +3\right)}{2k\left(2n - k + 1\right) } \ a_{k-1} .
\end{eqnarray}

The solution of this recursion relation is:
\begin{eqnarray}
a_k &=&
\left( - \right)^k \frac{n!}{\left(n-k\right)!k!}
\frac{\left(2n-k\right)! \left(2n+1\right)!! }{2^k \left(2n\right)!\left(2n-2k+1\right)!!} .
\end{eqnarray}

\end{subappendices}

\chapter{Twisted self-duality}

\label{Chap:tsdc}

We recalled that the equations of motion of the spin one (section \textbf{\ref{Sec:twistselfdual_1}}) and two (section \textbf{\ref{Sec:twistselfdual_2}}) are actually equivalent to the requirement that the dual of their gauge invariant curvature be itself the curvature of a dual field. This requirement is the so-called \textit{twisted self-duality condition}.

As we are going to see, the equations of motion of all higher spin can be rewritten as twisted self-duality conditions. Duality conditions, however, are better expressed in terms of Riemann tensor (introduced in section \textbf{\ref{Sec:Riemann}}), which is why we will begin by introducing a formalism in which no trace constraint is used to describe higher spin. We will then consider twisted self-duality conditions, first in a covariant form, before showing that a first order in time derivative subset of these equations is equivalent to the full set. We will then write down an action principle from which to extract these conditions, which will turn out to be precisely the Hamiltonian action derived earlier.

\section{Equations of motion in terms of Riemann tensor}

As we saw in section \textbf{\ref{Sec:Lag_s_bos}}, higher spin can be described by second order equations of motion provided trace constraints are imposed on both the field and the gauge parameters. On the other hand, we saw that (section \textbf{\ref{Sec:Riemann}}, if the gauge parameter is not subject to a trace constraint, the gauge invariant curvature of a spin-$s$ field contains $s$ derivatives of this field: it is the generalized Riemann tensor, which is simply computed by antisymmetrizing a derivative with each index of the field:
\begin{equation}
R_{\mu_1 \nu_1  \cdots \mu_s \nu_s} = 2^s \
\partial_{[\mu_1 \vert} \cdot \partial_{[\mu_s \vert} h_{\vert \nu_1] \cdots \vert \nu_s ]} .
\end{equation}

So, let us consider a symmetric tensor $h_{\mu_1 \cdots \mu_s}$ with a gauge variation:
\begin{equation} \label{bos_s_gauge_inv_no_trace_constr}
\delta h_{\mu_1 \cdots \mu_s} = s \ \partial_{(\mu_1} \xi_{\mu_2 \cdots \mu_s )},
\end{equation}
\noindent where $\xi_{\mu_2 \cdots \mu_s}$ is free of any tracelessness condition\footnote{This treatment was introduced in \cite{Bekaert:2003az,Damour:1987vm}.}.

In analogy with the spin-$2$ case, we are interested in the consequences of the vanishing of the trace of its Riemann tensor - the generalized \textit{Ricci tensor}. It is given by:
\begin{eqnarray}
R'_{\nu_1 \nu_2 \mu_3 \nu_3 \mu_4 \nu_4 \cdots \mu_s \nu_s} &\equiv&
\eta^{\mu_1 \mu_2} R_{\mu_1 \nu_1 \mu_2 \nu_2 \cdots \mu_s \nu_s}
\nonumber \\ &=&
2^{s-2} \ \partial_{[\mu_3 \vert} \cdot \partial_{[\mu_s \vert} \left\lbrace
\square h_{\nu_1 \nu_2 \vert \nu_3 ] \cdots \vert \nu_s ]}
\right.
\nonumber \\ && \qquad
\left.
\ - \ 2 \ \partial_{(\nu_1} \partial^{\rho} h_{\nu_2 ) \rho \vert \nu_3 ] \cdots \vert \nu_s ]}
\ + \ \partial_{\nu_1} \partial_{\nu_2} h'_{\vert \nu_3 ] \cdots \vert \nu_s ]}
\right\rbrace 
\nonumber \\ &=&
2^{s-2} \ \partial_{[\mu_3 \vert} \cdot \partial_{[\mu_s \vert} \mathcal{F}_{\nu_1 \nu_2 \vert \nu_3 ] \cdots \vert \nu_s ]} ,
\end{eqnarray}
\noindent where $\mathcal{F}_{\mu_1 \cdots \mu_s}$ is the Fronsdal tensor (defined by equation \eqref{Fronsdal_bos}), which is precisely the curvature that can be built for a description of the spin-$s$ field where the gauge parameter is traceless.

So, if we consider a field with an unconstrained gauge parameter, and we impose on it the following equations of motion:
\begin{equation}
0 = 
R'_{\nu_1 \nu_2 \mu_3 \nu_3 \mu_4 \nu_4 \cdots \mu_s \nu_s} ,
\end{equation}
\noindent we see that this is equivalent to:
\begin{equation}
0 = d_{(s)}^{s-2} \mathcal{F} ,
\end{equation}
\noindent in the language of the generalized differential introduced in Appendix \textbf{\ref{App:generalized_differential}}, which satisfies $d_{(s)}^{s+1} = 0$, and has empty cohomology when acting on a symmetric tensor with $s$ indices like Fronsdal tensor $\mathcal{F}$. So, this is also equivalent to:
\begin{equation}
\mathcal{F} = d_{(s)}^3 f ,
\end{equation}
\noindent for a symmetric tensor $f$ with $s-3$ indinces. Explicitly, we can rewrite this as:
\begin{equation}
\mathcal{F}_{\mu_1 \cdots \mu_s} = \partial_{(\mu_1} \partial_{\mu_2} \partial_{\mu_3} f_{\mu_4 \cdots \mu_s )} .
\end{equation}

We see that this expression has precisely the form of the variation of Fronsdal tensor under a gauge transformation whose gauge parameter has a trace proportional with $f$ (equation \eqref{Fronsdal_bos_var_trace_eps}). 

This shows that, starting from a field with unconstrained gauge parameter, imposing the vanishing of the trace of its Riemann tensor is the same as imposing the vanishing of its Fronsdal tensor \textit{up to a traceful gauge transformation}. Making a partial gauge fixing can then bring us into a gauge where Fronsdal tensor actually vanishes and leaves us with a residual gauge freedom containing only traceless parameters.\\

There are, in fact, a few more subtleties in this rewriting of the equations than is suggested here, and we refer the reader to \cite{Bekaert:2003az,Damour:1987vm} for a more detailed and rigorous treatment.

\section{Covariant twisted self-duality}

\label{Sec:cov_EMduality_s}

We have just seen that higher spin could be described by a symmetric tensor field $h_{\mu_1 \cdots \mu_s}$ endowed with the gauge invariance \eqref{bos_s_gauge_inv_no_trace_constr}, the gauge parameter $\xi_{\mu_2 \cdots \mu_s}$ being traceful. The equations of motion are of order $s$, and they set the trace of the Einstein tensor of the field to zero:
\begin{equation}
0 = R_{\nu_1\nu_2 \mu_3 \nu_3 \cdots \mu_s \nu_s}'\left[ h \right] .
\end{equation}

Now, if we define the dual of the Riemann tensor of $h$ as the Hodge dual of the Riemann over one of its pairs of indices:
\begin{equation}
\ast R_{\mu_1\nu_1 \cdots \mu_s \nu_s} \left[ h \right] \equiv \frac{1}{2} \ \epsilon_{\mu_1 \nu_1 \rho_1 \sigma_1} R^{\rho_1 \sigma_1}_{\phantom{\rho_1 \sigma_1} \mu_2 \nu_2 \cdots \mu_s\nu_s} \left[ h \right] ,
\end{equation}
\noindent we can make the immediate observation that the tracelessness of the Riemann tensor is equivalent to its dual satisfying the cyclic identity (Bianchi algebraic identity) over its first three indices:
\begin{equation}
0 = R'_{\nu_1 \nu_2 \mu_3 \nu_3 \mu_4 \nu_4 \cdots \mu_s \nu_s}\left[h\right]
\Leftrightarrow
0 = \ast R_{[\mu_1 \nu_1 \mu_2 ] \nu_2 \cdots \mu_s \nu_s} \left[h\right].
\end{equation}

However, since $\ast R$ is already antisymmetric in each of its pairs of indices $(\mu_i , \nu_i )$\footnote{And since it already satisfies the cyclic identity on each triplet of indices not including the first two, $\mu_1\nu_2$.}, its satisfying the cyclic identity is equivalent to its having the same symmetry type as the Riemann tensor:
\begin{equation}
0 = R'_{\nu_1 \nu_2 \mu_3 \nu_3 \mu_4 \nu_4 \cdots \mu_s \nu_s}\left[h\right]
\Leftrightarrow
\ast R_{\mu_1\nu_1 \cdots \mu_s \nu_s}\left[h\right] \thicksim \overbrace{\yng(10,10)}^{\text{$s$ boxes}} . \label{S_Young_on_shell}
\end{equation}

On the other hand, by its definition, the tensor $\ast R$ already satisfies the differential Bianchi identity on each of its pairs of indices, except the first. But, on-shell, since $\ast R$ becomes invariant under a permutation of its pairs of indices (which is a consequence of its having the symmetry type \eqref{S_Young_on_shell}), it also satisfies Bianchi identity on its first pair of indices ($\mu_1\nu_1$). In other words, the dual of the Riemann tensor satisfies:
\begin{equation}
0 = d_{(s)} \ast R\left[h\right] ,
\end{equation}
\noindent where we used the generalized differential defined in Appendix \textbf{\ref{App:generalized_differential}}. $\ast R$ being a ``well-filled" tensor, this is equivalent to:
\begin{equation}
\ast R\left[h\right] = d_{(s)}^s f ,
\end{equation}
\noindent for some symmetric field $f_{\mu_1\cdots \mu_s}$. This is precisely the definition (up to a factor $2^s$) of the Riemann tensor of $f$.

In conclusion, we see that the vanishing of the Ricci tensor of $h$ is equivalent to the dual of the Riemann tensor of $h$ being itself the Riemann tensor of some other field $f$:
\begin{equation}
\ast R\left[h\right] = R\left[f\right] . \label{cov_tsc_s_no_index}
\end{equation}

Thanks to the invariance of the Riemann tensor under generalized linearized diffeomorphism, the tensor $f$ has, of course, the same gauge invariance as $h$: the one given by \eqref{bos_s_gauge_inv_no_trace_constr}.

This form of the equations is completely equivalent to the original form $R'\left[h\right]=0$, since we have seen that the equation $R'\left[h\right]=0$ implies (\ref{cov_tsc_s_no_index}).  And conversely, if (\ref{cov_tsc_s_no_index}) holds, then both  $h$ and $f$ obey $R'[h] = 0$, $R'[f] =0$, i.e., fulfill the spin-$s$ equations of motion.  Furthermore,  the two spin-$s$ fields are not independent since $f$ is completely determined by $h$ up to a gauge transformation and therefore carries no independent physical degrees of freedom.\\

Since the Hodge dual squares to minus one in a lorentzian four-dimensional space-time, the equation $\ast R \left[ h \right] = R \left[f \right]$ is equivalent to $R\left[h\right] = - \ \ast R\left[ f \right]$. So, if we gather our fields into a two-vector $h^a \equiv \left(h,f\right)$, we can rewrite the equations \eqref{cov_tsc_s_no_index} as:
\begin{equation}
\ast R^a_{\mu_1 \nu_1 \cdots \mu_s\nu_s} = \epsilon^a_{\phantom{a}b} R^b_{\mu_1 \nu_1 \cdots \mu_s\nu_s} ,
\end{equation}
\noindent where $R^a$ is the Riemann tensor of $h^a$\footnote{Following \cite{Cremmer:1998px}, one refers to (\ref{cov_tsc_s_no_index}) as the twisted self-dual formulation of the spin-$s$ theory.}. More explicitly, this means:
\begin{equation} \label{cov_tsc_matrix_s}
\left(\begin{array}{c}
\ast R_{\mu_1 \nu_1 \cdots \mu_s\nu_s} \left[h\right]\\
\ast R_{\mu_1 \nu_1 \cdots \mu_s\nu_s}\left[f\right]
\end{array}
\right)
= 
\left(\begin{array}{c}
 R_{\mu_1 \nu_1 \cdots \mu_s\nu_s} \left[f\right]\\
- \ R_{\mu_1 \nu_1 \cdots \mu_s\nu_s}\left[h\right]
\end{array}
\right) .
\end{equation}

This is sometimes written in the more compact form that gathers the curvatures of the field and its dual into a single object:
\begin{equation}
\mathcal{R} \equiv 
\left(\begin{array}{c}
\ast R_{\mu_1 \nu_1 \cdots \mu_s\nu_s} \left[h\right]\\
\ast R_{\mu_1 \nu_1 \cdots \mu_s\nu_s}\left[f\right]
\end{array}
\right),
\end{equation}
\noindent and the matrix $\varepsilon \equiv \left(\begin{array}{cc}
0 & 1 \\ - \ 1 & 0
\end{array}
\right)$, so that the equations of motion become:
\begin{equation}
\ast \mathcal{R} = \varepsilon \mathcal{R} .
\end{equation}

\section{Electric and magnetic fields}

\label{Sec:non_cov_tsc_s}

The covariant form of the twisted self-duality conditions \eqref{cov_tsc_s_no_index} actually contains many superfluous equations, and a small subset of them only containing first order time derivatives is now going to be identified. This subset will not be manifestly covariant, and it will involve generalized electric and magnetic fields, which we will now define.

\subsection{Definitions}

Let us first define the electric and magnetic components of the Weyl tensor, which coincides on-shell with the Riemann tensor.  It would seem natural to define the electric components as the components of the Weyl tensor with the maximum number of indices equal to zero (namely $s$), and the magnetic components as the components with the maximum number minus one of indices equal to zero (namely $s-1$). By the tracelessness conditions of the Weyl tensor, the electric components can be related to the components with no zeroes when $s$ is even, like for gravity, or just one zero when $s$ is odd, like for Maxwell.  It turns out to be more convenient for dynamical purposes to define the electric and magnetic components starting from the other end, i.e., in terms of components with one or no zero.  Now, it would be cumbersome in the general analysis to have a definition of the electric and magnetic components that would depend on the spin.  For that reason, we shall adopt a definition which is uniform for all spins, but which coincides with the standard conventions given above only for even spins.  It makes the Schwarzschild field ``electric", but the standard electric field of electromagnetism is viewed as ``magnetic".   Since the electric (magnetic) components of the curvature of the spin-$s$ field are the magnetic (electric) components of the curvature of the dual spin-$s$ field, this is just a matter of convention, but this convention may be confusing when confronted with the standard Maxwell terminology.

Before providing definitions, we recall that the curvature $R_{i_1 j_1\cdots i_s j_s}$ of the three-dimensional ``spin-$s$ field" $h_{i_1 \cdots i_s}$ given by the spatial components  of the spacetime spin-$s$ field $h_{\mu_1 \cdots \mu_s}$ is completely equivalent to its Einstein tensor (introduced in section \textbf{\ref{Sec:Einstein_s}}) defined as:
\be
G^{i_1 \cdots i_s} = \frac{1}{2^s} \epsilon^{i_1 j_1 k_1} \cdots \epsilon^{i_s j_s k_s} R_{j_1 k_1\cdots j_s k_s} .
\ee

This tensor is completely symmetric and identically conserved,
\be
\partial_{i_1} G^{i_1 i_2 \cdots i_s} = 0
\ee
In the sequel, when we shall refer to the Einstein tensor of the spin $s$ field, we shall usually mean this three-dimensional Einstein tensor (the four-dimensional Einstein tensor vanishes on-shell).

We now define precisely the electric and magnetic fields off-shell as follows:
\begin{itemize} 
\item The electric field ${\mathcal E}^{i_1 \cdots i_s}$ of the spin-$s$ field $h_{\mu_1 \cdots \mu_s}$ is equal to the Einstein tensor $G^{i_1 \cdots i_s}$ of its spatial components $h_{i_1 \cdots i_s}$,
\be {\mathcal E}^{i_1 \cdots i_s} = G^{i_1 \cdots i_s}
\ee
By construction, the electric field fully captures the spatial curvature and involves only the spatial components of the spin-$s$ field.  It is completely symmetric and conserved,
\be
 {\mathcal E}^{i_1 \cdots i_s} =  {\mathcal E}^{(i_1 \cdots i_s)}, \; \; \; \partial_{i_1} {\mathcal E}^{i_1 i_2 \cdots i_s} = 0
\ee
\item The magnetic field ${\mathcal B}^{i_1 \cdots i_s}$ of the spin-$s$ field $h_{\mu_1 \cdots \mu_s}$ is equivalent to the components with one zero of the space-time curvature tensor and is defined through
\be {\mathcal B}_{i_1 \cdots i_s} = \frac{1}{2^{s-1}} R_{0i_1}^{\; \; \; \; \;  j_2 k_2 \cdots j_s k_s} \epsilon_{i_2 j_2 k_2} \cdots \epsilon_{i_s j_s k_s}
\ee
It contains one time (and $s-1$ space) derivatives of the spatial components $h_{i_1 \cdots i_s}$, and $s$ derivatives of the mixed components $h_{0i_2 \cdots i_s}$.  The magnetic field is symmetric in its last $s-1$ indices.  It is also transverse on each index,
\be
\partial_{i_1} {\mathcal B}^{i_1 i_2 \cdots i_s} = 0, \; \; \; \partial_{i_2} {\mathcal B}^{i_1 i_2 \cdots i_s} =0,
\ee
and traceless on the first index and any other index,
\be
\delta_{i_1 i_2} {\mathcal B}^{i_1 i_2 \cdots i_s} = 0.
\ee
It is useful to make explicit the dependence of the magnetic field -- or equivalently, $R_{0i_1 j_2k_2 \cdots j_s k_s}$ -- on $h_{0i_2 \cdots i_s}$.   One finds
\be
R_{0i_1 j_2k_2 \cdots j_s k_s}  = \partial_{i_1} \left(d_{(s-1)}^{s-1} N \right)_{j_2 k_2 \cdots j_s k_s} + ``more" \label{ambiguity}
\ee
where ``$more$" involves only spatial derivatives of $\dot{h}_{i_1 \cdots i_s}$ and where $N_{i_1 \cdots i_{s-1}}$ stands for $h_{0i_1 \cdots i_{s-1}}$, i.e., $N_{i_1 \cdots i_{s-1}} \equiv h_{0i_1 \cdots i_{s-1}}$.
\end{itemize}

Similar definitions apply to the dual spin-$s$ field $f_{\lambda_1 \cdots \lambda_s}$.

The electric and magnetic fields possess additional properties on-shell.  First, the electric field is traceless as a result of the equation $R'^0_{\; \; 0 i_5 \cdots i_s} - \frac12 \delta^0_0 R''_{i_5 \cdots i_s} = 0$,
\be
\delta_{i_1 i_2} {\mathcal E}^{i_1 i_2 \cdots i_s} = 0.
\ee
Second, the magnetic field is symmetric as a result of the equation $R'_{0i_4 i_5 \cdots i_s} =0$,
\be
 {\mathcal B}^{i_1 \cdots i_s} =  {\mathcal B}^{(i_1 \cdots i_s)} = 0.
\ee

We also note that there are no other independent components of the space-time curvature tensor on-shell, since components with more than one zero can be expressed in terms of components with one or no zero through the equations of motion.

\subsection{Twisted self-duality in terms of electric and magnetic fields}
It is clear that the twisted self-duality conditions (\ref{cov_tsc_s_no_index}) with all indices being taken to be spatial read
\begin{equation}
 \begin{pmatrix} {\mathcal E}^{i_1 i_2 \cdots i_s}[h] \\ {\mathcal E}^{i_1 i_2 \cdots i_s}[f] \\ \end{pmatrix} = \begin{pmatrix} {\mathcal B}^{i_1 i_2 \cdots i_s}[f] \\ -{\mathcal B}^{i_1 i_2 \cdots i_s}[h] \\ \end{pmatrix} . \label{key3}
\end{equation}
It turns out that these equations are completely equivalent to the full set of twisted self-duality conditions.  This is not surprising since the components of the curvature tensor with two or more zeroes are not independent on-shell from the components with one or no zero.  The fact that (\ref{key3}) completely captures all the equations of motion will be an automatic consequence of our subsequent analysis and so we postpone its proof to later (Section \textbf{\ref{Sec:action_tsc}} below, where the action principle from which these equations derive is shown to be the same as the one obtained earlier by Hamiltonian analysis).

\subsection{Getting rid of the Lagrange multipliers}
While a generic component of the curvature may contain up to $s$ time derivatives, the twisted self-duality conditions (\ref{key3}) contain only the first-order time derivatives $\dot{h}_{i_1 \cdots i_s}$ and $\dot{f}_{i_1 \cdots i_s}$.   One can give the fields $h_{i_1 \cdots i_s}$ and $f_{i_1 \cdots i_s}$ as Cauchy data on the spacelike hypersurface $x^0 = 0$.  The subsequent values of these fields are determined by the twisted self-duality conditions up to gauge ambiguities.  The Cauchy data $h_{i_1 \cdots i_s}$ and $f_{i_1 \cdots i_s}$ cannot be taken arbitrarily but must be such that their respective electric fields are both traceless since this follows from ${\mathcal E} = \pm {\mathcal B}$ and the fact that the magnetic field is traceless.  The constraints are equivalent to the condition that the traces of the Einstein tensors of both $h$ and $f$ should be zero,
\be
\bar{G}^{i_1 \cdots i_{s-2}} [h] = 0, \; \; \; \; \bar{G}^{i_1 \cdots i_{s-2}} [f] = 0 \label{constraints0}
\ee

The twisted self-duality conditions involve also the mixed components $h_{0i_2 \cdots i_s}$ and $f_{0i_2 \cdots i_s}$.  These are pure gauge variables, which act as Lagrange multipliers for constraints in the Hamiltonian formalism.  It is useful for the subsequent discussion to get rid of them.  Since they occur only in the magnetic fields, and through a gradient, this can be achieved by simply taking a curl on the first index.    Explicitly, from the twisted self-duality conditions (\ref{key3}) rewritten as
\be
{\mathcal E}^{a \, i_1 \cdots i_s} = \epsilon^{a}_{\; \; b} \, {\mathcal B}^{b \, i_1 \cdots i_s} \label{key4}
\ee
(${\mathcal E}^{a \, i_1 \cdots i_s}  \equiv {\mathcal E}^{ i_1 \cdots i_s} [h^a]$, ${\mathcal B}^{a \, i_1 \cdots i_s}  \equiv {\mathcal B}^{ i_1 \cdots i_s} [h^a]$, $a=1,2$, $(h^a) = (h,f)$, $\epsilon_{ab} = - \epsilon_{ba}$, $\epsilon_{12} = 1$), follows obviously the equation
\be
\epsilon_{jki_1} \partial^{k} {\mathcal E}^{a \, i_1 \cdots i_s} = \epsilon^{a}_{\; \; b} \, \epsilon_{jki_1} \partial^{k}{\mathcal B}^{b \, i_1 \cdots i_s} \label{key5}
\ee
which does not involve the mixed components $h_{0i_2 \cdots i_s}$ or $f_{0i_2 \cdots i_s}$ any more.

The equations (\ref{key5}) are physically completely equivalent to (\ref{key4}).  Indeed, it follows from (\ref{key5}) that 
\be
{\mathcal E}^{a \, i_1 \cdots i_s} = \epsilon^{a}_{\; \; b} \, \tilde{ {\mathcal B}}^{b \, i_1 \cdots i_s}
\ee
where $\tilde{ {\mathcal B}}^{b \, i_1 \cdots i_s}$ differs from the true magnetic field ${\mathcal B}^{b \, i_1 \cdots i_s}$ by an arbitrary gradient in $i_1$, or, in terms of the corresponding curvature components 
\be
\tilde{R}^a_{0i_1 j_2 k_2 \cdots j_s k_s} = R^a_{0i_1 j_2 k_2 \cdots j_s k_s} + \partial_{i_1} \mu^a_{j_2 k_2 \cdots j_s k_s}
\ee
for some arbitrary $\mu^a_{j_2 k_2 \cdots j_s k_s}$ with Young symmetry type
$$
\overbrace{\yng(9,9)}^{\text{$s-1$ boxes}}.
$$
Now, the cyclic identity fulfilled by the curvature implies $\partial_{[i_1} \mu^a_{j_2 k_2] \cdots j_s k_s} = 0$, i.e., in index-free notation, $d_{(s-1)} \mu^a =0$, and this yields $\mu^a = d_{(s-1)}^{s-1} \nu^a$ for some symmetric $\nu^a_{j_2 \cdots j_s}$ (see the Appendix \textbf{\ref{App:generalized_differential}}).  Comparing with (\ref{ambiguity}), we see that this is just the ambiguity in $R^a_{0i_1 j_2 k_2 \cdots j_s k_s}$ due to the presence of $h^a_{0j_2 \cdots j_s}$.  Therefore, one can absorb $\mu^a_{j_2 k_2 \cdots j_s k_s}$ in a redefinition of the pure gauge variables $h^a_{0j_2 \cdots j_s}$ and get thereby the equations (\ref{key4}).

It is in the form (\ref{key5}) that we shall derive the twisted self-duality conditions from a variational principle.

\section{Variational principle}
\label{VarPri}

\subsection{Prepotentials}
The searched-for variational principle involves as basic dynamical variables not the fields $h^a_{i_1 \cdots i_s}$, which are constrained, but rather ``prepotentials" that solve the constraints (\ref{constraints0}) and can be varied freely in the action.  The general solution of the constraint equation $\bar{G}^{a \, i_1 \cdots i_{s-2}} = 0$ was, of course, worked out in the previous chapter, since it is exactly the Hamiltonian constraint solved in section \textbf{\ref{Sec:solv_ham_constr_s}}: it implies the existence of prepotentials $Z^a_{i_1 \cdots i_s}$ from which $h^a_{i_1 \cdots i_s}$ derives, such that the Einstein tensor $G^{a\,  i_1 \cdots i_s}$ of $h^a_{i_1 \cdots i_s}$ is equal to the Cotton tensor $B^{a\, i_1 \cdots i_s}$ of $Z^a_{i_1 \cdots i_s}$.

The Cotton tensor $B^{a\,  i_1 \cdots i_s}$ was defined by equation \eqref{Cotton_s_def}. It involves $2s-1$ derivatives of the prepotentials and possesses the property of being invariant under spin-$s$ diffeomorphisms and Weyl symmetries,
\be
\delta Z^a_{i_1 \cdots i_s} = s \partial_{(i_1} \rho^a_{i_2 \cdots i_s)} + \frac{s(s-1)}{2} \delta_{(i_1 i_2} \sigma^a_{i_3 \cdots i_s)}. \label{gaugeZ}
\ee
It is symmetric, transverse and traceless.

Because of the gauge symmetries, the solution of the equation $G^{a\,  i_1 \cdots i_s}[h]= B^{a\,  i_1 \cdots i_s}[Z]$ for $h^a_{i_1 \cdots i_s}$ involves ambiguities.  To any given solution $h^a[Z]$ one can add an arbitrary variation of $h^a_{i_1 \cdots i_s}$ under spin-$s$ diffeomorphisms.   Furthermore $Z^a_{i_1 \cdots i_s}$ and $Z^a_{i_1 \cdots i_s} + \delta Z^a_{i_1 \cdots i_s}$ (with $\delta Z^a_{i_1 \cdots i_s}$ given by (\ref{gaugeZ})) yield $h^a[Z]$'s that differ by a spin-$s$ diffeomorphism. 

The expression for the spin-$s$ fields $h^a_{i_1 \cdots i_s}$ in terms of the prepotentials $Z^a_{i_1 \cdots i_s}$ was computed earlier. It is given in the Appendix \textbf{\ref{App:even_spin_prepot}} for even spin and in the Appendix \textbf{\ref{App:odd_spin_prepot}} for odd spin.

\subsection{Twisted self-duality and prepotentials}
In terms of the prepotentials, the electric fields are given by:
\be
{\mathcal E}^{a \, i_1 \cdots i_s} = B^{a i_1 \cdots i_s}[Z],
\ee
\noindent while the magnetic fields have the property 
\be
\epsilon_j^{\; \;  i_1k} \partial_{k}{\mathcal B}^{a \, ji_2 \cdots i_s} = \dot{B}^{a i_1 \cdots i_s}[Z] .
\ee

It follows that the twisted self-duality conditions \eqref{key5} take the form
\be
\epsilon^{i_1}_{\; \; jk} \partial^{j} B^{a \, k i_2 \cdots i_s}[Z] = \epsilon^{a}_{\; \; b} \dot{B}^{b i_1 \cdots i_s}[Z]  \label{key6}
\ee in terms of the prepotentials: the curl of the Cotton tensor of one prepotential is equal to ($\pm$) the time derivative of the other.  

\subsection{Action}

\label{Sec:action_tsc}

In their form (\ref{key6}), the twisted self-duality conditions are easily checked to derive from the following variational principle:
\be
S[Z] = \int \ d^4 x \ \left[  \frac{1}{2}\varepsilon_{ab} \dot{Z}^{a\, i_1 \cdots i_s} B^b_{i_1 \cdots i_s} \ - \ H \right] \label{action_s_prepotbis}
\ee
where the Hamiltonian is:
\begin{equation}
H = \delta_{ab} \left(\sum_{k=0}^{[\frac{s}{2}]} a_k  G^{[k] a \, i_1 \cdots i_{s-2k}}G^{[k] b}_{\; \; \; \; \; \; \; i_1 \cdots i_{s-2k}}  \right) .
\end{equation}

This is precisely the action obtained by the Hamiltonian analysis of Fronsdal covariant Lagrangian action for the spin-$s$ (see section \textbf{\ref{Sec:action_prepot_bos_s}}). The coefficient $a_k$ have the same value as then, being uniquely determined (up to a global factor) by the gauge invariance of the prepotentials, \eqref{gaugeZ}. The kinetic term indeed produces the right-hand side of \eqref{key6}, and the global factor of the Hamiltonian can be checked by computing one term the left-hand side of \eqref{key6} and comparing it with the corresponding term in the variation of the Hamiltonian.\\

Since we have started this chapter with Fronsdal description of higher spins and end up with this action as an equivalent description of its field, we can conclude that the action proposed in section \textbf{\ref{Sec:action_prepot_bos_s}} was indeed the Hamiltonian action of the spin-$s$ field.

We will examine again in Appendix \textbf{\ref{App:EMduality_s}} exactly how the symmetries constraint this action. In particular, we will see that this action enjoys, in addition to the gauge invariance of its fields (the prepotentials), an invariance under rigid rotations in the plane of its two prepotentials, which is precisel the off-shell form of \textit{electric-magnetic duality rotations}, hereby generalized to all higher spin.

\begin{subappendices}

\section{Higher dimensions and twisted self-duality}
In higher space-time dimension $D$,  the equations of motion can also be reformulated as twisted self-duality conditions on the curvatures of the spin-$s$ field and its dual. What is new is that the dual of a spin-$s$ field is not given by a symmetric tensor, but by a tensor of mixed Young symmetry type
$$
^\text{$D-3$ boxes}  \overbrace{\left\{ \yng(10,1,1,1,1,1,1) \right.}^{\text{$s$ boxes}}. 
$$
Consequently, the curvature tensor and its dual are also tensors of different types.  Nevertherless, the electric (respectively, magnetic) field of the spin-$s$ field is a spatial tensor of the same type as the magnetic (respectively, electric) field of its dual and  the twisted self-duality conditions again equate them (up to $\pm$ similarly to Eq. (\ref{key3})).  
The electric and magnetic fields are subject to tracelessness constraints that can be solved in terms of appropriate prepotentials, which are the variables for the variational principle from which the twisted self-duality conditions derive.  Again, this variational principle is equivalent to the Hamiltonian variational principle.

We have not worked out the specific derivation for all spins in higher dimensions $D$, but the  results of \cite{Bunster:2013oaa} for the spin-$2$ case, together with our above analysis, make us confident that this derivation indeed goes through as described here.

\newpage

\section{Electric-magnetic duality}

\label{App:EMduality_s}

We saw earlier that the equations of motion of the spin one (section \textbf{\ref{Sec:EMduality_1}}) and two (section \textbf{\ref{Sec:EMduality_2}}) possess a duality invariance under $SO(2)$ rotations in the plane of their (generalized) electric and magnetic fields\footnote{This can also be described - and were introduced in section \textbf{\ref{Sec:EMduality_1}} and \textbf{\ref{Sec:EMduality_2}} - as rotation in the plane of the curvature tensor and its Hodge dual.}. However, these rotations - called \textit{electric-magnetic duality} transformations - are not an apparent symmetry of their Lagrangian action - it is not even clear how to express these duality transformations at the level of the field variables, $A_{\mu}$ or $h_{\mu\nu}$. In order to make the off-shell form of these duality rotations explicit, it was necessary to go to the Hamiltonian formalism, where they appeared to be simply an $SO(2)$ rotation in the plane of the two prepotentials (obtained by solving the momentum and Hamiltonian constraints), under which the first order action was indeed invariant.

The first order description developed in this chapter for all higher spin - which is equivalent to their Hamiltonian formalism, developed in the last chapter - precisely offers the possibility to extend these observations to all spin.

In $D=4$ spacetime dimensions, the spin-$s$ field $h_{\mu_1 \cdots \mu_s}$ and its dual $f_{\mu_1 \cdots \mu_s}$ are tensors of the same type, as we have seen (section \textbf{\ref{Sec:cov_EMduality_s}}).  The equations enjoy then $SO(2)$ electric-magnetic duality invariance that rotates the field and its dual in the internal two-dimensional space that they span\footnote{The equations of motion \eqref{cov_tsc_matrix_s} are obviously invariant under rotations in the $h-f$ plane.}.  This comes over and above the twisted self-duality reformulation.

When one goes to the non-covariant formalism in which twisted self-duality conditions become first order in time (section \textbf{\ref{Sec:non_cov_tsc_s}}), the form \eqref{key3} of the equations of motion obviously conserve this invariance under rotations in the $h-f$ plane. Once the constraint part of these equations is solved and the prepotentials are introduced, it clearly appears that the $SO(2)$ electric-magnetic duality invariance amounts to perform rotations in the internal space of the prepotentials $Z^a_{i_1 \cdots i_s}$ - since the prepotentials are such that their respective Cotton tensors are equal to the Einstein tensors of the spatial components of the $h$ and $f$ fields (or, equivalently, to the electric fields of the $h$ and $f$ fields). Again, the two prepotentials have the same type in $D=4$: they are fully symmetric (spatial) tensors.

We saw that the equations of motion in terms of the prepotentials \eqref{key6} derived from the action principle \eqref{action_s_prepotbis}, which is also the Hamiltonian action of the spin-$s$. It is clear that this action is invariant under $SO(2)$ rotations of the prepotentials $Z^a$.

Thus, the prepotential reformulation makes it obvious that $SO(2)$ electric-magnetic duality invariance is a manifest off-shell symmetry, and not just a symmetry of the equations of motion.

\subsection{How the symmetries fix the action}

The Hamiltonian action principle of the spin-$s$ in terms of its prepotentials - or the action from which the first order in time twisted self-duality conditions can be extracted - is actually uniquely fixed by its invariance under gauge transformations of the prepotentials (which combine spatial generalized linearized diffeomorphism with traceful parameters and generalized linearized Weyl rescaling) and under electric-magnetic duality transformations, or $SO(2)$ rotations in the plane of the two prepotentials\footnote{These gauge and duality invariance only fix the action provided the quadratic presence of the prepotentials and the number of spatial and time-like derivatives present are fixed. The global factors of the Hamiltonian and kinetic terms remain, of course, undetermined.}

This action is: 
\be
S[Z] = \int \ d^4 x \ \left[  \frac{1}{2} \ \varepsilon_{ab} B^{a\, i_1 \cdots i_s} \dot{Z}^b_{i_1 \cdots i_s} \ - \ H \right] \label{action_s_prepot_app}
\ee
where the Hamiltonian is:
\begin{equation}
H = \delta_{ab} \left(\sum_{k=0}^{[\frac{s}{2}]} a_k  G^{[k] a \, i_1 \cdots i_{s-2k}}G^{[k] b}_{\; \; \; \; \; \; \; i_1 \cdots i_{s-2k}}  \right) .
\end{equation}

The gauge invariance of the prepotentials is:
\begin{equation}
\delta Z^a_{i_1 \cdots i_s} = s \ \partial_{(i_1} \rho^a_{i_2 \cdots i_s)} \ + \ \frac{s(s-1)}{2} \ \delta_{(i_1 i_2} \sigma^a_{i_3 \cdots i_s)}.\label{app_dualEM_gaugeZ}
\end{equation}

The kinetic term in the action is manifestly invariant under the gauge symmetries of the prepotentials, because the Cotton tensor is both invariant under \eqref{app_dualEM_gaugeZ} and traceless. The Hamiltonian is  manifestly invariant under the spin-$s$ diffeomorphisms, since it involves the Einstein tensors of the prepotentials.  It is also invariant under spin-$s$ Weyl transformations up to a surface term, since its coefficients were precisely adjusted (and uniquely fixed, up to a global factor) in this purpose. 

The action is furthermore manifestly invariant under $SO(2)$ electric-magnetic duality rotations in the internal plane of the prepotentials (we go back to the different naming of the two prepotentials $(Z^a_{i_1 \cdots i_s}) \equiv (P_{i_1 \cdots i_s}, \Phi_{i_1 \cdots i_s})$):
\begin{eqnarray}
\Phi'_{i_1 \cdots i_s} &=& \cos \theta \Phi_{i_1 \cdots i_s} - \sin \theta P_{i_1 \cdots i_s} ,
\\
P'_{i_1 \cdots i_s} &=& \cos \theta P_{i_1 \cdots i_s} + \sin \theta\Phi_{i_1 \cdots i_s} ,
\end{eqnarray}
since it involves only the $SO(2)$ invariant tensors $\varepsilon_{ab}$ and $\delta_{ab}$.  As recalled in the introduction, exhibiting duality symmetry in the case of spin two was in fact the main motivation of \cite{Henneaux:2004jw} for solving the constraints and introducing the prepotentials.

The gauge symmetries combined with duality invariance constrain the form of the action in a very powerful way. Indeed, the most general invariant quadratic kinetic term involving $2s$ derivatives of the prepotentials, among which one is a time derivative,  is a multiple of the above kinetic term.  Similarly, the most general invariant quadratic Hamiltonian involving $2s$ spatial derivatives of the prepotentials is  a mutiple of the above Hamiltonian\footnote{$SO(2)$ electric-magnetic duality invariance forces us to contract the prepotentials with $SO(2)$ invariant tensors, either $\varepsilon_{ab}$ or $\delta_{ab}$, and it is easily seen that we need an antisymmetric tensor in the kinetic term and a symmetric one in the Hamiltonian, other choices leading to boundary terms.}. By rescaling appropriately the time if necessary, one can therefore bring the action to the above form, which is consequently the most general gauge and duality invariant quadratic action with the required number of derivatives.

\end{subappendices}

\part{Toward supersymmetry}

\chapter{Hypergravity and electric-magnetic duality}

\label{Chap:hygra}

We have, up to this point, only considered free bosonic fields over flat space, whose dynamic and geometry we have systematically studied: based on our investigation of higher spin three dimensional Euclidean conformal geometry (chapter \textbf{\ref{Chap:confgeom}}), we have obtained the Hamiltonian formulation of these theories (chapter \textbf{\ref{Chap:hambos}}) and expressed their equations of motion as twisted self-duality conditions (chapter \textbf{\ref{Chap:tsdc}}).

However, a large part of the physical interest of these bosonic theories comes from the possibility of giving them a supersymmetric extension. We will not be able to explore thoroughly this vast and fascinating field in the present work, and we will limit ourselves to the opening of two pathways that could lead to a better understanding of theories of significant interest.

In this chapter, we will begin to include fermions into the picture developed in the previous part in a very straightforward way, by considering the Hamiltonian analysis of a supermultiplet. The simplest supermultiplet including higher spin fields is the so-called \textit{hypergravity}\footnote{The ``hyper" here refers to the fact that the fermionic superpartern has spin superior to the bosonic one. This termionology can be misleading, since hypersymmetry is a term often used to describe a symmetry whose parameter is a spinor-vector, as opposed to supersymmetry, for which it is a spinor.}, which pairs the graviton with a spin-$5/2$ field\footnote{This theory has originally been studied in \cite{Aragone:1979hx,Berends:1979wu,Berends:1979kg}.}, as opposed to the well-known supergravity, in which the superpartner of the spin-$2$ is a spin-$3/2$ field. Hypergravity has been much less studied than supergravity for the basic reason that the introduction of interactions seems impossible in the former, due to well-known no-go theorems \footnote{See \cite{Aragone:1979hx,Aragone:1980rk}. Interactions can still be introduced in three spacetime dimensions, see \cite{Aragone:1983sz}}.

Nevertheless, more recent investigations have led to the promising construction of possible interacting theories incorporating the spin-$2$ and $5/2$ gauge fields, in addition to an infinite number of other ones \footnote{This is the work of Lebedev school, see \cite{Fradkin:1986ka,Fradkin:1986qy,Fradkin:1987ks,Vasiliev:1990en,Vasiliev:1999ba}. For a recent review: \cite{Bekaert:2010hw,Didenko:2014dwa}}, so that the dynamic and symmetries of hypergravity might after all deserve some attention.

We will particularly be interested, here, into the way the electric-magnetic duality invariance of the spin two theory (reviewed in section \textbf{\ref{Sec:EMduality_2}}) can be extended into a symmetry of hypergravity that commutes with the supersymmetric (rigid) transformations that rotate the spin-$2$ and $5/2$ fiels into each other. Since we saw that electic-magnetic duality only becomes a manifest symmetry of the action in first order formalism, this suggests to make the Hamiltonian analysis of the spin-$5/2$ theory, which will show that the properties of the bosonic systems seem to still be present in the fermionic ones: the solving of the Hamiltonian constraints leads to prepotentials whose gauge symmetry includes both diffeomorphism and Weyl rescalings. 

Remarkably, supersymmetry will map the two prepotentials of the graviton (the ``momentum" and ``field" prepotentials) into the single prepotential of the spin-$5/2$ in such a way that an electric-magnetic duality rotation of the graviton will be mapped by supersymmetry onto a chiral rotation of the fermion. In particular, this will allow us to combine an electric-magnetic duality transformation of the boson with a chirality rotation of the fermion into a symmetry commuting with supersymmetry. This generalizes similar results well-established for the ``classic" supergravity\footnote{See \cite{Bunster:2012jp}.} \\

We will begin by recalling the covariant formulation of  hypergravity.  We then turn in Section \ref{HamiltonianForm} to the Hamiltonian formulation of the theory as this is necessary to manifestly exhibit duality.  Section \ref{Solving} provides the solution of the constraints of the Hamiltonian formulation in terms of the prepotentials.  In Section \ref{SUSYPrep}, we write the supersymmetry transformations for the prepotentials. In Section \ref{FormOfSusyAction}, we give the expression of the action in terms of the prepotentials.  The more technical derivations are relegated to Appendix \ref{Technical}.  

For comparison, we have also included an appendix \ref{132Mult} which pursues the similar duality analysis for the spin-$(1, 3/2)$ system.
\section{Covariant form of hypergravity}
\label{FreeLS}
\subsection{Action}
The action describing the combined system of a free spin-2 massless field, described by a symmetric tensor field  $h_{\mu \nu} = h_{\nu \mu}$, with a free spin-5/2 massless field, described by a symmetric spinor-rensor field $\psi_{\mu \nu} = \psi_{\nu \mu}$, is the sum of the actions of these two fields, introduced in equations \eqref{action_bos_full} and \eqref{action_fermi_full} for an arbitrary spin: 
\begin{equation}
S[h_{\mu \nu}, \psi_{\mu \nu}]=\int d^{4}x \left({\mathcal L}_2 + {\mathcal L}_{\frac52} \right) \label{Action0}
\end{equation}
where
\begin{eqnarray}
{\mathcal L}_2&=&
- \ \frac{1}{2} \ \left\lbrace
\partial_{\rho} h_{\mu\nu} \partial^{\rho} h^{\mu\nu}
\ - \ 2 \ \partial^{\rho} h_{\rho\mu}\partial_{\nu} h^{\nu\mu}
\ + \ 2 \ \partial^{\mu} h_{\mu\nu} \partial^{\nu} h'
\ - \ \partial_{\mu} h' \partial^{\mu} h'
\right\rbrace ,
\end{eqnarray}
and
\begin{eqnarray}
{\mathcal L}_{\frac52} &=& 
- \ i \   \left\lbrace
 \bar{\psi}^{\mu\nu} \slashed{\partial} \psi_{\mu\nu}
\ - \  4 \ \bar{\psi}^{\mu\rho}\gamma_\rho \partial^{\nu} \psi_{\mu\nu}
\ + \  2 \ \bar{\psi}^{\mu \rho}\gamma_\rho \slashed{\partial} \slashed{\psi}_{\mu}
\ + \ 2 \  \bar{\psi}^{\mu \rho}\gamma_\rho\partial_{\mu} \psi'
\ - \   \frac{1}{2} \ \bar{\psi}'\slashed{\partial}\psi ' 
\right\rbrace . \nonumber \\   \label{action_lag_5_2}
\end{eqnarray}

This fermionic action is not exactly the action introduced in \eqref{action_fermi_full}, because we are now using a spin-$5/2$ which is a ``real" field (since we pair it with a real bosonic field into a supermultiplet): $\psi_{\mu\nu}$ is taken to be a Majorana spinor-tensor. This allows us to rewrite a certain number of terms in \eqref{action_fermi_full} (through partial integrations), in order to get to \eqref{action_lag_5_2}.

\subsection{Gauge symmetries}
The action (\ref{Action0}) is invariant under linearized diffeomorphisms
\begin{equation}
\delta_{\text{gauge}} h_{\mu\nu} = \partial_{\mu} \xi_{\nu} \ + \ \partial_{\nu} \xi_{\mu} .
\end{equation}
and gauge transformations of the spin-$5/2$ field,
\begin{equation}
\delta_{\text{gauge}} \psi_{\mu\nu} = \partial_{\mu} \zeta_{\nu} \ + \ \partial_{\nu} \zeta_{\mu} ,
\end{equation}
where the ``spin-$3/2$" gauge parameter $\zeta_\mu$ is $\gamma$-traceless,
\begin{equation}\slashed{\zeta} = 0 .
\end{equation}
Because of the tracelessness condition, one can express $\zeta_0$ in terms of $\zeta_k$,
\begin{equation}
\zeta_0 = - \gamma_0 \gamma^k \zeta_k.
\end{equation}
There are three independent fermionic gauge symmetries.
\subsection{Supersymmetry}
The action is also invariant under rigid supersymmetry, which reads,
\begin{eqnarray}
\delta_{\text{SUSY}} h_{\mu\nu} &=& 8i \ \bar{\epsilon}\psi_{\mu\nu}, \label{susy_h}
\\
\delta_{\text{SUSY}} \psi_{\mu\nu} &=& 
\left( \partial_{\mu} h_{\nu\rho} \ + \ \partial_{\nu} h_{\mu\rho} \right) \gamma^{\rho} \epsilon
\ - \ 2 \ \partial_{\rho} h_{\mu\nu} \gamma^{\rho} \epsilon
\nonumber \\ &&
 +  \ \left( \epsilon_{\mu \lambda \sigma \rho} \partial^{\lambda} h_{\nu}^{\phantom{\nu} \sigma} 
\ + \ \epsilon_{\nu \lambda \sigma \rho} \partial^{\lambda} h_{\mu}^{\phantom{\nu} \sigma} \right)
\gamma^{\rho} \gamma_5 \epsilon
. \qquad \label{susy_psi}
\end{eqnarray}
The supersymmetry parameter $\epsilon$ is a constant spinor.
\subsection{Equations of motion}
The equations of motion can be written as
\begin{eqnarray}
&&R_{\mu \nu}= 0  \label{Einstein}\\
&& \mathcal{F}_{\mu\nu} \label{EOM52}= 0
\end{eqnarray}
where $R_{\mu \nu}$ is the linearized Ricci tensor of the spin-$2$\footnote{Note that $R_{\mu\nu}$ is both the trace of the linearized Riemann tensor of the spin-$2$ and its Fronsdal tensor; see section \textbf{\ref{Sec:low_bos}}} (see section \textbf{\ref{Sec:low_bos}}) and $\mathcal{F}_{\mu \nu}$ the Fronsdal tensor of the spin-$5/2$ (see section \textbf{\ref{Sec:Lag_s_fermi}}). These tensors are given by:
\begin{eqnarray}
R_{\mu\nu} &=&
\square h_{\mu\nu}
\ - \ 2 \ \partial_{(\mu} \partial^{\rho} h_{\nu )\rho}
\ + \ \partial_{\mu} \partial_{\nu} h' ,
\\
\mathcal{F}_{\mu\nu} &=&
\slashed{\partial} \psi_{\mu\nu}
\ - \ 2 \ \partial_{(\mu} \slashed{\psi}_{\nu )} .
\end{eqnarray}

The equations of motion and the action are invariant under electric-magnetic duality rotations in the internal two-dimensional space spanned by the Riemann tensor $R_{\mu\nu\rho\sigma} = 4 \ \partial_{[\mu \vert } \partial_{[\rho} h_{\sigma ] \vert \nu ]}$ of the spin-$2$ and its dual (see section \textbf{\ref{Sec:EMduality_2}}).  

They are also invariant under chirality rotations of the spinor fields:
\begin{equation}
\delta_{\text{chiral}} h_{\mu \nu} = 0, \; \; \; \; \delta_{\text{chiral}}  \psi_{\mu \nu} =  \lambda \gamma_5 \psi_{\mu \nu}
\end{equation}
That the action is invariant under chirality rotations is manifest. To display explicitly its electric-magnetic duality invariance, one must go to the first-order formalism and introduce prepotentials for the spin-two field. Supersymmetry forces one to then introduce  prepotentials for the spin-five-half partner, as we now discuss.  We will also see that a definite chirality rotation must accompany a duality transformation if duality is to commute with supersymmetry.
 
\section{Hamiltonian form}
 \label{HamiltonianForm}
 
 \subsection{Spin-2 part}
 As we recalled in section \textbf{\ref{Sec:Grav_ham}}, the action of the graviton takes the canonical form:
\begin{eqnarray}
S_2 \left[Z^a_{kl} \right] &=&
\frac{1}{2} \ \int \ d^4 x \ \left\lbrace
\epsilon_{ab} \dot{Z}^{a kl} B^b_{kl}
\ - \  \delta_{ab} \left[G^{a kl} G^b_{kl} \ - \ \frac{1}{2} \ \bar{G}^a \bar{G}^b \right]
\right\rbrace , 
\end{eqnarray}
\noindent where the prepotentials $Z^a_{kl}$ are related to the field $h_{kl}$ and its conjugate momenta $\Pi^{kl}$ (the variables $h_{0\mu}$ are Lagrange multipliers and do not have independent conjugate momenta; in this prepotential formalism, they are absent) through:
\begin{eqnarray}
P_{kl} &=& \varepsilon_{kmp} \varepsilon_{lnq} \partial^m \partial^n \Phi^{pq} ,
\\
h_{kl} &=&
\frac{1}{2} \ \varepsilon_{(k\vert mn} \partial^m \phi^{n}_{\phantom{n}\vert l )} 
\ + \ 2 \ \partial_{(k} \xi_{l)} ,
\end{eqnarray}
\noindent where the two prepotentials were renamed according to $Z^a_{kl} \equiv \left(P_{kl} , \phi_{kl} \right)$. 

The gauge invariance of these prepotentials is given by:
\begin{equation}
\delta Z^a_{kl} = 2 \ \partial_{(k} \lambda_{l)}
\ + \ \delta_{kl} \mu^a .
\end{equation}
The $\xi_k$ are pure gauge.

 \subsection{Spin-5/2 part}
 We turn now to the spin-5/2 action.  Its Hamiltonian formulation has been performed in \cite{Aragone:1979hw,Borde:1981gh} but our treatment differs from these earlier works in that we do not fix the gauge at any stage and solve the (first-class) constraints through the introduction of  prepotentials on which the fields depend locally. 
 
The action for the spin-five-half field, being already of first order, is almost in canonical form (see section \textbf{\ref{Sec:1st_order_gen}} for the general formalism).  Because there are gauge symmetries, some of the components of $\psi_{\mu \nu}$ are Lagrange multipliers for the corresponding first class constraints, while the other components define the phase space of the system.  What differentiates the Lagrange multipliers  from the standard phase space variables is that the gauge variations of the former contain the time derivatives $\dot{\zeta}_m$ of the gauge parameters $\zeta_m$, while the gauge variations of the latter do not involve $\dot{\zeta}_m$.  
 
To bring the action for the spin-five-half field to canonical form, we make the redefinition
\begin{eqnarray} 
&&\Xi = \psi_{00} - 2\ \gamma^{0} \gamma^{k} \psi_{0k} , \label{IV_variablechange}\\
&&\psi_{0k} = \psi_{0k}, \\
&&\psi_{mn} = \psi_{mn}.
\end{eqnarray}
One then has the following transformation rules,
\begin{eqnarray}
\delta_{\text{gauge}}\Xi &=& - \ 2\ \gamma^{k} \gamma^{l} \partial_{k} \zeta_{l} \label{IV_gaugetransf_00'} \\
\delta_{\text{gauge}}\psi_{0k} &=& \partial_{0} \zeta_{k} + \ \gamma^{0} \gamma^{l} \partial_{k} \zeta_{l} \label{IV_gaugetransf_0k}\\
\delta_{\text{gauge}} \psi_{kl} &=& \partial_{k}\zeta_{l} + \partial_{l}\zeta_{k}, \label{IV_gaugetransf_mn}
\end{eqnarray}
from which one immediately identifies the $\psi_{0k}$ components as the Lagrange multipliers.  
Indeed, the action takes the form:
\begin{eqnarray}
S_{\frac52}[\Xi, \psi_{mn}, \psi_{0k}] &=& \int \ d^{4} x \  \lbrace  \Theta^{A} ( \psi_{B} ) \dot{\psi}_{A}
\ - \  \mathcal{H}\left( \psi_{B} \right)  
\ + \  \psi_{0k}^T \mathcal{F}_{k}\left( \psi_{B} \right)  \rbrace, \qquad \hspace{.5cm}
\label{IV_action_newform}
\end{eqnarray}
with $(\psi_A) \equiv (\Xi, \psi_{mn})$ and where the Hamiltonian $\mathcal{H}$,  the constraint functions $\mathcal{F}_k$ and the symplectic potential $\Theta^A$ are explicitly given by,
\begin{eqnarray}
\mathcal{H} &=& 
- \ \frac{3i}{4} \ \bar{\Xi} \gamma^{k} \partial_{k} \Xi
\ + \ \frac{i}{2} \ \bar{\psi}^{kl} \gamma^{m} \partial_{m} \psi_{kl}
\ - \ 2i\ \bar{\psi}^{kl} \gamma_k \partial^{m} \psi_{lm} 
\ - \ i \ \bar{\psi}^{kl} \gamma_{k} \partial_{l} \Xi
\nonumber \\ &&
 + \ i \ \bar{\psi}^{kl} \gamma_{k} \partial_{l} \psi
\ + \ \frac{i}{2} \ \bar{\Xi} \gamma^{k} \partial_{k} \psi
\ - \ \frac{i}{4} \ \bar{\psi} \gamma^{l} \partial_{l}\psi
\ + \ i \ \bar{\psi}_{kl} \gamma^{l} \gamma^{m} \gamma^{n} \partial_{m} \psi^k_{\ n}
, \ \ \ \ \ \label{IV_hamiltonian} 
\\
\mathcal{F}_k &=&
 i \  \left[
\partial_{k} \Xi
\ - \ 2 \ \partial^{l} \psi_{kl}
\ + \ \partial_{k} \psi
\ + \ \gamma_{k} \gamma^{l} \partial_{l} \Xi 
\ + \ 2 \ \gamma^{l} \gamma^{m} \partial_{l} \psi_{km}
\right.
\nonumber \\ &&
\left. \qquad 
\ - \ \gamma_{k} \gamma^{l} \partial_{l} \psi 
 + \ 2 \ \gamma_k \gamma^l \partial^m \psi_{lm} \right] , \ \ \ \ \ \ \label{IV_constraints}
\\
\Theta_{\Xi} &=&
 \frac{i}{4}\ \Xi^{T}
\ - \ \frac{i}{2} \ \psi^{T}, \qquad \qquad \label{IV_theta_00}
\\
\Theta_{kl} &=&
\frac{i}{2} \ \psi^{T}_{kl}
\ - \ \frac{i}{4} \ \psi^{T} \delta_{kl}
\ - \ \frac{i}{2} \ \psi^{T}_{km} \gamma^{m} \gamma_{l}
\ - \ \frac{i}{2} \ \psi^T_{lm} \gamma^{m} \gamma_{k}.  \label{IV_theta_kl}
\end{eqnarray}
Here and from now on, $\psi$ denotes the {\it spatial} trace $\psi = \psi^k_{\; \; k}$.
The action (\ref{IV_action_newform}) is the searched-for action in canonical form because the symplectic form $\omega = d \Theta$ is non-degenerate and yields the following brackets between the canonical variables,
\begin{eqnarray}
\left\lbrace \Xi \left( \vec{x} \right)  , \Xi  \left( \vec{x}' \right) \right\rbrace_D &=& 
- \ \frac{5i}{4} \ \delta \left(  \vec{x} - \vec{x}' \right) ,\label{IV_diracbracket_0000}
\\
\left\lbrace \Xi\left( \vec{x} \right)  , \psi_{kl}  \left( \vec{x}' \right) \right\rbrace_D &=& 
\frac{i}{4} \ \delta_{kl} \ \delta \left(  \vec{x} - \vec{x}' \right) , \label{IV_diracbracket_00kl}
\end{eqnarray}
\begin{eqnarray}
\left\lbrace \psi_{kl} \left( \vec{x} \right)  , \psi_{mn}  \left( \vec{x}' \right) \right\rbrace_D &=& 
\delta \left(  \vec{x} - \vec{x}' \right)
\left[ - \ \frac{i}{4} \ ( \delta_{km} \delta_{ln} + \delta_{kn} \delta_{lm} )
\ + \ \frac{i}{4} \ \delta_{kl} \delta_{mn} \right.
\nonumber \\ && \ \ \
 + \left. \ \frac{i}{8} \ \left( \delta_{km} \gamma_{ln} + \delta_{kn} \gamma_{lm} + \delta_{lm} \gamma_{kn} + \delta_{ln} \gamma_{km} \right) \right] .
 \nonumber \\ &&   \hspace{1.2cm}
 \label{IV_diracbracket_klmn}
\end{eqnarray}
These brackets would appear as Dirac brackets had one introduced conjugate momenta for the fermionic variables $\psi_A$ and eliminated the corresponding second class constraints that express these momenta in terms of the $\psi_A$ through the Dirac bracket procedure - hence the notation $\lbrace, \rbrace_D$.
The variables $\psi_{0k}$ are the Lagrange multipliers for the first-class constraints $\mathcal{F}_k= 0$, which are easily verified to generate the fermionic gauge transformations (\ref{IV_gaugetransf_00'}) and (\ref{IV_gaugetransf_mn}) through the Dirac bracket.
\section{Solving the constraints - Prepotentials}
\label{Solving}
The fermionic constraint (\ref{IV_constraints}) can be rewritten in the simpler form
\begin{equation}
\partial_k \Xi + \partial_k \psi + 2\gamma^{ij} \partial_i \psi_{kj} = 0 \label{C5/2}
\end{equation}
Indeed, if one multiplies (\ref{C5/2}) with $k$ replaced by $l$ by the invertible operator $i \left(\delta^l_k + \gamma_k \gamma^l \right)$, one gets ${\mathcal F}_k = 0$.
The equation (\ref{C5/2}) itself can be rewritten as $\partial_i \left(\delta^i_k \Xi + \delta^i_k \psi + 2 \gamma^{ij} \psi_{kj} \right) = 0$, which by virtue of the Poincar\'e lemma implies
\begin{equation}\delta^i_k \Xi + \delta^i_k \psi + 2 \gamma^{ij} \psi_{kj} = \partial_r A^{[ir]}_{\; \; \; \; k} \label{KeyEq0} \end{equation}
for some tensor $A^{[ir]}_{\; \; \; \; k} = - A^{[ri]}_{\; \; \; \; k}$, which can be decomposed into irreducible components as 
$$ A^{[ir]}_{\; \; \; \; k} = \delta^i_k a^r - \delta^r_k a^i + \epsilon^{irm} m_{mk} $$ with $m_{mk} = m_{km}$.  The equation (\ref{KeyEq0}) can be solved to yield  $\psi_{ij}$ and $\Xi$  in terms of $a^k$ and $m_{ij}$ as
\begin{eqnarray}
&& 2 \psi_{lk} = \frac12 \gamma_l \gamma_m \left(\partial_k a^m - \frac13 \delta^m_k \partial_q a^q + \epsilon^{qms} \partial_q m_{sk} \right) \nonumber \\
&& \hspace{1cm}  - \partial_k a_l + \frac13 \delta_{lk} \partial_q a^q + \epsilon_l^{\; \; qs} \partial_q m_{sk} \label{psiam}\\
&& \Xi = - \frac23 \partial_q a^q - \psi \label{Xiam}
\end{eqnarray}
Now, the right-hand side of (\ref{psiam}) is not symmetric in general, while $\psi_{lk}$ is symmetric.  The system of equations (\ref{KeyEq0}) is overdetermined.  A symmetric solution of (\ref{KeyEq0}) exists if and only if $A^{[ir]}_{\; \; \; \; k}$ -- or equivalently, $a^q$ and $m_{ij}$ --  fulfills the constraints expressing that  $\psi_{lk} = \psi_{kl}$.  This condition is a differential constraint which can be rewritten, by virtue of the Poincar\'e lemma, as
\begin{eqnarray}
&& a_k \delta^q_l - a_l \delta^q_k + \frac12 \left(\gamma_l \gamma_m \delta^q_k - \gamma_k\gamma_m \delta^q_l \right) a^m - \frac13 \gamma_{lk} a^q \nonumber \\
&& \; \; \; + \epsilon_l^{\; \; qs} m_{sk}  - \epsilon_k^{\; \; qs} m_{sl} + \frac12 \gamma_l \gamma_m \epsilon^{qms} m_{sk} - \frac12 \gamma_k \gamma_m \epsilon^{qms} m_{sl}\nonumber \\ && = \partial_p B_{lk}^{qp} \hspace{1cm} \label{KeyEq1}
\end{eqnarray}
for some tensor $B_{lk}^{qp}$ antisymmetric in $l,k$ and $q,p$,  $B_{lk}^{qp} = -B_{kl}^{qp} = -B_{lk}^{pq}$.  Trading each antisymmetric pair for one single index using the Levi-Civita tensor, this tensor $B_{lk}^{qp}$ is equivalent to a tensor $B_{ij}$, $B^{lkqp} \sim \epsilon^{lki} \epsilon^{qpj} B_{ij}$, which can be decomposed into a symmetric part $\Sigma_{ij}$ and an antisymmetric part $\epsilon_{ijk} V^k$.  
The system (\ref{KeyEq1}) is a system of 9 linear spinorial equations for the 9 spinorial unknows $a^q$ and $m_{ij} = m_{ji}$.  We leave it to the reader to verify that this system can be solved uniquely for  $a^q$ and $m_{ij}$ in terms of the first derivatives of $\Sigma_{ij}$ and $V^k$, which are therefore unconstrained.  Next, one substitutes the resulting expressions in (\ref{psiam}) and (\ref{Xiam}), allowing in addition for a general gauge transformation term in which one absorbs all gauge transformation terms involving  $\Sigma_{ij}$ and $V^k$ through redefinitions.  Then one finally gets the searched-for expressions
\begin{eqnarray}
&& \Xi = \triangle \Sigma + \partial^a \partial^b \Sigma_{ab}- 2 \gamma^k \gamma^l \partial_k v_l \label{XiSigma1}\\
&& \psi_{ij} = \delta_{ij} \triangle \Sigma  - \delta_{ij} \partial^a \partial^b \Sigma_{ab} - 2 \triangle \Sigma_{ij} 
 - \left( \gamma_{ia} \partial^a \partial^b \Sigma_{jb} + \gamma_{ja} \partial^a \partial^b \Sigma_{ib}
\right)  \nonumber \\
&& \hspace{1.5cm} + \left( \gamma_{ia} \triangle \Sigma^{a}_{\; \; j} + \gamma_{ja} \triangle \Sigma^{a}_{\; \; i}\right)  +\partial_i v_j + \partial_j v_i \label{psiSigma1}
\end{eqnarray}
of the fields $\Xi$ and $\psi_{ij}$ in terms of a spin 5/2 prepotential $\Sigma_{ij} = \Sigma_{ji}$  and a spin 3/2 prepotential $v_i$ (in which $V^k$ has been completely absorbed).
As a consistency check, it is easy to verify that these expressions identically fulfill the spinorial constraint (\ref{C5/2}), as they should. Note that the prepotential $v_i$ drops out from gauge-invariant expressions and plays for that reason a less fundamental role.  It is similar to the prepotential $u_i$ for the graviton.   
One can easily verify that the expressions are invariant under the gauge symmetries of the prepotential $\Sigma_{ij}$
\begin{equation}
\delta \Sigma_{ij} = \partial_i \mu_j + \partial_j \mu_i +\gamma_i \eta_j + \gamma_j \eta_i \label{TransSigma}
\end{equation}
where $\mu_i$ and $\eta_i$ are arbitrary. Indeed, one finds $\delta \Xi = 0$ and $\delta \psi_{ij} = 0$ provided one transforms simultaneously the other prepotential $v_i$ as
\begin{equation} \delta v_i = 2 \triangle \mu_i + \gamma_{ia} \partial^a \partial^b \mu_b - \gamma_{ia} \triangle \mu^a - \gamma^a \partial_a \eta_i + \gamma_i \partial^a \eta_a.\end{equation}
Note that the special choice $\eta_i = \gamma_i \phi$ yields the Weyl rescaling $2 \phi \delta_{ij}$ of the prepotential $\Sigma_{ij}$, which is therefore contained among the gauge symmetries.  The gauge transformations (\ref{TransSigma}) are redundant since one gets $\delta \Sigma_{ij} = 0$ for $\mu_i = \gamma_i \epsilon$ and $\eta_i = - \partial_i \epsilon$.  There are thus $3 \times 4$ (components of the vector-spinor $\mu_i$) plus $3 \times 4$ (components of the vector-spinor $\eta_i$) minus $1 \times 4$ (reducibility relations) $=  5 \times 4 = 20$ independant gauge parameters.
The gauge transformations of the prepotentials $v_i$ are mere shifts $ v_i \rightarrow v_i + \omega_i$ where $\omega_i$ are arbitrary vector-spinors.  These prepotentials contain therefore no degree of freedon and can be set for instance equal to zero.
The gauge transformations  (\ref{TransSigma}) are the only gauge symmetries of the prepotential $\Sigma_{ij}$, as can be seen by mere counting.  There are $6 \times 4 = 24$ components of the tensor-spinor $\Sigma_{ij}$.  These are unconstrained, but the gauge freedom removes as we have just seen $20$ components, leaving $4$ independent arbitrary functions needed to describe the physical helicities  of a massless spin-$5/2$ field. 
Another way to reach the same conclusion goes as follows.  One may rewrite (\ref{psiSigma1}) in terms of the Schouten tensor $S_{ij}[\Sigma]$ of $\Sigma_{ij}$,  
\begin{eqnarray}
&& S_{ij}[\Sigma] = \frac{1}{2} \left( \partial_i \partial^m \Sigma_{mj} + \partial_j \partial^m \Sigma_{mi}  - \triangle \Sigma_{ij} - \partial_i \partial_j \Sigma \right) \nonumber \\
&& \hspace{2cm} - \frac{1}{4} \left( \partial^m \partial^n \Sigma_{mn} - \triangle \Sigma \right) \delta_{ij}, \nonumber
\end{eqnarray}
as
\begin{equation}
\psi_{ij} =4 S_{ij} - 2 \gamma_i^{\;  a} S_{aj} - 2 \gamma_j^{\;  a} S_{ai} + \partial_i \bar{v}_j + \partial_j \bar{v}_i
\end{equation}
with
\begin{equation}
\bar{v}_i = v_i  - 2 \partial^m \Sigma_{mi} + \partial_i \Sigma + \gamma_i^{\; a} \partial^b \Sigma_{ab} - \gamma_i^{\; a} \partial_a \Sigma
\end{equation}
which transforms as
\begin{equation}
\delta \bar{v}_i =  \gamma ^k (\partial_i \eta_k - \partial_k \eta_i) + 3 \gamma_{i}^{\; \, km} \partial_m \eta_k 
\end{equation}
under (\ref{TransSigma}).  Similarly, one finds
\begin{equation}
\Xi = 4 S - 2 \gamma^k \gamma^l \partial_k \bar{v}_l.
\end{equation}
The Schouten tensor $S_{ij}[\Sigma]$ is invariant under linearized diffeomorphisms and ``sees" only the $\gamma$-transformations.  One finds explicitly 
\begin{eqnarray}
\delta S_{ij}[\Sigma] &=& \left( - \partial_i \partial_j + \frac12 \triangle \right) \gamma^k \eta_k \nonumber \\
&& \left( \frac12 (\gamma_i \partial_j + \gamma_j \partial_i) - \frac12 \delta_{ij} \gamma^m \partial_m \right) \partial^k \eta_k \nonumber \\
&& + \frac12 \gamma^k \partial_k \left( \partial_i \eta_j + \partial_j \eta_i \right) \nonumber  \\ && - \frac12 \left(\gamma_i \triangle \eta_j + \gamma_j \triangle \eta_i \right) \label{TransSchouten}
\end{eqnarray}
under (\ref{TransSigma}).
Now, if $\psi_{ij}$ and $\Xi$ are both equal to zero, one finds that the Schouten tensor is given by
\begin{eqnarray}
8 S_{ij} &=& - 2 \partial^m\bar{v}_m  \delta_{ij}+ \partial_m(\gamma_j^{\; \, m} \bar{v}_i + \gamma_i^{\; \, m} \bar{v}_j) \nonumber \\ && + \partial_i(\gamma_j^{\; \, m} \bar{v}_m +  \bar{v}_j) +\partial_j(\gamma_i^{\; \, m} \bar{v}_m +  \bar{v}_i) \nonumber \\ \label{FormOfS}
\end{eqnarray}
with $\bar{v}_i$ constrained by
\begin{equation}
\gamma^k \gamma^l \partial_k \bar{v}_l +\partial^k \bar{v}_k = 0; \label{ConstraintV}
\end{equation}
By virtue of the Poincar\'e lemma, the general solution of (\ref{ConstraintV}) is given by 
\begin{equation}
\bar{v}_k = - 7 \gamma^m\partial_k \eta_m + 8 \gamma_k \partial^m \eta_m - 9 \gamma^m \partial_m\eta_k + 5 \gamma_k^{\; \, st} \partial_s \eta_t 
\end{equation} for some arbitrary $\eta_k$. 
But then, (\ref{FormOfS}) has exactly the form (\ref{TransSchouten}) of a $\gamma$-gauge transformation.  This means that one can set the Schouten tensor $S_{ij}[\Sigma]$ -- and hence also the Riemann tensor of $\Sigma_{ij}$ since the number of dimensions is three -- equal to zero by a $\gamma$-gauge transformation.  This implies that in that $\gamma$-gauge,  the tensor $\Sigma_{ij}$ itself is given by $\partial_i \mu_j + \partial_j \mu_i$,  or, in general, by (\ref{TransSigma}) if one does not impose any $\gamma$-gauge condition.  This shows that (\ref{TransSigma}) exhausts indeed the most general gauge freedom.
In parallel to what was found for the spin-2 prepotentials, the gauge symmetries of the spin-$\frac52$ prepotential take the same form as the gauge symmetries of the free conformal spin 5/2 field theory in 4 dimensions \cite{Fradkin:1985am}. To make the comparison it should be borne in mind that in \cite{Fradkin:1985am}, the redundancy in the gauge parameters is fixed by imposing the condition $\gamma^i \mu_i = 0$.
Of course, two physically different theories may have the same gauge symmetries. It is nevertheless quite remarkable that gauge conformal invariance should emerge automatically when one requires electric-magnetic duality invariance to be manifest. This is yet another evidence for the subtle interplay between duality invariance and spacetime symmetry.
\section{Supersymmetry transformations in terms of the prepotentials}
\label{SUSYPrep}
A somewhat tedious but direct computation shows that the supersymmetry transformations of the bosonic prepotentials are given by
\begin{equation}
\delta\Phi_{ij}= -8i\bar{\epsilon}\chi_{ij} \label{SUSYPHI}
\end{equation}
and 
\begin{equation}
\delta P_{ij}=-8i  \bar{\epsilon} \gamma_5 \chi_{ij}. \label{SUSYP}
\end{equation}
Here, 
\begin{eqnarray}
\chi_{ij} &=& \frac{1}{2}(\epsilon_{jmn}\partial^{m}\Sigma_{i}^{\ n}+\epsilon_{imn}\partial^{m}\Sigma_{j}^{\ n} \nonumber \\ && -\epsilon_{lmn}\gamma_{j}^{\ l}\partial^{m}\Sigma_{i}^{\ n}-\epsilon_{lmn}\gamma_{i}^{\ l}\partial^{m}\Sigma_{j}^{\ n})
\end{eqnarray}
 transforms under gauge transformations of $\Sigma_{ij}$ in the same way as the bosonic prepotentials, i.e.,
$$\delta \chi_{ij} = \partial_i \alpha_j + \partial_i \alpha_j  + \delta_{ij} \beta, $$  for some $\alpha_i$ and $\beta$. 
The details are given in  Appendix \ref{Technical}.   Similarly, one finds,
\begin{equation}
\delta \Sigma_{ij}=2\gamma_{5}\gamma_{0}(\Phi_{ij}-\gamma_{5}P_{ij})\epsilon , \label{SUSYS}
\end{equation}
where again the gauge transformations match.
Formulas (\ref{SUSYPHI}), (\ref{SUSYP}) and (\ref{SUSYS}) are the analogs for the spin-$(2,5/2)$ multiplet of the formulas (\ref{SusyPrepA10}), (\ref{SusyPrepA20}) and (\ref{SusyPrepChi0}) giving the supersymmetry transformations of the prepotentials of the spin-$(1, 3/2)$ multiplet.  Even though the relationship between the original fields and the prepotentials is rather different for the two systems - they involve only first-order derivatives for the lower spin multiplet, but also second-order derivatives for the higher spin multiplet -, the final formulas giving the supersymmetry transformation rules are remarkably similar and simple. 
In both cases, the transformation of the fermionic prepotential involves the combination $M_{ij} \equiv \Phi_{ij}-\gamma_{5}P_{ij}$ (or $M_i \equiv A^2_{i}-\gamma_{5}A^1_{i}$) of the two bosonic prepotentials that transforms under duality as
$$ \delta_{\text{dual}} M_{ij} = - \ \alpha \ \gamma_5 M_{ij} $$
(or $ \delta_{\text{dual}} M_{i} = - \ \alpha \ \gamma_5 M_{i} $).  At the same time,  the transformation of the first bosonic prepotential $\Phi_{ij}$ (or $A^1_i$) involves the function $\chi_{ij}$ (or $\psi_i$) of the fermionic prepotential that has identical gauge transformation properties as $\Phi_{ij}$ (or $A^1_i$), while the transformation of the second prepotential involves $-\gamma_5$ times it.
\section{Action}
\label{FormOfSusyAction}
The action for the combined spin-$(2,5/2)$ system can be written in terms of the prepotentials. One finds by direct substitution,
\be
S[Z^a_{ij}, \Sigma_{ij}] = \int dt \left[ \int d^3 \, x \left( \frac{1}{2} \ \varepsilon^{ab} B^{ij}_a \dot{Z} _{bij}  \ + \ \Theta^{A}  \dot{\psi}_{A}\right) - H \right], \label{FinalAction}
\ee
with 
\be
H =\int d^3x \left[ \frac{1}{2} \ \left(G^{a kl} G^b_{kl} \ - \ \frac{1}{4} \ \bar{G}^a \bar{G}^b \right) \delta_{ab} + \mathcal{H} \right]. \label{HFinal}
\ee
Here, $B_a^{\; ij} \equiv B^{\; ij}[Z_a]$ (respectively $G_a^{\; ij} \equiv G^{\; ij}[Z_a]$) is the co-Cotton tensor (respectively, the Ricci tensor) constructed out of the prepotential $Z_{aij}$ \cite{Bunster:2012km}, whereas $\Theta^{A}(\Xi(\Sigma), \psi(\Sigma))$ and $\mathcal{H}(\Xi(\Sigma),\psi(\Sigma))$ are the functions of the second and third order derivatives of the prepotential $\Sigma_{ij}$ obtained by merely substituting (\ref{XiSigma1}) and (\ref{psiSigma1}) in (\ref{IV_theta_00}), (\ref{IV_theta_kl}) and (\ref{IV_hamiltonian}).
The action (\ref{FinalAction}) is manifestly separately invariant under duality rotations of the bosonic prepotentials $Z^a_{ij}$:
\begin{equation}
\delta_{\text{dual}} Z^1_{ij} = \alpha Z^2_{ij}, \; \; \; \; \;  \delta_{\text{dual}} Z^2_{ij} = - \alpha Z^1_{ij}
\end{equation}
and chirality rotations of the prepotential $\Sigma_{ij}$,
\begin{equation}
\delta_{\text{chiral}} \Sigma_{ij} = \lambda \gamma_5 \Sigma_{ij},
\end{equation}
which implies $\delta_{\text{chiral}} \psi_{ij} = \lambda \gamma_5 \psi_{ij}$.
As such, neither the duality rotations (with untransforming fermions) nor the chirality transformations (with untransforming bosons) commute with supersymmetry. One can, however,  extend the duality transformation to the fermionic prepotential in such a way that duality and supersymmetry commute.  This is done by combining the duality rotations of the bosonic superpotentials with a chirality transformation of the fermionic prepotential of amplitude $ - \alpha$, 
\begin{eqnarray}
\delta_{\text{dual}} \Sigma_{jk} &=&  - \alpha \ \gamma_5 \Sigma_{jk} .
\end{eqnarray}

\begin{subappendices}

\section{Technical derivations}
\label{Technical}

\subsection{Bosonic prepotential $Z_{2}^{mn}=\phi^{mn}$}
The supersymmetry transformation rule of the graviton is
\bea
\delta h_{ij}&=&8i\bar{\epsilon}\psi_{ij}=8i\bar{\epsilon}\left[\delta_{ij}\Delta \Sigma-\delta_{ij}\partial^{a}\partial^{b}\Sigma_{ab}\right.\nonumber\\
&-&\left.2\Delta\Sigma_{ij}-\gamma_{ia}\partial^{a}\partial^{b}\Sigma_{jb}-\gamma_{ja}\partial^{a}\partial^{b}\Sigma_{ib}\right.\nonumber\\
&+&\left.\gamma_{ja}\Delta \Sigma^{a}_{\ i} + \gamma_{ia}\Delta \Sigma^{a}_{\ j} + \partial_{i}\epsilon_{j}+\partial_{j}\epsilon_{i}\right]\label{susyh}
\eea
On the other hand
\begin{equation}
\delta h_{ij}=\partial^{r}\epsilon_{irs}\delta\phi^{s}_{\ j}+\partial^{r}\epsilon_{jrs}\delta\phi^{s}_{\ i}+\partial_{i}\delta v_{j}+\partial_{j}\delta v_{i} .\end{equation}
In order to compare these two expressions and find out the form of the supersymmetry transformation of the graviton prepotential, $\delta\phi_{ij}$, it is useful to recast (\ref{susyh}) in the form 
\begin{equation}
\delta h_{ij}=-8i\bar{\epsilon}\left[\partial^{r}\epsilon_{irs}\chi^{s}_{\ j}+\partial^{r}\epsilon_{jrs}\chi^{s}_{\ i}+\partial_{i}\eta_{j}+\partial_{j}\eta_{i}\right]
\end{equation}
This can be accomplished by setting 
\begin{eqnarray}
\chi_{js}&=&\frac{1}{2}\left[\epsilon_{jmn}\partial^{m}\Sigma^{n}_{\ s}+\epsilon_{smn}\partial^{m}\Sigma^{n}_{\ j} -\epsilon_{lmn}\gamma_{j}^{\ l}\partial^{m}\Sigma_{s}^{\ n}-\epsilon_{lmn}\gamma_{s}^{\ l}\partial^{m}\Sigma_{j}^{\ n}\right] ,\label{chi}
\end{eqnarray}
and
\begin{eqnarray}
\epsilon_{j}&=&\eta_{j}-\frac{1}{2}\partial_{j}\Sigma+\frac{3}{2}\partial^{r}\Sigma_{jr}+\frac{1}{2}\partial^{l}\gamma_{jl}\Sigma  +\frac{1}{2}\gamma^{al}\partial_{l}\Sigma_{ja}-\frac{1}{2}\partial_{r}\gamma_{ja}\Sigma^{ra} .
\end{eqnarray}
Thus one concludes that, up to a gauge transformation,
\begin{eqnarray}
\delta\Phi_{ij}&=&-8i\bar{\epsilon}\chi_{ij}\\
&=& -4i\bar{\epsilon}(\epsilon_{jmn}\partial^{m}\Sigma_{i}^{\ n}+\epsilon_{imn}\partial^{m}\Sigma_{j}^{\ n} -\epsilon_{lmn}\gamma_{j}^{\ l}\partial^{m}\Sigma_{i}^{\ n}-\epsilon_{lmn}\gamma_{i}^{\ l}\partial^{m}\Sigma_{j}^{\ n})
\hspace{0.5cm} \end{eqnarray}
and
\begin{eqnarray}
\delta v_{i}&=&-8i\bar{\epsilon}\eta_{i} \\
&=&-8i\bar{\epsilon}(\epsilon_{i}+\frac{1}{2}\partial_{i}\Sigma-\frac{3}{2}\partial^{n}\Sigma_{in}-\frac{1}{2}\partial^{r}\gamma_{ir}\Sigma  \qquad +\frac{1}{2}\gamma_{is}\partial_{r}\Sigma^{sr}-\frac{1}{2}\partial^{r}\gamma_{sr}\Sigma_{i}^{\ s}) 
\hspace{0.5cm}\end{eqnarray}
Note that the field $\chi_{ij}$ defined by (\ref{chi}) transforms as 
\begin{equation}
\delta \chi_{ij} = \partial_i \alpha_j + \partial_j \alpha_i  + \delta_{ij} \beta
\end{equation}
with
\begin{equation}
\alpha_i = \frac12 \left(\epsilon_{imn} - \epsilon_{lmn} \gamma_i^{\; \, l} \right)\partial^m \mu^n + \tilde{\gamma}_5 \eta_i
\end{equation}
and
\begin{equation}
\beta = \epsilon_{lmn} \gamma^l \partial^m \eta^n
\end{equation}
under the gauge transformations (\ref{TransSigma}) of the prepotential $\Sigma_{ij}$.
\subsection{Bosonic prepotential $Z_{1}^{mn}=P^{mn}$}
Using $\delta h_{ij}=8i\bar{\epsilon}\psi_{ij}$ and the equation of motion of the hypergraviton
\be
\gamma_{\rho}\partial^{\rho}\psi_{\mu\nu}-\partial_{\mu}\gamma^{\rho}\psi_{\rho\nu}-\partial_{\nu}\gamma^{\rho}\psi_{\rho\mu}=0
\ee
one may write the supersymmetry transformation of the extrinsic curvature $K_{ij}=-\frac{1}{2}(\partial_{0}h_{ij}-\partial_{i}h_{0j}-\partial_{j}h_{0i})$ as
\be
\delta K_{ij}= -2 i \bar{\epsilon}\epsilon^{mkl} \gamma_5 \gamma_{kl}  \left[\partial_m \psi_{ij} - \partial_i \psi_{mj} - \partial_j \psi_{mi} \right]. \label{exc}
\ee
Substituting the expression of $\psi_{ij}$ in terms of the fermionic prepotential $\Sigma_{ij}$, one then 
finds
\begin{eqnarray}
\delta K_{ij} &=&
- \ 4i \bar{\epsilon} \gamma^0 \left[
\delta_{ij}\gamma_k \partial^k \left(\Delta \Sigma 
\ - \ \partial^m \partial^n \Sigma_{mn} \right)
 \ - \ 2 \ \gamma_k \partial^k \Delta _{ij}
 \right.
\nonumber \\ &&
\left. 
\ + \ 2 \ \gamma_{m} \partial_i\partial^m \partial^n \Sigma_{nj} 
\ + \ 2 \ \gamma_{m} \partial_j\partial^m \partial^n \Sigma_{ni} 
\ + \ \gamma_{kim} \partial^k\Delta \Sigma^{m}_{\phantom{m}j}
\right.
\nonumber \\ &&
\left. 
\ + \ \gamma_{kjm} \partial^k\Delta \Sigma^{m}_{\phantom{m}i}
\ + \ \gamma_{kjm} \partial^m \partial_n \partial_i \Sigma^{nk}
\ + \ \gamma_{kim} \partial^m \partial_n \partial_j \Sigma^{nk}
\right] . 
\end{eqnarray}
As a corrolary, we get:
\begin{eqnarray}
\delta K &=&
- \ 4i \bar{\epsilon} \gamma^0 \left[ \gamma_k \partial^k \Delta \Sigma
\ + \ \gamma_{m} \partial^i\partial^m \partial^n \Sigma_{ni}
\right] . 
\end{eqnarray}
and thus 
\begin{eqnarray}
\delta \Pi_{ij} &=& - \ \delta K_{ij} \ + \ \delta_{ij} \delta K
\nonumber \\ &=&
4i \bar{\epsilon} \gamma^0 \left[
\ - \ 2 \ \delta_{ij}\gamma_k \partial^k  \partial^m \partial^n \Sigma_{mn}
\ - \ 2 \ \gamma_k \partial^k \Delta \Sigma_{ij}
\ + \ 2 \ \gamma_{m} \partial_i\partial^m \partial^n \Sigma_{nj}
\right.
\nonumber \\ &&
\left. 
\ + \ 2 \ \gamma_{m} \partial_j\partial^m \partial^n \Sigma_{ni}
\ - \ \epsilon_{ikm} \gamma^0 \gamma_5 \partial^k\Delta \Sigma^{m}_{\phantom{m}j}
\ - \ \epsilon_{jkm} \gamma^0 \gamma_5 \partial^k\Delta \Sigma^{m}_{\phantom{m}i}
\right.
\nonumber \\ &&
\left. 
\ + \ \epsilon_{ikm} \gamma^0\gamma_5 \partial^k \partial_n \partial_j \Sigma^{nm}
\ + \ \epsilon_{jkm} \gamma^0 \gamma_5 \partial^k \partial_n \partial_i \Sigma^{nm}
\right]   \label{DeltaPi007}
\end{eqnarray}
Now, knowing the supersymmetry transformation $\delta\Phi_{ij}=-8i\bar{\epsilon}\chi_{ij}$ of the prepotential $\Phi_{ij}$ and comparing with the spin-1-spin-3/2 system, it is natural to guess that the supersymmetry transformation of the other bosonic prepotential $P_{ij}$ is simply 
\be
\delta P_{ij}=8i\bar{\epsilon}\gamma_5\chi_{ij} \label{DeltaChi007}
\ee
To prove this claim, it suffices to compute $\delta \Pi_{ij}$ from its definition in terms of $P_{ij}$ and (\ref{DeltaChi007}),  and verify that the resulting expression coincides with (\ref{DeltaPi007}).  The computation is direct.  One finds
\begin{eqnarray}
\delta \Pi_{ij} &=&
- \ 8i\bar{\epsilon}\gamma_5 \ \epsilon_{ikl} \epsilon_{jmn} \partial^k \partial^m \chi^{ln}
\nonumber \\ &=&
- \ 4 i\bar{\epsilon}\gamma_5 \ \left[ \epsilon_{ikl} \epsilon_{jmn}\epsilon^{lrs} \partial^k \partial^m \partial_r \Sigma_s^{\phantom{s}n}
\ + \ \epsilon_{ikl} \epsilon_{jmn}\epsilon^{nrs} \partial^k \partial^m \partial_r \Sigma_s^{\phantom{s}l}
\right.
\nonumber \\ && 
\left. \qquad
\ - \ \epsilon_{ikl} \epsilon_{jmn} \epsilon_{prs} \gamma^{np} \partial^k \partial^m \partial^r \Sigma^{sl}
\ - \ \epsilon_{ikl} \epsilon_{jmn}\epsilon_{prs} \gamma^{lp} \partial^k \partial^m \partial^r \Sigma^{sn}
\right] 
\nonumber \\ &=&
- \ 4i\bar{\epsilon}\gamma_5 \ \left[ 
 \epsilon_{jmn} \partial^k \partial^m \partial_i \Sigma_k^{\phantom{s}n}
\ + \ \epsilon_{ikl}  \partial^k \partial^m \partial_j \Sigma_m^{\phantom{m}l} 
\ - \ \epsilon_{jmn}  \partial^m \Delta \Sigma_i^{\phantom{s}n}
\ - \ \epsilon_{ikl} \partial^k \Delta \Sigma_j^{\phantom{s}l}
\right.
\nonumber \\ && 
\left. \ \ \
\ - \ 2 \ \delta_{ij} \epsilon_{prs} \gamma^{np} \Delta \partial^r \Sigma^{s}_{\phantom{s}n}
\ + \ 2 \ \delta_{ij} \epsilon_{prs} \gamma^{np} \partial_n \partial_m \partial^r \Sigma^{sm}
\ - \ \epsilon_{prs} \gamma^{np} \partial_n \partial_i \partial^r \Sigma^{s}_{\phantom{s}j}
\right. \nonumber \\ && \left. \ \ \
\ - \ \epsilon^{prs} \gamma_{jp} \partial^n \partial_i \partial_r \Sigma_{sn}
\ - \ \epsilon^{prs} \gamma_{ip} \partial_j \partial^m \partial_r \Sigma_{sm}
\ - \ \epsilon_{prs} \gamma^{mp} \partial_j \partial_m \partial^r \Sigma^{s}_{\phantom{s}i}
\right.
\nonumber \\ && 
\left. \ \ \
\ + \ \epsilon^{prs} \gamma_{ip} \Delta \partial_r \Sigma_{sj}
\ + \ \epsilon^{prs} \gamma_{jp} \Delta \partial_r \Sigma_{si} 
\ + \ \epsilon_{prs} \gamma^{np} \partial_j \partial_i \partial^r \Sigma^{s}_{\phantom{s}n}
\right. \nonumber \\ && \left. \ \ \
\ + \ \epsilon_{prs} \gamma^{np} \partial_j \partial_i \partial^r \Sigma^{s}_{\phantom{s}n}
\right] . \qquad \qquad
\end{eqnarray}
Using the identity $\gamma^{np} = \gamma^0 \gamma_5 \epsilon^{npq} \gamma_q$, this expression becomes
\begin{eqnarray}
\delta \Pi_{ij} &=&
- \ 4i \bar{\epsilon}\gamma^0 \ \left[ 
\ - \ \epsilon_{jmn}\gamma^0 \gamma_5 \partial^k \partial^m \partial_i \Sigma_k^{\phantom{s}n}
\ - \ \epsilon_{ikl} \gamma^0 \gamma_5 \partial^k \partial^m \partial_j \Sigma_m^{\phantom{m}l}
\ + \ \epsilon_{jmn} \gamma^0 \gamma_5 \partial^m \Delta \Sigma_i^{\phantom{s}n}
\right. \nonumber \\ && \left. \qquad
\ + \ \epsilon_{ikl} \gamma^0 \gamma_5\partial^k \Delta \Sigma_j^{\phantom{s}l}
\ - \ 2 \ \delta_{ij} \gamma_r  \partial^r \left(\Delta \Sigma
\ - \ \partial_n \partial_m\Sigma^{nm}\right)
\right. \nonumber \\ && \left. \qquad
\ - \ 2 \ \gamma_r\partial_j \partial^m \partial^r \Sigma_{im}
\ - \ 2 \ \gamma_r\partial^n \partial_i \partial^r \Sigma_{nj}
\ + \ 2 \ \gamma_r \Delta \partial^r \Sigma_{ij}
\ + \ 2 \  \gamma_r\partial_j \partial_i \partial^r \Sigma
\right] . 
\end{eqnarray}
The term $2\gamma_r\partial_j \partial_i \partial^r \Sigma - 2 \delta_{ij} \gamma_r  \partial^r \Delta \Sigma$ is a gauge transformation of $\Pi_{ij}$, and can be removed, giving:
\begin{eqnarray}
\delta \Pi_{ij} &=&
- \ 4i\bar{\epsilon}\gamma^0 \ \left[ 
\ - \ \epsilon_{jmn}\gamma^0 \gamma_5 \partial^k \partial^m \partial_i \Sigma_k^{\phantom{s}n}
\ - \ \epsilon_{ikl} \gamma^0 \gamma_5 \partial^k \partial^m \partial_j \Sigma_m^{\phantom{m}l}
\ + \ \epsilon_{jmn} \gamma^0 \gamma_5 \partial^m \Delta \Sigma_i^{\phantom{s}n}
\right. \nonumber \\ && \left. \qquad 
\ + \ \epsilon_{ikl} \gamma^0 \gamma_5\partial^k \Delta \Sigma_j^{\phantom{s}l} 
\ + \ 2 \ \delta_{ij} \gamma_r  \partial^r \partial_n \partial_m\Sigma^{nm}
\ + \ 2 \ \gamma_r \Delta \partial^r \Sigma_{ij}
\right. \nonumber \\ && \left. \qquad
\ - \ 2 \ \gamma_r\partial_j \partial^m \partial^r \Sigma_{im}
\ - \ 2 \ \gamma_r\partial^n \partial_i \partial^r \Sigma_{nj}
\right] , 
\end{eqnarray}
which is exactly (\ref{DeltaPi007}).  This proves the correctness of (\ref{DeltaChi007}).
\subsection{Supersymmetric variation of the fermionic prepotential $\Sigma_{ij}$}
One can rewrite the supersymmetry transformation for $\psi_{\mu\nu}$ as:
\bea
\delta\psi_{ij}&=&(\partial_{i}h_{j\rho}
+\partial_{j}h_{i\rho})\gamma^{\rho}\epsilon-2 \ \partial_{\rho}h_{ij}\gamma^{\rho}\epsilon
+(\epsilon_{i\lambda\sigma\rho}\partial^{\lambda}h_{j}^{\ \sigma} +\epsilon_{j\lambda\sigma\rho}\partial^{\lambda}h_{i}^{\ \sigma})\gamma^{\rho}\gamma_{5}\epsilon\nonumber\\
&=&\delta_{\Pi}\psi_{ij}+\delta_{h}\psi_{ij} ,\label{psi}
\eea
where:
\bea
\delta_{\Pi}\psi_{ij}&=&(\partial_{i}h_{j0}+\partial_{j}h_{i0})\gamma^{0}\epsilon-2 \ \partial_{0}h_{ij}\gamma^{0}\epsilon
+(\epsilon_{i0km}\partial^{0}h_{j}^{\ k}+\epsilon_{j0km}\partial^{0}h_{i}^{\ k}\nonumber \\ &&\ \ \ -\epsilon_{i0km}\partial^{k}h_{j}^{\ 0}-\epsilon_{j0km}\partial^{k}h_{i}^{\ 0})\gamma^{m}\gamma_{5}\epsilon , \nonumber\\
\delta_{h}\psi_{ij}&=&(\partial_{i}h_{jk}+\partial_{j}h_{ik})\gamma^{k}\epsilon \ - \ 2 \ \partial_{k}h_{ij}\gamma^{k}\epsilon
 +(\epsilon_{ikl0}\partial^{k}h_{j}^{\ l}+\epsilon_{jkl0}\partial^{k}h_{i}^{\ l})\gamma^{0}\gamma_{5}\epsilon .
\eea
We compute $\delta_{\Pi}\psi_{ij}$ and $\delta_{h}\psi_{ij}$ separately.
\subsubsection{Term depending on the prepotential $Z^{1}_{mn}=P_{mn}$}
Let us first focus  on the term involving the extrinsic curvature. Adding to it the gauge transformation $\delta\psi_{ij}=(\partial_{i}h_{j0}+\partial_{j}h_{i0})\gamma^{0}\epsilon-(\epsilon_{i0km}\partial_{j}h^{k0}+\epsilon_{j0km}\partial_{i}h^{k0})\gamma^{m}\gamma_{5}\epsilon$ yields:
\bea
\delta_{\Pi}\psi_{ij}&=&2 \ (\partial_{i}h_{j0}+\partial_{j}h_{i0}-\partial_{0}h_{ij})\gamma^{0}\epsilon+(\epsilon_{i0km}\partial^{0}h_{j}^{\ k}
-\epsilon_{i0km}\partial^{k}h_{j}^{\ 0}
-\epsilon_{i0km}\partial_{j}h^{k0}
\nonumber \\ && 
+\epsilon_{j0km}\partial^{0}h_{i}^{\ k}
-\epsilon_{j0km}\partial^{k}h_{i}^{\ 0}
-\epsilon_{j0km}\partial_{i}h^{k0})\gamma^{m}\gamma_{5}\epsilon\nonumber\\
&=& 4 \ K_{ij}\gamma^{0}\epsilon
-2\ \epsilon_{0ikm}K_{j}^{\ k}\gamma^{m}\gamma_{5}\epsilon
-2\ \epsilon_{0jkm}K_{i}^{\ k}\gamma^{m}\gamma_{5}\epsilon\nonumber\\
&=& 4 \ (- \ \Pi_{ij}
+\frac{\Pi}{2}\delta_{ij})\gamma^{0}\epsilon+2\epsilon_{0ikm}\Pi_{j}^{\ k}\gamma^{m}\gamma_{5}\epsilon
+2 \ \epsilon_{0jkm}\Pi_{i}^{\ k}\gamma^{m}\gamma_{5}\epsilon .
\eea
Using:
\bea
 - \ 4 \ \Pi_{ij} \ + \ 2 \ \Pi\delta_{ij}
&=&
- \ 4 \ \epsilon_{iab}\epsilon_{jcd}\partial^{a}\partial^{c}P^{bd}
+2 \ \delta_{ij}\epsilon_{mab}\epsilon^{m}_{\ cd}\partial^{a}\partial^{c}P^{bd}\nonumber\\
&=& - \ 4 \ \epsilon_{iab}\epsilon_{jcd}\partial^{a}\partial^{c}P^{bd}
+\epsilon_{ixy}\epsilon_{jxy}\epsilon_{mab}\epsilon^{m}_{\ cd}\partial^{a}\partial^{c}P^{bd}\nonumber\\
&=& - \ 2 \ \epsilon_{iab}\epsilon_{jcd}\partial^{a}\partial^{c}P^{bd}
-\epsilon_{jmb}\epsilon^{m}_{\ cd}\partial_{i}\partial^{c}P^{bd}
-\epsilon_{imb}\epsilon^{m}_{\ cd}\partial_{j}\partial^{c}P^{bd}
\nonumber \\ && \ \ \
+\epsilon_{jma}\epsilon^{m}_{\ cd}\partial^{a}\partial^{c}P_{i}^{\ d} +\epsilon_{ima}\epsilon^{m}_{\ cd}\partial^{a}\partial^{c}P_{j}^{\ d} ,
\eea
one finds:
\bea
\delta_{\Pi}\psi_{ij}&=&
-\left[2\epsilon_{iab}\epsilon_{jcd}\partial^{a}\partial^{c}P^{bd}
-\epsilon_{jma}\epsilon^{m}_{\ cd}\partial^{a}\partial^{c}P_{i}^{\ d} 
-\epsilon_{ima}\epsilon^{m}_{\ cd}\partial^{a}\partial^{c}P_{j}^{\ d}\right]\gamma^{0}\epsilon\nonumber\\
&&+2\left[\epsilon_{0jnm}\epsilon_{iab}\epsilon^{n}_{\ cd}\partial^{a}\partial^{c}P^{bd}
 +\epsilon_{0inm}\epsilon_{jab}\epsilon^{n}_{\ cd}\partial^{a}\partial^{c}P^{bd}\right]\gamma^{m}\gamma_{5}\epsilon\nonumber\\
&&-\epsilon_{jmb}\epsilon^{m}_{\ cd}\partial_{i}\partial^{c}P^{bd}\gamma^{0}\epsilon
-\epsilon_{imb}\epsilon^{m}_{\ cd}\partial_{j}\partial^{c}P^{bd}\gamma^{0}\epsilon , \label{psi-pi}
\eea
and by dropping again a gauge transformation for $\psi_{ij}$ the following supersymmetry transformation rule may be derived for $\chi_{ij}$:
\be
\delta\chi_{j}^{\ b}=2\partial_{m}P_{j}^{\ b}\gamma^{m}\gamma_{5}\epsilon-\epsilon_{jcd}\partial^{c}P^{bd}\gamma^{0}\epsilon-\epsilon^{bcd}\partial_{c}P_{jd}\gamma^{0}\epsilon . \label{chi1}
\ee
\subsubsection{Term depending on the $Z^{2}_{mn}=\phi_{mn}$ prepotential}
We turn now to $\delta_{h}\psi_{ij}$. Expressing it in terms of the prepotential $\phi_{mn}$ gives:
\bea
\delta_{h}\psi_{ij}&=&
(\partial_{i}h_{jk}+\partial_{j}h_{ik})\gamma^{k}\epsilon
-2 \ \partial_{k}h_{ij}\gamma^{k}\epsilon
+(\epsilon_{ikl0}\partial^{k}h_{j}^{\ l}+\epsilon_{jkl0}\partial^{k}h_{i}^{\ l})\gamma^{0}\gamma_{5}\epsilon
\nonumber\\
&=&\partial_{i}(\partial^{l}\epsilon_{jlm}\phi^{m}_{\ k}+\partial^{l}\epsilon_{klm}\phi^{m}_{\ j})\gamma^{k}\epsilon
+\partial_{j}(\partial^{l}\epsilon_{ilm}\phi^{m}_{\ k}+\partial^{l}\epsilon_{klm}\phi^{m}_{\ i})\gamma^{k}\epsilon
\nonumber\\
&&-2 \ \partial_{k}(\partial^{l}\epsilon_{ilm}\phi^{m}_{\ j}+\partial^{l}\phi^{m}_{\ i})\gamma^{k}\epsilon
 +\epsilon_{ikl0}\partial^{k}(\partial^{m}\epsilon_{jmn}\phi^{nl}+\partial^{m}\epsilon^{l}_{\ mn}\phi^{n}_{\ j})\gamma^{0}\gamma_{5}\epsilon\nonumber\\
&&+\epsilon_{jkl0}\partial^{k}(\partial^{m}\epsilon_{imn}\phi^{nl}+\partial^{m}\epsilon^{l}_{\ mn}\phi^{n}_{\ i})\gamma^{0}\gamma_{5}\epsilon \nonumber \\ 
\eea
which, up to a gauge transformation, contributes to the supersymmetry transformation rule of $\chi_{ij}$ as follows:
\be
\delta\chi_{jm}=-2\partial_{k}\phi_{jm}\gamma^{k}\epsilon-(\partial^{n}\epsilon_{jnp}\phi^{p}_{\ m}+\partial^{n}\epsilon_{mnp}\phi^{p}_{\ j})\gamma^{0}\gamma_{5}\epsilon\label{chi2}
\ee
Finally, by adding up (\ref{chi1}) and (\ref{chi2}) the complete supersymmetry transformation rule for the fermionic prepotential $\chi_{ij}$ is obtained:
\begin{eqnarray}
\delta\chi_{ij} &=&
-2 \ \partial_{k}(\phi_{ij}
+P_{ij}\gamma_{5})\gamma^{k}\epsilon
+\epsilon_{icd}\partial^{c}(\phi^{d}_{\ j}\gamma_{5}
-P_{j}^{\ d})\gamma^{0}\epsilon
\nonumber \\ &&
+\epsilon_{jcd}\partial^{c}(\phi^{d}_{\ i}\gamma_{5}
-P_{i}^{\ d})\gamma^{0}\epsilon
\end{eqnarray}
Using the identity $\epsilon_{lxy}\gamma_{i}^{\ l}=(\delta_{ix}\gamma_{y}-\delta_{iy}\gamma_{x})\gamma_{0}\gamma_{5}$ in (\ref{chi}), one can see immediatly that this implies the following transformation rule for the prepotential $\Sigma_{ij}$,
\be
\delta \Sigma_{ij} = \ 2 \ \gamma_{5}\gamma_{0}(\phi_{ij}-\gamma_{5}P_{ij})\epsilon
\ee

\newpage

\section{The spin-$(1,\frac32)$ multiplet}
\label{132Mult}
\subsection{Spin 1}
The covariant action of the Maxwell field 
\begin{equation}
S_{1} [A_\mu]
=
- \ \frac{1}{4} \ \int \ d^4 x \ F_{\mu\nu} F^{\mu\nu}
 \label{action_lag_1}
\end{equation}
can be recast in a manifestly duality-invariant form by going to the first-order formalism and introducing a second vector potential $A^2_i$ through the resolution of Gauss' contraint \cite{Deser:1976iy} (see section \textbf{\ref{Sec:EM_ham}}).  One finds, in duality covariant notations:
\begin{equation}
I = \frac{1}{2} \int dx^0 d^3x \left( \varepsilon_{ab} \vec{B}^a \cdot \dot{\vec{A}}^b - \delta_{ab} \vec{B}^a \cdot \vec{B}^b \right). \label{TwoPotential}
\end{equation}
Here, $\varepsilon_{ab}$ is given by $\varepsilon_{ab} = - \varepsilon_{ba}$, $\varepsilon_{12} = +1$ and 
$$
\vec{B} ^a  = \vec{\nabla} \times \vec{A}^a,
$$
with $A_i^1 \equiv A_i$.
The action (\ref{TwoPotential}) is invariant under rotations in the $(1,2)$ plane (``electric-magnetic duality rotations") ,
\be
\begin{pmatrix} \vec{A}^1 \\ \vec{A}^2  \end{pmatrix} \equiv \vec{\mathbf{A}} \; \;  \longrightarrow \; \; e^{\alpha \varepsilon} \vec{\mathbf{A}} \label{E-MDuality}
\ee
because $\varepsilon_{ab}$ and $\delta_{ab}$ are invariant tensors.  In infinitesimal form, 
\begin{eqnarray}
\delta_{\text{dual}} A^1_k &=& \alpha \ A^2_k , \label{dual_1}
\\
\delta_{\text{dual}} A^2_k &=& - \ \alpha \ A^1_k . \label{dual_2}
\end{eqnarray}
The action (\ref{TwoPotential}) is also invariant under $U(1) \times U(1)$ gauge transformations,
$$
A^a_k \; \;  \longrightarrow \; \; A_k^a + \partial_k \Lambda^a.
$$
\subsection{Spin 3/2}
The covariant action of the spin $3/2$ is given by the expression (see section \textbf{\ref{Sec:low_fermi}}):
\begin{eqnarray}
S_{3/2} &=& 
i \ \int \ d^4 x \ 
\bar{\psi}_{\mu} \gamma^{\mu \nu \rho} \partial_{\nu} \psi_{\rho} \label{action_lag_3_2}
\end{eqnarray}
which is invariant under the gauge transformation $\delta \psi_{\mu} = \partial_{\mu} \epsilon $.
This first-order action is already in canonical form, with $\psi_k$ being self-conjugate canonical variables and $\psi_0$ the Lagrange multiplier for the constraint
\begin{eqnarray}
0 &=&
\gamma^{kl} \partial_k \psi_l . \label{constraint_psi}
\end{eqnarray}
The general solution of the constraint (\ref{constraint_psi}) (see section \textbf{\ref{Sec:spin_3_2_ham}}) reads \cite{Bunster:2012jp}:
\begin{eqnarray}
\psi_k &=& 
- \ \frac{1}{2} \ \epsilon^{lmn} \gamma_l \gamma_k \partial_m \chi_n , \label{def_chi}
\end{eqnarray}
\noindent where $\chi_k$ is a vector-spinor, which is the prepotential for the spin-3/2 field. The ambiguity in $\chi_k$  is given by \cite{Bunster:2012jp},
\begin{eqnarray}
\delta_{\text{gauge}} \chi_k &=& \partial_k \eta \ + \ \gamma^0 \gamma_5 \gamma_k \epsilon ,
\end{eqnarray}
\noindent where $\eta$ and $\epsilon$ are arbitrary spinor fields.  As observed in \cite{Bunster:2012jp}, these are the same gauge symmetries as those of a conformal spin-3/2 field.
\subsection{Supersymmetry}
The supersymmetry transformations for the $(1, 3/2)$-multiplet  read 
\begin{eqnarray}
\delta_{\text{SUSY}} A_{\mu} &=& i \ \bar{\epsilon}\psi_{\mu}, \label{susy_A}
\\
\delta_{\text{SUSY}} \psi_{\mu} &=& 
\frac{1}{4} \ F_{\mu \nu} \gamma^{\nu} \epsilon 
\ + \ \frac{1}{4} \ \tilde{F}_{\mu \nu} \gamma_5  \gamma^{\nu} \epsilon
. \qquad \label{susy_psi0}
\end{eqnarray}
and are easily verified to leave the covariant action $S_1 \ + \ S_{3/2}$ invariant.
The supersymmetry transformations can be rewritten in terms of the prepotentials $(A^a_k, \chi_k)$. 
From the supersymmetry transformation of the photon field (\ref{susy_A}), one immediately deduces that:
\begin{eqnarray}
\delta_{\text{SUSY} }A^{1}_{k} &=& i \ \bar{\epsilon} \psi_k , \label{SusyPrepA10}
\end{eqnarray}
where $\psi_k$ is now to be viewed as the function (\ref{def_chi}) of the prepotential $\chi_k$.  Similarly, a direct computation shows that the supersymmetry transformation of the momentum $\Pi^k$ conjugate to $A_k \equiv A^1_k$ ( minus the original electric field) is 
\begin{eqnarray}
\delta_{\text{SUSY}} \Pi^k &=&
- \ \frac{i}{2} \ \epsilon^{klm} \bar{\epsilon} \gamma_5 \partial_l \psi_m
\ + \ \frac{i}{2} \  \bar{\epsilon}\gamma^0  \gamma^{klm} \partial_l \psi_m 
\nonumber \\ &=&
- \ i \ \epsilon^{klm} \bar{\epsilon} \gamma_5 \partial_l \psi_m .
\end{eqnarray}
Since $\Pi^k = \epsilon^{klm} \partial_l A^2_m$, one gets
\begin{equation}
\delta_{\text{SUSY}} A^2_k =
- \ i \ \bar{\epsilon} \gamma_5 \psi_k  \label{SusyPrepA20}
\end{equation}
(up to a gauge transformation that can be set to zero).
Finally, one easily derives from  (\ref{susy_psi0})
\begin{eqnarray}
\delta_{\text{SUSY}} \psi_k &=& 
- \ \frac{1}{4} \ W_k \gamma^{0} \epsilon 
\nonumber \\ &&
\ + \ \frac{1}{4} \ \epsilon_{klm} W^m \gamma_5  \gamma^{l} \epsilon .
\end{eqnarray}
where we have defined $W_k \equiv \Pi_k \ - \ B_k \gamma_5$.  It follows that, again up to a gauge transformation of the prepotential that can be set to zero, 
\begin{eqnarray}
\delta_{\text{SUSY}}\chi^k &=&
\frac{1}{2} \ \left(A^2_k \ - \ A^1_k \gamma_5 \right) \gamma^0  \epsilon \nonumber \\
&=& \frac{1}{2} \ M_k \gamma^0  \epsilon \label{SusyPrepChi0}
\end{eqnarray}
with 
\begin{equation}
M_k \equiv A^2_k \ - \ A^1_k \gamma_5 \, .
\end{equation}
The transformations (\ref{SusyPrepA10}), (\ref{SusyPrepA20}) and (\ref{SusyPrepChi0}) are the searched-for supersymmetry transformations in terms of the prepotentials.
We close by noting that the duality rotations  (\ref{dual_1}) and (\ref{dual_2}) with $\delta \chi = 0$, and the chirality rotations
\begin{eqnarray}
\delta_{\text{chiral}} \chi_{k} &=&  \lambda \ \gamma_5 \chi_{k}  \label{chirality32}
\end{eqnarray}
with $\delta A^1_k = \delta A^2_k = 0$, separately leave the action invariant.  None of these transformations commutes with supersymmetry. However, the combined duality-chirality transformation (\ref{dual_1}), (\ref{dual_2}) and (\ref{chirality32}) with $\lambda = - \alpha$ commutes with supersymmetry. This is because $M_k$ transforms as
\begin{eqnarray}
\delta_{\text{dual}} M_k &=& - \ \alpha \ \gamma_5 M_k 
\end{eqnarray}
under duality.  In the case of extended supersymmetry,  duality acting only on the vector fields will not commute either with supersymmetry, but again it can be redefined to do so with one of the supersymmetries.

\end{subappendices}

\chapter{Beyond fully symmetric tensors: type $(2,2)$}

\label{Chap:chiral_mix}

In this chapter, we will consider a bosonic theory which is of particular interest regarding supersymmetry, because it describes a constituent field of a maximally extended supersymmetric theory, the so-called $(4,0)$ theory in a six-dimensional space-time. This supermultiplet has been the object of considerable hopes and speculations lately\footnote{See \cite{Hull:2000rr}.}. Its most remarkable element, the $(2,2)$-\textit{Curtright field}\footnote{See \cite{Curtright:1980yk}. Curtright was the first to study gauge fields of mixed symmetry of any type.}, is a bosonic field $T$ of a type that has no equivalent in lower dimension: it is described by a tensor field $T_{\mu\nu\rho\sigma}$ having the same algebraic symmetries as the Riemann tensor of the graviton, that is, its type is:
\begin{equation}
T_{\mu\nu\rho\sigma} \thicksim \yng(2,2) .
\end{equation}

We are going to see that this theory exhibits features very similar to those already observed in the higher spin cases: beginning with a Lagrangian manifestly covariant theory in which the field has a gauge invariance generalizing spin-$s$ diffeomorphism, the Hamiltonian formalism and the resolution of the constraints that it naturally produces lead to the introduction of prepotentials enjoying a conformal gauge invariance. Moreover, the Hamiltonian equations of motion can be read as self-duality conditions\footnote{In $D=4$, we had twisted self-duality conditions, because Hodge duality squares to minus one in this dimension. In $D=4$, it squares to one, so that self-duality is authorized.} on the curvature of the prepotentials.

We will also explicitly consider the dimensional reduction of this theory, because of the relevance of the dimensional reduction of the complete $(4,0)$ theory to the study of supergravity.

\section{Covariant description}

The $(2,2)$-Curtright field $T_{\mu\nu\rho\sigma}$ satisfies\footnote{This theory was introduce in \cite{Curtright:1980yk}; see also \cite{Boulanger:2004rx} for a recent review.}:
\begin{eqnarray}
T_{\mu\nu\rho\sigma} &=&
T_{[\mu\nu ]\rho\sigma}
= T_{\mu\nu [\rho\sigma]} ,
\\
0 &=& T_{[\mu\nu\rho ]\sigma} .
\end{eqnarray}
This can be seen either as the generalization of a spin-two field (which is described by a symmetric tensor with two indices) or as the generalization of a two-form (which is described by an antisymmetric tensor with two indices), and it is usually referred to as the later.

Its gauge transformations are given by:
\begin{equation}
\delta T_{\mu\nu\rho\sigma} = 
\frac{1}{2} \left(\alpha_{\mu\nu [\rho , \sigma ]}
\ + \ \alpha_{\rho \sigma [\mu , \nu ]}\right) , \label{cov_gauge_curtright}
\end{equation}
\noindent where the gauge parameter $\alpha_{\mu\nu\rho}$ is of Young type $(2,1)$ - the so-called ``hook diagram":
\begin{equation}
\alpha_{\mu\nu\rho} \thicksim \yng(2,1) .
\end{equation}
In other words, it satisfies:
\begin{eqnarray}
\alpha_{\mu\nu\rho} &=& \alpha_{[\mu\nu ] \rho},
\\
0 &=& \alpha_{[\mu\nu\rho ]} .
\end{eqnarray}
As we see, the gauge variation \eqref{cov_gauge_curtright} is simply the projection of the tensorial product of a derivative with the gauge parameter onto the tensor type of the field $T$:
\begin{equation}
\delta T_{\mu\nu\rho\sigma} = 
\mathbb{P}_{(2,2)} \left(\partial_{\sigma} \alpha_{\mu\nu\rho} \right) .
\end{equation}
In terms of the generalized exterior derivative introduced in appendix \textbf{\ref{App:generalized_differential}}, we also have:
\begin{equation}
\delta T = d_{(2)} \alpha ,
\end{equation}
\noindent where $d_{(2)}$ acts on tensors whose Young diagram has all rows containing $2$ boxes, with the possible exception of the last one (which can contain one or two boxes). Its cube vanishes:
\begin{equation}
0 = d_{(2)}^3 ,
\end{equation}
\noindent and, according to the theorem \textbf{\ref{Theorem_cohomlogy}} of appendix \textbf{\ref{App:generalized_differential}}\footnote{See \cite{DuboisViolette:1992ye,DuboisViolette:1999rd}.}, the cohomology of $d_{(2)}$ acting on a tensor such as $T$ - a ``well-filled" tensor - is empty, so that a necessary and sufficient condition for the field $T$ to be pure gauge is for its second generalized exterior derivative to vanish. This suggests to introduce the following gauge invariant curvature:
\begin{equation}
R \equiv d_{(2)}^2 T .
\end{equation}
In components, this gives:
\begin{equation}
R_{\mu\nu\rho\sigma\lambda\tau}  \equiv
\partial_{[\mu} T_{\nu\rho ] [ \sigma \lambda , \tau ]} .
\end{equation}
This will generally be referred to as the \textit{Riemann tensor}, since it is plainly the generalization of the spin-$2$ corresponding tensor (just as the gauge variation \eqref{cov_gauge_curtright} is the natural generalization of a spin-$2$ diffeomorphism). 

The symmetry type of the Riemann tensor is obvisously:
\begin{equation}
R_{\mu\nu\rho\sigma\lambda\tau} \thicksim \yng(2,2,2) ,
\end{equation}
or, in component form, it satisfies:
\begin{eqnarray}
R_{\mu\nu\rho\sigma\lambda\tau} &=& 
R_{[\mu\nu\rho ]\sigma\lambda\tau}
= R_{\mu\nu\rho[\sigma\lambda\tau ]} ,
\\
0 &=& R_{[\mu\nu\rho\sigma ]\lambda\tau} .
\end{eqnarray}
It is also subject to a Bianchi identity which is a consequence of the nilpotency of $d_{(2)}$: 
\begin{equation}
0 = d_{(2)} R .
\end{equation}
Explicitly, this is:
\begin{equation}
0 = \partial_{[\mu} R_{\nu\rho\sigma ] \lambda \tau \theta} .
\end{equation}
Finally, let us observe that, thanks to the cohomological properties of $d_{(2)}$, any tensor with the algebraic and differential properties of $R$ is the Riemann tensor of some field $T$.

An obvious choice of gauge invariant equations of motion to impose on the $(2,2)$-Curtright field is the vanishing of its \textit{Ricci tensor}, which is simply the trace of its Riemann tensor:
\begin{equation}
0 = R'_{\mu\nu\rho\sigma} ,
\end{equation}
\noindent where the Ricci tensor is defined as:
\begin{equation}
R'_{\mu\nu\rho\sigma} \equiv R^{\lambda}_{\phantom{\lambda}\mu\nu\lambda\rho\sigma} .
\end{equation}
Its Young symmetry type is:
\begin{equation}
R'_{\mu\nu\rho\sigma} \thicksim \yng(2,2) . \label{EOM_Curt}
\end{equation}

These equations of motion can actually be seen to derive from a variational principle, in which the gauge invariant Lagrangian is:
\begin{equation}
\mathcal{L} = - \ \frac{5}{2} \ \delta^{\mu_1 \dots \mu_5}_{\nu_1 \dots \nu_5} \, M\indices{^{\nu_1\nu_2\nu_3}_{\mu_1\mu_2}} \, M\indices{_{\mu_3\mu_4\mu_5}^{\nu_4\nu_5}},
\end{equation}
where $M_{\mu\nu\rho \sigma\tau} = \partial_{[\mu}T_{\nu\rho]\sigma\tau}$ and $\delta^{\mu_1 \dots \mu_5}_{\nu_1 \dots \nu_5} = \delta^{\mu_1}_{[\nu_1} \dotsm \delta^{\mu_5}_{\nu_5]}$.\\

The construction outlined above offers additional possibilities of development in space-time dimension $D = 2 + 2p$ (p even) where chiral $p$-forms exist. In our case, $p = 2$ and $D = 6$ is of interest, since our field generalizes a $2$-form. Its Riemann tensor can naturally be subjected to a self-chirality condition, if one defines its dual as:
\begin{equation}
\ast R_{\mu\nu\rho\sigma\lambda\tau} \equiv
\frac{1}{3!} \ \epsilon_{\mu\nu\rho \alpha\beta\gamma} R^{\alpha\beta\gamma}_{\phantom{\alpha\beta\gamma}\sigma\lambda\tau} .
\end{equation}
This dual tensor $\ast R$ is not necessarily of the same symmetry type as $R$. However, a necessary and sufficient condition for it to be so is for the trace of $R$ to vanish, $R' = 0$. In this case, $\ast R$, since it would also satisfy Bianchi identity, shall be the Riemann tensor of a field such as $T$. So, we can reformulate the equations of motion \eqref{EOM_Curt} as demanding that the dual of the Riemann tensor of $T$ be itself the dual tensor of some field having the same algebraic symmetries and gauge invariance as $T$, which shall be the dual of $T$.

Since in $D = 6$ duality squares to one, $\left(\ast \right)^2 = 1$, we can actually identify this dual field be simply equal to $T$, and the equations of motion then take the elegant form of \textit{self-duality conditions}:
\begin{equation}
R = \ast R . \label{SD22}
\end{equation}

There is a mismatch between the number of equations (\ref{SD22}), namely $175$, and the number of components of the $(2,2)$-tensor field,  namely $105$.  But the equations (\ref{SD22}) are not all independent.  

We are now going to identify a complete subset of these equations which shall actually be of first order in time derivative, but whose Lorentz invariance shall not be manifest. We will then search for a variational principle from which to extract them. 

It is of course sufficient that the searched-for variational principle yields a system of equations equivalent to (\ref{SD22}).
 
 \section{Electric and magnetic fields}
 
 To identify such a subset derivable from a variational principle, we introduce the electric and magnetic fields.  The electric field contains the components of the curvature tensor with the maximum number of indices equal to the time direction $0$, namely, two, 
 ${\mathcal E}^{ijkl}  \sim R^{0ij0kl} $,  or what is the same on-shell, the components of the curvature with no index equal to zero.  
 
However, in 5 dimensions, the curvature tensor $R_{pijqkl}$ is completely determined by the Einstein tensor, which is its spatial double Hodge dual: 
\begin{eqnarray}
G\indices{^{ij}_{kl}} &=& \frac{1}{(3!)^2} \ R\indices{^{abc def}} \varepsilon\indices{_{abc}^{ij}} \varepsilon_{defkl}
\\  &=& \bar{R}\indices{^{ij}_{kl}} \ - \ 2 \ \delta^{[i}_{[k} \bar{\bar{R}}\indices{^{j]}_{l]}} \ +\ \frac{1}{3} \ \delta^i_{[k} \delta^j_{l]} \bar{\bar{\bar{R}}} ,
\end{eqnarray}
(the Weyl tensor identically vanishes).  

One then  defines explicitly the electric field as:
 \be {\mathcal E}^{ijkl} \equiv G^{ijkl}. \ee
 Here, $\bar{R}\indices{^{ij}_{kl}} = R\indices{^{mij}_{mkl}} $, $\bar{\bar{R}}\indices{^{j}_{l}} = R\indices{^{mij}_{mil}}$ and 
$\bar{\bar{\bar{R}}} = R\indices{^{mij}_{mij}} $ are the successive traces.  Similar conventions will be adopted below for the traces of the tensors that appear.
 The electric field has the $(2,2)$ Young symmetry and is identically transverse, $ \partial_i {\mathcal E}^{ijkl} = 0$.  It is also traceless on-shell,
 \be
 \bar{{\mathcal E}}^{ik} \equiv {\mathcal E}^{ijkl}\delta_{jl} = 0 . \label{trace0}
 \ee
 
 The magnetic field contains the components of the curvature tensor with only one index equal to $0$,
 \be
 {\mathcal B}_{ijkl}  \equiv \frac{1}{3!} \ R\indices{_{0ij}^{abc}} \varepsilon_{abckl} .
 \ee
It is identically traceless, $\bar{{\mathcal B}}^{jl} \equiv {\mathcal B}^{ijkl} \delta_{ik} = 0$, and transverse on the second pair of indices, $ \partial_k {\mathcal B}^{ijkl} = 0$.  On-shell, it has the $(2,2)$ Young symmetry. \\
 
The self-duality equation (\ref{SD22}) implies:
\be
{\mathcal E}^{ijrs} - {\mathcal B}^{ijrs} = 0 . \label{E=B}
\ee
Conversely, the equation (\ref{E=B}) implies all the components of the self-duality equation (\ref{SD22}). This is verified in appendix \ref{app:eom} by repeating the argument of \cite{Bunster:2012km} (recalled here in section \textbf{\ref{Sec:twistselfdual_2}}) given there for a $(2)$-tensor, which is easily adapted to a $(2,2)$-tensor. 
We have thus replaced the self-duality conditions (\ref{SD22}) by a smaller, equivalent, subset.  One central feature of this subset is that it is expressed in terms of spatial objects.  

Note that the trace condition (\ref{trace0}) directly follows by taking the trace of (\ref{E=B}) since the magnetic field is traceless. It appears as a constraint on the initial conditions because it does not involve the time derivatives of $T_{ijrs}$.  There is no analogous constraint in the $p$-form case.

Since the number of components of the electric field is equal to the number of spatial components $T_{ijrs}$ of the $(2,2)$-tensor $T_{\alpha \beta \lambda \mu}$, one might wonder whether the equations (\ref{E=B}) can be derived from an action principle in which the basic variables would be the $T_{ijrs}$.   This does not work, however. 
Indeed, while the electric field involves only the spatial components $T_{ijrs}$ of the gauge field, the magnetic field involves also the gauge component $T_{0jrs}$, through an exterior derivative.  One must therefore get rid of $T_{0jrs}$. 

To get  equations that involve only the spatial components $T_{ijrs}$, we proceed as in the $2$-form case and take the curl of (\ref{E=B}), i.e.
\be
\epsilon^{mnijk}\partial_{k}\left({\mathcal E}_{ij}^{\; \; \; \; rs} - {\mathcal B}_{ij}^{\; \; \; \; rs} \right)=0 , \label{E2=B2}
\ee
eliminating thereby  the gauge components $T_{0jrs}$.   We also retain the equation (\ref{trace0}), which is a consequence of (\ref{E=B}) involving only the electric field.  There is no loss of physical information in going from (\ref{E=B}) to the system (\ref{trace0}), (\ref{E2=B2}). Indeed, as shown in appendix \ref{app:eom}, if (\ref{trace0}) and (\ref{E2=B2}) are fulfilled, one recovers (\ref{E=B}) up to a term that can be absorbed in a redefinition of $T_{0jrs}$.  The use of (\ref{trace0}) is crucial in the argument.  It is in the form (\ref{trace0}), (\ref{E2=B2}) that the self-duality equations can be derived from a variational principle.

\section{Prepotentials - Action}

To achieve the goal of constructing the action for the chiral tensor, we first solve the constraint (\ref{trace0}) by introducing a prepotential $Z_{ijrs}$ for $T_{ijrs}$.   Prepotentials were defined for gravity in \cite{Henneaux:1988gg} (see section \textbf{\ref{Sec:Grav_ham}}) and generalized to arbitrary symmetric tensor gauge fields in \cite{Henneaux:2015cda,Henneaux:2016zlu} (see section \textbf{\ref{sec:spins}}). 
The introduction of a prepotential for the mixed tensor $T_{ijrs}$ proceeds along similar lines. 

Explicitly, the prepotential $Z_{ijrs}$ provides a parametrization of the most general $(2,2)$ tensor field $T_{ijrs}$ that solves the constraint (\ref{trace0}). One has:
\be 
T_{ijrs} = {\mathbb P}_{(2,2)} \left( \frac{1}{3!} \epsilon_{ij}^{\; \; \;   k mn} \partial_k Z_{mn rs} \right) + \hbox{gauge transf.}, \label{TZ}
\ee
which is a direct generalisation of the formula given in \cite{Henneaux:2004jw} for a $(2)$-tensor (see our equation \eqref{2_h_Z}).   The prepotential is determined up to the gauge symmetries:
\be
\delta Z_{ijrs}  = {\mathbb P}_{(2,2)} \left(\partial_{i} \xi_{rs j} + \lambda_{ir}\delta_{js}\right) \label{Weyl}
\ee
where $\xi_{rs j}$ is a $(2,1)$-tensor parametrizing the ``linearized spin-$(2,2)$ diffeomorphisms" of the prepotential and $\lambda_{ir}$ a symmetric tensor parametrizing its ``linearized spin-$(2,2)$ Weyl rescalings".

Because the Weyl tensor of a $(2,2)$-tensor identically vanishes, the relevant tensor that controls Weyl invariance is the ``Cotton tensor", defined as
\be B_{ij kl} = \frac{1}{3!} \ \varepsilon_{ijabc} \partial^{a} S\indices{^{bc}_{kl}},
\ee
where: 
\be
S\indices{^{ij}_{kl}} \equiv G\indices{^{ij}_{kl}} - 2 \ \delta^{[i}_{[k} \bar{G}\indices{^{j]}_{l]}} + \frac{1}{3} \ \delta^i_{[k} \delta^j_{l]} \bar{\bar{G}}\ee is the ``Schouten tensor", which has the key property of transforming as:
\be \delta S\indices{^{ij}_{kl}} = -\frac{4}{27} \partial^{[j}\partial_{[k} \lambda\indices{^{i]}_{l]}}\ee under Weyl rescalings. In terms of our generalized exterior derivative (see appendix \textbf{\ref{App:generalized_differential}}), this means:
\begin{equation}
\delta S = d_{(2)} \lambda ,
\end{equation}
 \noindent where the tensors are of the symmetry type:
\begin{equation}
\lambda \thicksim \yng(2) , \qquad \qquad S \thicksim \yng(2,2) .
\end{equation}
The complete set of gauge invariants under \eqref{Weyl} is, tanks to the cohomological properties of $d_{(2)}$, the tensor:
\begin{equation}
C \equiv d_{(2)} S ,
\end{equation}
\noindent or, in components:
\begin{equation}
C_{abckl} \equiv \partial_{[a} S_{bc]kl} .
\end{equation}
Its symmetry type is:
\begin{equation}
C \thicksim \yng(2,2,1) .
\end{equation}

Since we are in a five dimensional space, we can naturally dualize it over its first three indices, getting to the aforementioned Cotton tensor $B$. This is a straightforward generalization of what was done for completely symmetric tensors in section \textbf{\ref{Sec:Weyl_s}}.

The Cotton tensor $B_{ij kl}$  is a $(2,2)$-tensor which is gauge invariant under (\ref{Weyl}), as well as identically transverse (this is a consequence of Bianchi identity $d_{(2)} C = 0$) and traceless (this is a consequence of the Young symmetry of $C$), $\partial_i B^{ij rs} = 0 = B^{ij rs}  \delta_{js}$.  Furthermore, a necessary and sufficient condition for $Z_{ijrs}$ to be pure gauge is that its Cotton tensor vanishes.

The relation (\ref{TZ}) implies that 
\be
\mathcal{E}^{ij rs} [T[Z]] \equiv G^{ij rs} [T[Z]] = B^{ij rs} [Z].
\ee
The relation (\ref{TZ}) gives the most general solution for $T_{ij rs}$ subject to the constraint that $\mathcal{E}^{ij rs}$ is traceless (this is proved in   \cite{Henneaux:2015cda} - see our section \textbf{\ref{Sec:solv_ham_constr_s}} - for general higher spins described by completely symmetric tensors, and is easily extended to tensors with mixed Young symmetry). \footnote{We note that in three dimensions, the analogous relations on the Cotton tensor for symmetric gauge fields have a nice supersymmetric interpretation \cite{Kuzenko:2016bnv,Kuzenko:2016qdw,Kuzenko3}. It would be of interest to explore whether a similar interpretation holds here.}

It follows from (\ref{TZ}) that 
\be
\frac12 \epsilon^{mnijk}\partial_{k}{\mathcal B}_{ij}^{\; \; \; \; rs}= \dot{B}^{mn rs} [Z]
\ee
and therefore, in terms of the prepotential $Z_{ijrs}$, the self-duality condition (\ref{E2=B2}) reads
\be
\frac12 \epsilon^{mnijk}\partial_{k}B_{ij}^{\; \; \; \; rs}[Z]  - \dot{B}^{mn rs} [Z]=0,
\ee
an equation that we can rewrite as
\be L^{mnrs \vert ijpq} Z_{ijpq} = 0 \label{E3=B3}
\ee
where the differential operator $L^{mnrs \vert ijpq}$ contains four derivatives and can easily be read off from (\ref{E3=B3}).  The operator $L^{mnrs \vert ijpq}$ is {\em symmetric}, so that one can form the action 
\begin{eqnarray} 
&& S[Z] = \frac12 \int d^6x  Z_{mnrs} \left(L^{mnrs \vert ijpq} Z_{ijpq}\right) \label{Action} \\
&&= \frac12 \int d^6x  Z_{mnrs} \left(\dot{B}^{mn rs} [Z] - \frac12\epsilon^{mnijk}\partial_{k}B_{ij}^{\; \; \; \; rs}[Z]\right)  \nonumber
\end{eqnarray}
which yields (\ref{E3=B3}) as equations of motion.  
Given that $Z \sim \partial^{-1} T$,  this action contains the correct number of derivatives of $T$, namely two,  and has therefore the correct dimension.

\section{Chiral and non-chiral actions}

The action (\ref{Action}) is our central result.  Although not manifestly so, it is  covariant.  One way to see this is to observe that (\ref{Action}) can be derived from the manifestly covariant $(2,2)$-Curtright action for a $(2,2)$-field \cite{Curtright:1980yk,Boulanger:2004rx} rewritten in Hamiltonian form.  As explained in appendix \ref{app:ham} , this action involves the spatial components $T_{ijrs}$ and their conjugate momenta $\pi^{ijrs}$ as canonically conjugate dynamical variables, while the temporal components $T_{0ijk}$ and $T_{0i0j}$ play the role of Lagrange multipliers for the ``momentum constraint": \be
\mathcal{C}^{ijk} \equiv \partial_l \pi^{ijlk} \approx 0 \ee and the ``Hamiltonian constraint": \be \mathcal{C}^{ij} \equiv \mathcal{E}\indices{^{ikj}_k}[T] \approx 0. \ee  

These constraints can be solved by introducing two prepotentials $Z^{(1)}_{ijrs}$ and $Z^{(2)}_{ijrs}$.   As for a chiral $2$-form \cite{Bekaert:1998yp}, the linear change of variables: \be (Z^{(1)}_{ijrs}, Z^{(2)}_{ijrs}) \rightarrow (Z^+_{ijrs}= Z^{(1)}_{ijrs} + Z^{(2)}_{ijrs}, Z^-_{ijrs} = Z^{(1)}_{ijrs} - Z^{(2)}_{ijrs})\ee  splits the action as a sum of two independent terms, one for $Z^+_{ijrs}$ and one for $Z^-_{ijrs}$.   The Poincar\'e generators also split similarly,   one for $Z^+_{ijrs}$ and one for $Z^-_{ijrs}$, which transform separately.  The action (\ref{Action}) is the action for $Z^+_{ijrs}$ obtained though this decomposition procedure, with the identification $Z^+_{ijrs} \equiv Z_{ijrs}$.  This second method for obtaining the action for a chiral $(2,2)$ tensor  shows as a bonus how a  non-chiral $(2,2)$-tensor dynamically splits as the sum of a chiral $(2,2)$-tensor  and an  anti-chiral $(2,2)$-tensor.

\section{Dimensional reduction}

Upon reduction from $5+1$ to $4+1$ dimensions, the prepotential $Z_{ijrs}$ decomposes into a $(2,2)$-tensor, a $(2,1)$-tensor and a $(2)$-tensor.  Using part of the Weyl symmetry, one can set the $(2)$-tensor equal to zero, leaving  one with a  $(2,2)$-tensor and a $(2,1)$-tensor which are exactly the prepotentials of the pure Pauli-Fierz theory in $4+1$ dimensions \cite{Bunster:2013oaa}, with the same action and gauge symmetries (see appendix \ref{app:dimred}).  It is this remarkable connection between the $(2,2)$-self-dual theory in $6$ spacetime dimensions and pure (linearized) gravity in $5$ spacetime dimensions that is at the heart of the work \cite{Hull:2000zn,Hull:2000rr}.  We have shown here that the connection holds not just for the equations of motion, but also for the actions themselves.  

\section{Generalizations}

The extension to more general two-column Young symmetry tensors is direct. The ``critical dimensions" where one can impose self-duality conditions on the curvature are those where chiral $p$-forms exist. The first colum of the Young tableau characterizing the Young symmetry must have $p$ boxes, and the second column has then a number $q \leq p$ of boxes.  So, in $D=6$ spacetime dimensions, one has also the interesting case of $(2,1)$-tensors, also considered in \cite{Hull:2000zn,Hull:2000rr}.  This case is treated along lines identical to those described here.  For the next case -- $D=10$ spacetime dimensions --, the first column must have length 4, and the second colum has length $q \leq 4$, an interesting example being the  $(2,2,2,2)$-tensors.  Again, the extension to this two-column symmetry case is direct, as in all higher spacetime dimensions $D= 14, 18, 22, 26 , \cdots$.  

The extension to more than two column Young symmetries is more subtle but proceeds as in \cite{Henneaux:2016zlu}, by relying on the crucial property demonstrated in \cite{Bekaert:2003az}, where it was shown that the second-order Fronsdal-Crurtright type equations can be replaced by equations on the curvatures, which involves higher order derivatives.  The self-duality conditions can then be derived from an action principle involving the appropriate prepotentials.  The action is obtained by combining the above derivation with the methods of  \cite{Henneaux:2016zlu} for introducing prepotentials.

The present analysis can be developed in various directions.  First, following  \cite{Hull:2000zn,Hull:2000rr}, it would be of great interest to consider the supersymmetric extensions of the $6$-dimensional chiral theory and to determine how the fermionic prepotentials enter the picture \cite{Bunster:2012jp,Bunster:2014fca}. The attractive $(4,0)$-theory of \cite{Hull:2000zn,Hull:2000rr} deserves a particular effort in this respect.  Second, the inclusion of sources, which would be dyonic by the self-duality condition, and the study of the corresponding quantization conditions, would also be worth understanding \cite{Deser,Seiberg:2011dr,Bunster:2013era}. 

Finally, we note that we restricted the analysis to flat Minkowski space.  The trivial topology of $\mathbb{R}^n$ enabled us to integrate the differential equations for the prepotentials without encountering obstructions, using the Poincar\'e lemma of \cite{DuboisViolette:1992ye,DuboisViolette:1999rd}. The consideration of Minkowski space is not optional at this stage since the coupling of a single higher spin field to curved backgrounds is problematic.  It is known how to surpass the problems only in the context of the Vasiliev theory, which requires an infinite number of fields \cite{Vasiliev:1995dn,Vasiliev:2004cp,Bekaert:2010hw,Didenko:2014dwa,Metsaev:1993mj}.  Important ingredients to extend the analysis of the present article to nonlinear backgrounds are expected to include the cohomological considerations of \cite{Bekaert:1998yp}, the nonlinear extension of the higher spin Cotton tensors \cite{Linander:2016brv}, as well as duality in cosmological backgrounds \cite{Julia:2005ze,Julia2,Hortner:2016omi}.

\begin{subappendices}

\section{Equations of motion} \label{app:eom}

In this appendix, we show the equivalences between the different forms of the self-duality equations above, \eqref{SD22} $\Leftrightarrow$ \eqref{E=B} $\Leftrightarrow$ \eqref{trace0},\eqref{E2=B2}.

\eqref{SD22} $\Leftrightarrow$ \eqref{E=B}:
In components, the self-duality equation $R = \,^*\! R$ reads
\begin{align}
R_{0ijklm} &= \frac{1}{3!} \varepsilon\indices{_{ij}^{abc}}  R_{abcklm} \label{R0} \\
R_{0ij0kl} &= \frac{1}{3!} \varepsilon\indices{_{ij}^{abc}}  R_{abc0kl} \label{R00}.
\end{align}
The first of these equations is equivalent to \eqref{E=B} by dualizing on the $klm$ indices. Conversely,  we must show that \eqref{E=B} implies \eqref{R00} or, equivalently, that \eqref{R0} implies \eqref{R00}. To do so, we use the Bianchi identity $\partial_{[\alpha_1} R_{\alpha_2\alpha_3\alpha_4]\beta_1 \beta_2 \beta_3} = 0$ on the curvature, which imples
\begin{equation}
\partial_0 R_{ijk \beta_1 \beta_2 \beta_3} = 3 \partial_{[i} R_{jk]0 \beta_1 \beta_2 \beta_3} .
\end{equation}
Therefore, taking the time derivative of equation \eqref{R0} gives
\begin{equation}
\partial_{[k} R_{lm]00ij} = \frac{1}{3!} \partial_{[k} R_{lm]0 abc} \varepsilon\indices{_{ij}^{abc}}, \label{curlR00}
\end{equation}
which is exactly the curl of \eqref{R00}. Now, the tensor $R_{0lm0ij}$ has the $(2,2)$ symmetry, and so does $\frac{1}{3!} R_{0lm abc} \varepsilon\indices{_{ij}^{abc}} = \mathcal{B}_{lmij}$ because of equation \eqref{E=B} and the fact that $\mathcal{E}$ has the $(2,2)$ symmetry. Using the Poincaré lemma (our theorem \textbf{\ref{Theorem_cohomlogy}} of Appendix \textbf{\ref{App:generalized_differential}}\footnote{See \cite{DuboisViolette:1992ye,DuboisViolette:1999rd}.}) for rectangular Young tableaux, one recovers equation \eqref{R00} up to a term of the form $\partial_{[i} N_{j][k,l]}$ for $N_{jk}$ symmetric. This term can be absorbed in a redefinition of the $T_{0j0k}$ components appearing in $R_{0ij0kl}$. (In fact, the components $T_{0j0k}$ drop from equation \eqref{curlR00}, and this explains how one can get equation \eqref{R00} from \eqref{R0}, which does not contain $T_{0j0k}$ either.)

\eqref{E=B} $\Leftrightarrow$ \eqref{trace0},\eqref{E2=B2}:
Equation \eqref{E=B} obviously implies \eqref{E2=B2}. It also implies \eqref{trace0} because the magnetic field $\mathcal{B}$ is identically traceless. To prove the converse, we introduce the tensor $K_{ijk lm} = \varepsilon\indices{_{ijk}^{ab}} (\mathcal{E} - \mathcal{B})_{lmab}$. Equation \eqref{trace0} and the fact that $\mathcal{B}$ is traceless imply that $K$ has the $(2,2,1)$ symmetry, $K\sim  $ {\tiny $  \yng(2,2,1)$ }. Equation \eqref{E2=B2} states that the curl of $K$ on its second group of indices vanishes, $K_{ijk[lm,n]} = 0$. The explicit formula
\begin{equation}
K_{ijk lm} = \frac{1}{3} \left(\varepsilon_{lmpqr} \partial^{p} T\indices{^{qr}_{[ij,k]}} - \partial_{[0} T_{lm][ij,k]} \right)
\end{equation}
shows that the curl of $K$ on its first group of indices also vanishes, $\partial_{[i} K_{jkl]mn} = 0$. Using the generalized Poincaré lemma (our theorem \textbf{\ref{Theorem_cohomlogy}} of appendix \textbf{\ref{App:generalized_differential}}\footnote{See \cite{DuboisViolette:1992ye,DuboisViolette:1999rd}.}) for arbitrary Young tableaux, this implies that $K_{ijklm} = \partial_{[i} \lambda_{jk][l,m]}$, where $\lambda_{jkl}$ is a tensor with the $(2,1)$ symmetry that can be absorbed in a redefinition of $T_{0ijk}$. (Similarly to the previous case, those components actually drop from \eqref{E2=B2}.) One finally recovers equation \eqref{E=B} by dualizing again $K$ on its first group of indices.


\newpage

\section{Hamiltonian formulation} \label{app:ham}

The Lagrangian for a non-chiral $(2,2)$ tensor $T_{\mu\nu\rho\sigma}$ is given by \cite{Boulanger:2004rx}
\begin{equation}
\mathcal{L} = - \frac{5}{2} \,\delta^{\mu_1 \dots \mu_5}_{\nu_1 \dots \nu_5} \, M\indices{^{\nu_1\nu_2\nu_3}_{\mu_1\mu_2}} \, M\indices{_{\mu_3\mu_4\mu_5}^{\nu_4\nu_5}},
\end{equation}
where $M_{\mu\nu\rho \sigma\tau} = \partial_{[\mu}T_{\nu\rho]\sigma\tau}$ and $\delta^{\mu_1 \dots \mu_5}_{\nu_1 \dots \nu_5} = \delta^{\mu_1}_{[\nu_1} \dotsm \delta^{\mu_5}_{\nu_5]}$.
The associated Hamiltonian action is
\begin{equation}
S_H = \int \! dt \, d^5\! x \left( \pi_{ijkl} \dot{T}^{ijkl} - \mathcal{H} - n_{ijk} \,\mathcal{C}^{ijk} - n_{ij}\, \mathcal{C}^{ij} \right),
\end{equation}
where the Hamiltonian is
\begin{align}
\mathcal{H} &= \mathcal{H}_\pi + \mathcal{H}_T \\
\mathcal{H}_\pi &= 3 \left( \pi^{ijkl}\pi_{ijkl} - 2 \pi^{ij}\pi_{ij} + \frac{1}{3} \pi^2\right) \\
\mathcal{H}_T &= \frac{5}{2}\, \delta^{i_1 \dots i_5}_{j_1 \dots j_5} \, M\indices{^{j_1 j_2 j_3}_{i_1 i_2}} \, M\indices{_{i_3 i_4 i_5}^{j_4 j_5}} .
\end{align}
The components $n_{ijk} = - 4 T_{ij0k}$ and $n_{ij} = 6 T_{0i0j}$ of $T$ with some indices equal to zero only appear as Lagrange multipliers for the constraints
\begin{align}
\mathcal{C}^{ijk} &\equiv \partial_l \pi^{ijlk} = 0\\
\mathcal{C}^{ij} &\equiv \mathcal{E}\indices{^{ikj}_k}[T] = 0.
\end{align}
Those constraints are solved by introducing two prepotentials $Z^{(1)}_{ijkl}$ and $Z^{(2)}_{ijkl}$ through
\begin{align}
\pi^{ijkl} &= G^{ijkl}[Z^{(1)}] \\
T_{ijkl} &= \frac{1}{3} \,{\mathbb P}_{(2,2)} \left( \epsilon\indices{_{ij}^{abc}} \partial_a Z^{(2)}_{bckl} \right) .
\end{align}
In terms of prepotentials, we have up to a total derivative
\begin{align}
\pi_{ijkl} \dot{T}^{ijkl} &= 2\, Z^{(1)}_{ijkl} \dot{D}^{ijkl}[Z^{(2)}] \\
\mathcal{H}_\pi &= 3\, G_{ijkl}[Z^{(1)}]S^{ijkl}[Z^{(1)}] \\ \mathcal{H}_T &= 3\, G_{ijkl}[Z^{(2)}]S^{ijkl}[Z^{(2)}] .
\end{align}
Again up to a total derivative, one has $G_{ijkl}[Z]S^{ijkl}[Z] = \frac{1}{3!} Z_{ijkl} \epsilon^{ijabc}\partial_a D\indices{_{bc}^{kl}}[Z]$. Therefore, defining the prepotentials $Z^{\pm}_{ijkl} = Z^{(1)}_{ijkl} \pm Z^{(2)}_{ijkl}$, the action splits into two parts, $S[Z^+, Z^-] = S^+[Z^+] - S^-[Z^-]$. The action $S^+[Z^+]$ is exactly the action \eqref{Action} provided in the text for a chiral tensor, while $S^-[Z^-]$ is the analog action for an anti-chiral tensor (which differs from equation \eqref{Action} only by the sign of the second term).

\newpage
\section{Dimensional reduction} \label{app:dimred}

The prepotential decomposes into three tensors,
\begin{equation}
Z_{IJKL} \longrightarrow Z_{ijkl},\; Z_{ij k5},\; Z_{i5j5} .
\end{equation}
(For the purposes of this appendix, uppercase indices run from $1$ to $5$ while lowercase indices run from $1$ to $4$.)
The $(2,2)$-tensor and the $(2,1)$-tensor are identified with the two prepotentials $P_{ijkl}$ and $\Phi_{ijk}$ for linearized gravity in $4+1$ dimensions \cite{Bunster:2013oaa} as
\begin{equation}
Z_{ijkl} = 12 \sqrt{3} P_{ijkl}, \qquad Z_{ijk5} = -3 \sqrt{3} \Phi_{ijk}.
\end{equation}
The $(2)$-tensor $Z_{i5j5}$ transforms under the Weyl symmetries \eqref{Weyl} as $\delta Z_{i5j5} = \frac{1}{3} ( \lambda_{ij} + \delta_{ij} \lambda_{55} )$ and can therefore be set to zero. Remaining gauge transformations on $Z$ must respect this choice: this restricts the gauge parameters to $\lambda_{ij} = - \delta_{ij} \lambda_{55}$ and $\xi_{i55} = 0$. The surviving gauge parameters are then $\lambda_{55}$, $\lambda_{i5}$, $\xi_{ijk}$ and $\xi_{5ij}$. (Note that $\xi_{ij5} = - 2 \xi_{5[ij]}$ is not independent, due to the cyclic identity $\xi_{[IJK]}=0$.) The map with the gauge parameters of \cite{Bunster:2013oaa} is
\begin{align}
\chi_{ijk}&=-\frac{\xi_{ijk}}{24\sqrt{3}}, \quad S_{ij} = \frac{\xi_{5(ij)}}{12\sqrt{3}}, \quad A_{ij} = -\frac{\xi_{5[ij]}}{12\sqrt{3}} \nonumber \\
\xi &= \frac{2\lambda_{55}}{9\sqrt{3}},\quad B_i = \frac{2\lambda_{i5}}{9\sqrt{3}} .
\end{align}
This shows that field content and gauge symmetries match. For the comparison of the actions, one needs the following expressions for the reduction of the Cotton tensor:
\begin{align}
D\indices{_{ij}^{kl}} &= -\frac{2}{\sqrt{3}} \,\varepsilon_{ijab} \partial^a (E^{klb} + \delta^{b[k} E^{l]} ) \\
D\indices{_{ij}^{k5}} &= 2\sqrt{3} \,\varepsilon_{ijab} \partial^a ( R^{kb} - \frac{1}{3} \delta^{kb} R )\\
D\indices{_{i5}^{j5}} &= \frac{1}{\sqrt{3}} \,\varepsilon_{iabc} \partial^a ( E^{bcj} + \delta^{jb} E^c ) ,
\end{align}
where $R^{ij}[P]$ and $E^{ijk}[\Phi]$ are defined as
\begin{align}
R^{ij}[P] &= \frac{1}{(3!)^2} \varepsilon^{iabc} \varepsilon^{jdef} \partial_a \partial_f P_{bcde} \\
E^{ijk}[\Phi] &= \frac{1}{2.3!} \varepsilon^{ij de} \varepsilon^{k abc} \partial_a \partial_e \Phi_{bcd}
\end{align}
and the traces are $R = R\indices{^i_i}$, $E^i=E\indices{^{ij}_j}$.
Using these formulas, one recovers the action of \cite{Bunster:2013oaa} for linearized gravity in $4+1$ dimensions in the prepotential formalism. 

\end{subappendices}

\part{Surface charges on AdS}

\chapter{Bose fields}

\label{Chap:Bose}

\section{Introduction}\label{sec:intro_bose}

In theories with a gauge freedom, conserved quantities associated with the gauge symmetry are given by surface integrals (see section \textbf{\ref{Sec:Ham_surf_charges}}). We are now going to make a systematic investigation of these surface charges in the case of higher spin massless fields over AdS, through their Hamiltonian analysis. Similarly to the flat space case considered above, this will lead us to first class constraints generating gauge transformations.

These gauge transformations will have to be given a boundary term in order for their Poisson bracket with the fields to make sense. On-shell, the bulk part of these generators will identically vanish, since it is proportional to the constraints. On the other hand, the boundary term will precisely give us the value of higher spin charges, provided the values plugged into the fields and gauge parameters appearing in it are carefully chosen. The fields will have to be given their on-shell value, so that their asymptotic behaviour will have to be evaluated through a (asymptotic) resolution of Fronsdal equations of motion. As for the gauge parameters, they will have to correspond to variations of the fields preserving their on-shell asymptotic behaviour while changing their physical content.\\

This chapter is organised as follows: in order to highlight the key points of our analysis of higher-spin charges, we begin by discussing in section \textbf{\ref{sec:spin3}} the simplest example given by a spin-3 field. We detail the Hamiltonian description of the free dynamics and we provide boundary conditions on the canonical variables that secure finiteness of charges. In section \textbf{\ref{sec:spins}} we move to arbitrary spin: we first identify the Hamiltonian constraints that generate Fronsdal's gauge transformations and then the associated charges. We eventually present boundary conditions inspired by the behaviour at spatial infinity of the solutions of the free equations of motion (recalled in Appendix~\textbf{\ref{app:boundary}}) and we verify that they give finite charges. We conclude summarising our results and discussing their expected regime of applicability. Other appendices provide a summary of our conventions (Appendix~\textbf{\ref{app:conventions}}) and more details on various results used in the main text (appendices \textbf{\ref{app:fronsdal}}, \textbf{\ref{app:identities}} and \textbf{\ref{app:charges}}). \\

The next chapter is devoted to symmetric massless fermionic fields \cite{Campoleoni:2017}.

\section{Spin-3 example}\label{sec:spin3}

To compute surface charges within the canonical formalism, we first rewrite in Hamiltonian form the Fronsdal action for a spin-3 field on an Anti de Sitter background of dimension $d$. The charges are then identified with the boundary terms that enter the generator of gauge transformations and we propose boundary conditions on fields and deformation parameters that secure their finiteness. The final expression for the charges is given in section \textbf{\ref{sec:spin3-charges}}, where we also compare our outcome with other results in literature.

\subsection{Hamiltonian and constraints}\label{sec:spin3-H}

We begin with the manifestly covariant Fronsdal action \cite{fronsdal-AdS,fronsdal-AdS-D}
\be \label{fronsdal-action}
\begin{split}
S = \int\! d^{\,d}x\, \sqrt{-\bar{g}}\, \bigg\{ & -\frac{1}{2}\, \bar{\nabla}_{\!\m} \vf_{\n\r\s} \bar{\nabla}^{\m} \vf^{\n\r\s} + \frac{3}{2}\, \bar{\nabla}_{\!\m} \vf_{\n\r\s} \bar{\nabla}^{\n} \vf^{\m\r\s} - 3\, \bar{\nabla}\!\cdot \vf^{\m\n} \bar{\nabla}_{\m} \vf_{\n} \\
& + \frac{3}{2}\, \bar{\nabla}_{\!\m} \vf_{\n} \bar{\nabla}^{\m} \vf^{\n} + \frac{3}{4} \left( \bar{\nabla}\!\cdot \vf \right)^2 - \frac{2d}{L^2} \left( \vf_{\m\n\r} \vf^{\m\n\r} - \frac{3}{2}\, \vf_\m \vf^\m \right)\! \bigg\} \, .
\end{split}
\ee
Here $\bar{\nabla}$ denotes the AdS covariant derivative, $L$ is the AdS radius\footnote{All results of this subsection apply also to de Sitter provided that one maps $L \to i L$.} and omitted indices signal a trace, e.g.\ $\vf_\m = \vf_{\m\l}{}^\l$. If one parameterises the AdS$_d$ background with static coordinates
\be \label{AdS_bose}
ds^2 = - f^2(x^k) dt^2 + g_{ij}(x^k) dx^i dx^j \, ,
\ee 
one finds that the terms in the Lagrangian with two time derivatives are
\be \label{2dots}
\cL = \sqrt{g} \left\{ \frac{f^{-1}}{2}\, \dot{\vf}^{ijk} \left( \dot{\vf}_{ijk} - 3\, g_{ij} \dot{\vf}_k \right) + \frac{f^{-7}}{4} \left( \dot{\vf}_{000} - 3f^{2} \dot{\vf}_{0} \right)^2 + \cdots \right\} ,
\ee
where Latin indices take values along spatial directions, omitted indices denote from now on a spatial trace and $g$ is the determinant of the spatial metric. Integrating by parts one can also eliminate all time derivatives acting on $\vf_{00i}$ and on the remaining contributions in $\vf_{0ij}$.
In analogy with linearised gravity, the spatial components of the covariant field thus play the role of canonical variables. The novelty is that this role is played also by the combination $(\vf_{000} - 3f^{2} \vf_{0})$, while the remaining components of the symmetric tensor are Lagrange multipliers which enforce first class constraints. That $(\vf_{000} - 3f^{2} \vf_{0})$ is a dynamical variable, in much the same way as the purely spatial components $\vf_{ijk}$, can be also inferred from its gauge transformation, which does not involve time derivatives (see the discussion around \eqref{lambda}).

It is convenient to perform the  redefinitions
\be \label{defN}
\a \equiv f^{-3} \vf_{000} - 3f^{-1} \vf_{0i}{}^i \, , \qquad
\cN_i \equiv f^{-1} \vf_{00i} \, , \qquad
N_{ij} \equiv \vf_{0ij} \, .
\ee
Introducing then the conjugate momenta to $\vf_{ijk}$ and $\alpha$, 
\be \label{Pijk}
\begin{split}
\P^{ijk} \equiv \frac{\d \cL}{\d \vf_{ijk}} & = \frac{\sqrt{g}}{f}\, \bigg\{ \dot{\vf}^{ijk} -3\, g^{(ij} \dot{\vf}^{k)} - 3 \, \nabla^{(i} N^{jk)} + 3\,g^{(ij|}\! \left(\, 2\, \nd N^{|k)} + \pr^{|k)} N \,\right) \\
& + \frac{3f}{2}\, g^{(ij} g^{k)l}\! \left(\, \pr_l \a - \G^0{}_{0l} \a \,\right)  \! \bigg\} \, ,
\end{split}
\ee
\be \label{Pt}
\tilde{\P} \equiv \frac{\d \cL}{\d \a} = \frac{\sqrt{g}}{2f} \left\{ \dot{\a} + 3\, \nd \cN \right\} ,
\ee
one can equivalently rewrite the action \eqref{fronsdal-action} in Hamiltonian form as\footnote{The rewriting of Fronsdal's action in Hamiltonian form has been previously discussed in \cite{Metsaev:2011iz}, relying however on the Poincar\'e parameterisation of the AdS metric. See also \cite{Hframe} for another Hamiltonian description of the free higher-spin dynamics on AdS.}
\be \label{action}
S[ \vf_{ijk}, \alpha, \P^{ijk}, \tilde{\P}, \cN^i , N^{ij}]= \int\! d^{\,d}x\, \Big\{ \P^{ijk} \dot{\vf}_{ijk} + \tilde{\P} \dot{\a} - \cH - \cN^i \cC_i - N^{ij} \cC_{ij} \Big\} \, ,
\ee
where $\cH$, $\cC_i$ and $\cC_{ij}$ are functions only of $\vf_{ijk}$, $\a$ and of their conjugate momenta $\P^{ijk}$, $\tilde{\P}$. 
Here $\nabla$ denotes the Levi-Civita connection for the spatial metric $g_{ij}$, while the ``extrinsic'' Christoffel symbol depends on $g_{00}$ as
\be \label{christoffel}
\G^0{}_{0i} = f^{-1} \pr_i f \, .
\ee
Parentheses denote a symmetrisation of the indices they enclose, and dividing by the number of terms in the sum is understood.
While \eqref{Pijk} and \eqref{Pt} hold for any static metric, the rewriting of the Fronsdal action \eqref{fronsdal-action} in the form \eqref{action} requires that the metric \eqref{AdS_bose} be of constant curvature (see Appendix~\textbf{\ref{app:fronsdal-3}} for details).

The Hamiltonian in \eqref{action} reads explicitly
\begin{align}
& \cH = f \bigg\{ \frac{1}{\sqrt{g}} \! \left[\, \frac{1}{2}\, \P^{ijk}\! \left( \P_{ijk} - \frac{3}{d}\, g_{ij} \P_k \right) + \tilde{\P}^2 \,\right] + \frac{3}{2d}\, \P^i\! \left( \pr_i \a - \G^0{}_{0i}\a \right) \nn \\[5pt]
& + \sqrt{g}\, \bigg[\, \frac{1}{2}\, \nabla_{\!i} \vf_{jkl} \nabla^{i} \vf^{jkl} - \frac{3}{2}\, \nabla_{\!i} \vf_{jkl} \nabla^{j} \vf^{ikl} + 3\, \nabla\!\cdot \vf^{ij} \nabla_{\!i} \vf_{j} - \frac{3}{2}\, \nabla_{\!i} \vf_j \nabla^i \vf^j - \frac{3}{4} \left(\nabla\!\cdot\vf\right)^2 \nn \\[5pt]
&\qquad\ + \frac{2d}{L^2}\! \left( \vf_{ijk} \vf^{ijk} - \frac{3}{2}\, 
\vf_i \vf^i \right) + \G^0{}_{0i}\! \left( 3\,\vf^{ijk} \nabla_{\!j}\vf_k - \frac{3}{2}\, \vf^i \nabla\cdot \vf - \frac{9}{4}\,  \G^0{}_{0j} \vf^i \vf^j \right) \nn \\[5pt]
&\qquad\ + \frac{5d-3}{8d}\, \pr_i \a\, \pr^i \a + \frac{d}{L^2}\, \a^2 + \frac{3(d+1)}{8d}\, g^{ij} \G^0{}_{0i}\! \left( 2\,\a \pr_j \a - \G^0{}_{0j} \a^2 \right) \bigg]\bigg\} \, . \label{H3}
\end{align}
The constraints are a generalisation of the Hamiltonian constraint in linearised gravity (but note here the dependence on the additional momentum $\tilde{\P}$),
\be \label{Ci}
\begin{split}
\cC_i & = 3\, \bigg\{ \pr_{i} \tilde{\P} - \frac{\sqrt{g}}{2} \bigg[\, 2 \left( \D \vf_i - \nd\nd \vf_i \right) + \nabla_i \nd \vf \\
& - \frac{4d-1}{L^2}\, \vf_i - 3\, \G^0{}_{0j}\nabla_{\!i} \vf^j + 3\, \G^0{}_{0i}\G^0{}_{0j} \vf^j \bigg] \bigg\} \, ,
\end{split}
\ee
and a generalisation of the constraint that generates spatial diffeomorphisms,
\be \label{Cij}
\begin{split}
\cC_{ij} & = -\, 3\, \bigg\{ \nd \P_{ij} + \frac{\sqrt{g}}{2}\, g_{ij}\! \left( \D - \frac{d-1}{L^2} \right) \a \\
& + \sqrt{g} \left[\, \G^0{}_{0(i} \nabla_{\!j)} \a - \G^0{}_{0i} \G^0{}_{0j} \a + \frac{1}{2}\, g_{ij} g^{kl} \G^0{}_{0k} \left( \pr_l \a - \G^0{}_{0l} \a \right) \right] \!\bigg\} \, .
\end{split}
\ee
One can verify the absence of secondary constraints and that \eqref{Ci} and \eqref{Cij} are of first class. A simple way to convince oneself of these statements is to check that they provide the correct counting of local degrees of freedom (see e.g.\ \S~1.4.2 of \cite{Henneaux:1994pup}):
\be
\#\, \textrm{d.o.f.} = \underbrace{\frac{(d+1)!}{3!(d-2)!} + 1}_{\textrm{canonical variables/2}} - \underbrace{\left[ \frac{(d-1)d}{2} + (d-1) \right]}_{\textrm{first-class constraints}} = \frac{(d-3)(d-2)(d+2)}{3!} \, .
\ee
The right-hand side is the dimension of a representation of $so(d-2)$ labelled by a single-row Young tableau with three boxes, as it is appropriate to describe a massless spin-3 particle in $d$ space-time dimensions.

\subsection{Gauge transformations}\label{sec:spin3-gauge}

Being of first class, the constraints $\cC_i = 0$ and $\cC_{ij} = 0$ generate gauge transformations. These correspond to the variation
\be \label{cov-gauge}
\d \vf_{\m\n\r} = 3\,\bar{\nabla}_{\!(\m} \L_{\n\r)} 
\ee
that, in the covariant language, leaves the action \eqref{fronsdal-action} invariant provided that $\L^{\m\n}$ be traceless.

In the canonical formalism, the generator of gauge transformations is 
\be \label{G}
\cG[\xi^{ij},\l^i] = \int d^{\,d-1}x \left( \x^{ij} \cC_{ij} + 
\l^i \cC_i \right) + Q_1[\x^{ij}] + Q_2[\l^i] \, ,
\ee
where $Q_1$ and $Q_2$ are the boundary terms that one has to add in order that $\cG[\x,\l]$ admit well defined functional derivatives, i.e.\ that its variation be again a bulk integral:
\be \label{deltaG}
\d \cG = \int d^{\,d-1}x \left( A_{ijk} \d \P^{ijk} + B^{ijk} \d \vf_{ijk} + C \d \tilde{\P} + D \d\a \right) .
\ee
From \eqref{deltaG} one can read the gauge transformations of the canonical variables as
\begin{subequations} \label{gauge-canonical}
\begin{alignat}{3}
\d \vf_{ijk} & = \{ \vf_{ijk} , \cG[\xi,\l] \} = A_{ijk} \, , \qquad & \d \P^{ijk} & = \{ \P^{ijk} , \cG[\xi,\l] \} = - B^{ijk} \, , \\[5pt]
\d \a & = \{ \a , \cG[\x,\l] \} = C \, , \qquad & \d \tilde{\P} & = \{ \tilde{\P} , \cG[\x,\l] \} = - D \, .
\end{alignat}
\end{subequations}
The boundary terms $Q_1[\x]$ and $Q_2[\l]$ give the asymptotic charges (see e.g.~\cite{Regge,benguria-cordero}). Sensible boundary conditions on the canonical variables must then be chosen such that the charges be finite when evaluated on deformation parameters which generate transformations preserving the given boundary conditions. 

Inserting the definitions \eqref{Ci} and \eqref{Cij} of the constraints in \eqref{G} and taking into account that the background has constant curvature, one obtains
\begin{subequations}
\begin{align}
A_{ijk} & = 3\, \nabla_{\!(i} \x_{jk)} \, , \label{Aijk} \\[5pt]
B^{ijk} & = 3 \sqrt{g} \left\{ \nabla^{(i} \nabla^j \l^{k)}\! - g^{(ij|}\! \left[ \left( \D - \frac{2(d-1)}{L^2} \right)\! \l^{|k)} + \frac{1}{2} \left(\nabla^{|k)}\! + 3\, \G^0{}_{0}{}^{|k)} \right)\! \nd \l \right]\! \right\} , \label{Bijk} \\[5pt]
C & = -\, 3\, \nd \l \, , \label{C} \\[5pt]
D & = -\, \frac{3}{2}\sqrt{g} \left[ \left( \D - \frac{2d}{L^2} \right)\! \x - \G^0{}_{0i}\! \left( 2 \nd \x^i + \pr^i \x \right) \right] . \label{D}
\end{align}
\end{subequations}
The boundary terms generated by the integrations by parts putting the variation of $\cG[\x,\l]$ in the form \eqref{deltaG} must be cancelled by the variations $\d Q_1$ and $\d Q_2$ of the charges. Being linear in the fields, these variations are integrable and yield:\footnote{Here $d^{d-2}S_i \equiv d^{d-2} x\, \hat{n}_i$, where $\hat{n}_i$ and $d^{d-2} x$ are respectively the normal and the product of differentials of the coordinates on the $d-2$ sphere at infinity (e.g.\ $d^{2} x = d\th d\phi$ for $d=4$, so that $d^2S_i$ does not include the determinant of the intrinsic metric that appears in the full volume element).}
\begin{subequations} \label{spin3-charges}
\begin{align}
Q_1[\x^{ij}] & = 3\! \int\! d^{\,d-2}S_i \left\{ \x_{jk} \P^{ijk} + \frac{\sqrt{g}}{2} \left[\, \x \nabla^i \a - \a \nabla^i \x + \G^0{}_{0j}\! \left( 2\,\x^{ij} + g^{ij} \x \right)\! \a \,\right] \right\} , \label{Q1} \\[10pt]
Q_2[\l^i] & = 3\! \int\! d^{\,d-2}S_i\, \bigg\{ - \l^i \tilde{\P} + \frac{\sqrt{g}}{2} \left[\, 2\, \l_j \nabla^i \vf^j - 2\, \l_j \nd \vf^{ij} + \l^i \nd \vf \right. \nn \\
& \phantom{= 3\! \int\! d^{\,d-2}S_i\, \bigg\{}\, \left. - \, 2\, \vf_j \nabla^i \l^j + 2\, \vf^{ijk} \nabla_{\!j} \l_k - \nd \l \vf^i - 3\, \G^0{}_{0j} \l^i \vf^j \,\right]\! \bigg\} \, . \label{Q2}
\end{align}
\end{subequations}
In presenting $Q_1$ and $Q_2$ we also adjusted the integration constants so that the charges vanish for the zero solution.

Expanding \eqref{cov-gauge} in components and comparing with \eqref{Aijk} and \eqref{C}, one can also identify the deformation parameter $\x^{ij}$ with the spatial components of the covariant gauge parameter, while $\l^i$ is related to $\L^{\m\n}$ by
\be \label{lambda}
\l^i = -\, 2 f\, \L^{0i} \, .
\ee
The remaining component of the covariant gauge parameter, $\L^{00}$, is proportional to $g_{ij} \x^{ij}$ thanks to the Fronsdal constraint $g_{\m\n} \L^{\m\n} = 0$ that allows the cancellation of time derivatives in the gauge variation of $(\vf_{000} - 3f^{2} \vf_{0})$. The other components of \eqref{cov-gauge} give the gauge transformations of the Lagrange multipliers:
\begin{subequations}
\begin{align}
\d N_{ij} & = \dot{\x}_{ij} + f \left( \nabla_{\!(i} \l_{j)} - \G^0{}_{0(i}\l_{j)} \right) , \\[5pt]
\d \cN_i & = \dot{\l}_i + f \left( \pr_i \x - 2\,\G^0{}_{0i} \x - 2\, \G^0{}_{0j} \x_i{}^j \right) .
\end{align}
\end{subequations}
Substituting the previous variations in the definitions of the momenta one can finally check the consistency of \eqref{Bijk} and \eqref{D} with Fronsdal's gauge transformations.
Notice that time derivatives only appear in the gauge transformations of Lagrange multipliers, in agreement with the general results discussed e.g.\ in \S~3.2.2 of \cite{Henneaux:1994pup}.  This confirms once more our splitting of the covariant field into canonical variables and Lagrange multipliers. 

\subsection{Boundary conditions}\label{sec:spin3-bnd}

We now have to set boundary conditions on the canonical variables and restrict the deformation parameters to those that generate gauge transformations preserving them. Since we deal with the linearised theory, we can actually fully specify the space of solutions of the equations of motion. We then extract boundary conditions from the behaviour at spatial infinity of the solutions in a convenient gauge, with the expectation that the regime of applicability of both the charges \eqref{spin3-charges} and the following fall-off conditions will extend even beyond the linearised regime (see section \textbf{\ref{sec:conclusions}} for an ampler discussion of this strategy).

In Appendix~\textbf{\ref{app:boundary}} we recall the falloff at the boundary of the two branches of solutions of the second-order equations of motion imposed by the covariant action principle. In a coordinate system in which the AdS metric reads
\be \label{poincare3}
ds^2 = \frac{dr^2}{r^2} + r^2\, \h_{IJ} dx^Idx^J \, ,
\ee
the solutions in the \emph{subleading branch} behave at spatial infinity ($r \to \infty$) as 
\begin{subequations} \label{boundary}
\begin{align}
\vf_{IJK} & = r^{3-d}\, \cT_{IJK}(x^M) + \cO(r^{1-d}) \, , \label{dvIJK} \\[5pt]
\vf_{rIJ} & = \cO(r^{-d}) \, , \label{dvrIJ} \\[5pt]
\vf_{rrI} & = \cO(r^{-d-3}) \, , \label{dvrrI} \\[5pt]
\vf_{rrr} & = \cO(r^{-d-6}) \, . \label{dvrrr}
\end{align}
\end{subequations}
We remark that capital Latin indices denote all directions which are transverse to the radial one (including time) and that here and in the following we set the AdS radius to $L=1$. The field equations further impose that $\cT_{IJK}$ be conserved and traceless:
\be \label{current}
\pr^K \cT_{IJK} = \h^{JK} \cT_{IJK} = 0 \, . 
\ee
We take \eqref{boundary} and \eqref{current} as the definition of our boundary conditions.

In the case of spin $2$ included in the discussion of section \textbf{\ref{sec:bndH}}, for which the complete theory is known in closed form (AdS gravity), the boundary conditions generally considered in the literature agree with the behaviour of the solutions in the subleading branch \cite{HT,BrownHenneaux,AdS-generic}. Since in this case finiteness of the charges and consistency have been completely checked, we also adopt here boundary conditions defined by the subleading branch, which is the direct generalisation of these previous works.  It would be of interest to extend the analysis to more general asymptotics, as done for scalar fields in \cite{scalar1,scalar2,scalar3}, but we leave this question for future work. Furthermore, for $d=3$ the boundary conditions \eqref{boundary} agree with those in eq.~(3.9) of \cite{metric3D}, which have been derived from the Chern-Simons formulation of the full interacting theory of a spin-3 field with gravity.\footnote{Actually in \cite{metric3D} the trace of $\vf_{rIJ}$ is even $\cO(r^{-d-2})$ for the following reason: in the interacting theory --~at least in $d=3$~-- the fall-off conditions \eqref{boundary} are not preserved by asymptotic symmetries unless the fields also satisfy the  equations of motion up to an order in $r$ which depends on the highest spin appearing in the model. As we shall see, these additional specifications are anyway irrelevant to compute charges.}

From \eqref{boundary} one directly gets the behaviour at spatial infinity of half of the canonical variables. Denoting the coordinates that parameterise the $d-2$ sphere at infinity by Greek letters from the beginning of the alphabet, one obtains
\begin{subequations} \label{bnd-q}
\begin{align}
\vf_{\a\b\g} & = r^{3-d}\, \cT_{\a\b\g} + \cO(r^{1-d}) \, , \qquad \a = -\frac{2}{r^d}\, \cT_{000} + \cO(r^{-d-2})\, , \\[10pt]
\vf_{r\a\b} & = \cO(r^{-d}) \, , \qquad \vf_{rr\a} = \cO(r^{-d-3}) \, , \qquad \vf_{rrr} = \cO(r^{-d-6}) \, .
\end{align}
\end{subequations}
To fix the behaviour of $\a$ we used the trace constraint on $\cT_{IJK}$, which implies $\d^{\a\b}\cT_{0\a\b} = \cT_{000}$. The boundary conditions on the momenta follow from the substitution of \eqref{bnd-q} in \eqref{Pijk} and \eqref{Pt}:
\begin{subequations} \label{bnd-p}
\begin{align}
\P^{r\a\b} & = \frac{d+1}{r^4}\, \cT_0{}^{\a\b} + \cO(r^{-6}) \, , \qquad\qquad \tilde{\P} = \cO(r^{-4}) \, , \\[10pt]
\P^{\a\b\g} & = \cO(r^{-7}) \, , \qquad 
\P^{rr\a} = \cO(r^{-3}) \, , \qquad
\P^{rrr} = \cO(1) \, . 
\end{align}
\end{subequations}
In the formulae above we displayed explicitly only the terms which contribute to the charges (see section \textbf{\ref{sec:spin3-charges}}). We expressed everything in terms of the components of the conserved boundary current $\cT_{IJK}$, but $\cT_{\a\b\g}$, $\cT_{000}$ and $\cT_0{}^{\a\b}$ can be considered only as convenient labels to denote the boundary values of, respectively, $\vf_{\a\b\g}$, $\a$ and $\P^{r\a\b}$.  The covariant boundary conditions \eqref{boundary} would also fix the falloff of the Lagrange multipliers. The resulting conditions, however, would correspond to a particular choice of gauge, while --~as noticed in three space-time dimensions \cite{chemical,BH4}~-- the freedom in the choice of the Lagrange multipliers is instrumental in fitting some physically relevant solutions within the boundary conditions (see also \cite{BHregularity}). We postpone to future work a detailed analysis of this issue, especially because this freedom does not affect charges.

\subsection{Asymptotic symmetries}\label{sec:spin3-symm}

As a next step we identify asymptotic symmetries, that are the gauge transformations preserving boundary conditions. Our goal is to specify appropriate deformation parameters in the generator of gauge transformations \eqref{G}. We begin however by selecting covariant gauge transformations compatible with the fall-off conditions \eqref{boundary}, to later identify the corresponding deformation parameters using e.g.\ \eqref{lambda}.

Asymptotic symmetries contain at least the ``higher-rank isometries'' of the AdS space, i.e.\ the gauge transformations leaving the vacuum solution \mbox{$g_{\m\n} = g^{\textrm{AdS}}_{\m\n}$} and \mbox{$\vf_{\m\n\r} = 0$} invariant. These are generated by traceless AdS Killing tensors, satisfying
\be \label{exact-killing}
\bar{\nabla}_{\!(\m} \Lambda_{\n\r)} = 0 \, , \qquad\qquad \bar{g}_{\m\n} \Lambda^{\m\n} = 0 \, ,
\ee
where the latter condition is the usual Fronsdal constraint. These equations have been studied e.g.\ in \cite{killing,Bekaert:2005ka,HScharges}: they admit $\frac{(d-2)(d+1)(d+2)(d+3)}{12}$ independent solutions, obtained as traceless combinations of symmetrised products of AdS Killing vectors \cite{killing}. This guarantees that non-trivial asymptotic symmetries do exist. We shall also notice that --~as far as the free theory in $d > 3$ is concerned~-- asymptotic and exact Killing tensors only differ in terms that do not contribute to surface charges. It is nevertheless instructive to specify  asymptotic Killing tensors in a unified framework that applies for any $d \geq 3$.

In Appendix~\textbf{\ref{app:gauge}} we recall the behaviour at the spatial boundary of traceless AdS Killing tensors in the coordinates \eqref{poincare3}. We assume that asymptotic Killing tensors have the same leading behaviour at $r \to \infty$, that is
\begin{subequations} \label{boundary-gauge}
\begin{align}
\Lambda^{IJ} & = \chi^{IJ}(x^K) + \sum_{n\,=\,1}^\infty r^{-2n}\, \chi_{(n)}^{IJ} (x^K)\, , \label{chi-exp} \\
\Lambda^{rI} & = r\, w^I(x^K) + \sum_{n\,=\,1}^\infty r^{1-2n}\, w_{(n)}^{I} (x^K)\, , \label{w-exp} \\
\Lambda^{rr} & = r^2\, u(x^K) + \sum_{n\,=\,1}^\infty r^{2-2n}\, u_{(n)} (x^K) \, , \label{u-exp}
\end{align}
\end{subequations}
and we now analyse the constraints imposed by the preservation of boundary conditions. Notice that $\Lambda^{rr}$ and the trace of $\L^{IJ}$ are not independent: Fronsdal's constraint gives 
\be \label{trace-poincare}
\Lambda^{rr} = - r^4\, \h_{IJ} \L^{IJ} \, .
\ee
As a result $\chi^{IJ}$ in \eqref{chi-exp} is traceless, while the trace of $\chi^{IJ}_{(n+1)}$ is proportional to $u_{(n)}$.

For any value of the space-time dimension $d$, the variation $\d \vf_{IJK}$ induced by \eqref{boundary-gauge} decays slower at spatial infinity than the boundary conditions \eqref{boundary}:
\be \label{leading-var}
\d \vf_{IJK} = 3\, r^4 \left\{ \pr_{(I} \chi_{JK)} + 2\, \h_{(IJ} w_{K)} \right\} + \cO(r^2) \, .
\ee
The expression between parentheses must therefore vanish; this requires that $\chi^{IJ}$ be a \emph{traceless conformal Killing tensor} for the Minkowski metric $\h_{IJ}$ (see e.g.~\cite{algebra}),\footnote{For simplicity we fixed the coordinates \eqref{poincare3} such that the boundary metric is the flat Minkowski metric, but \eqref{kill-boundary} is invariant under conformal rescalings of the boundary metric. See e.g.~\cite{metric3D} for a discussion of this issue in the $d=3$ example, that can be extended verbatim to $d > 3$.}
\be \label{kill-boundary}
\pr_{(I} \chi_{JK)} - \frac{2}{d+1}\, \h_{(IJ\,} \prd \chi_{K)} = 0 \, , \qquad\qquad \h_{IJ} \chi^{IJ} = 0 \, ,
\ee
and that $w^I$ be fixed as
\be
w_I = -\, \frac{1}{d+1}\, \prd \chi_I \, .
\ee
As recalled in Appendix~\textbf{\ref{app:identities}}, when $d > 3$ the equations \eqref{kill-boundary} --~here defined in \mbox{$d-1$} dimensions~-- admit the same number of independent solutions as the equations \eqref{exact-killing}. It is therefore not surprising that their solutions can be extended to solutions of the Killing equation \eqref{exact-killing}, provided one fixes the subleading components of the gauge parameter in terms of $\chi^{IJ}$. As a result, in the linearised theory asymptotic Killing tensors coincide with traceless Killing tensors of the AdS space up to a certain order in $r$, except in $d = 3$ where the equations \eqref{kill-boundary} admit locally infinitely many solutions.

To prove the previous statement, let us look at the variations of the other components with at least one transverse index:
\begin{subequations} \label{var-with-I}
\begin{align}
\d \vf_{rIJ} & = \sum_{n\,=\,1}^\infty r^{3-2n} \left\{ -2n\, \chi^{(n)}_{IJ} - 2\, \h_{IJ} \chi^{(n)} + 2\, \pr_{(I}^{\phantom{(}} w_{J)}^{(n-1)} \right\} ,  \\
\d \vf_{rrI} & = \sum_{n\,=\,1}^\infty r^{-2n} \left\{ -4n\, w^{(n)}_I - \pr_I \chi^{(n)} \right\} .
\end{align}
\end{subequations}
The leading order must cancel for any $d$ and this requires 
\be
\chi^{(1)}_{IJ} = - \frac{1}{d+1} \left( \pr_{(I} \prd \chi_{J)} - \frac{1}{d}\, \h_{IJ} \prd\prd \chi \right) , \qquad
w^{(1)}_I = \frac{1}{4d(d+1)}\, \pr_I \prd\prd \chi \, .
\ee
Substituting in $\d \vf_{IJk}$ one discovers that the first subleading order in \eqref{leading-var} vanishes as well provided that 
\be \label{kill1}
\pr_{(I} \pr_{J} \prd \chi_{K)} - \frac{3}{2d}\, \h_{(IJ} \pr_{K)} \prd\prd \chi = 0 \, .
\ee
This identity is shown to follow from \eqref{kill-boundary} in Appendix~\textbf{\ref{app:identities}}.

The subleading orders can be analysed in a similar way. From \eqref{var-with-I} one sees that the boundary conditions \eqref{dvrIJ} and \eqref{dvrrI} are preserved provided that for $1 < n < \frac{d+1}{2}$ one has
\begin{subequations}
\begin{align}
\chi^{(n)}_{IJ} & = \frac{(-1)^n (d-1)!}{4^{n-1} n! (d+n-2)! (d+1)} \left(  \pr_I \pr_J - \frac{1}{d+n-1}\, \h_{IJ} \Box \right) \Box^{n-2} \prd\prd \chi , \\
w^{(n)}_I & = \frac{(-1)^{n+1} (d-1)!}{4^{n}n!(d+n-1)!(d+1)}\, \Box^{n-1} \pr_{I} \prd\prd \chi \, .
\end{align}
\end{subequations}
Computing three divergences of the conformal Killing equation \eqref{kill-boundary} one finds however
\be \label{kill3}
(d-1)\, \Box\, \prd\prd \chi = 0 \, ,
\ee
so that all $w^{(n)}_I$ with $n > 1$ and all $\chi^{(n)}_{IJ}$ with $n > 2$ vanish when $d > 1$. This also implies $u^{(n)} = 0$ for $n \geq 1$ thanks to the trace constraint \eqref{trace-poincare} and, in its turn, that the condition \eqref{dvrrr} is preserved ($\chi^{(2)}_{IJ}$ does not vanish but it is traceless).

The variation of the transverse component thus reduces to
\be \label{var-subleading}
\d \vf_{IJK} = \frac{1}{8d(d+1)}\, \pr_I \pr_J \pr_K \prd\prd \chi + \cO(r^{3-d}) \, .
\ee
The term written explicitly must vanish in all space-time dimensions except $d=3$ and, indeed, in Appendix~\textbf{\ref{app:identities}} we prove that \eqref{kill-boundary} implies
\be \label{kill2}
(d-3)\, \pr_I \pr_J \pr_K \prd\prd \chi = 0 \, .
\ee
In $d=3$ this identity is not available and one has two options: if one wants to solve the Killing equation \eqref{exact-killing} one has to impose the cancellation of the triple gradient of $\prd\prd\chi$ and the additional condition is satisfied only on a finite dimensional subspace of the solutions of the conformal Killing equation \eqref{kill-boundary}.  If one is instead interested only in preserving the boundary conditions \eqref{boundary}, which is the only option when the background is not exact AdS space, a shift of $\vf_{IJK}$ at $\cO(1)$ is allowed. The corresponding variation of the surface charges is at the origin of the central charge that appears in the algebra of asymptotic symmetries (see sect.~3.4 of \cite{metric3D}). 

To summarise: parameterising the AdS$_d$ background as in \eqref{poincare3}, linearised\footnote{A similar analysis has been performed in $d=3$ including interactions \cite{metric3D}. These introduce a dependence on the boundary values of the fields in \eqref{final-gauge_bose}, while asymptotic Killing tensors are still in one-to-one correspondence with the solutions of the conformal Killing equation.} covariant gauge transformations preserving the boundary conditions \eqref{boundary} are generated by
\begin{subequations} \label{final-gauge_bose}
\begin{align}
\Lambda^{IJ} & = \chi^{IJ} - \frac{r^{-2}}{d+1} \left( \pr^{(I} \prd \chi^{J)} - \frac{1}{d}\,\h^{IJ} \prd\prd\chi \right) + \frac{r^{-4}}{8d(d+1)}\, \pr^I \pr^J \prd\prd \chi \\
& + \cO(r^{-d-3}) \, , \nn \\
\Lambda^{rI} & = - \frac{r}{d+1}\, \prd \chi^I + \frac{r^{-1}}{4d(d+1)}\, \pr^I \prd\prd \chi + \cO(r^{-d-2}) \, , \\[2pt]
\Lambda^{rr} & = \frac{r^2}{d(d+1)}\, \prd\prd\chi + \cO(r^{-d-1}) \, , 
\end{align}
\end{subequations}
where $\chi^{IJ}$ satisfies the conformal Killing equation \eqref{kill-boundary}, whose general solution is recalled in \eqref{sol-conf}.
In section \textbf{\ref{sec:spin3-gauge}} we have seen that the deformation parameter $\xi^{ij}$ has to be identified with the spatial components of the covariant gauge parameter $\L^{\m\n}$, while the deformation parameter $\l^i$ is related to $\L^{\m\n}$ by \eqref{lambda}. Combining this information with \eqref{final-gauge_bose}, one sees that asymptotic symmetries are generated by deformation parameters behaving as 
\begin{subequations} \label{bnd-deform}
\begin{align}
\x^{\a\b} & = \chi^{\a\b} + \cO(r^{-2}) \, , \qquad
\l^\a  = -\,2r\, \chi^{0\a} + \cO(r^{-1}) \, , \\[10pt] 
\x^{r\a} & = \cO(r) \, , \qquad 
\x^{rr} = \cO(r^2) \, , \qquad
\l^r = \cO(r^{2}) \, .
\end{align}
\end{subequations}
As in \eqref{bnd-q}, Greek letters from the beginning of the alphabet denote coordinates on the $d-2$ sphere at infinity and we specified the dependence on $\chi^{IJ}$ only in the terms that contribute to surface charges. In particular, as for gravity, when $d > 3$ the charges are insensitive to the differences between asymptotic and exact Killing tensors, corresponding the unspecified subleading orders in \eqref{final-gauge_bose}.

\subsection{Charges}\label{sec:spin3-charges}

Having proposed boundary conditions on both canonical variables (see \eqref{bnd-q} and \eqref{bnd-p}) and deformation parameters (see \eqref{bnd-deform}), we can finally evaluate the asymptotic charges \eqref{spin3-charges}. In the coordinates \eqref{poincare3}, the normal to the $d-2$ sphere at infinity is such that $\hat{n}_r = 1$ and $\hat{n}_\a = 0$. At the boundary the charges thus simplify as
\begin{subequations} \label{Q-final}
\begin{align}
\lim_{r\to\infty}Q_1[\x^{ij}] & = 3\! \int\! d^{\,d-2}x \left\{ \x_{\a\b} \P^{r\a\b}\! + \frac{\sqrt{g}}{2}\, g^{rr} g_{\a\b}\! \left( \x^{\a\b} \pr_r \a - \a \nabla_{\!r} \x^{\a\b} + \G^0{}_{0r} \x^{\a\b} \a \right) \! \right\} , \label{Q1-final} \\[5pt]
\lim_{r\to\infty}Q_2[\l^{i}] & = 3\! \int\! d^{\,d-2}x\, \sqrt{g}\, g^{rr} g^{\a\b}\! \left\{ \l^\g \nabla_{\!r} \vf_{\a\b\g} + \l^\g \G^\d{}_{r\a} \vf_{\b\g\d} - \vf_{\a\b\g} \nabla_{\!r} \l^\g \right\} . \label{Q2-final}
\end{align}
\end{subequations}
The terms which survive in the limit give a finite contribution to the charges; one can make this manifest by substituting their boundary values so as to obtain
\begin{subequations} \label{Q-chi}
\begin{align}
\lim_{r\to\infty}Q_1 & = 3(d+1)\! \int\! d^{\,d-2}x \left( \chi^{\a\b} \cT_{0\a\b} + \chi^{00} \cT_{000} \right) , \\[5pt]
\lim_{r\to\infty}Q_2 & = 6(d+1)\! \int\! d^{\,d-2}x\, \chi^{0\a} \cT_{00\a} \, , 
\end{align}
\end{subequations}
where we used the trace constraints on both $\chi^{IJ}$ and $\cT_{IJK}$.
When one sums both charges, the result partly covariantises in the indices transverse to the radial direction:
\be \label{charge-cov_bose}
Q \equiv \lim_{r\to\infty}\! \left( Q_1 + Q_2 \right) = 3 (d+1) \!\int\! d^{\,d-2}x\, \chi^{IJ} \cT_{0IJ} \, .
\ee
The boundary charge thus obtained is manifestly conserved: it is the spatial integral of the time component of a conserved current since
\be
J_I \equiv \chi^{JK} \cT_{IJK} \quad \Rightarrow \quad \prd J = \pr^{(I} \chi^{JK)} \cT_{IJK} + \chi^{JK} \prd \cT_{JK} = 0 \, ,
\ee
where the conservation holds thanks to \eqref{current} and \eqref{kill-boundary}. We recovered in this way the standard presentation of the global charges of the boundary theories entering the higher-spin realisations of the AdS/CFT correspondence (see e.g.\ sect.~2 of \cite{review-Giombi} for a review).

In three space-time dimensions, the rewriting \eqref{charge-cov_bose} exhibits the chiral splitting of charges that one obtains in the Chern-Simons formulation with a suitable choice of the boundary value of the Lagrange multipliers\cite{HR,CS3,BH4}. Introducing the light-cone coordinates $x^\pm = t \pm \phi$, one obtains
\be \label{3Dcharge}
\begin{split}
Q_{d=3} & = 12 \!\int\! d\phi \left( \chi^{IJ} \cT_{+IJ} + \chi^{IJ} \cT_{-IJ} \right) = 12 \!\int\! d\phi \left( \chi^{++} \cT_{+++} + \chi^{--} \cT_{---} \right) \\
& = 12 \!\int\! d\phi \left( \chi(x^+) \cT(x^+) + \tilde{\chi}(x^-) \tilde{\cT}(x^-) \right) ,
\end{split}
\ee
where we took advantage of the form of the general solutions of \eqref{current} and \eqref{kill-boundary}, see e.g.~\eqref{sol-conf3}.
The separation in $Q_1$ and $Q_2$ does not correspond, however, to the splitting in left and right-moving components since
\begin{align}
\lim_{r\to\infty}Q_1 & = 6\! \int\! d\phi\, ( \chi + \tilde{\chi} ) ( \cT + \tilde{\cT} ) \, , \qquad
\lim_{r\to\infty}Q_2 = 6\! \int\! d\phi\, ( \chi - \tilde{\chi} ) ( \cT - \tilde{\cT} ) \, . 
\end{align}

The analysis of the linearised theory allowed us to recover the expression \eqref{3Dcharge} that also holds in the non-linear theory \cite{HR,CS3}, in line with our general expectations. Note, however, that in $d > 3$ the asymptotic symmetries \eqref{final-gauge_bose} leave the charges \eqref{charge-cov_bose} invariant, while in $d=3$ only a variation of \eqref{3Dcharge} which is independent of $\cT$ and $\tilde{\cT}$ is allowed. Since the variation of the charges is generated by the charges themselves \cite{Brown:1986ed} as
\be
\d_{\xi_2} Q[\x_1] = \{ Q[\xi_1] , Q[\xi_2] \} \, , 
\ee
one concludes that --~although the linearised theory suffices to identify the charges~-- their algebra does depend on interactions. As discussed in section \textbf{\ref{sec:conclusions}}, this phenomenon is anyway not a peculiarity of higher-spin theories. For an explicit example of the correlation between algebras of charges and interactions we refer to \cite{metric3D} where, assuming the expression \eqref{3Dcharge} for spin-3 charges in $d=3$, their algebra has been computed by including the first interaction vertices in a weak-field expansion in the spin-3 field.

We conclude this section by explicitly comparing our results with the higher-spin charges derived from Fronsdal's action in \cite{HScharges} within the covariant approach of \cite{covariant-charges1,covariant-charges2}.\footnote{See also Appendix~D of \cite{covariant-charges2}, where the methods of \cite{covariant-charges1,covariant-charges2} are applied to Hamiltonian actions to discuss the equivalence between canonical and covariant derivations of surface charges.} In the latter context, the spin-3 charges are given by the integral on a closed $d-2$ surface of the \emph{on-shell} closed $(d-2)$-form
\be \label{form}
k[\L] = \frac{1}{2(d-2)!}\, k^{[\m\n]}[\L]\, \e_{\m\n\r_3 \cdots \r_{d}} dx^{\r_3} \cdots dx^{\r_{d}} \, ,
\ee
with
\be \label{spin3-covariant}
\begin{split}
k^{[\m\n]}[\L] = \sqrt{-\bar{g}}\, & \Big\{\, \vf_{\r\s}{}^{[\m} \bar{\nabla}^{\n]} \Lambda^{\r\s} + \vf_\r \bar{\nabla}^{[\m} \Lambda^{\n]\r} + \Lambda^{\r\s} \bar{\nabla}^{[\m} \vf^{\n]}{}_{\r\s} \\
& + \L_\r{}^{[\m|}\! \left(\, 2\,\bar{\nabla}^{|\n]} \vf^\r + \bar{\nabla}^\r \vf^{|\n]} - 2\,\bar{\nabla}\!\cdot \vf^{|\n]\r} \,\right) \Big\} \, .
\end{split}
\ee
Here we went back to a manifestly covariant notation as in \eqref{fronsdal}, while square brackets denote an antisymmetrisation.
The closure of $k[\L]$ --~from which conservation of the charges follows~-- requires that the traceless parameter $\Lambda^{\m\n}$ satisfies the Killing tensor equation \eqref{exact-killing}.
By integrating \eqref{form} on a $d-2$ sphere at fixed time and radial distance from the origin and using \eqref{lambda}, one obtains
\be \label{compare-charges}
\int\! d^{\,d-2}S_i\, k^{0i}[\L] = -\, \frac{1}{3} \left( Q_1[\x] + Q_2[\l] \right) + \int\! d^{\,d-2}S_i\, \sqrt{g} \left( K_1^i + K_2^i \right) . 
\ee
One thus recovers the canonical charges \eqref{spin3-charges} plus a contribution that vanishes after integration since
\be \label{k1}
K_1^i = \nabla_{\!j}\! \left[\, \l^{[i}\vf^{j]} - f^{-1}\! \left( \l^{[i}\cN^{j]} - 4\, \x^{k[i}N^{j]}{}_k \right) \right]
\ee
is a total derivative on the sphere, while
\be \label{k2}
\begin{split}
K_2^i = & - \frac{3}{f} \left( g^{ij} N^{kl} - g^{kl} N^{ij} \right) \bar{\nabla}_{\!(j} \Lambda_{kl)} + \frac{3}{f} \left( \vf^{ijk} - g^{ij} \vf^k + f^{-1} g^{ij} \cN^k \right) \bar{\nabla}_{\!(0} \Lambda_{jk)} \\
& - \frac{3}{2f^2} \left( g^{ij} \a + 2f^{-1} \left( N^{ij} + g^{ij} N \right) \right) \bar{\nabla}_{\!(0} \Lambda_{0j)} + \cN^i\, \bar{\nabla}\!\cdot \L^0
\end{split}
\ee
vanishes identically for parameters that satisfy \eqref{exact-killing}. The deformation parameters in \eqref{spin3-charges} satisfy the Killing tensor equation \eqref{exact-killing} at leading order, so that eq.~\eqref{compare-charges} implies that the covariant and canonical expressions for the charges coincide asymptotically. Similarly, the constraint on the $\L^{\m\n}$ in \eqref{spin3-covariant} can be imposed only asymptotically to capture the infinitely many conserved charges that we exhibited in three space-time dimensions (see e.g.~\cite{covariant3d} for a discussion of $3d$ gravity in the covariant context).

\section{Arbitrary spin}\label{sec:spin_s}

In this section we extend most of the previous results to arbitrary spin. To simplify expressions we resort to an index-free notation: we omit indices and we denote e.g.\ the \mbox{$n$-th} trace of the field $\vf_{\m_1 \cdots \m_s}$ by $\vf^{[n]}$. Symmetrised gradients and divergences are denoted by $\nabla$ and $\nabla\cdot$, while in terms which are quadratic in the fields a contraction of all free indices is understood. A summary of our conventions is presented in Appendix~\textbf{\ref{app:conventions}}.

\subsection{Constraints and gauge transformations}\label{sec:gauge_cons}

In Fronsdal's formulation of the dynamics, a spin-$s$ particle is described by a symmetric tensor of rank $s$ with vanishing double trace \cite{fronsdal-AdS}. The action can be cast in the form
\begin{align}
S =\! \int & d^{\,d}x \sqrt{-\bar{g}}\, \bigg\{\! - \12\, \bar{\nabla}_{\!\m} \vf \bar{\nabla}^{\m} \vf + \frac{s}{2}\, \bar{\nabla}_{\!\m} \vf_{\n} \bar{\nabla}^\n \vf^{\m} - \binom{s}{2} \bar{\nabla}\!\cdot \vf^\m \bar{\nabla}_{\!\m} \vf^{[1]} + \12 \binom{s}{2} \bar{\nabla}_{\!\m} \vf^{[1]} \bar{\nabla}^{\m} \vf^{[1]} \nn \\
& + \frac{3}{4} \binom{s}{3} \bar{\nabla}\!\cdot \vf^{[1]} \bar{\nabla}\!\cdot \vf^{[1]} - \frac{(s-1)(d+s-3)}{L^2}\, \vf \left( \vf - \frac{1}{2}\binom{s}{2} g\, \vf^{[1]} \right)\! \bigg\} \, , \label{fronsdal-s}
\end{align}
where $\vf^{[1]}$ denotes a trace and omitted indices are understood to be contracted as e.g.\ in
\be
\bar{\nabla}_{\!\m} \vf_{\n} \bar{\nabla}^\n \vf^{\m} \equiv \bar{\nabla}_{\!\m} \vf_{\n\r_1 \cdots\, \r_{s-1}} \bar{\nabla}^\n \vf^{\m\r_1 \cdots\, \r_{s-1}} \, .
\ee 
The action \eqref{fronsdal-s} is invariant under the transformations
\be \label{fronsdal-gauge}
\d\vf = s\, \bar{\nabla} \L \, , \qquad\quad \L^{[1]} = 0 \, ,
\ee
where, as in \eqref{cov-gauge}, a symmetrisation of the indices carried by the gradient and the gauge parameter is understood.

By splitting the covariant field in time and spatial components one can solve the double-trace constraint; one can choose e.g.\ to encode the independent components of the spin-$s$ field in the traceful spatial tensors $\vf_{i_1 \cdots i_s}$, $\vf_{0i_1 \cdots i_{s-1}}$, $\vf_{00i_1 \cdots i_{s-2}}$ and $\vf_{000i_1 \cdots i_{s-3}}$. Within this set one has to distinguish between canonical variables and Lagrange multipliers. As discussed in section \textbf{\ref{sec:spin3-gauge}}, combinations whose variations \eqref{fronsdal-gauge} do not contain time derivatives are canonical variables having conjugate momenta such that the Legendre transformation is invertible. These are the spatial components $\vf_{i_1 \cdots i_s}$ of the field and
\be
\a_{i_1 \cdots i_{s-3}} \equiv f^{-3} \vf_{000i_1 \cdots i_{s-3}} - 3 f^{-1} g^{kl} \vf_{0i_1 \cdots i_{s-3}kl} \, ,
\ee
where $f$ denotes the lapse in a static parametrisation of the AdS metric as in \eqref{defN}. The remaining independent components of the covariant field are Lagrange multipliers (see their gauge transformations in \eqref{var-multipliers}). We denote them as
\be
\cN_{i_1 \cdots i_{s-2}} \equiv f^{-1} \vf_{00i_1 \cdots i_{s-2}} \, , \qquad N_{i_1 \cdots i_{s-1}} \equiv \vf_{0i_1 \cdots i_{s-1}} \, .
\ee
As showed in \cite{Metsaev:2011iz}, one can then rewrite the action \eqref{fronsdal-s} in the form
\be \label{action-s}
S = \int\! d^{\,d}x\, \Big\{ \P\,\dot{\vf} + \tilde{\P}\,\dot{\a} - \cH - \cN\,\cC_{s-2} - N\,\cC_{s-1} \Big\} \, ,
\ee
where $\cH$ and the tensorial densities $\cC_{s-1}$ and $\cC_{s-2}$ only depend on the canonical variables and their conjugate momenta. The latter are related to the other fields by
\begin{align}
\P & = \frac{\sqrt{g}}{f} \sum_{n=0}^{\left[\frac{s}{2}\right]} \!\binom{s}{2n} g^n \Big\{ (1-2n) \dot{\vf}^{[n]} + (2n-1)(s-2n) \nabla N^{[n]} + 2n(2n-1) \nabla\!\cdot\! N^{[n-1]} \nn \\
& + \frac{n(s-2n)}{2}\, f \left( \nabla \a^{[n-1]} - \G \a^{[n-1]} \right) + n(n-1) f \left( \nabla\!\cdot \a^{[n-2]} - \G\cdot \a^{[n-2]} \right) \Big\} \, , \label{p1} \\[5pt]
\tilde{\P} & = \frac{\sqrt{g}}{f} \sum_{n=1}^{\left[\frac{s-1}{2}\right]} \frac{n}{2} \binom{s}{2n+1} g^{n-1} \Big\{ \dot{\a}^{[n-1]} + (s-2n-1) \nabla \cN^{[n]} + (2n+1) \nabla\!\cdot \cN^{[n-1]} \Big\} \, , \label{p2}
\end{align}
where we denoted by $\G$ the ``extrinsic'' Christoffel symbol $\G^0{}_{0i}$ defined in \eqref{christoffel}. For instance, $\G \a^{[n-1]}$ stands for the symmetrisation of the index carried by $\G^0{}_{0i}$ with the free indices in the $(n-1)$-th trace of the tensor $\a$.

The computation of surface charges does not require knowledge of the Hamiltonian $\cH$. For this reason we refer to \cite{Metsaev:2011iz} for an account of the Hamiltonian form of Fronsdal's action (in Poincar\'e coordinates) and we focus on the constraints, which are relevant for our analysis. As in the \mbox{spin-3} case, the constraints $\cC_{s-1} = 0$ and $\cC_{s-2} = 0$ are of first class and there are no secondary constraints \cite{Metsaev:2011iz}.
Consequently, they can be reconstructed from the gauge transformations they generate, i.e.\ from \eqref{fronsdal-gauge}. More specifically, one can compute
\be \label{generator}
\cG[\x ,\l] = \int d^{d-1}x \left( \x\, \cC_{s-1} + \l\, \cC_{s-2} \right) + Q_1[\x] + Q_2[\l]
\ee
by integrating the variations
\be \label{var-canonical}
\d \vf = \frac{\d \cG[\x,\l]}{\d\P} \, , \quad
\d \P = - \frac{\d \cG[\x,\l]}{\d\vf} \, , \quad 
\d \a = \frac{\d \cG[\x,\l]}{\d \tilde{\P}} \, , \quad
\d \tilde{\P} = - \frac{\d \cG[\x,\l]}{\d \a} \, ,
\ee
that one can derive from \eqref{fronsdal-gauge}. In the previous expressions $\x$ and $\l$ are tensors of rank $s-1$ and $s-2$, like the constraints with whom they are contracted.

Following this procedure, one obtains for the time-like spin-$s$ diffeomorphisms,
\begin{align}
& \cC_{s-2} = -\, s\, \nd \P - \sqrt{g} \sum_{n=0}^{\left[\frac{s-1}{2}\right]} \binom{s}{2n+1}\, g^n\, \bigg\{\, n\, \bigg[\, \frac{(2n+1) \triangle + m_n}{2}\, \a^{[n-1]} \nn \\
& + (n-1)(2n+1) \nd\nd \a^{[n-2]} + (s-2n-1)\frac{4n+1}{2}\, \nabla \nd \a^{[n-1]} + \binom{s-2n-1}{2} \nabla^2 \a^{[n]} \nn \\
& + (2n+1)\, \G_{k} \left( (n-1) \nd \a^{[n-2]\,k} + \frac{1}{2}\,\nabla^k \a^{[n-1]} \right) + (s-2n-1) \left( \frac{2n+1}{2}\, \G\, \nd \a^{[n-1]} \right. \nn \\
& \left. + (n+1) \G_{k} \nabla \a^{[n-1]\,k} - \frac{4n+3}{2}\, \G\, \G\cdot \a^{[n-1]} \right) - (2n+1)(n-1)\, \G_{k} \G_{l}\, \a^{[n-2]\,kl} \nn \\
& - \frac{2n+1}{2} \left( \G \cdot \G\right) \a^{[n-1]} \bigg] + (n+1) \binom{s-2n-1}{2} \left(\, \G \nabla \a^{[n]} - \G^2 \a^{[n-1]} \,\right) \bigg\} \, , \label{C(s-2)}
\end{align}
where $\triangle \equiv g^{ij} \pr_i \pr_j$ and the mass coefficients in the first line read
\be
\begin{split}
m_n = & - (2n+1) \left[ (s-2n-1)(d+s-2n-5) + (d+4s-11) - 2n(2n-1) \right] \\
& + 2(n+1)(s-2n-1) \, .
\end{split}
\ee
In several contributions we displayed some indices in order to avoid ambiguities as e.g.\ in
\be \label{example}
g^n\, \G_{k} \nabla \a^{[n-1]\,k} \equiv \underbrace{g_{ii} \cdots g_{ii}}_{\textrm{$n$ times}}\, \G^0{}_{0k} \nabla_{\!i\,} \a^k{}_{i_{s-2n-2}\,l_{2n-2}} \underbrace{g^{ll} \cdots g^{ll}}_{\textrm{$n-1$ times}} \, .
\ee
On the right-hand side of \eqref{example}, we also denoted symmetrisations by repeated covariant or contravariant indices, while the indices carried by a tensor are denoted by a single label with a subscript indicating their total number.

The constraint which generalises the generator of spatial diffeomorphisms reads instead
\begin{align}
& \cC_{s-1} = 3\, \nabla \tilde{\P} + (s-3)\, g\, \nd \tilde{\P} - \sqrt{g} \sum_{n=1}^{\left[\frac{s}{2}\right]} n \binom{s}{2n} g^{n-1} \bigg\{ \left(\, n \triangle - m'_n \right) \vf^{[n]} \nn \\
& + (n-2)(2n-1) \nd\nd \vf^{[n-1]}  + \frac{(4n-3)(s-2n)}{2}\, \nabla \nd \vf^{[n]} + \binom{s-2n}{2} \nabla^2 \vf^{[n+1]} \nn \\
& - \frac{(2n+1)(s-2n)}{2}\, \G_{k} \nabla \vf^{[n]\,k} - \frac{s-2n}{2}\, \G \left[\, (s-2n-1) \nabla \vf^{[n+1]} + 2(n-1) \nd \vf^{[n]} \,\right] \nn \\
& - (n-1)\, \G_{k} \left[\, (2n-1) \nd \vf^{[n-1]\,k} + \nabla^k \vf^{[n]} \,\right] + \binom{s-2n}{2} \G^2 \vf^{[n+1]} \nn \\
& + \frac{(4n-1)(s-2n)}{2}\, \G\, \G\cdot \vf^{[n]} + (n-1) \left[\, (2n-1)\, \G_k \G_l \vf^{[n-1]\,kl} + \left( \G \cdot \G \right)  \vf^{[n]} \,\right] \bigg\} \, , \label{C(s-1)}
\end{align}
where the mass coefficients in the first line are
\be
m'_n = 2n(s-2n)(D+s-3)-4n^2(s-5)+2n(3D+6s-26)-(4D+5s-22) \, .
\ee
The gauge variations of coordinates and momenta that we used to derive \eqref{C(s-2)} and \eqref{C(s-1)} are collected in Appendix~\textbf{\ref{app:fronsdal-s}}.

The boundary terms in \eqref{generator} that cancel the contributions from the integrations by parts needed to rewrite the variation of the generator as a bulk integral are
\begin{subequations} \label{Qgen}
\be \label{Q1gen}
\begin{split}
Q_1[\x] & = \int d^{d-2}S_k \bigg\{ s \, \x\, \P^{k} + \sqrt{g} \sum_{n=1}^{\left[\frac{s-1}{2}\right]}\! \frac{n}{2} \binom{s}{2n+1} \Big\{ (2n+1)\, \x^{[n]} \big[ \nabla^k \a^{[n-1]} \\
& + 2(n-1) \nd \a^{[n-2]\,k} \big] - (2n+1)\, \a^{[n-1]} \big[ \nabla^k \x^{[n]} + 2(n-1) \nd \x^{[n-1]\,k} \big] \\[6pt]
& + (s-2n-1)\, \x^{[n]\,k} \big[ (4n+1) \nd \a^{[n-1]} + (s-2n-2) \nabla \a^{[n]} \big] \\[5pt]
& - (s-2n-1)\, \a^{[n-1]\,k} \big[ (4n+1) \nd \x^{[n]} + (s-2n-2) \nabla \x^{[n+1]} \big] \\[3pt]
& + (2n+1) \Big[\, \x^{[n]} \big[ \G^k \a^{[n-1]} \! + 2(n-1)\, \G\cdot \a^{[n-2]\,k} \big] + 2n\, \a^{[n-1]} \G\cdot \x^{[n-1]\,k} \Big] \\
& + (s-2n-1)\, \big[ (2n+1)\, \x^{[n]}  \G\, \a^{[n-1]\,k} + 2(n+1)\, \a^{[n-1]}  \G\, \x^{[n]\,k} \big] \Big\} \bigg\} \, ,
\end{split}
\ee
\be \label{Q2gen}
\begin{split}
Q_2[\l] & = \int d^{d-2}S_k \bigg\{ - 3\, \l^k\, \tilde{\P} - (s-3) \l^{[1]} \tilde{\P}^{k}  + \sqrt{g} \sum_{n=0}^{\left[\frac{s}{2}\right]} n\binom{s}{2n} \Big\{ \l^{[n-1]}\! \left[\, n\, \nabla^k \vf^{[n]} \right. \\
& \left. +\, (n-2)(2n-1) \nd \vf^{[n-1]\,k} \,\right] + \frac{s-2n}{2}\, \l^{[n-1]\,k} \left[\, (s-2n-1) \nabla \vf^{[n+1]} \right. \\[4pt]
& \left. +\, (4n-3) \nd \vf^{[n]} \,\right] - \vf^{[n]} \left[\, n\, \nabla^k \l^{[n-1]} + (n-1)(2n-1) \nd \l^{[n-2]\,k} \,\right] \\[3pt]
& - \frac{s-2n}{2n}\, \vf^{[n]\,k} \left[\, (n-1)(s-2n-1) \nabla \l^{[n]} + n(4n-3) \nd \l^{[n-1]} \,\right] \\[4pt]
& -(n-1)\, \l^{[n-1]} \left[\, (s-2n) \G\, \vf^{[n]\,k} + (2n-1) \G\cdot \vf^{[n-1]\,k} + \G^k \vf^{[n]} \,\right] \\[3pt]
& - \frac{s-2n}{2}\, \l^{[n-1]\,k} \left[\, (s-2n-1) \G\, \vf^{[n+1]} + (2n+1) \G\cdot \vf^{[n]} \,\right] \Big\} \bigg\} \, .
\end{split}
\ee
\end{subequations}
Again, as for spin-$3$, to obtain \eqref{Qgen} we took advantage of the linear dependence on the fields of the variations $\delta Q_1$ and $\delta Q_2$, which implies their integrability. We also fixed the integration constants to zero on the zero solution.\footnote{The latter operation requires some care in $d = 3$, where it is customary to assign the negative energy $M = - 1/8G$ to AdS$_3$. To this end, one can either introduce an integration constant in \eqref{Qgen} for $s=2$ or declare that the vacuum is identified by a corresponding non-trivial $\cT_{IJ}$ in the boundary conditions specified below in \eqref{boundary-s}. This modification of the background metric does not affect the final expression for the charges \eqref{qfin}, since it introduces only corrections that vanish at the boundary. It however allows a direct match with the conventions of the Chern-Simons formulation \cite{HR,CS3} (see e.g.~\eqref{total-charge}).}
The resulting $Q_1$ and $Q_2$ are the surface charges: in the following we shall introduce boundary conditions on the canonical variables and on deformation parameters generating asymptotic symmetries. In section \textbf{\ref{sec:charges}} we shall show how these conditions simplify asymptotically \eqref{Qgen} and we shall verify that the resulting spin-$s$ charges are finite and non-vanishing.

\subsection{Boundary conditions and asymptotic symmetries}\label{sec:bndH}

As in the spin-3 case, we derive boundary conditions on the canonical variables from the falloff at spatial infinity of the solutions of Fronsdal's equation in a convenient gauge, adopting the subleading branch. This is recalled in Appendix~\textbf{\ref{app:boundary}}: in the coordinates \eqref{poincare3}, the relevant solutions behave at spatial infinity ($r \to \infty$) as
\begin{subequations} \label{boundary-s}
\begin{align} \label{boundaryspins}
\varphi_{I_{1}\cdots I_{s}} & = r^{3-d}\, \cT_{I_{1}\cdots I_{s}}(x^M) + \cO(r^{1-d}) \, , \\[5pt]
\varphi_{r\cdots rI_{1}\cdots I_{s-k}} & = \cO(r^{3-d-3k}) \, ,
\end{align}
\end{subequations}
where capital Latin indices denote again directions transverse to the radial one. The symmetric boundary tensor $\cT_{I_1 \cdots I_s}$ is a \emph{traceless conserved current}:
\be \label{cons-curr-s}
\pr^J \vf_{JI_1 \cdots I_{s-1}} = \h^{JK} \vf_{JKI_1\cdots I_{s-2}} = 0 \, .
\ee
For $s=2$ the fall-off conditions on $h_{IJ}$ and $h_{rI}$ agree with those proposed for non-linear Einstein gravity in eq.~(2.2) of \cite{AdS-generic}. Our $h_{rr}$ decays instead faster at infinity; the mismatch can be interpreted anyway as a radial gauge fixing, as discussed in sect.~2.1 of \cite{metric3D}. For $d=3$, \eqref{boundary-s} also agrees with the boundary conditions in eq.~(5.3) of \cite{metric3D}, with the same provisos as those discussed for the spin-3 case in section \textbf{\ref{sec:spin3-bnd}}.

From \eqref{boundary-s} one obtains the boundary conditions on half of the canonical variables:
\begin{subequations}\label{boundary-spins-varphi}
\begin{align} 
\varphi_{\alpha_{1}\cdots \alpha_{s}} &= r^{3-d}\,{\cal{T}}_{\alpha_{1} \cdots \alpha_{s}}+{\cal{O}}(r^{1-d}) \, , \\[5pt]
\varphi_{r\cdots r\alpha_{1}\cdots \alpha_{s-k}} &= {\cal{O}}(r^{3-d-3k}) \, , \\[5pt]
\alpha_{\alpha_{1}\cdots \alpha_{s-3}} &= -2\,r^{-d}\,{\cal{T}}_{000\alpha_{1}\cdots \alpha_{s-3}}+{\cal{O}}(r^{-d-2}) \, , \\[5pt]
\alpha_{r\cdots r\alpha_{1}\cdots \alpha_{s-3-k}} &={\cal{O}}(r^{-d-3k}) \, ,
\end{align}
\end{subequations}
where Greek indices from the beginning of the alphabet denote angular coordinates on the $d-2$ sphere. The boundary conditions on the momenta are determined by making use of \eqref{boundary-s} in the definitions \eqref{p1} and \eqref{p2}:
\begin{subequations}\label{boundary-spins-varP}
\begin{align} \label{boundary-spins-Pi}
\Pi^{r\cdots r\alpha_{1} \cdots \alpha_{s-k}} &= \left\lbrace
\begin{array}{ll}
{\cal{O}}(r^{-1-2(s-k)})\, , \ & k=2m-2 \, , \\[5pt]
{\cal{O}}(r^{-2(s-k)})\, , \ & k=2m-1 \, ,
\end{array}
\right. \\[15pt]
\label{boundary-spins-P}
\tilde{\Pi}^{r\cdots r \alpha_{1} \cdots \alpha_{s-3-k}} &= \left\lbrace
\begin{array}{ll}
{\cal{O}}(r^{2-2(s-k)})\, , \quad & k=2m-2 \, , \\[5pt]
{\cal{O}}(r^{3-2(s-k)})\, , \quad & k=2m-1 \, ,
\end{array}
\right.
\end{align}
\end{subequations}
with $m$ a positive integer. For arbitrary spin, the components actually contributing to the charges are only $\P^{r\a_1\cdots\a_{s-1}}$ and $\tilde{\P}^{r\a_1\cdots\a_{s-4}}$, whose dependence on the boundary current $\cT_{I_1\cdots I_s}$ is detailed, respectively, in \eqref{Pr} and \eqref{Ptr}.

The deformation parameters that generate, via \eqref{generator} and \eqref{var-canonical}, gauge transformations preserving the boundary conditions behave at spatial infinity as
\begin{subequations}\label{bdn-deform-spins}
\begin{alignat}{5}
& \xi^{\alpha_{1}\cdots\alpha_{s-1}}=\chi^{\alpha_{1}\cdots\alpha_{s-1}}+{\cal{O}}(r^{-2}) \,, \qquad 
& \xi^{r\cdots r\alpha_{1}\cdots\alpha_{s-1-k}} & =  {\cal{O}}(r^{k}) \,, \\[10pt]
& \lambda^{\alpha_{1}\cdots\alpha_{s-2}}=-2\,r\,\chi^{0\alpha_{1}\cdots\alpha_{s-2}}r+{\cal{O}}(r^{-1}) \,, \qquad
& \lambda^{r\cdots r\alpha_{1}\cdots\alpha_{s-2-k}} & =  {\cal{O}}(r^{1+k}) \, ,
\end{alignat}
\end{subequations}
where $\chi^{I_1 \cdots I_{s-1}}$ is a \emph{traceless conformal Killing tensor} for the Minkowski metric $\h_{IJ}$ \cite{algebra}, satisfying
\be \label{conf-kill}
\pr_{(I_1} \chi_{I_2 \cdots I_{s})} - \frac{s-1}{d+2(s-2)-1}\, \h_{(I_1I_2\,} \prd \chi_{I_3 \cdots I_{s})} = 0 \, , \qquad \h_{JK} \chi^{JKI_1 \cdots I_{s-3}} = 0 \, .
\ee
The conditions \eqref{bdn-deform-spins} and \eqref{conf-kill} can be derived as for the spin-3 case discussed in section \textbf{\ref{sec:spin3-symm}}. We refrain from detailing the analysis of asymptotic symmetries also for arbitrary spin because only the information on the leading terms displayed in \eqref{bdn-deform-spins} is relevant for the computation of charges. The existence of non-trivial asymptotic symmetries --~in which part of the subleading terms in \eqref{bdn-deform-spins} are fixed in order to preserve \eqref{boundary-s}~-- is again guaranteed by the existence of traceless Killing tensors for the AdS background (see e.g.~\cite{killing,Bekaert:2005ka,HScharges}). The latter solve the equations
\be \label{kill-s}
\bar{\nabla}_{\!(\m_1} \Lambda_{\m_2 \cdots \m_s)} = 0 \, , \qquad\qquad \L^\r{}_{\r\m_1 \cdots \m_{s-3}} = 0 \, ,
\ee
and generate gauge transformations preserving the vacuum solution $\vf_{\m_1 \cdots\m_s} = 0$, which is trivially included in the boundary conditions \eqref{boundary-s}.

\subsection{Charges}\label{sec:charges}

The computation of asymptotic charges for arbitrary integer spin $s$ is performed following similar steps to those for the spin-3 case, although with the contribution of terms that vanish for $s=3$, in both charges and momenta. Using the coordinates \eqref{poincare3} and the asymptotic behaviour of the canonical variables (\ref{boundary-spins-varphi})--(\ref{bdn-deform-spins}) one sees that the only terms that contribute to (\ref{Qgen}) at the boundary are
\begin{subequations} \label{qfin}
\be\label{q1}
\begin{split}
\lim\limits_{r\rightarrow\infty}Q_1[\xi] & = s \!\int\! d^{d-2}x \bigg\{\, \xi\, \Pi^{r} + \sqrt{g} \sum_{n=1}^{\left[\frac{s-1}{2}\right]} \frac{n}{2} \binom{s-1}{2n} \Big\{\, \xi^{[n]} \left[ \nabla^r \a^{[n-1]} - \G^r \a^{[n-1]} \right. \\
& \left. +\, 2(n-1) \nabla\cdot \a^{[n-2]\,r} \right] - \a^{[n-1]} \left[ \nabla^r \xi^{[n]} + 2(n-1) \nabla\cdot \xi^{[n-1]\,r} \right] \Big\} \bigg\}
\end{split}
\ee
and
\begin{align}
& \lim\limits_{r\rightarrow\infty}Q_2[\lambda] = \!\int\! d^{d-2}x\bigg\{\! - (s-3) \lambda^{[1]} \tilde{\Pi}^{r}\! + \sqrt{g} \sum_{n=0}^{\left[\frac{s}{2}\right]} n\binom{s}{2n} \Big\{ \lambda^{[n-1]}\! \left[ (n-2)(2n-1) \nabla\!\cdot \varphi^{[n-1]\,r} \right. \nn \\
& \left. +\, n\, \nabla^r \varphi^{[n]}\,\right] -  \varphi^{[n]} \left[\, n\, \nabla^r \lambda^{[n-1]} + (n-1)\left(\, (2n-1) \nabla\cdot \lambda^{[n-2]\,r} + \G^r \lambda^{[n-1]} \,\right) \right] \Big\}\bigg\} \, , \label{q2}
\end{align}
\end{subequations}
where, as in \eqref{Q-final}, all omitted indices (including those involved in traces) take values in the $d-2$ sphere at infinity. 

The explicit form of these contributions in terms of the boundary current and of the boundary conformal Killing tensor that generate asymptotic symmetries is (see Appendix~\textbf{\ref{app:charges}} for details)
\begin{subequations} \label{Qgen-chi}
\begin{align}
\lim\limits_{r\rightarrow\infty}Q_{1} & = C \!\int\! d^{d-2}x\sum_{n=0}^{[(s-1)/2]}\binom{s-1}{2n} (\chi^{[n]})^{\a_{s-2n-1}} ({\cal{T}}^{\ [n]})_{0\a_{s-2n-1}} \, , \label{Qgen-chi1} \\
\lim\limits_{r\rightarrow\infty}Q_{2} & = C \!\int\! d^{d-2}x\sum_{n=1}^{[s/2]}\binom{s-1}{2n-1} (\chi^{[n-1]})^{0\a_{s-2n}} ({\cal{T}}^{[n]})_{\a_{s-2n}} \, , \label{Qgen-chi2}
\end{align}
\end{subequations}
with $C=s(d+2s-5)$. For better clarity, we specified the number of angular indices omitted in \eqref{qfin}, and we recall that traces are understood to follow from contractions with the metric $g_{\a\b}$ on the $d-2$ sphere. Taking advantage of the trace constraints on both ${\cal{T}}_{I_{s}}$ and ${\cal{\chi}}^{I_{s-1}}$, the sum of the two charges partly covariantises in the indices transverse to the radial direction as in \eqref{charge-cov_bose}:
\be \label{total-charge}
\begin{split}
Q & \equiv \lim\limits_{r\rightarrow\infty}(Q_{1}+Q_{2})=
C \!\int\! d^{d-2}x\sum_{n=0}^{s-1}\binom{s-1}{n}\chi^{\supsuboverbrac[n]{0 \cdots 0}\, \a_{s-n-1}} {\cal{T}}_{\supsuboverbrace[n+1]{0 \cdots 0}\,\a_{s-n-1}} \\
& = s(d+2s-5) \!\int\!d^{d-2}x\, \chi^{I_{s-1}}{\cal{T}}_{0I_{s-1}} \, .
\end{split}
\ee
The charge $Q$ is manifestly preserved by time evolution, since it is the integral of the time component of a current which is conserved thanks to \eqref{cons-curr-s} and \eqref{conf-kill}. The final rewriting of $Q$ also manifests a chiral splitting in three space-time dimensions as in \eqref{3Dcharge}.

\section{Summary and further developments}\label{sec:conclusions}

We identified surface charges in AdS Fronsdal's theory as the boundary terms which enter the canonical generator of gauge transformations. This gives the charges \eqref{spin3-charges} in the \mbox{spin-3} example and the charges \eqref{Qgen} in the \mbox{spin-$s$} case. As discussed at the end of section \textbf{\ref{sec:spin3-charges}}, these results are the analogues of the higher-spin charges identified by covariant methods in \cite{HScharges}. They are conserved when evaluated on-shell and on deformation parameters which generate gauge transformations leaving the AdS background strictly invariant. Both conditions can be however weakened: the field equations need to be fulfilled only asymptotically through suitable boundary conditions on the canonical variables and the residual gauge transformations need only to preserve these boundary conditions, without being exact Killing tensors everywhere.\footnote{Alternatively one can fix the gauge everywhere and not only asymptotically, and classify the gauge transformations preserving it; see sect.~2 of \cite{superrotations} for more comments on the relation between asymptotic conditions and gauge fixings.} This is the typical setup within the canonical approach we employed \cite{Regge,benguria-cordero}, and it allows one to discover infinitely many conserved charges in three space-time dimensions.  We specified boundary conditions in sects.~\ref{sec:spin3-bnd} and \ref{sec:bndH} and the resulting, greatly simplified, asymptotic expression for the charges are collected in \eqref{Q-final} for the \mbox{spin-3} example and in \eqref{qfin} for the \mbox{spin-$s$} case. We also showed that, with our choice of boundary conditions, the asymptotic charges can be rearranged so as to result from the integral of the time component of a conserved current, built from the contraction of a boundary conserved current with a conformal Killing tensor (see \eqref{Q-chi} and \eqref{Qgen-chi}). This form of our outcome allows a direct contact with the charges associated to the global symmetries of the boundary theory in AdS/CFT scenarios (compare e.g.\ with \cite{review-Giombi}).

Although we worked in the linearised theory, the fact that the conservation of the asymptotic charges \eqref{Q-final} or \eqref{qfin} only relies on the boundary conditions \eqref{boundary} or \eqref{boundary-s} suggests that both charges and boundary conditions may remain valid when switching on interactions, at least in certain regimes, -- the idea being that asymptotically the fields become weak and the linearised theory applies. This expectation is supported by notable examples: our charges and fall-off conditions coincide with those obtained within the canonical formulation of gravity \cite{AD1,HT,BrownHenneaux,AdS-generic} and within the Chern-Simons formulation of higher-spin gauge theories in three space-time dimensions \cite{HR,CS3,metric3D}. Color charges in Yang-Mills can be obtained through a similar procedure \cite{AD2}. See also e.g.~\cite{covariant-charges1} for more examples of models in which surface charges are linear in the fields. 

In spite of these reassuring concurrencies, one should keep two important facts in mind. 
\begin{itemize}
\item For AdS Einstein gravity coupled to scalar matter, the scalar field might have a back-reaction on the metric that is sufficiently strong to force a weakening of the boundary conditions \cite{scalar1,scalar2,scalar3,Henneaux:2002wm}.  This observation is relevant in the present context because all known interacting higher-spin theories in more than three space-time dimensions have a scalar field in their spectra. One can thus foresee both additional non-linear contributions to the charges and relaxed boundary conditions when considering backgrounds with non-trivial expectation values for scalar fields. For instance, the boundary conditions of the scalar greatly influence the nature of the boundary dual within the AdS/CFT correspondence \cite{review-Giombi,ABJ-HS}.  As the pure-gravity analysis played an important role in paving the way to the study of scalar couplings, we expect nevertheless that our results will be important to attack the more general analysis of higher-spin interacting theories, e.g.\ in a perturbative approach. We defer the study of these interesting questions to future work. 
\item Second, we focused on the subleading branch of the boundary conditions.  How to include the other branch, or a mixture of the two,  has been investigated in the Hamiltonian asymptotic analysis for scalar fields, where nonlinearities may arise \cite{scalar1,scalar2,scalar3}.   This question is of course of definite importance  for higher spin holography \cite{boundary-action,unfolding-holography,alternate-bnd-cond}.
\end{itemize}

Even when the surface integrals are determined by the linear theory without nonlinear corrections, the interactions play a crucial role  in the study of the asymptotic symmetry algebra. This can be seen in many ways.  For instance, the Poisson bracket of charges, that is their algebra, can be derived from their variation under gauge transformations preserving the boundary conditions. However, as shown in sects.~\ref{sec:spin3-symm} and \ref{sec:bndH}, in the linearised theory the charges are left invariant by asymptotic symmetries in all space-time dimensions except $d = 3$ (but there the variation only gives rise to a central term in the algebra). The fact that one must include interactions in order to derive the asymptotic symmetry algebra is not a higher-spin feature, though.  In general relativity, although the surface charges are correctly reproduced by the linear theory,  one has to know how the metric transforms under diffeomorphisms (and not only under their linearised version) in order to obtain the algebra of asymptotic symmetries (see e.g.~\cite{BrownHenneaux,Brown:1986ed}). A similar phenomenon arises for color charges in Yang-Mills theory.

This mechanism is also clearly displayed by the comparison between sects.~\ref{sec:spin3-symm} and \ref{sec:bndH} and the similar analysis performed in $d=3$ in \cite{metric3D}, where the effect of gravitational couplings and of the self-interactions of higher-spin fields manifests itself in the non-linear variations of the linear charges. Similarly, higher-spin Lie algebras emerge when considering commutators of gauge transformations induced by cubic vertices and restricted to gauge parameters that at linearised level leave the AdS background invariant \cite{cubic-symmetries-1,cubic-symmetries-2}. It will be interesting to compare this analysis with that of surface charges developed in this and the related papers \cite{HScharges,Campoleoni:2017}. Note that the analysis of asymptotic symmetries in \cite{ABJ-HS,review-strings} also involves ingredients that go beyond the linearised regime, like the full knowledge of the higher-spin gauge algebra on which the bulk theory is built upon.   

Let us finally point out other possible extensions.  Our results apply to any space-time dimension but they do not cover all possible bosonic higher-spin gauge theories. When $d > 4$, mixed symmetry fields (see e.g.~\cite{review-mixed} for a review) are a new possibility. In AdS they loose almost all gauge symmetries they display in flat space, but the remaining ones should give rise to conserved charges generalising those discussed in this chapter. An analysis along the lines we followed for symmetric fields may start from the actions of \cite{Campoleoni:2012th}, which constitutes at present the closest generalisation of the Fronsdal action known in closed form. Partially massless fields \cite{pm1,pm2,pm3} are other gauge fields which can be defined on constant curvature backgrounds and have less gauge symmetry than Fronsdal's fields.  This generalisation comprises gauge fields that do not propagate unitarily on AdS. It would be interesting to investigate whether one can define in this case boundary conditions allowing for non-trivial residual symmetries and supporting consequently non-trivial conserved charges. How such charges, if they exist, would fit with the classical instability of the partially massless theories would also deserve to be understood.



\begin{subappendices}

\section{Notation and conventions}\label{app:conventions}

We adopt the mostly-plus convention for the space-time metric and we denote by $d$ the dimension of space-time. The AdS radius $L$ is defined as
\be \label{commutator}
[ \bar{\nabla}_{\!\m} \,, \bar{\nabla}_{\!\n}] V_\r = \frac{1}{L^2}\! \left( g_{\n\r} V_\m - g_{\m\r} V_\n \right) .
\ee
In the static coordinates \eqref{AdS_bose} the spatial metric satisfies \eqref{commutator} as the full space-time metric provided one substitutes everywhere $\m,\n \to i,j$.

We distinguish between four types of indices, depending on whether the time and/or radial coordinates are included or not. Greek letters from the middle of the alphabet include all coordinates, small Latin letters include all coordinates except $t$, capital Latin letters include all coordinates except $r$, while Greek letters from the beginning of the alphabet denote angular coordinates on the unit $d-2$ sphere. In summary:
\begin{alignat}{3}
\m,\n,\ldots & \in \{t,r,\phi^1,\ldots,\phi^{d-2}\} \, , \qquad
& i,j,\ldots & \in \{r,\phi^1,\ldots,\phi^{d-2}\} \, , \nn \\[3pt]
I,J,\ldots & \in \{t,\phi^1,\ldots,\phi^{d-2}\} \, , \qquad
& \a,\b,\ldots & \in \{\phi^1,\ldots,\phi^{d-2}\} \, . \label{conventions-indices}
\end{alignat}

Indices between parentheses (or square brackets) are meant to be (anti)symmetrised with weight one, i.e.\ one divides the (anti)symmetrised expression by the number of terms that appear in it, so that the (anti)symmetrisation of a (anti)symmetric tensor is a projection.

In section \textbf{\ref{sec:spin3}} omitted indices denote a trace, whose precise meaning depends on the context: in covariant expressions a contraction with the full space-time metric $g_{\m\n}$ is understood, while contractions with the spatial metric $g_{ij}$ or with the boundary metric $g_{IJ}$ are understood in formulae displaying the corresponding types of indices.

In most of section \textbf{\ref{sec:spin_s}} we omit instead all indices: in linear expressions we elide all free indices, which are meant to be symmetrised, while in expressions quadratic in the fields omitted indices are meant to be contracted. In order to avoid ambiguities, however, in some terms we display explicitly a pair of contracted indices, continuing to elide the remaining ones. Traces, symmetrised gradients and divergences are denoted by
\be \label{notation}
\vf^{[n]} \equiv \vf_{\m_1 \cdots \m_{s-2n}\l_1 \cdots \l_n}{}^{\!\l_1 \cdots \l_n} \, , \qquad
\bar{\nabla} \vf \equiv \bar{\nabla}_{\!(\m_1} \vf_{\m_2 \cdots \m_{s+1})} \, , \qquad
\bar{\nabla}\cdot\vf \equiv \bar{\nabla}^\l \vf_{\l\m_1 \cdots \m_{s-1}} \, . 
\ee
We recalled here covariant expressions as an example, but contractions may also run only over spatial or transverse indices depending on the context. Similar shortcuts are adopted to denote symmetrisations or contractions with the ``extrinsic'' Christoffel symbol \eqref{christoffel}:
\be
\G \vf \equiv \G^0{}_{0(i_1} \vf_{i_2 \cdots i_{s+1})} \, , \qquad
\G\cdot \vf \equiv g^{kl} \G^0{}_{0k} \vf_{li_1 \cdots i_{s-1}} \, .
\ee

In Appendix \textbf{\ref{app:boundary}} and in some expressions of section \textbf{\ref{sec:spin_s}} we reinstate indices with the following convention: repeated covariant or contravariant indices denote a symmetrisation, while a couple of identical covariant and contravariant indices denotes as usual a contraction. Moreover, the indices carried by a tensor are substituted by a single label with a subscript indicating their total number. For instance, the combinations in \eqref{notation} may also be presented as
\be \label{example-repeated}
\vf^{[n]} = \vf_{\m_{s-2n}} \, , \qquad
\bar{\nabla} \vf = \bar{\nabla}_{\!\m} \vf_{\m_s} \, , \qquad
\bar{\nabla}\cdot\vf \equiv \bar{\nabla}^\l \vf_{\l\m_{s-1}} \, . 
\ee

Finally, our Hamiltonian conventions are $[q,p] = 1$ and $\dot{F} = [F, H]$.

\newpage
\section{Hamiltonian form of Fronsdal's action}\label{app:fronsdal}

This appendix collects some additional details on the rewriting of spin-$3$ Fronsdal's action in Hamiltonian form and on the derivation of first class constraints in the spin-$s$ case.

\subsection{Spin 3} \label{app:fronsdal-3}

Integrating by parts the time derivative, one can eliminate on any static background all terms with $\dot{\cN}_i$ and $\dot{N}_{ij}$ which appear in the expansion in components of the Fronsdal action \eqref{fronsdal-action}. This is because the lapse does not depend on time, so that this step of the computation works as in flat space. Eliminating time derivatives from $\cN_i$ and $N_{ij}$, however, does not suffice to identify them as Lagrange multipliers: one must also verify that they enter linearly the action. The cancellation of the quadratic terms requires instead that the metric \eqref{AdS_bose} be of constant curvature since in general
\be \label{S-generic}
\begin{split}
S & = S_{\textrm{Fronsdal}} + \int\! d^{\,d}x\, \frac{3\sqrt{g}}{2f}\, \bigg\{ f \left( R_{0ij}{}^0 - \frac{1}{L^2}\, g_{ij} \right)\! \left( 2 \cN_k \vf^{ijk} - \cN^i \vf^j - g^{ij} N \a \right) \\
& + 4 \left[ N^{ij} [ \nabla_{\!k} \,, \nabla_{\!i} ] N_j{}^k  + \frac{d}{L^2}\! \left( N_{ij}N^{ij} + N^2 \right) - R_{0ij}{}^0\! \left( N^i{}_k N^{jk} + 2 N^{ij} N + g^{ij} N^2 \right) \right] \\
& - \left[ \cN^i [ \nabla_{\!j} \,, \nabla_{\!i} ] \cN^j + \frac{2d}{L^2}\, \cN_i\cN^i - R_{0ij}{}^0\! \left( 3\cN^i\cN^j + g^{ij}\cN_i\cN^i \right) \right] \!\bigg\} \, ,
\end{split}
\ee
where $S_{\textrm{Fronsdal}}$ is the same action as \eqref{action}, with the same $\cH$, $\cC_i$ and $\cC_{ij}$, while $R_{\m\n\r}{}^\s$ denotes the Riemann tensor of the background. 
In obtaining this result we used only
\be
\G^0{}_{00} = \G^0{}_{ij} = \G^i{}_{0j} = 0 \, , \qquad \G^i{}_{00} = f^2 g^{ij} \G^0{}_{0j}
\ee
together with \eqref{christoffel}, which hold for a generic static metric.

The first line in \eqref{S-generic} would provide an additional contribution to the constraints, that vanishes on a constant curvature background on account of
\be \label{const-curvature}
R_{\m\n\r\s} = - \frac{1}{L^2} \left( g_{\m\r} g_{\n\s} - g_{\m\s} g_{\n\r} \right) .
\ee
The second and third lines must instead vanish and this requires \eqref{const-curvature}. The missing cancellation of the terms quadratic in $N_i$ and $N_{ij}$ is the counterpart of the loss of the gauge symmetry \eqref{cov-gauge} on arbitrary backgrounds. Correspondingly, in the canonical formalism one  obtains first class constraints only on constant curvature backgrounds. 

\subsection{Arbitrary spin} \label{app:fronsdal-s}

In order to reconstruct the canonical generator of gauge transformations \eqref{generator} by integrating \eqref{var-canonical}, one has to reconstruct the gauge transformations of the canonical variables from Fronsdal's covariant gauge transformation \eqref{fronsdal-gauge}. To this end, one has to set a map between the components of the covariant gauge parameter $\L_{\m_1 \cdots \m_{s-1}}$ and the deformation parameters $\x_{i_1 \cdots i_{s-1}}$ and $\l_{i_1 \cdots i_{s-2}}$ that enter \eqref{generator}. To obtain dimensions compatible with the action \eqref{action-s} and to agree with the spin-3 example, we choose
\be \label{comp-par}
\x_{i_1 \cdots i_{s-1}} = \L_{i_1 \cdots i_{s-1}} \, , \qquad \l_{i_1 \cdots i_{s-2}} = 2f^{-1} \L_{0\,i_1 \cdots i_{s-2}} \, .
\ee
The other components of the covariant gauge parameter are not independent from the ones above due to Fronsdal's constraint. 

Combining \eqref{fronsdal-gauge} with \eqref{comp-par} one obtains the following variations for the canonical variables and their traces:
\begin{subequations} \label{var-can-trace}
\begin{align}
\d \vf^{[n]} & = 2n\, \nd \x^{[n-1]} + (s-2n) \nabla \x^{[n]} \, , \label{dphi} \\[10pt]
\d \a^{[n]} & = -(2n+3) \nd \l^{[n]} - (s-2n-3) \nabla \l^{[n+1]} \, . \label{dalpha}
\end{align}
\end{subequations}
The gauge transformations of the Lagrange multipliers and their traces are obtained in a similar fashion:
\begin{subequations} \label{var-multipliers}
\begin{align}
\d N^{[n]} & = \dot{\x}^{[n]} + n f \left\{ \nd \l^{[n-1]} - \G\cdot \l^{[n-1]} \right\} + \frac{s-2n-1}{2} f \left\{ \nabla \l^{[n]} - \G\, \l^{[n]} \right\} , \label{dN1} \\[10pt]
\d \cN^{[n]} & = \dot{\l}^{[n]} + 2f \left\{ n\, \nd \x^{[n]} - (2n+1)\, \G\cdot \x^{[n]} \right\} + (s-2n-2)f \left\{ \nabla \x^{[n+1]} - 2\, \G\, \x^{[n+1]} \right\} . \label{dN2}
\end{align}
\end{subequations}
Substituting \eqref{var-can-trace} and \eqref{var-multipliers} in the definitions of the momenta \eqref{p1} and \eqref{p2} one obtains
\begin{align}
& \d \P = \sqrt{g} \sum_{n=0}^{\left[\frac{s}{2}\right]} \binom{s}{2n}\, g^n\, \bigg\{ (n-1)\binom{s-2n}{2} \nabla^2 \l^{[n]} + n\, \bigg[ (n-1)(2n-1) \nd\nd\l^{[n-2]} \nn \\[2pt]
& + \left( n\,\triangle - \frac{n(s-2n)(D+s-2n-3)+(2n-1)(D+2s-2n^2+4n-8)}{L^2} \right)\!\l^{[n-1]} \nn \\[2pt]
& + \frac{(4n-3)(s-2n)}{2}\, \nabla \nd \l^{[n-1]} + \frac{s-2n}{2}\, \G \left( (s-2n-1) \nabla \l^{[n]} + (2n+1) \nd \l^{[n-1]} \right) \nn \\[4pt]
& + (n-1)\, \G_{k} \left( (s-2n) \nabla \l^{[n-1]\,k} + (2n-1) \nd \l^{[n-2]\,k} + \nabla^k \l^{[n-1]} \right) \bigg] \bigg\} \, , \label{dp1}
\end{align}
\begin{align}
& \d\tilde{\P} = \sqrt{g} \sum_{n=1}^{\left[\frac{s-1}{2}\right]} \frac{n}{2} \binom{s}{2n+1}\, g^{n-1}\, \bigg\{\,  2(n-1)(2n+1) \nd\nd \x^{[n-1]} \nn \\[2pt]
& + (2n+1)\! \left[ \triangle - \frac{(s-2n-1)(D+s-2n-4) + 2(D+2s-2n^2+3n-7)}{L^2} \right] \x^{[n]} \nn \\[3pt]
& + (s-2n-1)\! \left[\, (4n+1) \nabla \nd \x^{[n]} + (s-2n-2) \nabla^2 \x^{[n+1]} \,\right] \nn \\[9pt]
& - (s-2n-1)\, \G \left[\, (s-2n-2) \nabla \x^{[n+1]} + 2(n+1) \nd \x^{[n]} \,\right] \nn \\[4pt]
& - (2n+1)\, \G_{k} \left[\, (s-2n-1) \nabla \x^{[n]\,k} + 2n\, \nd \x^{[n-1]\,k} + \nabla^k \x^{[n]} \,\right] \bigg\} \, . \label{dp2}
\end{align}
To derive \eqref{dp1} and \eqref{dp2} we used identities that are valid only on a constant-curvature space-time, as
\be \label{dGamma}
\nabla_{i} \G^0{}_{0j} + \G^0{}_{0i} \G^0{}_{0j} = \frac{1}{L^2}\, g_{ij}
\ee
which follows from \eqref{const-curvature}. We also commuted covariant derivatives using, for instance,
\be
[\, \nabla_{\!k} , \nabla \, ]\, \l^{[n-1]\,k} = - \frac{1}{L^2} \left\{ (D+s-2n-3) \l^{[n-1]} - (s-2n-1) g \l^{[n]} \right\} .
\ee

\newpage
\section{Covariant boundary conditions}\label{app:boundary}

In this appendix we recall the falloff at the boundary of AdS$_d$ of the solutions of Fronsdal's equations of motion (see also e.g.~\cite{Mikhailov,metsaev-solutions,boundary-action}). To achieve this goal we partially fix the gauge freedom, and we also exhibit the fall-off conditions on the parameters of the residual gauge symmetry (which include the traceless Killing tensors of AdS$_d$).

We set the AdS radius to $L=1$ and we work in the Poincar\'e patch parameterised as
\be \label{poincare-z_bose}
ds^2 = \frac{1}{z^2} \left( dz^2 + \h_{IJ} dx^I dx^J \right) 
\ee 
In these coordinates the spatial boundary is at $z \to 0$. All results can be easily translated in the coordinates \eqref{poincare3} used in the main body of the text ($z = 1/r$), in which the boundary is at $r \to \infty$. We denote by capital Latin indices all directions transverse to the radial one (including time). 

\subsection{Falloff of the solutions of the free equations of motion}\label{sec:eom}

We wish to study the space of solutions of the Fronsdal equation in AdS$_d$ \cite{fronsdal-AdS,fronsdal-AdS-D} which, in the index-free notation of section \textbf{\ref{sec:spin_s}}, reads
\be \label{fronsdal}
\Box \vf - s\, \bar{\nabla}\! \left( \bar{\nabla}\cdot \vf - \frac{s-1}{2}\, \bar{\nabla} \vf^{[1]} \right) - \frac{s^2 + (d-6)s - 2(d-3)}{L^2}\, \vf
- \frac{2}{L^2}\binom{s}{2}\, g\, \vf^{[1]} = 0
\ee
To this end it is convenient to partially fix the gauge freedom \eqref{fronsdal-gauge} by imposing $\bar{\nabla} \cdot \vf = 0$ and $ \vf^{[1]} = 0$.
This gauge is reachable on-shell,\footnote{In flat space one easily sees that one can reach the gauge $\pr \cdot \vf = \vf^{[1]} = 0$ only on-shell: imposing $\vf^{[1]} = 0$ implies $\prd \x = 0$, but the divergence of the field, which now transforms as $\d \prd \vf = \Box \x$, is not divergenceless. In the gauge $\vf^{[1]} = 0$ the solutions of the equations of motion satisfy however $\Box \vf - s\, \pr\prd \vf = 0$; taking a divergence one obtains $\prd\prd \vf = 0$ allowing to reach the desired gauge.} as it is proven e.g.\ in sect.~3.1 of \cite{Mikhailov}. The previous statement amounts to say that the space of solutions (modulo gauge transformations) of \eqref{fronsdal} is equivalent to the space of solutions of the Fierz system
\begin{subequations} \label{fierz}
\begin{align}
& \!\!\left[ \Box - \left( s^2 + (d-6)s - 2(d-3) \right) \right] \vf = 0 \, , \label{fierz-eom} \\[3pt] 
& \bar{\nabla} \cdot \vf = 0 \, , \label{fierz-div} \\[1pt]
& \vf^{[1]} = 0 \, , \label{fierz-trace}
\end{align}
\end{subequations}
which also describes the propagation of a free massless particle of spin $s$ in the AdS$_d$ background \cite{fierz}. Actually the conditions \eqref{fierz-div} and \eqref{fierz-trace} do not fix completely the gauge: the Fierz system admits gauge transformations still of the form $\d \vf = s \bar{\nabla} \L$, but with gauge parameters constrained as
\begin{subequations} \label{parameter}
\begin{align}
& \!\!\left[\, \Box - (s-1)(d+s-3) \,\right] \L = 0 \, , \label{par-eom} \\[3pt]
& \bar{\nabla} \cdot \L = 0 \, , \label{par-div} \\[1pt]
& \L^{[1]} = 0 \, . \label{par-trace}
\end{align}
\end{subequations}

To analyse the falloff at the boundary of the solutions of the conditions \eqref{fierz}, one has to treat separately field components with a different number of indices along the $z$ direction. We denote them as
\be \label{radial-label}
\vf_{z_kI_{s-k}} \equiv \vf_{z \cdots z\,I_1 \cdots I_{s-k}} \, .
\ee
The divergence constraint \eqref{fierz-div} then gives
\be \label{divergence}
\left( z\pr_z - (d-2) \right) \vf_{z\m_{s-1}} + z\, \pr^I \vf_{I\m_{s-1}} = 0 \, ,
\ee
where Greek letters denote indices that take values in all space-time dimensions (including $z$).
The trace constraint \eqref{fierz-trace} gives instead
\be \label{trace}
\vf_{zz\m_{s-2}} + \h^{IJ} \vf_{IJ\m_{s-2}} = 0 \, .
\ee
Using \eqref{divergence} and \eqref{trace} the components of the equation of motion \eqref{fierz-eom} read
\be \label{eom}
\begin{split}
& \left[ z^2 \pr_z^2 - \left( d-2(s-k+1) \right) z\pr_z - (d+k-3)(2s-k-2) \right] \vf_{z_k I_{s-k}}\\[5pt]
& + z^2\, \Box \vf_{z_k I_{s-k}} - 2(s-k)\, z\,\pr_I \vf_{z_{k+1}I_{s-k-1}} + (s-k)(s-k-1)\, \h_{II} \vf_{z_{k+2}I_{s-k-2}} = 0 \, ,
\end{split}
\ee
where now $\Box = \h^{IJ} \pr_I\pr_J$ and repeated covariant or contravariant indices denote a symmetrisation, see e.g.~\eqref{example-repeated}.

Eq.~\eqref{divergence} implies $\vf_{z_k I_{s-k}} \sim z^{\D + k}$; substituting this ansatz in \eqref{eom} the terms in the second line are subleading for $z \to 0$ and the first line vanishes provided that
\be \label{fall-off-field}
\vf_{z\cdots z\, I_1 \cdots I_{s-k}} \sim z^{\D_\pm+\,k} \quad \textrm{with} \ \
\left\{
\begin{array}{l}
\D_+ = d-3 \\[5pt]
\D_- = 2-2s
\end{array}
\right. .
\ee
For $s = 0$ one recovers the asymptotic behaviour of the conformally coupled scalar of mass $m^2 = - 2(d-3)$ that enters the Vasiliev equations. For higher spins the subleading $\D_+$ branch gives the boundary conditions usually considered within the AdS/CFT correspondence \cite{review-Giombi,review-strings} and adopted in the text, while the $\D_-$ branch has been considered in a holographic setup in \cite{boundary-action,unfolding-holography,alternate-bnd-cond}.

\subsection{Residual gauge symmetry}\label{app:gauge}

The constraints \eqref{parameter} force the gauge parameters to have a precise fall off at the boundary, which can be determined as above. The divergence and trace constraints give again
\begin{subequations}
\begin{align}
& \left( z\pr_z - (d-2) \right) \L_{z\m_{s-2}} + z\, \pr^I \L_{I\m_{s-2}} = 0 \, , \label{par-div-open} \\[5pt]
& \L_{zz\m_{s-3}} + \h^{IJ} \L_{IJ\m_{s-3}} = 0 \, . \label{par-trace-open}
\end{align} 
\end{subequations}
Eq.~\eqref{par-div-open} implies $\L_{z_k I_{s-k-1}} \sim z^{\Theta + k}$ and using the relations above in \eqref{par-eom} one gets
\be \label{par-eom-open}
\begin{split}
& \left[ z^2 \pr_z^2 - \left( d-2(s-k) \right) z\pr_z - (d+k-1)(2s-k-2) + z^2 \Box \right] \L_{z_k I_{s-k-1}}\\[5pt]
& - 2(s-k-1)\, z\pr_I \L_{z_{k+1}I_{s-k-2}} + (s-k-1)(s-k-2)\, \h_{II} \L_{z_{k+2}I_{s-k-3}} = 0 \, .
\end{split}
\ee
This equation is identical to \eqref{eom}, apart from the shift $s \to s-1$ and a modification in the mass terms which implies that the first line vanishes asymptotically when
\be \label{fall-off-x}
\x_{z\cdots z\, I_1 \cdots I_{s-k-1}} \sim z^{\Theta_\pm+\,k} \quad \textrm{with} \ \
\left\{
\begin{array}{l}
\Theta_+ = d-1 \\[5pt]
\Theta_- = 2-2s
\end{array}
\right. .
\ee

We solved a second order equation and, as a result, we obtained two allowed asymptotic behaviours for the gauge parameters. On the contrary, the Killing equation \eqref{kill-s} is of first order and Killing tensors admit only a given boundary falloff. To fix it, notice that the AdS background is left invariant by the same set of gauge transformations in both Fronsdal's and Fierz's formulations: the gauge parameters are traceless in both setups and a traceless Killing tensor is also divergenceless thanks to
\be
g^{\a\a} \bar{\nabla}_{\!\a} \L_{\a\m_{s-2}} = - \frac{s-2}{2}\, g^{\a\a} \bar{\nabla}_{\!\m} \L_{\a\a\m_{s-2}} = 0 \, .
\ee
Traceless Killing tensors must therefore display one of the two boundary behaviours above. The Killing equation $\bar{\nabla}_{\!\m} \x_{\m_{s-1}} = 0$ branches in components as
\be \label{killingeq}
\left[ z\pr_z + (2s-k-2) \right] \L_{z_k I_{s-k-1}} = \frac{s-k-1}{k+1} \left[ -z\, \pr_I \L_{z_{k+1} I_{s-k-2}} + (s-k-2) \h_{II} \L_{z_{k+2} I_{s-k-3}} \right] \, .
\ee
The right-hand side is subleading, and one thus sees that Killing tensors belong to the $\Theta_-$ branch of \eqref{fall-off-x}.

\subsection{Initial data at the boundary}\label{sec:leading}

In this subsection we recall the constraints on the initial data at the boundary imposed by divergence and trace constraints, and how the number of independent components is further reduced by the residual gauge symmetry. To this end we denote the leading contributions in the $\D_+$ branch by
\begin{subequations} \label{branch-vev}
\begin{align}
\vf_{I_s} & = z^{\D_+} \cT_{I_s}(x^J) + \cO\!\left(z^{\D_+ + 2}\right) , \label{vev} \\[5pt]
\vf_{z_k I_{s-k}} & = z^{\D_+ + k}\, t^{(k)}_{I_{s-k}}(x^J) + \cO\!\left(z^{\D_+ + k + 2}\right) , \quad 1\leq k \leq s \, ,
\end{align}
\end{subequations}
and the leading contributions in the $\D_-$ branch by
\begin{subequations} \label{branch-source}
\begin{align}
\vf_{I_s} & = z^{\D_-} \Phi_{I_s}(x^J) + \cO\!\left(z^{\D_- + 2}\right) , \label{source} \\[5pt]
\vf_{z_k I_{s-k}} & = z^{\D_- + k}\, \phi^{(k)}_{I_{s-k}}(x^J) + \cO\!\left(z^{\D_- + k + 2}\right) , \quad 1\leq k \leq s \, .
\end{align}
\end{subequations}
The subleading terms can be expressed in terms of the leading ones via the field equations (see e.g.~\cite{metsaev-solutions}).

The tensors $\cT_{I_s}$ and $\Phi_{I_s}$ are boundary fields of conformal dimensions, respectively, $\Delta_c = d + s - 3$ and $\Delta_s = 2 - s$. They thus correspond to the \emph{conserved currents} and \emph{shadow fields} of \cite{shadows}.\footnote{If one performs a dilatation $x^\m \to \l x^\m$ the tensor $\vf_{I_1 \cdots I_s}$ transforms as $\vf_{I'_s} = \l^{-s} \vf_{I_s}$, while on the right-hand side of \eqref{vev} or \eqref{source} one has $z'^{\D_\pm} = \l^{\D_\pm} z^{\D_\pm}$. As a result, $\cT$ and $\Phi$ must transform as $\cT_{I'_s} = \l^{- ( \Delta_+ + s )} \cT_{I_s}$ and $\Phi_{I'_s} = \l^{- ( \Delta_- + s )} \Phi_{I_s}$, from where one reads the conformal dimensions. To obtain a direct matching between the exponents $\D_\pm$ and the conformal dimensions, one can contract all indices with the (inverse) vielbein $e_\m{}^M= z^{-1}\, \d_\m{}^M$ as e.g.\ in \cite{boundary-action}. }
Note that $\D_c + \D_s = d -1$, i.e.\ the sum of conformal dimensions is equal to the dimension of the boundary. Therefore the coupling $\Phi^{I_s} \cT_{I_s}$ is conformally invariant. In the AdS/CFT jargon $\Phi_{I_s}$ is a \emph{source} and $\cT_{I_s}$ is the corresponding \emph{vev}.

The trace constraint \eqref{trace} then implies that all tensors on the right-hand side of \eqref{branch-vev} and \eqref{branch-source} are traceless, since the trace of $\vf_{z_k I_{s-k}}$ is subleading with respect to the traceless part. The divergence constraint \eqref{divergence} has instead different consequences in the two branches: in the $\D_+$ branch of vevs one obtains
\begin{subequations}
\begin{align}
& \prd \cT_{I_{s-1}} = 0 \, , \label{conservation} \\[5pt]
& t^{(k+1)}_{I_{s-k-1}} = - \frac{1}{k}\, \prd t^{(k)}_{I_{s-k-1}} \, , \quad 1 \leq k \leq s \, ,
\end{align}
\end{subequations}
while in the $\D_-$ branch of sources one obtains
\be
\phi^{(k+1)}_{I_{s-k-1}} = \frac{1}{d+2s-k-5}\, \prd \phi^{(k)}_{I_{s-k-1}} \, , \quad 0 \leq k \leq s \, ,
\ee
where $\phi^{(0)}_{I_s} \equiv \Phi_{I_s}$ and where divergences are meant to involve sums only over indices transverse to $z$. Eq.~\eqref{conservation} shows that, as expected from its conformal weight, $\cT$ is a conserved current. Up to this point $t^{(1)}$ is instead an arbitrary tensor (but we still have to consider the residual gauge symmetry), while all other tensors in \eqref{branch-vev} are not independent. In the other branch the only independent tensor is instead $\Phi$, whose number of independent components is the same as those of $\cT$ plus $t^{(1)}$. 

The residual gauge symmetry is generated by
\be \label{par+}
\L_{z_k I_{s-k-1}} = z^{\Th_+ + k}\, \x^{(k)}_{I_{s-k-1}}(x^J) + \cO\!\left(z^{\Th_+ + k + 2}\right) , 
\ee
or
\be \label{par-}
\L_{z_k I_{s-k-1}} = z^{\Th_- + k}\, \ve^{(k)}_{I_{s-k-1}}(x^J) + \cO\!\left(z^{\Th_- + k + 2}\right) .
\ee
In the following we shall often denote $\x^{(0)}$ by $\x$ and $\ve^{(0)}$ by $\ve$.
As for the fields, the trace constraint \eqref{par-trace-open} imposes that all tensors in \eqref{par+} and \eqref{par-} be traceless. For gauge parameters the divergence constraint \eqref{par-div-open} has instead the same form in both branches:
\begin{subequations}
\begin{align}
& \x^{(k+1)}_{I_{s-k-2}} = - \frac{1}{k+2}\, \prd \x^{(k)}_{I_{s-k-2}} \, , \quad 0 \leq k \leq s-1 \, , \\[5pt]
& \ve^{(k+1)}_{I_{s-k-2}} = \frac{1}{d+2s-k-5}\, \prd \ve^{(k)}_{I_{s-k-2}} \, , \quad 0 \leq k \leq s-1 \, . \label{div-ve}
\end{align}
\end{subequations}
This means that only the tensors $\x_{I_{s-1}}$ and $\ve_{I_{s-1}}$ are independent.

The components of the field transform as
\be \label{var-residual}
\d \vf_{z_k I_{s-k}} = k \left[ \pr_z + \frac{2s-k-1}{z} \right]\! \L_{z_{k-1}I_{s-k}} + (s-k)\! \left[ \pr_I \L_{z_k I_{s-k-1}}\! - \frac{s-k-1}{z}\, \h_{II} \L_{z_{k+1}I_{s-k-2}} \right] .
\ee
In the $\Th_+$ branch the first contribution is leading and gives
\be
\d \vf_{z_k I_{s-k}} = k (d+2s-3) z^{d-3+k} \x^{(k-1)}_{I_{s-k}} + \cO(z^{d-1+k}) \, .
\ee
These gauge transformations act naturally on vevs, where they induce
\be
\d \cT_{I_s} = 0 \, , \qquad
\d t^{(1)}_{I_{s-1}} = (d+2s-3) \L_{I_{s-1}} \, .
\ee
The tensor $t^{(1)}$ is thus a Stueckelberg field that can be eliminated using the residual symmetry generated by $\x$ and the solution is fully specified by the conserved current $\cT$.

In the $\Th_-$ branch the coefficient of the leading contribution vanishes, and one obtains
\be
\begin{split}
\d \vf_{z_k I_{s-k}} & = \frac{z^{2-2s+k}}{d+2s-5} \Big[\, k\, \Box \ve^{(k-1)}_{I_{s-k}} + (s-k)(d+2(s-k)-5)\, \pr_I \ve^{(k)}_{I_{s-k-1}} \\
& - (s-k)(s-k-1)(d+2s-k-5)\, \h_{II} \ve^{(k+1)}_{I_{s-k-2}} \,\Big] + \cO(z^{4-2s+k}) \, ,
\end{split}
\ee
where we used the second-order equation \eqref{par-eom-open} to express the subleading contributions in $\L_{z_{k-1}I_{s-k}}$ in terms of the leading ones. These gauge transformations act naturally on sources, where they induce
\be \label{var-source}
\d \Phi_{I_s} = s\, \pr_{I} \ve_{I_{s-1}} - \frac{s(s-1)}{(d-1)+2(s-2)}\, \h_{II} \prd\ve_{I_{s-2}}
\ee
and similar transformations on the $\phi^{(k)} = \phi^{(k)}(\Phi)$.  To obtain \eqref{var-source} we used \eqref{div-ve} to eliminate $\ve^{(1)}$. This gauge freedom reduces the number of independent components of $\Phi_{I_s}$ such that it becomes identical to that of the conserved current $\cT_{I_s}$. Moreover, the gauge transformations \eqref{var-source} leave invariant the coupling $\Phi^{I_s} \cT_{I_s}$.

To summarise, the solutions of the field equations are specified asymptotically either by a traceless and conserved current $\cT_{I_s}$ or by a traceless tensor $\Phi_{I_s}$ subjected to the gauge symmetry \eqref{var-source}. These tensors enter the components transverse to $z$ as in \eqref{vev} and \eqref{source}, while all other components of the spin-$s$ field are expressed in terms of them via the field equations (or set to zero by the residual gauge symmetry \eqref{var-residual}). 

\newpage
\section{Conformal Killing tensors}\label{app:identities}

In this appendix we recall the structure of the general solution of the conformal Killing equations \eqref{conf-kill} and we present it explicitly in the rank-2 case. We also prove the identities \eqref{kill1} and \eqref{kill2}, that we used in the analysis of asymptotic symmetries in the spin-3 example.

\subsection*{General solution of the rank-2 conformal Killing equation}

When $d-1 > 2$, the solutions of the conformal Killing tensor equation \eqref{conf-kill} are in one-to-one correspondence with rectangular traceless Young tableaux with two rows of \mbox{$s-1$} boxes in $d+1$ dimensions \cite{algebra}, that we denote by $\{s,s\}$. For instance, a generic conformal Killing vector, $v_I = a_I - \o_{I|J} x^J + \l\, x_I + b_{K}\! \left( 2\,x_Ix^K - \d_I{}^{K} x^2 \right)$,
can be cast in the form \mbox{$v^I = \Phi_{\und{M}} V^{\und{M}|\und{N}} \,\Psi^I{}_{\!\und{N}}$}, where underlined indices take value in the $(d+1)$-dimensional ambient space $\mathbb{R}^{2,d-1}$ and
\be
V^{\und{I}|\und{J}} = 
\left(\begin{array}{ccc}
0 & 2b^J & \l \\
-2b^I & \omega^{I|J} & -a^I \\
-\l & a^J & 0
\end{array}\right) , \quad
\Phi_{\und{I}} = 
\left(\begin{array}{c}
-x^2/2 \\ x_I \\ 1
\end{array}\right) , \quad
\Psi^{I}{}_{\!\und{J}} =
\left(\begin{array}{c}
-x^I \\ \d^I{}_{\!J} \\ 0
\end{array}\right) .
\ee
A similar characterisation of traceless conformal Killing tensors exists for any value of the rank \cite{algebra}. 
In particular, the pattern of tensors in $d-1$ dimensions that specify the solution follows from the decomposition of a traceless $\{s,s\}$ tensor in $d+1$ dimensions (using the branching rules discussed e.g.\ in \S~8.8.A of \cite{Barut}). In the rank-2 case one obtains
\be
\{2,2\}_{o(d+1)} = \left( \{2,2\} + 2\,\{2,1\} + 3\,\{2\} + \{1,1\} + 2\,\{1\} + 1 \right){}_{\!o(d-1)} \, .  
\ee
Correspondingly, the general solution of \eqref{kill-boundary} is
\begin{align}
& \chi_{IJ} = a_{IJ} + \left( b_{(I}x_{J)} - \frac{1}{d-1}\, \h_{IJ}\, b\cdot x\right) + \o_{IJ|K}\,x^K + \l \left( x_I x_J - \frac{1}{d-1}\, \h_{IJ}\,x^2 \right) \nn \\
& + \r_{K|(I}x_{J)}x^K + \left(  2\, c_{K(I} x_{J)}x^K  - c_{IJ}\, x^2 - \frac{2}{d-1}\, \h_{IJ} c_{KL} x^K x^L \right) + \O_{IJ|KL}\,x^K x^L \nn \\
& + \left( 2\,\tilde{b}_K x_Ix_Jx^K \! - \tilde{b}_{(I} x_{J)}x^2 - \frac{1}{d-1}\, \h_{IJ} (b\cdot x) x^2 \right) + \left( 2\,\tilde{\o}_{KL|(I}x_{J)}x^Kx^L \! + \tilde{\o}_{IJ|K}x^Kx^2 \right) \nn \\[5pt]
& + \tilde{c}_{KL} \left( 4\,x_Ix_Jx^Kx^L - 4\, \d^K{}_{\!(I} x_{J)} x^L x^2 + \d^K{}_{\!I} \d^L{}_{\!J}\, x^4  \right) . \label{sol-conf} 
\end{align}
All tensors in this expression are irreducible and traceless, so that e.g.\ $\o_{IJ|K}$ is symmetric in its first two indices and satisfies $\o_{(IJ|K)} = 0$.

When $d-1 = 2$, the general solution depends instead on two chiral functions for any value of the rank. Introducing the light-coordinates $x^\pm = t \pm \phi$, eqs.~\eqref{conf-kill} are solved by 
\be \label{sol-conf3}
\chi^{+ \cdots +} = \chi(x^+) \, , \qquad
\chi^{+\cdots+-\cdots-} = 0 \, , \qquad
\chi^{- \cdots -} = \tilde{\chi}(x^-) \, .
\ee

\subsection*{Proof of eqs.~\eqref{kill1} and \eqref{kill2}}

We wish to prove that the rank-2 conformal Killing equation \eqref{kill-boundary} implies the identities \eqref{kill1} and \eqref{kill2}, which entail the cancellation of the first two subleading orders in $\d \vf_{IJK}$. The following proof is independent on the space-time dimension, but when $d > 3$ one could also verify these identities by acting with the differential operators they involve on \eqref{sol-conf}.

As a first step, one can act with a derivative on the conformal Killing equation, and rewrite the result as follows
\be \label{1st-id}
\begin{split}
0 & = \pr_M \!\left( \pr_{(I} \chi_{JK)} - \frac{2}{d+1}\, \h_{(IJ\,} \prd \chi_{K)} \right) = 2 \left( \pr_{(M} \pr_I \chi_{JK)} - \frac{2}{d+1}\, \h_{(MI\,} \pr_J \prd \chi_{K)} \right) \\
& - \, \pr_{(I} \pr_J\chi_{K)M} - \frac{1}{d+1} \left(\, \h_{(IJ|} \pr_M \prd\chi_{|K)} - 2\, \h_{M(I} \pr_J \prd \chi_{K)} - \h_{(IJ} \pr_{K)} \prd \chi_M \right) .\end{split}
\ee
The first term on the right-hand side vanishes since it is the symmetrisation of the left-hand side. As a result, one discovers that the second line vanishes as well.

To prove \eqref{kill1} one needs another identity obtained in a similar fashion:
\begin{align}
0 & = 3\, \pr_M \pr_N \!\left( \pr_{(I} \chi_{JK)} - \frac{2}{d+1}\, \h_{(IJ\,} \prd \chi_{K)} \right) \nn \\
& = 10 \left( \pr_{(M} \pr_N \pr_I \chi_{JK)} - \frac{2}{d+1}\, \h_{(MN\,} \pr_I \pr_J \prd \chi_{K)} \right) - \pr_{I} \pr_J \pr_K \chi_{MN} \label{2nd-id} \\
& + \frac{2}{d+1}\, \Big\{ \h_{(IJ|} \pr_M \pr_N \prd\chi_{|K)} - 2 \left( \h_{M(I} \pr_{J|} \pr_N \prd \chi_{|K)} + \h_{N(I} \pr_{J|} \pr_M \prd \chi_{|K)} \right) \nn \\
& - \h_{(IJ} \pr_{K)} \pr_{(M} \prd \chi_{N)} + 2 \left( \h_{M(I} \pr_J \pr_{K)} \prd \chi_N + \h_{N(I} \pr_J \pr_{K)} \prd \chi_M \right) + \h_{MN} \pr_{(I} \pr_J \prd \chi_{K)} \Big\} . \nn
\end{align}
The first term on the right-hand side --~with a symmetrisation over five indices~-- vanishes again because it is the symmetrisation of the left-hand side. To reach this expression we also used the identity derived from \eqref{1st-id}.

One can finally contract the result with $\h^{MN}$ obtaining (recall that $\chi_{MN}$ is traceless): 
\be \label{3rd-id}
(d-1)\, \pr_{(I} \pr_J \prd \chi_{K)} - \h_{(IJ|} \!\left\{ \pr_{|K)} \prd\prd\prd \chi - \Box\, \prd\chi_{|K)} \right\} = 0 \, .
\ee
By computing two divergences of the conformal Killing equation \eqref{kill-boundary} one also obtains
\be
\Box\, \prd \chi_I = - \frac{d-3}{2d}\, \pr_{I\,} \prd\prd \chi \, ,
\ee
so that \eqref{3rd-id} implies \eqref{kill1} for $d>1$.

To prove the identity \eqref{kill2}, it is convenient to compute two gradients of \eqref{kill1} and to manipulate the resulting expression as in the previous subsection:
\begin{align}
0 & = 3\, \pr_M \pr_N \!\left( \pr_{(I} \pr_J \prd \chi_{K)} - \frac{3}{2d}\, \h_{(IJ} \pr_{K)} \prd\prd \chi \right) \nn \\
& = 5 \left( \pr_{(M} \pr_N \pr_I \pr_J \prd \chi_{K)} - \frac{3}{2d}\, \h_{(MN} \pr_I \pr_J \pr_{K)} \prd\prd \chi \right) - 2\, \pr_I \pr_J \pr_K \pr_{(M} \prd\chi_{N)} \\
& + \frac{3}{4d} \left\{ \h_{MN} \pr_I \pr_J \pr_K + 3 \left( \h_{M(I} \pr_J \pr_{K)} \pr_N + \h_{N(I} \pr_J \pr_{K)} \pr_M \right) - 3\, \h_{(IJ} \pr_{K)} \pr_M \pr_N \right\} \prd\prd\chi \, . \nn
\end{align}
The first term on the right-hand side vanishes. Contracting the remaining addenda with $\h^{MN}$ and taking into account \eqref{kill3} one obtains \eqref{kill2}.

\newpage
\section{Spin-$s$ charges}\label{app:charges}

This appendix is dedicated to provide the reader with some details of the computation of the asymptotic charges in the general case of spin $s$. As mentioned in section \textbf{\ref{sec:spin_s}}, taking into account the boundary conditions on canonical variables and deformation parameters (\ref{boundary-spins-varphi})--(\ref{bdn-deform-spins}) one sees that the finite contributions to the charges come from the terms
\be \label{breakdown-charges}
\lim\limits_{r\rightarrow\infty}Q_1[\xi] = \!\int\! d^{d-2}x\, \Big\{A_{1}(\Pi)+B_{1}(\alpha)\Big\} \, ,\quad
\lim\limits_{r\rightarrow\infty}Q_2[\lambda] =  \!\int\! d^{d-2}x\, \Big\{A_{2}(\tilde{\Pi})+B_{2}(\varphi) \Big\} \, ,
\ee
where
\begin{subequations}
\begin{align}
A_{1}(\Pi)&\equiv s \, \xi \Pi^{r} \, , \label{A1} \\
B_{1}(\alpha)&\equiv s\,\sqrt{g} \sum_{n=1}^{\left[\frac{s-1}{2}\right]}\! \frac{n}{2} \binom{s-1}{2n} \Big\{ \xi^{[n]} \left[ \nabla^r \a^{[n-1]} + 2(n-1) \nabla\cdot \a^{r[n-2]} \right] \nn \\
& - \a^{[n-1]} \left[ \nabla^r \xi^{[n]} + 2(n-1) \nabla\cdot \xi^{r[n-1]} - \G^r \xi^{[n]} \right] \Big\} \, , \\[8pt]
A_{2}(\tilde{\Pi})&\equiv (3-s)\, \lambda^{[1]} \tilde{\Pi}^{r} \, , \\[1pt]
B_{2}(\varphi)&\equiv \sqrt{g} \sum_{n=0}^{\left[\frac{s}{2}\right]} n \binom{s}{2n} \Big\{ \lambda^{[n-1]} \left[\, n\, \nabla^r \varphi^{[n]} + (n-2)(2n-1) \nabla\cdot \varphi^{r[n-1]} \,\right] \nonumber\\
& -  \varphi^{[n]} \left[\, n\, \nabla^r \lambda^{[n-1]} + (n-1) \left( (2n-1) \nabla\cdot \lambda^{r[n-2]} + \G^r \lambda^{[n-1]} \right)\right] \Big\} \, .
\end{align}
\end{subequations}
At this stage, differently from \eqref{qfin}, the omitted indices in the expressions above still include all coordinates except time as in \eqref{Qgen}. Along the way we shall show that the contributions from radial components are actually subleading; eventually all omitted indices can thus be considered to be valued on the $d-2$ sphere at infinity as indicated in section \textbf{\ref{sec:charges}}.

The contribution of the terms in $B_{1}$ and $B_{2}$ is computed in a straightforward fashion using the boundary conditions (\ref{boundary-spins-varphi}) and (\ref{bdn-deform-spins}). As an example, let us consider the first term in $B_{1}$, that is \mbox{$\sqrt{g}\, \xi^{[n]} \nabla^r \a^{[n-1]}$}. Displaying explicitly the free indices on each tensor as in Appendix~\textbf{\ref{app:boundary}}, one obtains
\be\label{Ck}
\sqrt{g}\,g^{rr}\xi^{i_{s-2n-1}}\nabla_{\!r}\alpha_{i_{s-2n-1}} = \sqrt{g}\,g^{rr} \sum_{k=0}^{s-2n-1} \binom{s-2n-1}{k}\, \xi^{r_k\a_{s-k-1}}\nabla_{\!r}\alpha_{r_k\a_{s-k-1}} \, .
\ee
We recall that small Latin indices include all coordinates except time, while Greek indices from the beginning of the alphabet do not include neither time nor radial directions. We also resorted to a collective notation for the radial indices as in \eqref{radial-label}.
Expanding the covariant derivative and using the boundary conditions one gets
\be
\nabla_{\!r}\alpha_{r_k\a_{s-k-1-2n}} \sim -2(3-d-k-s)\,r^{1-d-3k-2n} \, ,
\ee
and so
\be \label{contraction-example}
\sqrt{g}\,g^{rr}\xi^{r_k\a_{s-k-1-2n}}\nabla_{\!r}\alpha_{r_k\a_{s-k-1-2n}} \sim -2(3-d-k-s)\,r^{-2k} \, .
\ee
Clearly the only finite contribution when $r\to\infty$ comes from the term $k=0$ in (\ref{Ck}). This is a general feature that one also encounters in the analysis of the other contributions in $B_1$ (and more generally in \eqref{breakdown-charges}). One can compute them along similar lines and find
\be\label{B1}
B_{1}=C\sum_{n=1}^{[\frac{s-1}{2}]} \frac{n(s-1)!}{(2n)!(s-2n-1)!}\, \chi^{\supsuboverbrac[2n]{0\cdots0}\a_{s-2n-1}}{\cal{T}}_{0\supsuboverbrace[2n]{0\cdots0}\a_{s-2n-1}} \, ,
\ee
with $C\equiv s(d+2s-5)$.

Now let us turn to $A_{1}$: from \eqref{boundary-spins-varP} and \eqref{bdn-deform-spins} one sees that, similarly to e.g.~\eqref{contraction-example}, the only finite contribution in $\xi \Pi^{r}$ comes from the term $\xi_{\a_{s-1}} \Pi^{r\a_{s-1}}$. Its computation is however less direct since one has to take into account the definition \eqref{p1} of the momentum. To illustrate this point, consider the first term in $\Pi_{r \a_1\cdots \a_{s-1}}$:
\be \label{my-rewriting}
\begin{split}
\Pi_{r \a_{s-1}} & = \frac{\sqrt{g}}{f} \sum_{n=0}^{\left[\frac{s}{2}\right]} \binom{s}{2n}(2n-1)(s-2n)(g_{\a\a})^n \Big[\nabla_{\!r} N_{\a_{s-2n-1}} \\ & + (s-2n-1)\nabla_{\!\a} N_{r\a_{s-2n-2}} \big] D_{1}(n) + \cdots \, ,
\end{split}
\ee
where we have introduced the degeneracy factor $D_{1}(n)\equiv\frac{1}{(2n+1)!!}\binom{s}{2n+1}^{-1}$, which counts the number of equivalent terms after the symmetrisation over free indices.
Expanding the covariant derivative and taking advantage of the trace constraint in \eqref{conf-kill} one finds
\begin{subequations}
\begin{align}
\nabla_{\!r}N_{\a_{s-2n-1}} & \sim (4-d-s)\,r^{2-d-2n}\,\cT_{0\supsuboverbrace[2n]{0\cdots0}\,\a_{s-2n-1}} \, , \\
\nabla_{\!\a}N_{r\a_{s-2n-2}} & \sim -\,r^{2-d-2n}\,\cT_{0\supsuboverbrace[2n]{0\cdots0}\,\a_{s-2n-2}} \, ,
\end{align}
\end{subequations}
so that
\be
\Pi_{r \a_{s-1}} \sim \sum_{n=0}^{\left[\frac{s}{2}\right]} \binom{s}{2n}(2n-1)(s-2n) D_{1}(n)(5-d-2s+2n)\,r^{-2}\,(g_{\a\a})^n {\cal{T}}_{0\supsuboverbrace[2n]{0\cdots0}a_{s-2n-1}} \, .
\ee
Repeating the analysis for the other relevant terms in $\Pi_{r \a_1\cdots \a_{s-1}}$ one gets
\be \label{Pr}
\Pi_{r\a_{s-1}} \sim -\,s(d+2s-5)\, r^{-2}\sum_{n=0}^{[s/2]}D_{1}(n)\binom{s-1}{2n}(n-1)\,(g_{\a\a})^n {\cal{T}}_{0\supsuboverbrace[2n]{0\cdots0}\a_{s-2n-1}}
\ee
and after contracting with the deformation parameter 
\be\label{xiP}
s\, \xi_{\a_{s-1}}\Pi^{r \a_{s-1}} \sim -\,C\sum_{n=0}^{[s/2]}D_{1}(n)I_{1}(n)\binom{s-1}{2n}(n-1)\,\chi^{\supsuboverbrac[2n]{0\cdots0}\a_{s-2n-1}}{\cal{T}}_{0\supsuboverbrace[2n]{0\cdots0}\a_{s-2n-1}} \, ,
\ee
where $C$ is the same factor as in \eqref{B1}, while $I_{1}(n) = \binom{s-1}{2n}(2n-1)!!$ is a different degeneracy factor that takes into account the number of non-equivalent terms in $\Pi^{r\a_{s-1}}$ that give the same contribution after contraction with $\xi_{\a_{s-1}}$ --~owing to the complete symmetric character of the latter. Then by adding up (\ref{B1}) and (\ref{xiP}) one gets
\be
\lim\limits_{r\rightarrow\infty}Q_{1} = C \!\int\! d^{d-2}x\sum_{n=0}^{[(s-1)/2]} \!\binom{s-1}{2n}\! \left[ n -(n-1)\frac{2^{n}n!(2n-1)!!}{(2n)!}\right]\chi^{\supsuboverbrac[2n]{0\cdots0}\a_{s-2n-1}}{\cal{T}}_{0\supsuboverbrace[2n]{0\cdots0}\,\a_{s-2n-1}} \\
\ee
thus recovering \eqref{Qgen-chi1} thanks to the identity \mbox{$(2n)! = 2^n n! (2n-1)!!$} (and taking into account the trace constraints defined in \eqref{cons-curr-s} and \eqref{conf-kill}).

A similar analysis yields 
\be \label{Ptr}
\tilde{\Pi}^{r\alpha_{s-4}} \sim -\,\frac{C}{2}\,r^{-2(s-4)-1}\sum_{n=2}^{[\frac{s+1}{2}]}D_{2}(n)\binom{s-1}{2n-1}\frac{n-1}{s-2n+1}(g^{\alpha\alpha})^{n-2}({\cal{T}}^{[n]})^{\alpha_{s-2n}}
\ee
from where one obtains
\be \label{lambda-Pi}
\lambda_{i_{s-4}}\tilde{\Pi}^{ri_{s-4}}\sim C\sum_{n=2}^{[(s+1)/2]}\binom{s-1}{2n-1}D_{2}(n)I_{2}(n)\frac{(n-1)}{s-2n+1}\,\chi^{\supsuboverbrac[2n-1]{0\cdots0}\a_{s-2n}}{\cal{T}}_{\supsuboverbrace[2n]{0\cdots0}\,\a_{s-2n}} \, ,
\ee
In the expressions above we have introduced the combinatorial factors
\be
D_{2}(n) =\frac{2^{n-2}(n-2)!(s-2n+1)!}{(s-3)!} \, , \qquad
I_{2}(n) = \binom{s-4}{2n-4}(2n-5)!! \, ,
\ee
and \eqref{lambda-Pi} allows one to finally derive \eqref{Qgen-chi2}, again taking into account the trace constraints in \eqref{cons-curr-s} and \eqref{conf-kill}.

\end{subappendices}

\chapter{Fermi fields}

\label{Chap:Fermi}

We are now going to extend the results of the previous chapter to fermions. In a completely similar fashion, we first need to cast the Fang-Fronsdal action in Hamiltonian form. Given that the action is of first order in the derivatives, this step essentially requires to distinguish the dynamical variables from the Lagrange multipliers enforcing the first-class constraints that generate the gauge symmetry. In flat space this analysis has been performed in \cite{Aragone:1979hw} for fields of arbitrary half-integer spins and revisited in the spin-$5/2$ case in \cite{Borde:1981gh} and more recently in \cite{Bunster:2014fca} (see our chapter \textbf{\ref{Chap:hambos}}). Hamiltonian actions involving a different field content --~inspired by the frame formulation of general relativity~-- have also been considered for higher-spin fermions in both flat \cite{Aragone:1980rk} and (A)dS backgrounds \cite{Hframe}. Here we extend the presentation of the spin-$5/2$ action in \cite{Aragone:1979hw,Bunster:2014fca} to AdS and (partly) to higher spins.

Once the constraints have been determined, we build the canonical generators of gauge transformations, which contain boundary terms. These boundary terms are non-vanishing in the case of ``improper gauge transformations''  \cite{benguria-cordero} -- i.e., transformations that take the form of gauge transformations but produce a non-trivial effect on the physical system because they do not go to zero fast enough at infinity -- and are identified with the higher-spin surface charges. Improper gauge transformations are determined by boundary conformal Killing spinor-tensors (precisely defined in the text by definite equations), up to proper gauge transformations. Hence, to each conformal Killing spinor-tensor of the boundary is associated a well-defined higher-spin surface charge.

To illustrate the logic of the procedure, in section \textbf{\ref{sec:example}} we first detail the rewriting in canonical form of the AdS Fang-Fronsdal action for a spin-5/2 Dirac field. We then provide boundary conditions on the dynamical variables and on the parameters of the gauge transformations preserving them. We finally use this information to evaluate the charges at spatial infinity and we conclude by discussing the peculiarities of the three-dimensional case. In section \textbf{\ref{sec:arbitrary}} we move to arbitrary half-integer spin: first of all we observe that, for the sake of computing charges, it is sufficient to know the form of the constraints in flat space. We therefore bypass a detailed Hamiltonian analysis of the AdS theory and we build surface charges from flat-space constraints. We then present boundary conditions on the dynamical variables inspired by the behaviour at spatial infinity of the solutions of the free equations of motion (recalled in Appendix~\textbf{\ref{app:boundary_fermi}}) and we verify that they guarantee finiteness of the charges. We conclude with a summary of our results and with a number of appendices. Appendix \textbf{\ref{app:conventions_fermi}} recalls our conventions while Appendix~\textbf{\ref{app:conformal-killing}} discusses the conformal Killing spinor-tensors which play a crucial role in the study of charges and asymptotic symmetries.

\section{Spin-5/2 example}\label{sec:example}

To apply the techniques developed within the canonical formalism to compute surface charges, we begin by rewriting the AdS Fang-Fronsdal action in a manifestly Hamiltonian form. The charges are then identified with the boundary terms that enter the canonical generator of improper gauge transformations. Finally, we propose boundary conditions on the dynamical variables and on the deformation parameters that give finite asymptotic charges.

\subsection{Hamiltonian and constraints}\label{sec:example-H}

Our starting point is the Fang-Fronsdal action\footnote{Actions for spin-$5/2$ gauge fields in four dimensional Minkowski space have also been presented independently in \cite{Schwinger:1970xc,Berends:1979wu}.} for a massless spin-5/2 Dirac field on  AdS$_d$ \cite{fronsdal-AdS_fermi}, described by a complex-valued spinor-tensor $\psi_{\m\n}^\a$ which is symmetric in its base-manifold indices $\m, \n$ (see also e.g.\ the review \cite{Francia:2002aa} and our section \textbf{\ref{Sec:fermi_lag_AdS}}):\footnote{The sign in front of the mass-like term is conventional. One can change it provided one also changes the sign of the $L^{-1}$ terms in the gauge transformation \eqref{cov-var}, consistently with the option to send \mbox{$\g^\m \to - \g^\m$}. See section \textbf{\ref{sec:3d}} for a discussion of the effect of this transformation (when $d=3$).}
\begin{align}
S & = -
i \!\int\! d^d x \sqrt{-\bar{g}} \left\{ \bar{\psi}^{\mu\nu} \slashed{D} \psi_{\mu\nu}
+ 2\, \bar{\psi}_{\m\n} \gamma^\n \gamma^\l \gamma^\r D_\l \psi_{\r}{}^\m - \frac{1}{2} \, \bar{\psi}\slashed{D}\psi 
+ \bar{\psi}_{\m\n} \g^\m D^\n \psi + \bar{\psi} D\!\cdot \slashed{\psi} \right. \nn \\
& - 2 \left( \bar{\psi}^{\m\n} D_{\m} \slashed{\psi}{}_\n + \bar{\psi}_{\m\n} \g^\m D\!\cdot \psi^\n \right) + \left. \frac{d}{2L} \left( \bar{\psi}^{\m\n} \psi_{\m\n} - 2\, \bar{\psi}^{\m}{}_{\n} \g^\n \g^\r \psi_{\m\r} - \frac{1}{2}\, \bar{\psi} \psi \right) \right\} . \label{action_lag_5_2_charge}
\end{align}
Spinor indices will always be omitted, while $D$ stands for the AdS covariant derivative \eqref{cov-fermi}, $L$ is the AdS radius,\footnote{All results of this subsection apply also to de Sitter provided that one maps $L \to iL$.} slashed symbols denote contractions with $\g^\m$ and omitted indices signal a trace, so that e.g.\ $\slashed{\psi}{}_\m = \g^\n \psi_{\m\n}$ and $\psi = g^{\m\n} \psi_{\m\n}$. In the previous formulae we employed ``curved'' $\g$ matrices, which are related to ``flat'' ones as $\g^\m = e^\m{}_{\!A} \hat{\g}^A$, where $e^\m{}_{\!A}$ is the inverse vielbein. Our conventions are also detailed in Appendix \textbf{\ref{app:conventions_fermi}}. The action \eqref{action_lag_5_2_charge} is invariant under the gauge transformations
\be \label{cov-var}
\d \psi_{\m\n} = 2 \left( D_{(\m} \e_{\n)} + \frac{1}{2L}\, \g_{(\m} \e_{\n)} \right)
\ee
generated by a $\g$-traceless spinor-vector. In \eqref{cov-var} and in the following, parentheses denote a symmetrisation of the indices they enclose and dividing by the number of terms in the sum is understood.

Being of first order, the action \eqref{action_lag_5_2_charge} is almost already in canonical form. However, one would like to distinguish the actual phase-space variables from the Lagrange multipliers that enforce the first-class constraints associated to the gauge symmetry \eqref{cov-var}. In flat space the rewriting in canonical form of the Fang-Fronsdal action for a spin-5/2 Majorana field in $d = 4$ has been presented in \cite{Aragone:1979hw, Borde:1981gh, Bunster:2014fca} (see our chapter \textbf{\ref{Chap:hygra}}); to extend it to Dirac fields on Anti de Sitter backgrounds of generic dimension, we parameterise the AdS metric with the static coordinates
\be \label{AdS}
ds^2 = - f^2(x^k) dt^2 + g_{ij}(x^k) dx^i dx^j \, .
\ee 
We also choose the local frame such that the non-vanishing components of the vielbein and the spin connection are
\be \label{local-frame}
e^0 = f dt \, , \quad e^i = e_j{}^i dx^j \, , \quad \o^{0i} = e^{ij} \pr_j f dt \, .
\ee
To separate dynamical variables and Lagrange multipliers within the components of $\psi_{\m\n}$, we recall that time derivatives of the gauge parameter can only appear in the gauge variations of Lagrange multipliers (see \textit{e.g.} \S~3.2.2 of \cite{Henneaux:1994pup}, and our section \textbf{\ref{Sec:1st_order_gen}}). This criterion allows one to identify the dynamical variables with the spatial components $\psi_{ij}$ of the covariant field and with the combination
\be \label{def-xi}
\Xi = f^{-2} \psi_{00} - 2\, \g^0 \g^i \psi_{0i} \, .
\ee
The remaining components of $\psi_{\m\n}$ play instead the role of Lagrange multipliers. The covariant gauge variation \eqref{cov-var} breaks indeed into
\begin{subequations} \label{var-can}
\begin{align}
\d \psi_{ij} & = 2 \left( \nabla_{\!(i} \e_{j)} + \frac{1}{2L}\, \g_{(i} \e_{j)} \right) , \label{var-psi} \\
\d \Xi & = -\, 2\, \slashed{\nabla} \slashed{\e} + \frac{d+1}{L}\, \slashed{\e} \, , \label{var-xi} \\
\d \psi_{0i} & = \dot{\e}_i + f^2 \g^0 \left( \nabla_{\!i} \slashed{\e} - \frac{1}{2}\, \slashed{\G} \e_i - \G_i\, \slashed{\e} - \frac{1}{2L} \left( \e_i + \g_i \slashed{\e} \right) \right) ,
\end{align}
\end{subequations}
where we remark that Latin indices are restricted to spatial directions, while $\nabla$ is the covariant derivative for the spatial metric $g_{ij}$ and $\G_i$ denotes the Christoffel symbol $\G^0_{0i}$. The latter depends on $g_{00}$ as
\be
\G_i = f^{-1} \pr_i f \, .
\ee 
Moreover, from now on slashed symbols denote contractions involving only spatial indices, e.g.\ $\slashed{\e} = \g^i \e_i$.
The cancellation of time derivatives in \eqref{var-xi} follows from Fronsdal's constraint on the gauge parameter:\footnote{The rewriting of the gauge variations in \eqref{var-can} also relies on the identities $\o_t{}^{0k} \hat{\g}_k = f\, \G_k \g^k$ and \mbox{$\G^k_{00} = f^2 g^{kl} \G_l$}, that hold thanks to \eqref{AdS} and \eqref{local-frame}, while $\G^0_{00} = \G^0_{kl} = \G^k_{0l} = 0$.}
\be \label{gamma-constr}
\g^\m \e_\m = 0 \quad \Rightarrow \quad \e_0 = f^2 \g^0 \g^i \e_i \, .
\ee

The previous splitting of the fields is confirmed by the option to cast the action \eqref{action_lag_5_2_charge} in the following canonical form:  
\be \label{can-action_5-2}
S = \!\int\! d^dx \left\{ \frac{1}{2}\! \left( \Psi^\dagger{}_{\!\!\!A}\, \o^{AB} \dot{\Psi}_B - \dot{\Psi}^\dagger{}_{\!\!\!A}\, \o^{AB} \Psi_B \right) - \psi^\dagger_{0k}\, \cF^k[\Psi] - \cF^\dagger_k[\Psi^\dagger]\, \psi_{0}{}^k - \cH[\Psi,\Psi^\dagger] \right\} ,
\ee
where we collected the phase-space variables by defining
\be \label{def_symplectic}
\Psi_A = 
\begin{pmatrix}
\psi_{kl} \\
\Xi
\end{pmatrix} , 
\qquad
\o^{AB} = 
\begin{pmatrix}
\o^{kl|mn} & \o^{kl|\bullet} \\
\o^{\bullet|mn} & \o^{\bullet|\bullet}
\end{pmatrix} .
\ee
The kinetic term is specified by the symplectic 2-form\footnote{With respect to section \textbf{\ref{Sec:1st_order_gen}} -- which recalls some general facts about first-order Grassmanian actions like \eqref{can-action_5-2} -- the symplectic 2-form has here implicit spinor indices and should incorporate a spatial delta function: $\O^{AB}\left(\vec{x},\vec{x}'\right) = \o^{AB} \d \left( \vec{x} - \vec{x}'\right)$.} $\o^{AB}$ with components 
\begin{subequations} \label{2_form_5_2}
\begin{align}
\o^{kl|mn} & = i\, \sqrt{g} \left( g^{k(m}g^{n)l} - 2\, \g^{(k} g^{l)(m} \g^{n)} - \frac{1}{2}\, g^{kl} g^{mn} \right) ,  \\
\o^{kl|\bullet} & = \o^{\bullet|kl} = -\, \frac{i}{2}\, \sqrt{g}\, g^{kl} \, , \\
\o^{\bullet|\bullet} & = \frac{i}{2}\,\sqrt{g} \, , 
\end{align}
\end{subequations}
where $\sqrt{g}$ involves only the determinant of the spatial metric. The symplectic 2-form satisfies $\o^{AB} = - \left( \o^{BA} \right)^{\dagger}$ and its inverse $\o_{AB}$ --~which enters the definition of the Dirac brackets given below~-- reads 
\begin{subequations} \label{inverse-omega}
\begin{align}
\o_{kl|mn} & = \frac{i}{\sqrt{g}} \left( -\, g_{k(m}g_{n)l} + \frac{2}{d}\, \g_{(k} g_{l)(m} \g_{n)} + \frac{1}{d}\, g_{kl} g_{mn} \right) , \\
\o_{kl|\bullet} & = \o_{\bullet|kl} = \frac{i}{d\sqrt{g}}\, g_{kl} \, , \\
\o_{\bullet|\bullet} & = -\, \frac{i}{\sqrt{g}}\, \frac{d+1}{d} \, .
\end{align}
\end{subequations}
The constraints enforced by the Lagrange multipliers $\psi_{0k}$ are instead
\be \label{constr}
\begin{split}
\cF_k = i\,\sqrt{g}\, \Big\{& 2 \left( \nd \psi_k - \g_k \nd \slashed{\psi} - \slashed{\nabla} \slashed{\psi}{}_k \right) - \nabla_{\!k} \psi + \g_k \slashed{\nabla} \psi - \nabla_{\!k} \X - \g_k \slashed{\nabla} \X \\
& + \frac{d}{2L} \left( 2\, \slashed{\psi}{}_k + \g_k \psi - \g_k \X \right) \Big\} \, ,
\end{split}
\ee
while the Hamiltonian reads 
\be \label{H}
\begin{split}
\cH & = i\,f\sqrt{g} \left\{ \left( \bar{\psi}_{kl} \slashed{\nabla} \psi^{kl} + \frac{1}{2}\, \G_m \bar{\psi}_{kl} \g^m \psi^{kl} \right) - \frac{3}{2} \left( \bar{\X} \slashed{\nabla} \X + \frac{1}{2}\, \G_k  \bar{\X} \g^k \X \right) \right. \\
& + 2 \left( \bar{\psi}_{kl} \g^k \g^m \g^n \nabla_{\!m} \psi_{n}{}^l + \frac{1}{2}\, \G_m \bar{\psi}_{kl} \g^k \g^m \g^n \psi_n{}^l \right) - \frac{1}{2} \left( \bar{\psi} \slashed{\nabla} \psi + \frac{1}{2}\, \G_k \bar{\psi} \g^k \psi \right) \\
& - 2 \left( \bar{\psi}^{kl} \nabla_{\!k} \slashed{\psi}{}_l + \bar{\psi}_{kl} \g^k \nd \psi^l + \G_m \bar{\psi}_{kl} \g^k \psi^{lm} \right) + \left( \bar{\psi}_{kl} \g^k \nabla^l \psi + \bar{\psi}\, \nd \slashed{\psi} + \G_k \bar{\psi} \slashed{\psi}{}^k \right) \\ 
& - \left( \bar{\X}\, \nd \slashed{\psi} + \bar{\psi}_{kl} \g^k \nabla^l \X + \G_k \bar{\X} \slashed{\psi}{}^k \right) + \frac{1}{2} \left( \bar{\X} \slashed{\nabla} \psi + \bar{\psi} \slashed{\nabla} \Xi + \G_k \bar{\Xi} \g^k \psi \right) \\ 
& \left. -\, \frac{3}{4}\, \G_k \left( \bar{\Xi} \g^k \psi - \bar{\psi} \g^k \X \right) + \frac{d}{4L} \left( 2\, \bar{\psi}_{kl} \psi^{kl} - 4\, \bar{\slashed{\psi}}{}_k \slashed{\psi}{}^k - \bar{\psi} \psi - 3\, \bar{\X} \X + \bar{\X} \psi + \bar{\psi} \X \right) \right\} .
\end{split}
\ee
Note that integrating by parts within $\cH$ generates contributions in $\G_k$ due to the overall dependence on $f(x^k)$. The terms collected within each couple of parentheses in \eqref{H} give an hermitian contribution to the action thanks to this mechanism. 

Following the steps outlined in section \textbf{\ref{Sec:1st_order_gen}} (to which we refer for more details), the knowledge of the symplectic 2-form allows to derive the Dirac brackets between fields:
\begin{subequations}\label{Dirac_5_2}
\begin{align}
\{ \psi^{\phantom{\dagger}}_{kl}(\vec{x}) , \psi^\dagger_{mn}(\vec{x}^{\,\pe}) \}_D & = \frac{i}{\sqrt{g}} \left( -\,g_{k(m} g_{n)l} + \frac{2}{d}\, \g_{(k} g_{l)(m} \g_{m)} + \frac{1}{d}\, g_{kl} g_{mn} \right) \d(\vec{x} - \vec{x}^{\,\pe}) \, , \\
\{ \psi^{\phantom{\dagger}}_{kl}(\vec{x}) , \Xi^\dagger(\vec{x}^{\,\pe}) \}_D & = \{ \Xi(\vec{x}) ,  \psi^\dagger_{kl}(\vec{x}^{\,\pe}) \}_D =    \frac{i}{d \sqrt{g}}\, g_{kl}\, \d(\vec{x} - \vec{x}^{\,\pe}) \, , \\
\{ \Xi(\vec{x}) , \Xi^\dagger(\vec{x}^{\,\pe})  \}_D & = -\,\frac{i}{\sqrt{g}}\, \frac{d+1}{d}\, \d(\vec{x} - \vec{x}^{\,\pe}) \, .
\end{align}
\end{subequations}
These are the same expressions as in flat space (compare e.g.\ with sect.~3 of \cite{Bunster:2014fca} or with our Chapter \textbf{\ref{Chap:hygra}}). This was to be expected since the AdS action differs from the Minkowski one only through its mass term and its covariant derivatives (which are modified only by the addition of algebraic terms), neither of which modifies the kinetic term, containing one time derivative.

\subsection{Gauge transformations}\label{sec:example-gauge}

The action \eqref{action_lag_5_2_charge} is invariant under \eqref{cov-var} for a $\g$-traceless $\epsilon_{\m}$ and this induces the variations \eqref{var-psi} and \eqref{var-xi} for the variables which are dynamically relevant. The constraints \eqref{constr} that we have just obtained are of first class on AdS (see \eqref{1st-class} below), and they generate these gauge transformations through their Dirac brackets with the fields.

The canonical generator of gauge transformations is indeed 
\begin{equation} \label{generator_5_2}
\mathcal{G} [ \lambda^k , \lambda^{\dagger\, l} ] = 
\int d^{d-1} x \left( \lambda^{\dagger\, k} \mathcal{F}_k + \mathcal{F}^{\dagger}{}_{\!k}\, \lambda^k \right) + Q[ \lambda^k , \lambda^{\dagger\, l} ] \, ,
\end{equation}
where $Q$ is the boundary term one has to add in order that $\mathcal{G}$ admit well defined functional derivatives, i.e.\ that its variation be again a bulk integral \cite{benguria-cordero}:
\begin{equation} \label{deltaG_fermi}
\d \mathcal{G} = \int d^{d-1} x \left(\d \psi^{\dagger}{}_{\!\!kl}  A^{kl} + \d \Xi^{\dagger} B  + A^{\dagger kl} \d \psi_{kl} + B^{\dagger} \d \Xi \right) .
\end{equation}
The gauge variations of the dynamical variables are recovered from the Dirac brackets with the constraint, including its surface addition, as follows:
\begin{subequations} \label{var-bracket}
\begin{align}
\d \psi_{kl} &=
\left\lbrace \psi_{kl} , \mathcal{G} \right\rbrace_D = \omega_{kl\vert mn} A^{mn}  + \omega_{kl\vert \bullet} B \,, \\[5pt]
\d \Xi &= 
\left\lbrace \Xi , \mathcal{G} \right\rbrace_D = \omega_{\bullet\vert kl} A^{kl} + \omega_{\bullet\vert \bullet} B \,,
\end{align}
\end{subequations}
where the $\o_{AB}$ are the components of the inverse of the symplectic 2-form, given in \eqref{inverse-omega}.
Inserting into (\ref{generator_5_2}) the constraints (\ref{constr}), one obtains
\begin{align}
A^{kl} &= i  \sqrt{g} \left\lbrace
2 \left(
\nabla^{(k} \lambda^{l)} - \g^{(k} \nabla^{l)} \slashed{\lambda} - \g^{(k} \slashed{\nabla} \lambda^{l)}
\right) 
+ g^{kl}\! \left( \slashed{\nabla} \slashed{\lambda} - \nabla\!\cdot\!\lambda \right) - \frac{d}{2L}\! \left( 2\, \g^{(k} \lambda^{l)} + g^{kl} \slashed{\lambda} \right)
\right\rbrace , \nn \\
B &= i  \sqrt{g} \left\lbrace - \nabla \cdot \lambda - \slashed{\nabla} \slashed{\lambda} + \frac{d}{2L} \slashed{\lambda} \right\rbrace .
\end{align}
Substituting the values of $\o_{AB}$ from \eqref{inverse-omega}, one gets back the gauge transformations (\ref{var-can}) (for $\psi_{kl}$ and $\Xi$) with the identification $\lambda^k = \epsilon^k$. 

The variations \eqref{var-bracket} leave the constraints and the Hamiltonian invariant up to the constraints themselves:
\be \label{1st-class}
\d \cF_i = 0 \, , \qquad\qquad
\d \cH = - \left( \d \psi^\dagger_{0k} - \dot{\e}^\dagger_k\right) \cF^k - \cF^{\dagger\, k}_{\phantom{0k}} 
\left( \d \psi^{\phantom{\dagger}}_{0k} - \dot{\e}^{\phantom{\dagger}}_k \right) \, .
\ee
On the one hand, when combined with the variation of the kinetic term, these relations just reflect the gauge invariance of the Fang-Fronsdal action \eqref{action_lag_5_2_charge} on AdS.\footnote{The variation of the constraints vanishes provided that the spatial metric be of constant curvature. To reproduce the variation of $\cH$ in \eqref{1st-class} one has instead to impose that the full space-time metric be of constant curvature.} On the other hand, given the link between gauge transformations and Dirac brackets recalled above, they also imply that both the constraints and the Hamiltonian are of first class and that there are no secondary constraints. This is confirmed by the associated counting of local degrees of freedom (see e.g.\ \S~1.4.2 of \cite{Henneaux:1994pup} or our chapter \textbf{\ref{Sec:Ham_form_constr}}):
\be
\#\, \textrm{d.o.f.} = 2^{\left[\frac{d}{2}\right]} \bigg(\! \underbrace{\frac{(d-1)d}{2} + 1}_{\textrm{dynamical variables}} -\ \,2 \!\!\!\!\!\!\underbrace{(d-1)\!\!\!\!\phantom{\frac{1}{2}}}_{\textrm{1st class constr.}} \!\!\!\!\bigg) = 2^{\left[\frac{d-2}{2}\right]} (d-3)(d-2) \, .
\ee
In $d=4$ the right-hand side is equal to four as expected, and in arbitrary $d$ it reproduces the number of degrees of freedom of a spin-5/2 Dirac fermion (compare e.g.\ with \cite{modave1}).

The boundary term $Q[\lambda^k , \lambda^{\dagger\, l}]$ will be of crucial importance in the following, since it gives the asymptotic charges. Its variation has to cancel the boundary terms generated by the integrations by parts putting the variation of $\cG[\lambda^k , \lambda^{\dagger\, l}]$ in the form \eqref{deltaG_fermi}. Being linear in the fields, these variations are integrable and yield:\footnote{Here $d^{d-2}S_k \equiv d^{d-2} x\, \hat{n}_k$, where $\hat{n}_k$ and $d^{d-2} x$ are respectively the normal and the product of differentials of the coordinates on the $d-2$ sphere at infinity (e.g.\ $d^{2} x = d\th d\phi$ for $d=4$).}
\be \label{charge_5-2}
\begin{split}
Q[\lambda^k , \lambda^{\dagger\, l}] = -
 i \int d^{d-2} S_k \sqrt{g}\,  \Big\lbrace
& 2\,  \lambda^{\dagger j} \psi_j{}^k - 2
\, \lambda^{\dagger}{}_{\!\!j} \g^j \g^l 
 \psi_l{}^k - 2\, \lambda^{\dagger j} \g^k \g^l \psi_{jl} 
- \lambda^{\dagger k} \psi \\  
& + \lambda^{\dagger}{}_{\!\!j} \g^j \g^k \psi
- \lambda^{\dagger k}\, \X - \lambda^{\dagger}{}_{\!\!j} \g^j \g^k\, \X  
\,\Big\rbrace + \textrm{h.c.}
\end{split}
\ee
In the definition of $Q$, we also adjusted the integration constant so that the charge vanishes for the zero solution. Note that, since the constraint \eqref{constr} contains a single derivative, the expression above for the boundary term on AdS is the same as that one we would have obtained in Minkowski. For clarity, in this example we displayed the complete Hamiltonian form of the AdS Fang-Fronsdal action; still knowledge of the constraints in flat space suffices to compute charges. 
In section \textbf{\ref{sec:arbitrary}} we shall follow this shortcut when dealing with arbitrary half-integer spins.

\subsection{Boundary conditions}\label{sec:example-bnd}

The previous considerations remain a bit formal in the sense that the surface integrals (\ref{charge_5-2}) might diverge.  This is where boundary conditions become relevant.  In fact, for generic theories, the problem of cancelling the unwanted surface terms that appear in the variation of the Hamiltonian and the problem of defining boundary conditions are entangled and must be considered simultaneously, because it is only for some appropriate boundary conditions that the requested charges are integrable and that one can perform the cancellation. The reason why we got above (formal) integrability of the charges without having to discuss boundary conditions is that the constraints are linear.  One can then construct formal expressions for the charges first since integrability is automatic.

To go beyond this somewhat formal level and  to evaluate the asymptotic charges, however, we have to set boundary conditions on the dynamical variables. In analogy with the strategy we employed for Bose fields in the previous chapter, we propose to use as boundary conditions the falloffs at spatial infinity of the solutions of the linearised field equations in a convenient gauge. We check in section \textbf{\ref{sec:example-charges}} that these conditions make the charges finite.

In Appendix~\textbf{\ref{app:boundary_fermi}} we recall the behaviour at the boundary of AdS of the solutions of the Fang-Fronsdal equations of motion; in spite of being of first order, these equations admit two branches of solutions, related to different projections that one can impose asymptotically on the fields.\footnote{The existence of two branches of solutions associated to different projections on the boundary values of the fields is not a peculiarity of higher spins. For spin-$3/2$ fields on AdS it has been noticed already in \cite{HT}, while for spin-$1/2$ fields it has been discussed e.g.\ in \cite{AdS/CFT-spinors,boundary-dirac}.} In a coordinate system in which the AdS metric reads
\be \label{poincare}
ds^2 = \frac{dr^2}{r^2} + r^2\, \h_{IJ} dx^I dx^J \, ,
\ee
the solutions in the \emph{subleading branch} behave at spatial infinity ($r \to \infty$) as
\begin{subequations} \label{fall-off_5-2}
\begin{align}
\psi_{IJ} & = r^{\frac{5}{2}-d}\, \cQ_{IJ}(x^K) + \cO(r^{\frac{3}{2}-d}) \, , \\[5pt]
\psi_{rI} & = \cO(r^{-d-\frac{1}{2}}) \, , \\[5pt]
\psi_{rr} & = \cO(r^{-d-\frac{7}{2}}) \, .
\end{align}
\end{subequations}
We remark that capital Latin indices denote all directions which are transverse to the radial one (including time) and that here and in the following we set the AdS radius to $L = 1$. The field equations further impose that $\cQ_{IJ}$ satisfies the following conditions:
\begin{subequations} \label{bnd-constr_5-2}
\begin{align}
\pr^J \cQ_{IJ} = \hat{\g}^J \cQ_{IJ} = 0 \, , \label{div-gamma} \\[5pt]
\left( 1 + \hat{\g}^r \right) \cQ_{IJ} = 0 \, , \label{proj-Q}
\end{align}
\end{subequations}
where a hat indicates ``flat'' $\g$ matrices, that do not depend on the point where the expressions are evaluated. For instance, $\hat{\g}^r = \delta^r{}_{\!A} \hat{\g}^A$. Eqs.~\eqref{fall-off_5-2} and \eqref{bnd-constr_5-2} define our boundary conditions.

In the case of spin $3/2$ included in the discussion of section \textbf{\ref{sec:bnd}}, the boundary conditions dictated by the subleading solution of the linearised e.o.m.\ agree with those considered for $\cN = 1$ AdS supergravity in four dimensions \cite{HT}.\footnote{This is actually true up to a partial gauge fixing allowing one to match the falloffs of the radial component. We refer to section \textbf{\ref{sec:bnd}} for more details.} This theory is known in closed form, and finiteness of the charges and consistency have been completely checked. Moreover, the agreement in the spin-3/2 sector extends a similar matching between the subleading falloffs of linearised solutions and the boundary conditions generally considered in literature for gravity (see the previous chapter). These are our main motivations to adopt the boundary conditions defined by subleading linearised solutions for arbitrary values of the spin. See also section \textbf{\ref{sec:3d}}, where we show how the conditions above allow one to match results previously obtained in the Chern-Simons formulation of three-dimensional higher-spin gauge theories \cite{susy_AdS-CFT_1,3d-fermi-asymptotics_1,3d-fermi-asymptotics_2,finite-superalgebras_1,finite-superalgebras_2,3d-fermi-asymptotics_3}.

Since the action is of first order, the constraints \eqref{constr} only depend on the dynamical variables $\psi_{ij}$ and $\Xi$, that somehow play both the role of coordinates and momenta. The boundary conditions \eqref{fall-off_5-2} and \eqref{bnd-constr_5-2}, obtained from the covariant field equations, can therefore be easily converted in boundary conditions on the canonical variables:
\begin{subequations} \label{Hboundary_5-2}
\begin{alignat}{5}
\psi_{\a\b} & = r^{\frac{5}{2}-d} \cQ_{\a\b} + \cO(r^{\frac{3}{2}-d})  \, , \qquad 
& \Xi & = -\, r^{\frac{1}{2}-d} \cQ_{00} + \cO(r^{-d-\frac{1}{2}}) \, , \\[5pt]
\psi_{r\a} & = \cO(r^{-d-\frac{1}{2}}) \, , \qquad 
& \psi_{rr} & = \cO(r^{-d-\frac{7}{2}}) \, .
\end{alignat}
\end{subequations}
In the formulae above we displayed explicitly only the terms which contribute to the charges (see section \textbf{\ref{sec:example-charges}}) and we used Greek letters from the beginning of the alphabet to indicate the coordinates that parameterise the $d-2$ sphere at infinity. Furthermore, we used the $\g$-trace constraint \eqref{div-gamma} to fix the boundary value of $\Xi$.

\subsection{Asymptotic symmetries}\label{sec:example-symm}

In order to specify the deformation parameters that enter the charge \eqref{charge_5-2}, we now identify all gauge transformations preserving the boundary conditions of section \textbf{\ref{sec:example-bnd}}. We begin by selecting covariant gauge transformations compatible with the fall-off conditions \eqref{fall-off_5-2}, and then we translate the result in the canonical language. 

Asymptotic symmetries contain at least gauge transformations leaving the vacuum solution $\psi_{\m\n} = 0$ invariant. These are generated by $\g$-traceless Killing spinor-tensors of the AdS background, which satisfy the conditions
\be \label{killing-ads}
D_{(\m} \e_{\n)} + \frac{1}{2}\, \g_{(\m} \e_{\n)} = 0 \, , \qquad\qquad
\g^\m \e_\m = 0 \, ,
\ee
and generalise the Killing spinors that are considered in supergravity theories (see \cite{AdS-killing-spinors} for a discussion of the Killing spinors of AdS$_d$ along the lines we shall follow for higher spins). We are not aware of any classification of $\g$-traceless Killing spinor-tensors of AdS spaces of arbitrary dimension, but they have been discussed for $d = 4$ \cite{ABJ-HS} and they are expected to be in one-to-one correspondence with the generators of the higher-spin superalgebras classified in \cite{HSsuperalgebras}. In section \textbf{\ref{app:conformal-killing}} we shall also show that the number of independent solutions of \eqref{killing-ads} is the same as that of its flat limit, whose general solution is given by \eqref{sol-flat-killing}. These arguments indicate that --~as far as the free theory is concerned~-- non-trivial asymptotic symmetries exist in any space-time dimension and we are going to classify them from scratch in the current spin-$5/2$ example. Along the way we shall observe that, when $d > 3$, asymptotic and exact Killing spinor-tensors only differ in terms that do not contribute to surface charges.

To identify the gauge transformations that preserve the boundary conditions \eqref{fall-off_5-2}, one has to analyse separately the variations of components with different numbers of radial indices. In the coordinates \eqref{poincare}, if one fixes the local frame as
\be \label{localframe-text}
e_r{}^A = - \frac{1}{r}\, \d_r{}^A \, , \qquad 
e_I{}^J = \o_I{}^{rJ} = r\, \d_I{}^J \, , \qquad
\o_r^{\m\n} = \o_I{}^{JK} = 0 \, ,
\ee 
one obtains the conditions
\begin{align}
\d \psi_{IJ} & = 2 \left( \pr_{(I} \e_{J)} + \frac{r}{2}\, \hat{\g}_{(I|} \left( 1-\hat{\g}_r \right) \e_{|J)} \right) + 2\, r^3\, \h_{IJ} \e_r = \cO(r^{\frac{5}{2}-d}) \, , \label{dpsi-ij} \\[7pt]
\d \psi_{rI} & = \frac{1}{r} \left( r\pr_r - \frac{4+\hat{\g}_r}{2} \right) \e_I + \left( \pr_I + \frac{r}{2}\, \hat{\g}_I \left( 1 - \hat{\g}_r \right) \right) \e_r = \cO(r^{-d-\frac{1}{2}}) \, , \label{dpsi-ri} \\[5pt]
\d \psi_{rr} & = \frac{2}{r} \left( r\pr_r + \frac{2-\hat{\g}_r}{2} \right) \e_r = \cO(r^{-d-\frac{7}{2}}) \, . \label{dpsi-rr}
\end{align}
Fronsdal's $\g$-trace constraint $\slashed{\e} = 0$ implies instead
\be \label{gmu-emu}
\hat{\g}^r \e_r = r^{-2}\, \hat{\g}^I \e_I \, ,
\ee
thus showing that the radial component of the gauge parameter, $\e_r$, is not independent. It is anyway convenient to start analysing the conditions above from \eqref{dpsi-rr}, which is a homogeneous equation solved by
\be \label{e-r}
\e_r = r^{-\frac{1}{2}}\, \l^{+}(x^k) + r^{-\frac{3}{2}}\, \l^{-}(x^k) + \cO(r^{-d-\frac{5}{2}})\, , 
\quad \textrm{with} \quad \hat{\g}^{r} \l^\pm = \pm\, \l^\pm \, .
\ee
Substituting in \eqref{dpsi-ri} one obtains
\be \label{e-i}
\e_I = r^{\frac{5}{2}}\, \z^+_I + r^{\frac{3}{2}}\, \z^-_I + \frac{r^{\frac{1}{2}}}{2} \left( \pr_I \l^+ + \hat{\g}_{I} \l^- \right) + \frac{r^{-\frac{1}{2}}}{2}\, \pr_I \l^- + \cO(r^{\frac{1}{2}-d}) \, , 
\ee
where the new boundary spinor-vectors that specify the solution satisfy
\be \label{proj-z}
\hat{\g}^{r} \z_I^\pm = \pm\, \z_I^\pm \, .
\ee

The gauge parameter is further constrained by \eqref{dpsi-ij} and \eqref{gmu-emu}. The latter equation implies
\be \label{g-traces}
\slashed{\z}^+ \equiv \hat{\g}^I \z^+_I = 0 \, , \qquad\
\l^+ = \slashed{\z}^- \, ,
\ee
and the differential conditions
\be \label{diff-constr}
\slashed{\pr} \l^+ = - (d+1) \l^- \, , \qquad\
\slashed{\pr} \l^- = 0 \, . 
\ee
As shown in Appendix~\textbf{\ref{app:independent}}, the relations \eqref{diff-constr} are however not independent from the constraints imposed by \eqref{dpsi-ij}, so that we can ignore them for the time being. We stress that in the equations above and in the rest of this subsection, contractions and slashed symbols only involve sums over transverse indices and flat $\g$ matrices, so that \mbox{$\slashed{\pr} \l^\pm = \hat{\g}^I \pr_I \l^\pm$}.
Eq.~\eqref{dpsi-ij} implies instead
\be \label{dpsi-ij-exp}
\begin{split}
\d \psi_{IJ} & = 2\, r^{\frac{5}{2}} \left( \pr_{(I} \z_{J)}{}^{\!\!+} + \hat{\g}_{(I} \z_{J)}{}^{\!\!-} + \h_{IJ} \l^+ \right) + 2\, r^{\frac{3}{2}} \left( \pr_{(I} \z_{J)}{}^{\!\!-} + \h_{IJ} \l^- \right) \\[5pt]
& + r^{\frac{1}{2}} \left( \pr_I\pr_J \l^+ + 2\, \hat{\g}_{(I} \pr_{J)} \l^- \right) + r^{-\frac{1}{2}}\, \pr_I\pr_J \l^- = \cO(r^{\frac{5}{2}-d}) \, .
\end{split}
\ee
The cancellation of the two leading orders requires
\be
\l^+ = - \frac{1}{d-1} \left( \prd \z^+ + \slashed{\z}^- \right) =  - \frac{1}{d}\, \prd \z^+ \, , \qquad\
\l^- = - \frac{1}{d-1}\, \prd \z^- \, ,
\ee
plus the differential conditions presented below in \eqref{conformal-killing}. In Appendix~\textbf{\ref{app:independent}} we prove that these constraints also force the cancellation of the second line in \eqref{dpsi-ij-exp} when $d > 3$. At the end of this subsection we shall instead comment on how to interpret the additional terms that one encounters when $d=3$.

To summarise: parameterising the AdS$_d$ background as in \eqref{poincare} and fixing the local frame as in \eqref{localframe-text}, linearised covariant gauge transformations preserving the boundary conditions \eqref{fall-off_5-2} are generated by 
\begin{subequations} \label{final-gauge}
\begin{align}
\e^I & = r^{\frac{1}{2}}\, \z^{+I} + r^{-\frac{1}{2}}\, \z^{-I} - \frac{r^{-\frac{3}{2}}}{2d}\! \left( \pr^I \prd \z^+ + \frac{d}{d-1}\, \hat{\g}^{I} \prd \z^- \right) - \frac{r^{-\frac{5}{2}}}{2(d-1)}\, \pr^I \prd \z^- \quad \nn \\
& + \cO(r^{-\frac{3}{2}-d}) \, , \\[5pt]
\e^r & = - \frac{r^{\frac{3}{2}}}{d}\, \prd \z^+ - \frac{r^{\frac{1}{2}}}{d-1}\, \prd \z^- + \cO(r^{-d-\frac{1}{2}}) \, , \label{xir}
\end{align}
\end{subequations}
where the spinor-vectors $\z^\pm_I$ are subjected the chirality projections \eqref{proj-z} and satisfy\footnote{These conditions also allow \eqref{final-gauge} to satisfy the $\g$-trace constraint \eqref{gmu-emu}, which is not manifest in the parameterisation of the solution we have chosen.}
\begin{subequations} \label{conformal-killing}
\begin{align}
& \pr_{(I} \z_{J)}{}^{\!\!+} - \frac{1}{d-1}\,\h_{IJ}\, \prd \z^+ = -\, \hat{\g}_{(I} \z_{J)}{}^{\!\!-} + \frac{1}{d-1}\, \h_{IJ}\, \hat{\g}\cdot \z^- \, , \label{conf2+} \\
& \pr_{(I} \z_{J)}{}^{\!\!-} - \frac{1}{d-1}\,\h_{IJ}\, \prd \z^- = 0 \, , \label{conf2-}\\[5pt]
& \hat{\g}\cdot \z^+ = 0 \, , \label{constr2+} \\[4pt]
& \hat{\g}\cdot \z^- = - \frac{1}{d}\, \prd \z^+ \, . \label{constr2-}
\end{align}
\end{subequations}
The left-hand sides of \eqref{conf2+} and \eqref{conf2-} have the same structure as the bosonic conformal Killing-vector equation in $d-1$ space-time dimensions. For this reason we call here the solutions of \eqref{conformal-killing}  ``conformal Killing spinor-vectors''. When $d > 3$ there are \mbox{$2^{\left[\frac{d-1}{2}\right]}(d-2)(d+1)$} independent solutions that we display in Appendix~\textbf{\ref{app:conformal-killing}}, while when $d = 3$ the space of solutions actually becomes infinite dimensional (see section \textbf{\ref{sec:3d}}). 

As discussed in section \textbf{\ref{sec:example-gauge}}, Fang-Fronsdal's gauge parameters coincide with the deformation parameters entering the charge \eqref{charge_5-2}. Asymptotic symmetries are therefore generated by deformation parameters behaving as
\be \label{deform_5-2}
\l^\a = r^{\frac{1}{2}}\, \z^{+\a} + \cO(r^{-\frac{1}{2}}) \, , \qquad 
\l^r = \cO(r^{\frac{3}{2}}) \, .
\ee
As in \eqref{Hboundary_5-2}, Greek letters from the beginning of the alphabet denote coordinates on the $d - 2$ sphere at infinity and we specified only the terms that contribute to surface charges.

To conclude this subsection, we remark that an infinite number of solutions is not the unique peculiarity of the three-dimensional setup: in this case one indeed obtains
\be \label{3d-var}
\d \psi_{IJ} = - \frac{r^{-\frac{1}{2}}}{2}\, \pr_I \pr_J \prd \z^- + \cO(r^{-\frac{3}{2}}) 
\ee 
even considering gauge parameters that satisfy \eqref{final-gauge} and \eqref{conformal-killing}. One can deal with this variation in two ways: if one wants to solve the Killing equation \eqref{killing-ads}, one has to impose the cancellation of $\pr_I \pr_J \prd \z^-$ and the additional condition is satisfied only on a finite dimensional subspace of the solutions of \eqref{conformal-killing}. If one is instead interested only in preserving the boundary conditions \eqref{Hboundary_5-2}, which is the only option when the background is not exact AdS space, a shift of $\psi_{IJ}$ at $\cO(r^{-\frac{1}{2}})$ is allowed. In analogy with what happens for Bose fields \cite{metric3D} (see the previous chapter), the corresponding variation of the surface charges is at the origin of the central charge that appears in the algebra of asymptotic symmetries. In $d=3$ the spinor-vector $\z^{-I}$ entering the variation \eqref{3d-var} indeed depends on the spinor-vector $\z^{+I}$ entering the charges (see section \textbf{\ref{sec:3d}}).

\subsection{Charges}\label{sec:example-charges}

Having proposed boundary conditions on both canonical variables (see \eqref{Hboundary_5-2}) and deformation parameters (see \eqref{deform_5-2}), we can finally obtain the asymptotic charges. In the coordinates \eqref{poincare}, the normal to the $d-2$ sphere at infinity is such that $\hat{n}_r = 1$ and $\hat{n}_\a = 0$. At the boundary the charge \eqref{charge_5-2} thus simplifies as
\be
\lim_{r\to\infty} Q[\lambda^k , \lambda^{\dagger\, l}]
 =  i \int d^{d-2} x \sqrt{g} \left\lbrace
2\, \lambda^{\dagger \alpha}\g^r \g^{\beta}\psi{}_{\alpha\beta}
- \lambda^{\dagger \alpha}\g_{\alpha} \g^r \left( \psi - \X  \right) \right\rbrace +\, \textrm{h.c.}
\ee
The terms which survive in the limit give a finite contribution to the charge; one can make this manifest by substituting their boundary values (where we drop the label $+$ on $\zeta^{+ I}$ to avoid confusion) so as to obtain 
\be \label{charge-cov}
Q = 2i \int d^{d-2} x  \left\lbrace
\bar{\zeta}^{I} \mathcal{Q}_{0I} + \bar{\mathcal{Q}}_{0I}\, \zeta^{I}
\right\rbrace .
\ee
This presentation of $Q$ relies on the $\g$-trace constraints on both $\mathcal{Q}_{IJ}$ and $\zeta^{I}$ and on the chirality conditions $(1+\hat{\g}^r) \cQ_{IJ} = 0$ and $(1-\hat{\g}^r) \z^I = 0$.
Remarkably, the result partly covariantises in the indices transverse to the radial direction.
The boundary charge thus obtained is manifestly conserved: it is the spatial integral of the time component of a conserved current since
\be
\cJ_I \equiv 2i\, \bar{\mathcal{Q}}_{IJ} \zeta^{ J} + \textrm{h.c.} \quad \Rightarrow \quad \prd \cJ = 2i \left( \pr^I
\bar{\cQ}_{IJ} \z^J + \bar{\cQ}_{IJ} \pr^{(I} \z^{J)} \right) + \textrm{h.c.} = 0 \, ,
\ee
where conservation holds thanks to \eqref{div-gamma} and \eqref{conf2+}. Eq.~\eqref{charge-cov} naturally extends the standard presentation of the bosonic global charges of the boundary theories entering the higher-spin realisations of the AdS/CFT correspondence (see e.g.\ sect.~2.5 of \cite{Campoleoni:2016uwr}).

\subsection{Three space-time dimensions}\label{sec:3d}

We conclude our analysis of the spin-$5/2$ example by evaluating the charge \eqref{charge-cov} in three space-time dimensions, where several peculiarities emerge and we can compare our findings with the results obtained in the Chern-Simons formulation of supergravity \cite{CS-sugra1,CS-sugra2} and higher-spin theories \cite{susy_AdS-CFT_1,3d-fermi-asymptotics_1,3d-fermi-asymptotics_2,finite-superalgebras_1,finite-superalgebras_2,3d-fermi-asymptotics_3}. 

To proceed, it is convenient to introduce the light-cone coordinates $x^\pm = t \pm \phi$ on the boundary. Two inequivalent representations of the Clifford algebra, characterised by $\hat{\g}^0 \hat{\g}^1 = \pm \hat{\g}^2$, are available in $d=3$. Since we conventionally fixed e.g.\ the relative sign in the gauge variation \eqref{cov-var}, we shall analyse them separately by choosing
\be \label{clifford}
\hat{\g}^\pm = \begin{pmatrix}
0 & 2 \\
0 & 0 
\end{pmatrix} ,
\quad
\hat{\g}^\mp = \begin{pmatrix}
0 & 0 \\
-2 & 0 
\end{pmatrix} ,
\quad
\hat{\g}^r = \begin{pmatrix}
1 & 0 \\
0 & -1 
\end{pmatrix} .
\ee
Equivalently, one could fix the representation of the Clifford algebra once for all and analyse the effects of a simultaneous sign flip in the gauge variation \eqref{cov-var} and in the chirality projections \eqref{proj-Q} and \eqref{proj-z}.  

One can exhibit the peculiar form of surface charges in $d=3$ by studying the general solutions of the conditions \eqref{bnd-constr_5-2} and \eqref{conformal-killing} on $\cQ_{IJ}$ and $\z^I$. The constraints on $\cQ_{IJ}$ are solved by
\be
\cQ_{\mp\mp} = \begin{pmatrix}
0 \\
\cQ(x^\mp)
\end{pmatrix} ,
\qquad
\cQ_{+-} = \cQ_{\pm\pm} = \begin{pmatrix}
0 \\
0
\end{pmatrix} ,
\ee
where the signs are selected according to the conventions in \eqref{clifford}. Note that the divergence constraint is satisfied by suitable left or right-moving functions as for bosons, but the interplay between the $\g$-trace constraint \eqref{div-gamma} and the chirality projection \eqref{proj-Q} forces one of the two chiral functions to vanish. Similarly, the constraints on $\z^{\pm I}$ are solved by
\be
\z^{(+)\mp} = \begin{pmatrix}
\z(x^\mp) \\
0
\end{pmatrix} ,
\qquad
\z^{(-)\mp} = \frac{1}{3}\begin{pmatrix}
0 \\
\pr_\mp \z(x^\mp)
\end{pmatrix} ,
\qquad
\z^{(+)\pm} = \z^{(-)\pm} = \begin{pmatrix}
0 \\
0
\end{pmatrix} ,
\ee
where we enclosed between parentheses the label denoting the chirality, which appears in the covariant $\z^{\pm}_I$ of section \textbf{\ref{sec:example-symm}}. 

When considering the representation of the Clifford algebra with $\hat{\g}^0 \hat{\g}^1 = \pm \hat{\g}^2$, the charge \eqref{charge-cov} therefore takes the form 
\be
Q_{d=3} = 2i \int d\phi\, \z^*(x^\mp) \cQ(x^\mp) + \textrm{c.c.} \, ,
\ee
so that a single Fang-Fronsdal field is associated to infinitely many asymptotic conserved charges, corresponding to the modes of an arbitrary function which is either left \emph{or} right-moving. On the contrary, a bosonic Fronsdal field is associated to both left \emph{and} right-moving charges \cite{metric3D,Campoleoni:2016uwr}. From the Chern-Simons perspective, the counterpart of this observation is the option to define supergravity theories with different numbers of left and right supersymmetries \cite{extended-3d-sugra}. One can similarly define higher-spin gauge theories in $AdS_3$ by considering the difference of two Chern-Simons actions based on different supergroups (modulo some constraints on the bosonic subalgebra -- see e.g.\ sect.~2.1 of \cite{Wlambda}), while models with identical left and right sectors are associated to an even numbers of Fang-Fronsdal fields. 

\section{Arbitrary spin}\label{sec:arbitrary}

We are now going to generalise the results of the previous section to a spin $s + 1/2$ Dirac field, omitting the details that are not necessary to compute surface charges. As discussed at the end of section \textbf{\ref{sec:example-gauge}}, these can be directly computed from the flat-space Hamiltonian constraints (which differ from the AdS$_d$ ones through algebraic terms) and the boundary conditions of fields and gauge parameters. We will focus on these two elements.

\subsection{Constraints and gauge transformations}\label{sec:gauge}

Our starting point is again the Fang-Fronsdal action for a massless spin $s + 1/2$ Dirac field on AdS$_d$ \cite{fronsdal-AdS_fermi}:
\be \begin{split} \label{fronsdal-action_fermi}
S = - i \int d^d x \sqrt{-\bar{g}}
& \left\{ \frac{1}{2}\,  \bar{\psi} \slashed{D} \psi
+ \frac{s}{2}\, \bar{\slashed{\psi}} \slashed{D} \slashed{\psi} - \frac{1}{4}\bin{s}{2} \bar{\psi}' \slashed{D} \psi'  
+ \bin{s}{2} \bar{\psi}' D\cdot\slashed{\psi} - s\, \bar{\psi} D \slashed{\psi} \right. \\
 & \left. +\, \frac{d+2(s-2)}{4L} \left( \bar{\psi} \psi - s\, \bar{\slashed{\psi}} \slashed{\psi} - \frac{1}{2}\bin{s}{2}\, \psi' \psi'
 \right) \right\} + \textrm{h.c.} 
\end{split}
\ee
The conventions are the same as in the spin-$5/2$ example (we will use the same static parametrisation of AdS$_d$, etc., from section \textbf{\ref{sec:example}}), except that we will now leave all indices (tensorial and spinorial) implicit. The field is a complex-valued spinor-tensor \mbox{$\psi^{\a}_{\m_1 \cdots \m_s} = \psi^{\a}_{(\m_1 \cdots \m_s)}$}, which is fully symmetric in its $s$ base-manifold indices $\m_1 , \ldots , \m_s$. It will generally be denoted by $\psi$ and its successive traces will be indicated by primes or by an exponent in brackets: $\psi^{[k]}$ is the $k$th trace, while we often use $\psi'$ to denote a single trace.

The novelty of the general case is that there is an algebraic constraint on the field in addition to that on the gauge parameter: its triple $\g$-trace is required to vanish, $\slashed{\psi}{}^{\pe} = 0$. The only algebraically independent components of the field have therefore zero, one or two tensorial indices in the time direction: $\psi_{k_1 \cdots k_s}$, $\psi_{0 k_2 \cdots k_s}$ and $\psi_{00 k_3 \cdots k_s}$.

The action \eqref{fronsdal-action_fermi} is invariant under gauge transformations
\be \label{gauge-fermi}
\d \psi = s \left( D \e + \frac{1}{2L}\, \g\, \e \right)
\ee
generated by a $\g$-traceless spinor-tensor of rank $s-1$. In \eqref{gauge-fermi} and in the following, a symmetrisation of all free indices is implicit, and dividing by the number of terms in the sum is understood.

Similarly to the spin-$5/2$ analysis, we note that the action \eqref{fronsdal-action_fermi} is almost already in canonical form. To identify the Lagrange multipliers which enforce the first-class constraints associated to the gauge symmetry \eqref{gauge-fermi}, one can again select combinations whose gauge variation contains time derivatives of the gauge parameter. This leads to identify the dynamical variables with the spatial components $\psi_{k_1 \ldots k_s}$ of the covariant field and with the combination 
\be
\Xi_{k_1 \cdots k_{s-2}} = f^{-2} \psi_{00 k_1 \cdots k_{s-2}} - 2\, \g^0 \g^j \psi_{0 j k_1 \cdots k_{s-2}} \, .
\ee
The remaining independent components of the covariant fields, $\psi_{0 k_1 \cdots k_{s-1}} \equiv N_{k_1 \cdots k_{s-1}}$, play instead the role of Lagrange multipliers. The covariant gauge variation \eqref{gauge-fermi} breaks indeed into (with all contractions and omitted free indices being from now on purely spatial):
\begin{subequations}
\begin{align} 
\d \psi &=
s \left( \nabla \epsilon + \frac{1}{2L}\, \gamma \epsilon \right) , \label{spin_s_non_cov_gauge_transf}
\\
\d \Xi &= -\, 2 \slashed{\nabla} \slashed{\epsilon} - \left(s - 2\right) \nabla \epsilon' + \frac{1}{2L} \left[\, 2 \left(d + 1 + 2 \left(s - 2\right)\right) \slashed{\epsilon} - \left(s - 2\right) \g \e' \,\right] , \label{xi_non_cov_gauge_transf}
\\
\d N &=
\dot{\e} + f^2 \g^0 \left[ \left(s - 1\right) \nabla \slashed{\e} - \frac{1}{2}\, \slashed{\G} \e - \left(s - 1\right) \G \, \slashed{\e} - \frac{1}{2L} \left( \e + \left(s - 1\right) \g \slashed{\e} \right) \right] .
\end{align}
\end{subequations}
This choice of variables is further confirmed by injecting it back into the action \eqref{fronsdal-action_fermi}, and using the Fronsdal constraint that sets the triple $\g$-trace of the field to zero, which leads to the following identities 
\begin{subequations}
\begin{align}
\psi_{0 \cdots 0 k_{2n+1} \cdots k_s} &=
f^{2n} \left[ n \, \Xi^{[n-1]}_{k_{2n+1} \cdots k_s}
+ 2n \, \g^0 \slashed{N}^{[n-1]}_{k_{2n+1} \cdots k_s}
- \left(n-1\right) \psi^{[n]}_{k_{2n+1} \cdots k_s}\right] ,
\\
\psi_{0 \cdots 0 k_{2n+2} \cdots k_s} &=
f^{2n} \left[ n \, \g^0  \slashed{\Xi}^{[n-1]}_{k_{2n+2} \cdots k_s}
 +  \left(2n + 1\right)  N^{[n]}_{k_{2n+2} \cdots k_s}
 -  n \, \g^0  \slashed{\psi}^{[n]}_{k_{2n+2} \cdots k_s}\right] .
\end{align}
\end{subequations}
This brings the action into the canonical form 
\be \label{can_action_s}
S = \!\int\! d^dx \left\{ \frac{1}{2}\! \left( \Psi^\dagger{}_{\!\!\!A}\, \o^{AB} \dot{\Psi}_B - \dot{\Psi}^\dagger{}_{\!\!\!A}\, \o^{AB} \Psi_B \right) - N^\dagger \cF[\Psi] - \cF^\dagger[\Psi^\dagger]\, N - \cH[\Psi,\Psi^\dagger] \right\} ,
\ee
where we collected the phase-space variables by defining
\be \label{def_symplectic_s}
\Psi_A = 
\begin{pmatrix}
\psi_{k_1 \ldots k_s} \\
\Xi_{k_3 \ldots k_s}
\end{pmatrix} , 
\qquad
\o^{AB} = 
\begin{pmatrix}
\o^{k_1 \cdots k_s| l_1 \cdots l_s} & \o^{k_1 \cdots k_s |\bullet\, i_3 \cdots i_s} \\
\o^{\bullet j_3 \ldots j_s | l_1 \ldots l_s} & \o^{\bullet j_3 \ldots j_s|\bullet\, i_3 \ldots i_s}
\end{pmatrix} .
\ee
We will not exhibit all terms (symplectic 2-form, Hamiltonian, etc.) of this action, but only those which are necessary to compute surface charges, that is the constraints. These have $s-1$ implicit spatial indices symmetrised with weight one and read
\be \label{F_s}
\begin{split}
\mathcal{F} & =
-i\, \frac{\sqrt{g}}{2} \sum_{n = 0}^{[s/2]} \bin{s}{2n} \bigg\lbrace
2n\, \g \,  g^{n-1} \Big[\, 
\slashed{\nabla} \Xi^{[n-1]} 
+ \left(s - 2n\right) \nabla \slashed{\Xi}^{[n-1]}
+ 2\left(n - 1\right) \nabla\!\cdot \slashed{\Xi}^{[n-2]} \\ 
& 
- \slashed{\nabla} \psi^{[n]}
+ \left(s-2n\right) \nabla \slashed{\psi}^{[n]}
+ 2n\, \nabla \cdot \slashed{\psi}^{[n-1]}
\Big] +
\left(s-2n\right)  g^{n} \Big[\,  2n\, \nabla \!\cdot \Xi^{[n-1]} \\
& +\left(s - 2n - 1\right) \nabla \Xi^{[n]}
+ 2\left(n - 1\right) \nabla\!\cdot \psi^{[n]} + \left(s-2n - 1\right) \nabla \psi^{[n+1]}
+ 2\, \slashed{\nabla} \slashed{\psi}^{[n]}
\Big]
\bigg\rbrace + \cdots \!\!\!\!
\end{split}
\ee
The dots stand for algebraic contributions coming from the mass term in the action and possible contributions in $\G_i$, which do not contribute to the surface charge.

Through their Dirac brackets (built from the inverse of the symplectic 2-form), these constraints generate the gauge transformations \eqref{spin_s_non_cov_gauge_transf} and \eqref{xi_non_cov_gauge_transf}, under which the Hamiltonian and the constraints are invariant (which confirms that they are first class). The canonical generator of gauge transformations is again
\begin{equation} \label{generator_s}
\mathcal{G} [ \lambda , \lambda^{\dagger} ] = 
\int d^{d-1} x \left( \lambda^{\dagger} \mathcal{F} + \mathcal{F}^{\dagger} \lambda \right) + Q[ \lambda , \lambda^{\dagger} ] \, ,
\end{equation}
where $Q$ is the boundary term one has to add in order that $\mathcal{G}$ admits well defined functional derivatives, i.e.\ that its variation be again a bulk integral:
\begin{equation} \label{deltaG_s}
\d \mathcal{G} = \int d^{d-1} x \left(\d \psi^{\dagger}{}  A + \d \Xi^{\dagger} B  + A^{\dagger} \d \psi + B^{\dagger} \d \Xi \right) .
\end{equation}

To compute surface charges we are only interested in $Q$ (whose expression is independent of the terms we omitted in the constraint \eqref{F_s}). Its variation has to cancel the boundary terms generated by the integrations by parts putting the variation of $\cG$ in the form \eqref{deltaG_s}. Being linear in the fields, these variations are integrable and yield (we display explicitly the index $k$ contracted with $d^{d-2}S_k$.):
\be \label{charge_s}
\begin{split}
& Q[\lambda , \lambda^{\dagger}] = \frac{i}{2} \!\int\! d^{d-2} S_k \sqrt{g} \sum_{n = 0}^{[s/2]} \bin{s}{2n}
\bigg\lbrace 
2n  \Big[\, 
\slashed{\lambda}^{\dagger [n-1]}  \g^k \left( \Xi^{[n-1]} - \psi^{[n]} \right) \\ 
& + \left(s-2n\right) \slashed{\lambda}^{\dagger [n-1] k} \left( \slashed{\psi}^{[n]} + \slashed{\Xi}^{[n-1]} \right)
+ 2\left(n - 1\right) \slashed{\lambda}^{\dagger [n-1]} \slashed{\Xi}^{[n-2]  k}  
+ 2n\, \slashed{\lambda}^{\dagger [n-1]} \slashed{\psi}^{[n-1] k}
\Big] \\ 
& +
\left(s-2n\right) \Big[  \left(s - 2n - 1\right)  \lambda^{\dagger [n]k} \left( \Xi^{[n]} + \psi^{[n+1]} \right)
+ 2n\, \lambda^{\dagger [n]}\Xi^{[n-1]k}
\\
& + 2 \left(n - 1\right)\lambda^{\dagger [n]} \psi^{[n] k} 
+ 2\, \lambda^{\dagger [n]}\g^k \slashed{\psi}^{[n]}
\Big]
\bigg\rbrace + \textrm{h.c.}  \end{split}
\ee
In the definition of $Q$, we again adjusted the integration constant so that the charge vanishes for the zero solution.

\subsection{Boundary conditions and asymptotic symmetries}\label{sec:bnd}

As in the spin-$5/2$ example, we derive boundary conditions on the dynamical variables from the falloff at spatial infinity of the solutions of the Fang-Fronsdal equations in a convenient gauge, adopting the subleading branch. As shown in Appendix \textbf{\ref{app:boundary_fermi}}, with the parameterisation \eqref{local-frame} of the local frame the relevant solutions behave at spatial infinity ($r \to \infty$) as
\begin{subequations} \label{cond_field_s}
\begin{align}
\psi_{I_1 \cdots I_s} & = r^{\frac{5}{2} - d} \cQ_{I_1 \cdots I_s}(x^M) + \cO(r^{\frac{3}{2}-d}) \, , \\[5pt]
\psi_{r \cdots r I_1 \cdots I_{s-n}} & = \cO(r^{\frac{5}{2} - d - 3n}) \, , \label{radial-psi}
\end{align}
\end{subequations}
where capital Latin indices denote directions transverse to the radial one as in section \textbf{\ref{sec:example-bnd}}. From now on we also set again $L = 1$. The boundary spinor-tensor $\cQ_{I_1 \cdots I_s}$ is fully symmetric as the Fang-Fronsdal field and satisfies
\begin{subequations} \label{bnd-constr}
\begin{align}
\prd \cQ = \slashed{\cQ} = 0 \, , \label{constr-s} \\[5pt]
\left( 1 + \hat{\g}^r \right) \cQ = 0 \, , \label{projQ-s}
\end{align}
\end{subequations}
where we omitted free transverse indices. Eqs.~\eqref{cond_field_s} and \eqref{bnd-constr} define our boundary conditions. For $s=1$ and $d=4$ the requirements on $\psi_I$ agree with those proposed for non-linear $\cN = 1$ supergravity in eq.~(V.1) of \cite{HT}. Our $\psi_r$ decays instead faster at infinity, but the leading term that we miss in \eqref{radial-psi} can be eliminated using the residual gauge freedom parameterised by the function $a(t,\th,\phi)$ in eq.~(V.5) of \cite{HT}. In conclusion, on a gravitino our boundary conditions agree with those considered in non-linear supergravity up to a partial gauge fixing that does not affect the charges.

The covariant boundary conditions \eqref{cond_field_s} fix the behaviour at spatial infinity of the dynamical variables as
\begin{subequations} \label{canonical-bnd}
\begin{align}
\psi_{\a_1 \cdots \a_s} & = r^{\frac{5}{2} - d} \cQ_{\a_1 \cdots \a_s} + \cO(r^{\frac{3}{2}-d}) \, , \\
\psi_{r \cdots r \a_1 \cdots \a_{s-n}} & = \cO(r^{\frac{5}{2} - d - 3n}) \, , \\[5pt]
\Xi_{\a_1 \cdots \a_{s-2}} & = -\, r^{\frac{1}{2}-d} \cQ_{00\a_1 \cdots \a_{s-2}} + \cO(r^{-d-\frac{1}{2}}) \, , \\
\Xi_{r \cdots r \a_1 \cdots \a_{s-n-2}} & = \cO(r^{\frac{1}{2} - d - 3n}) \, ,
\end{align}
\end{subequations}
where Greek indices from the beginning of the alphabet denote angular coordinates in the $d-2$ sphere at infinity. Moreover, we displayed only the dependence on the boundary values of the fields in the terms that actually contribute to surface charges.

The next step in the procedure we illustrated in section \textbf{\ref{sec:example}} requires to identify all gauge transformations that do not spoil the boundary conditions \eqref{canonical-bnd}. We are now going to provide necessary conditions for the preservation of the asymptotic form of the fields, which generalise those given for $s=2$ in \eqref{final-gauge} and \eqref{conformal-killing}. A proof that they are also sufficient (along the lines of the proof presented for $s=2$ in Appendix~\textbf{\ref{app:independent}}) will be given elsewhere. We stress, however, that the rank-$s$ counterparts of \eqref{final-gauge} and \eqref{conformal-killing} also characterise the exact $\g$-traceless Killing spinor-tensors of AdS$_d$, which satisfy
\be \label{killing-ads-gen}
D \e + \frac{1}{2}\, \g\, \e = 0 \, , \qquad \slashed{\e} = 0 \, . 
\ee
The general solution of these equations is provided in Appendix~\textbf{\ref{app:conformal-killing}} for $s = 2$. It shows that the number of independent solutions is the same as in the flat-space limit, where Killing spinor-tensors are easily obtained (see \eqref{killing-flat}). In the following we assume that this concurrence holds for arbitrary values of the spin and, hence, that the conditions we are going to present admit as many independent solutions as integration constants in \eqref{killing-flat}.

To characterise the gauge parameters which generate asymptotic symmetries, one has to analyse separately the variations of components with different numbers of radial indices. We continue to omit transverse indices and we denote them as
\be
\psi_n \equiv \psi_{r \cdots r I_1 \cdots I_{s-n}} \, .
\ee
Similarly, we denote by $\e_n$ the component of the gauge parameter with $n$ radial indices. With the choice \eqref{local-frame} for the local frame, the variations of the field components must satisfy
\be \label{var-psi-s}
\begin{split}
\d \psi_n & = \frac{n}{r} \left( r\pr_r - \frac{2(2s-3n+1)+ \hat{\g}^r}{2} \right) \e_{n-1} + (s-n) \left( \pr + \frac{r}{2}\, \hat{\g}\, (1 - \hat{\g}^r ) \right) \e_n \\
& + 2 \bin{s-n}{2} \, r^3 \h\, \e_{n+1} = \cO(r^{\frac{5}{2} - d - 3n}) \, .
\end{split}
\ee
The constraint $\slashed{\psi}{}^{\pe} = 0$ actually implies that one can focus only on the variations of $\psi_0$, $\psi_1$ and $\psi_2$, since all other components are not independent. One also has to consider the constraint $\slashed{\e} = 0$, which implies
\be \label{gamma-trace-symm}
\hat{\g}^r \e_{n+1} = r^{-2} \slashed{\e}{}_{n}
\ee
and shows that the only independent component of the gauge parameter is the purely transverse one, that is $\e_0 \equiv \e_{I_1 \cdots I_{s-1}}$.

The equations \eqref{var-psi-s} require\footnote{This can be shown e.g.\ by considering the redundant variation of $\psi_s \equiv \psi_{r \cdots r}$, which gives a homogeneous equation for $\e_{s-1} \equiv \e_{r \cdots r}$ as in section \textbf{\ref{sec:example-symm}}. One can then fix recursively the $r$-dependence of all other components of the gauge parameter.}
\be \label{gauge0-gen}
\e_0 = r^{2(s-1)} \left( r^{\frac{1}{2}} \z^+ + r^{-\frac{1}{2}} \z^- \right) + \sum_{k=1}^{s-1} r^{2(s-k-1)} \left(  r^{\frac{1}{2}} \a_k + r^{-\frac{1}{2}} \b_k \right) + \cO(r^{\frac{1}{2}-d})
\ee
together with the chirality projections
\be
(1\mp\hat{\g}^r) \z^\pm = 0
\ee
and similar restrictions on the subleading components: $(1-\hat{\g}^r)\a_k = (1+\hat{\g}^r)\b_k = 0$. The $\g$-trace constraint on the gauge parameter is then satisfied by
\begin{subequations}
\begin{align}
\e_{2n} = (-1)^n \sum_{k=n}^{s-n-1} r^{2(s-2n-k-1)} \left(  r^{\frac{1}{2}} \a_k^{[n]} + r^{-\frac{1}{2}} \b_k^{[n]} \right) + \cO(r^{\frac{1}{2}-d-6n}) \, , \\
\e_{2n+1} = (-1)^n \sum_{k=n}^{s-n-2} r^{2(s-2n-k)-5} \left(  r^{\frac{1}{2}} \slashed{\b}_k^{[n]} - r^{-\frac{1}{2}} \slashed{\a}_k^{[n]} \right) + \cO(r^{-d-\frac{5}{2}-6n}) \, .
\end{align}
\end{subequations}
Substituting these expressions in \eqref{var-psi-s} one obtains
\begin{subequations} \label{varpsi-gen}
\begin{align}
\d \psi_0 & = s\, r^{2(s-1)} \left\{ r^{\frac{1}{2}} \left[ \pr \z^+ + \g \z^- + (s-1)\, \h\, \slashed{\z}^{-} \right] + r^{-\frac{1}{2}} \left[ \pr \z^- - (s-1)\, \h\, \slashed{\a}{}_1 \right] \right\} \nn \\ 
& \!\!\!+ s \sum_{k=1}^{s-1} r^{2(s-k-1)}\! \left\{ r^{\frac{1}{2}}\! \left[ \pr \a_k + \g \b_k + (s-1)\, \h\, \slashed{\b}{}_k \right] + r^{-\frac{1}{2}}\! \left[ \pr \b_k - (s-1)\, \h\, \slashed{\a}_{k+1} \right] \right\} , \label{varpsi_0} \\[5pt]
\d \psi_1 & = \sum_{k=1}^{s-1} r^{2(s-k)-3}\! \left\{ r^{\frac{1}{2}}\! \left[ -2k\,\a_k + (s-1) \left( \pr \slashed{\b}{}_{k-1} - \g \slashed{\a}{}_k \right) - (s-1)(s-2)\, \h\, \a^{\pe}_k \right] \right. \nn \\
& \!\!\!\left. +\, r^{-\frac{1}{2}}\! \left[ -2k\, \b_k - (s-1) \left( \pr \slashed{\a}_k + (s-2) \h\, \b^{\,\pe}_k \right) \right] \right\} . \label{varpsi_1}
\end{align}
\end{subequations}
In complete analogy with the analysis of section \textbf{\ref{sec:example-symm}}, the cancellation of the first line in \eqref{varpsi_0} requires
\begin{subequations} \label{generic-conf-kill}
\begin{align}
\pr \z^+ - \frac{s-1}{d+2s-4}\, \h\, \prd \z^+ + \hat{\g}\, \z^- & = 0 \, , \label{conf1} \\[5pt]
\pr \z^- - \frac{s-1}{d+2s-5}\, \h\, \prd \z^- & = 0 \, , \label{conf2} \\[5pt]
\hat{\g}\cdot \z^+ & = 0 \, , \label{conf3} \\[5pt]
\hat{\g}\cdot\z^- + \frac{1}{d+2s-4}\, \prd\z^+ & = 0 \, . \label{conf4}
\end{align}
\end{subequations}
These conditions generalise the conformal Killing equations \eqref{conformal-killing} to arbitrary values of the rank. Their general solution is given below in \eqref{sol3d-killing} when $d = 3$, while for $d > 3$ it will be given elsewhere. Still, as anticipated, the detailed analysis of the $s=2$ case presented in Appendix~\textbf{\ref{app:conformal-killing}} makes us confident that the equations \eqref{generic-conf-kill} admit a number of independent solutions equal to the number of integration constants in \eqref{killing-flat}.

The subleading orders in \eqref{varpsi-gen} allow instead to fix the subleading components of the gauge parameter in terms of $\z^\pm$. For instance, one can manipulate these expressions to obtain the recursion relations presented in Appendix~\textbf{\ref{app:independent}}.
Let us stress that, in analogy with what we observed for $s = 2$, \eqref{varpsi_0} and \eqref{varpsi_1} provide a set of equations that are compatible only if one takes into account the constraints \eqref{generic-conf-kill}. We do not have yet a proof of the latter statement, but the analysis given for $s=2$ in Appendix~\textbf{\ref{app:conformal-killing}} gives strong indications that this is a robust assumption.

The deformation parameters that generate gauge transformations preserving the boundary conditions can then be related to Fang-Fronsdal's gauge parameters by comparing the Lagrangian field equations with their rewriting in \eqref{eom1o} and \eqref{gauge1o}. In particular, the Dirac brackets with the constraints can be inferred from the terms with Lagrange multipliers contained in the equations expressing the time derivatives of the dynamical variables in terms of the spatial derivatives of $\psi_{\m\n}$. This shows that the gauge parameter can be identified with the canonical deformation parameter also for arbitrary values of the spin. As a result, the latter behaves at spatial infinity as
\be \label{def-par}
\l^{\a_1 \cdots \a_{s-1}} = r^{\frac{1}{2}} \z^{+\a_1 \cdots \a_{s-1}} + \cO(r^{-\frac{1}{2}}) \, , \qquad
\l^{r \cdots r \a_1 \cdots \a_{s-n-1}} = \cO(r^{\frac{1}{2}+n}) \, ,
\ee
where $\z^+$ satisfies the conformal Killing equations \eqref{generic-conf-kill} and, as in \eqref{canonical-bnd}, Greek indices from the beginning of the alphabet denote angular coordinates in the $d-2$ sphere at infinity.

\subsection{Charges}\label{sec:charges_fermi}

Having proposed boundary conditions on both canonical variables (see \eqref{canonical-bnd}) and deformation parameters (see \eqref{def-par}), we can finally evaluate the asymptotic charges. The charge \eqref{charge_s} simplifies at the boundary as
\be \label{asymptotic-charge}
\begin{split}
\lim_{r\to\infty} Q[\lambda , \lambda^{\dagger}] =
i \!\int\! d^{d-2} x \sqrt{g}\, 
\sum_{n = 0}^{[s/2]} 
\bin{s}{2n} \Big\lbrace & n\, \slashed{\lambda}^{\dagger [n-1]} \g^r \left( \Xi^{[n-1]} - \psi^{[n]} \right) \\
& + \left(s-2n\right) \lambda^{\dagger [n]}\g^r \slashed{\psi}^{[n]} \Big\rbrace + \textrm{h.c.} \, ,
\end{split}
\ee
where all implicit indices are now purely spatial and transverse and we dropped the label $+$ on $\zeta^{+}$ to avoid confusion.

The terms which survive in the limit give a finite contribution to the charge; one can make this manifest by substituting their boundary values so as to obtain 
\be \label{charge-cov-gen}
Q = s\,i \int d^{d-2} x  \left\lbrace
\bar{\zeta}^{I_2 \ldots I_s} \mathcal{Q}_{0 I_2 \ldots I_s} + \bar{\mathcal{Q}}_{0 I_2 \ldots I_s}\, \zeta^{I_2 \ldots I_s}
\right\rbrace ,
\ee
where we used the $\g$-trace constraints on both $\mathcal{Q}$ and $\zeta$ and the chirality conditions $(1+\hat{\g}^r) \cQ = 0$ and $(1-\hat{\g}^r) \z = 0$.
The result partly covariantises in the indices transverse to the radial direction as in the spin-$5/2$ example, thus making the conservation of the charge manifest. It is indeed the spatial integral of the current \mbox{$\cJ_I = \bar{\zeta}^{K_2 \ldots K_{s}} \mathcal{Q}_{I K_2 \ldots K_s} + \textrm{h.c.}$}, which is conserved thanks to \eqref{constr-s} and \eqref{generic-conf-kill}.

In three space-time dimensions the charge \eqref{charge-cov-gen} is actually given by a left or right-moving function also for arbitrary half-integer values of the spin. As in section \textbf{\ref{sec:3d}}, one can deal with the two inequivalent representations of the Clifford algebra by choosing the $\g$ matrices as in \eqref{clifford}. The constraints on $\cQ$ are then solved by
\be
\cQ_{\mp \cdots \mp} = \begin{pmatrix}
0 \\
\cQ(x^\mp)
\end{pmatrix} ,
\qquad 
\cQ_{\pm \cdots \pm} = \cQ_{+ \cdots + - \cdots -} = \begin{pmatrix}
0 \\
0
\end{pmatrix} ,
\ee
where different signs correspond to the choices $\g^0 \g^1 = \pm \g^2$. Similarly, the conformal Killing equations \eqref{generic-conf-kill} are solved by
\begin{subequations} \label{sol3d-killing}
\begin{align}
\z^{(+)\mp \cdots \mp} & = \begin{pmatrix}
\z(x^\mp) \\
0
\end{pmatrix} ,
\qquad
\z^{(-)\mp \cdots \mp} = \frac{1}{2s-1}\begin{pmatrix}
0 \\
\pr_\mp \z(x^\mp)
\end{pmatrix} , \\
\z^{(\pm)\pm \cdots \pm} & = \z^{(\pm) + \cdots + - \cdots -} = \begin{pmatrix}
0 \\
0
\end{pmatrix} ,
\end{align}
\end{subequations}
where we enclosed again between parentheses the label denoting the chirality of the spinor-tensors. The charge \eqref{charge-cov-gen} takes therefore the form
\be
Q_{d=3} = s\, i \int d\phi\, \z^*(x^\mp) \cQ(x^\mp) + \textrm{c.c.} \, ,
\ee
which generalises the result for the spin-$5/2$ case discussed in section \textbf{\ref{sec:3d}}.

\section{Summary and further developments}\label{sec:conclusions_fermi}

We have explicitly constructed higher-spin charges for fermionic gauge fields of arbitrary spin on AdS backgrounds in any number of spacetime dimensions, extending our analogous work on bosonic gauge fields \cite{Campoleoni:2016uwr} (see the previous chapter). We have followed Hamiltonian methods. The charges appear as the surface integrals that must be added to the terms proportional to the constraints in order to make the generators of gauge transformations well-defined as phase-space generators. These integrals are finite with the boundary conditions that we have given, which crucially involve chirality-type projections generalising those of \cite{HT}. Improper gauge transformations --~associated to non-vanishing surface integrals~-- are determined by conformal Killing spinor-tensors of the boundary, and the corresponding charges take a simple, boundary-covariant expression in terms of them -- even though the intermediate computations are sometimes rather involved. While bosonic higher-spin charges have been derived also following other approaches \cite{HScharges,unfolded-charges1,unfolded-charges2}, to our knowledge our treatment provides the first presentation of fermionic higher-spin charges that applies to any number of space-time dimensions

We confined our analysis to the linearised theory, which suffices to derive the charges. In this context, however, the charges are abelian and their Dirac brackets vanish (modulo possible central extensions in $d=3$). To uncover a non abelian algebra, one must evaluate the brackets in the non-linear theory, since the bulk terms do play a role in that computation. A similar situation occurs for Yang-Mills gauge theories, where the surface terms giving the charges coincide with those of the abelian theory. The non-abelian structure appears when one computes the algebra of the charges, a step which involves the full theory (or, generically, at least the first non-linear corrections in a weak field expansion \cite{metric3D}). 

By working within the linearised theory, we have been able to associate conserved charges to any gauge field of given spin, although the spectra of interacting higher-spin theories are typically very constrained. We also worked with Dirac fields, that can be defined for any $d$, but the Majorana and/or Weyl projections that one may need to consistently switch on interactions can be easily implemented in our approach. In general, we have not found obstructions to define non-trivial higher-spin charges in any number of space-time dimensions and for arbitrary multiplicities of any value of the spin, consistently with the chance to define higher-spin algebras with fermionic generators in any dimension \cite{ABJ-HS,HSsuperalgebras}. Possible constraints could emerge from interactions, but  let us point out that once one starts considering half-integer higher spins, more exotic options than standard supersymmetry may become available. For instance, one can define higher-spin theories with increasing multiplicities for the fermionic fields without introducing any obvious pathology --~apart from difficulties in identifying a superconformal subalgebra within their algebra of asymptotic symmetries (see e.g.\ \cite{ABJ-HS,3d-fermi-asymptotics_3})~-- or try to define ``hypersymmetric'' theories (see e.g.\ \cite{Bunster:2014fca}) --~with fermionic gauge symmetries but without any gravitino at all. To analyse these phenomena it will be very interesting to combine our current results with those obtained for Bose fields \cite{Campoleoni:2016uwr}, and to analyse the effect of interactions on the algebra of surface charges, e.g.\ introducing them perturbatively in a weak field expansion (see \cite{cubic-symmetries-1,cubic-symmetries-2,metric3D} for related work restricted to bosonic models).


\begin{subappendices}

\section{Notation and conventions}\label{app:conventions_fermi}

We adopt the mostly-plus signature for the space-time metric $g_{\m\n}$ and we often distinguish among time and spatial components by breaking space-time indices as $\m = (0,i)$. Tangent-space indices are collectively denoted by capital Latin letters, but when we separate time and spatial directions we use the same letters as for the indices on the base manifold, i.e.\ $A = (0,i)$. The $\g$ matrices then satisfy
\be
\{ \hat{\g}^A , \hat{\g}^B \} = \h^{AB} \, , \quad
(\hat{\g}^0)^\dagger = -\, \hat{\g}^0 \, , \quad
(\hat{\g}^i)^\dagger = \hat{\g}^i \, ,
\ee
where the hat differentiates them from their curved counterparts involving the inverse vielbein:
\be
\g^\m = e^\m{}_{\!A}\, \hat{\g}^A \, .
\ee
For instance, with the choice \eqref{local-frame} for the local frame one has $\hat{\g}^0 = f \g^0$. Similarly, the Dirac conjugate is defined as $\bar{\psi} = \psi^\dagger \hat{\g}^0$, while the $\g$ matrices displayed explicitly in the Fronsdal action (see e.g.\ \eqref{action_lag_5_2_charge} or \eqref{fronsdal-action_fermi}) are curved ones. 

The space-time covariant derivative acts on a spin $s+1/2$ field as
\be \label{cov-fermi}
D_{\r} \psi_{\m_1 \cdots \m_s} = \pr_\r \psi_{\m_1 \cdots \m_s} + \frac{1}{8}\, \o_\r{}^{AB} [ \hat{\g}_A , \hat{\g}_B ] \psi_{\m_1 \cdots \m_s} - s\, \G^\l{}_{\r(\m_1} \psi_{\m_2 \cdots \m_s)\l} \, ,
\ee
and it satisfies $D_\m \g_\n = 0$. In the definition we omitted spinor indices as in the rest of the chapter. Moreover, indices between parentheses are meant to be symmetrised with weight one, i.e.\ one divides the symmetrised expression by the number of terms that appears in it. The spatial covariant derivative $\nabla$ is defined exactly as in \eqref{cov-fermi}, but with indices constrained to take values only along spatial directions. It also satisfies $\nabla_{\!i} \g_j = 0$.

On an (A)dS background the commutator of two covariant derivatives reads
\be
[ D_\m , D_\n ] \psi_{\r_1 \cdots \r_s} = \frac{s}{L^2} \left( g_{\n(\r_1} \psi_{\r_2 \cdots \r_s)\m} - g_{\m(\r_1} \psi_{\r_2 \cdots \r_s)\n} \right) - \frac{1}{2L^2}\, \g_{\m\n} \psi_{\r_1 \cdots \r_s} \, ,
\ee
where $\g_{\m\n} = \frac{1}{2} [ \g_\m , \g_\n ]$. This relation defines the (A)dS radius $L$ and suffices to fix the mass term in the Fronsdal action \eqref{fronsdal-action_fermi}.

When expanding tensors in components, we actually distinguish among four types of indices, depending on whether the time and/or radial coordinates are included or not. Greek letters from the middle of the alphabet include all coordinates, small Latin letters include all coordinates except $t$, capital Latin letters include all coordinates except $r$, while Greek letters from the beginning of the alphabet denote the angular coordinates on the unit $d-2$ sphere. In summary:
\begin{alignat}{3}
\m,\n,\ldots & \in \{t,r,\phi^1,\ldots,\phi^{d-2}\} \, , \qquad
& i,j,\ldots & \in \{r,\phi^1,\ldots,\phi^{d-2}\} \, , \nn \\
I,J,\ldots & \in \{t,\phi^1,\ldots,\phi^{d-2}\} \, , \qquad
& \a,\b,\ldots & \in \{\phi^1,\ldots,\phi^{d-2}\} \, .
\end{alignat}

Slashed symbols always denote a contraction with a $\g$ matrix, whose precise meaning depends on the context: the contraction may be with the full $\g^\m$ or with its spatial counterpart $\g^i$. In section \textbf{\ref{sec:example}} omitted indices denote a trace that, similarly, may result from a contraction with the full space-time metric $g_{\m\n}$ or with the spatial metric $g_{ij}$. In most of section \textbf{\ref{sec:arbitrary}} we omit instead all indices, which are always assumed to be fully symmetrised according to the conventions given above. Traces are instead denoted by an exponent between square brackets, so that, for instance,
\be \label{notation_bose}
\psi^{[n]} \equiv \psi_{\m_1 \cdots \m_{s-2n}\l_1 \cdots \l_n}{}^{\!\l_1 \cdots \l_n} \, , \qquad
D \psi \equiv D_{(\m_1} \psi_{\m_2 \cdots \m_{s+1})} \, , \qquad
\g\, \psi \equiv \g_{(\m_1} \psi_{\m_2 \cdots \m_{s+1})} \, . 
\ee
In Appendix~\textbf{\ref{app:boundary_fermi}} we reinstate indices with the following convention: repeated covariant or contravariant indices denote a symmetrisation, while a couple of identical covariant and contravariant indices denotes as usual a contraction. Moreover, the indices carried by a tensor are substituted by a single label with a subscript indicating their total number. For instance, the combinations in \eqref{notation_bose} may also be presented as
\be \label{example-repeated_bose}
\psi^{[n]} = \psi_{\m_{s-2n}} \, , \qquad
D \psi = D_{\m} \psi_{\m_s} \, , \qquad
\g\,\psi \equiv \g_\m \psi_{\m_{s}} \, . 
\ee

\newpage
\section{Covariant boundary conditions}\label{app:boundary_fermi}

In this appendix we recall the falloff at the boundary of $AdS_d$ of the solutions of the Fang-Fronsdal equations of motion (see also \cite{shadows-fermi}). To achieve this goal we partially fix the gauge freedom, and we also exhibit the falloffs of the parameters of the residual gauge symmetry (which include the $\g$-traceless Killing spinor-tensors of $AdS_d$).

We set the AdS radius to $L=1$ and we work in the Poincar\'e patch parameterised as
\be \label{poincare-z}
ds^2 = \frac{1}{z^2} \left( dz^2 + \h_{IJ} dx^I dx^J \right) .
\ee 
We also fix the local frame as
\be
e_\m{}^A = \frac{1}{z}\, \d_\m{}^A \, , \quad \o_I{}^{zJ} = \frac{1}{z}\, \d_I{}^J \, , \quad \o_z{}^{\m\n} = \o_I{}^{JK} = 0 \, ,
\ee
where we take advantage of the form of the vielbein to identify ``flat'' and ``curved'' indices.
In these coordinates the spatial boundary is at $z \to 0$. All results can be easily translated in the coordinates \eqref{AdS} used in the main body of the text, in which the boundary is at $r \to \infty$. We denote by capital Latin indices all directions transverse to the radial one (including time).

\subsection{Falloff of the solutions of the free equations of motion}\label{sec:eom-fermi}

We wish to study the solutions of the Fang-Fronsdal equation on a constant-curvature background of dimension $d$ \cite{fronsdal-AdS_fermi} which, in the index-free notation of section \textbf{\ref{sec:arbitrary}}, reads
\be \label{fang-fronsdal}
i \left( \slashed{D} \psi - s\, D \slashed{\psi} + \frac{d+2(s-2)}{2}\, \psi + \frac{s}{2}\, \g\, \slashed{\psi} \right) = 0 \, .
\ee
To this end it is convenient to partially fix the gauge freedom \eqref{gauge-fermi} by setting to zero the $\g$-trace of the field (see \cite{dWF} or sect.~2.2 of the review \cite{modave1} for a discussion of this partial gauge fixing in flat space). This leads to the system of equations
\begin{subequations} \label{fierz-fermi}
\begin{align} 
& i \left( \slashed{D} + \frac{d+2(s-2)}{2} \right) \psi = 0 \, , \label{eom-fermi} \\
& \slashed{\psi} = 0 \, . \label{gamma}
\end{align}
\end{subequations}
These conditions also imply that the divergence of the field vanishes: taking the $\g$-trace of the first equation one indeed obtains
\be \label{div-fermi}
0 = \g^\m \slashed{D} \psi_{\m} = - \slashed{D} \slashed{\psi} + 2\, D\cdot \psi \, ,
\ee
that implies $D\cdot \psi = 0$ thanks to the second equation.
Imposing \eqref{gamma} does not fix completely the gauge freedom: eqs.~\eqref{fierz-fermi} admit a residual gauge symmetry with parameters constrained as
\begin{subequations} \label{fierz-gauge-fermi}
\begin{align}
& \! \left( \slashed{D} + \frac{d+2(s-1)}{2} \right) \e = 0 \, , \label{gauge-eom-fermi} \\
& D\cdot \e = 0 \, , \label{gauge-div-fermi} \\
& \slashed{\e} = 0 \, , \label{gauge-gamma}
\end{align}
\end{subequations}
where the cancellation of the divergence follows from the other two conditions as above.\footnote{Eqs.~\eqref{fierz-gauge-fermi} manifestly guarantee that gauge transformations of the form \eqref{gauge-fermi} preserve the $\gamma$-trace constraint \eqref{gamma}. The divergence constraint \eqref{div-fermi} is also preserved thanks to
\[
\d\, D\cdot \psi = \left( \Box + \frac{1}{2} \slashed{D} - \frac{(s-1)(2d+2s-5)}{2} \right) \e = \left( \slashed{D} - \frac{d+2s-3}{2} \right) \! \left( \slashed{D} + \frac{d+2(s-1)}{2} \right) \e \, .
\]
}

To analyse the falloff at $z \to 0$ of the solutions of \eqref{fierz-fermi} one has to treat separately field components with a different number of indices along the $z$ direction. We denote them as
\be
\psi_{z_n I_{s-n}} \equiv \psi_{z \cdots z I_1 \cdots I_{s-n}} \, .
\ee
The $\g$-trace constraint \eqref{gamma} then gives
\be \label{gamma-exp}
\hat{\g}^z\, \psi_{z\m_{s-1}} + \hat{\g}\cdot \psi_{\m_{s-1}} = 0 \, ,
\ee
where here and below contractions only involve indices transverse to $z$.
Using \eqref{gamma-exp}, the components of the equation of motion \eqref{eom-fermi} read
\be \label{eom-exp-fermi}
\begin{split}
& \hat{\g}^z \left( z\,\pr_z - \frac{d-2(s-n)-1}{2} \right) \psi_{z_n I_{s-n}} + \frac{d+2(s-2)}{2}\, \psi_{z_n I_{s-n}}  \\
& + z\, \hat{\g}^J \pr_J \psi_{z_n I_{s-n}} - (s-n)\, \hat{\g}_I \psi_{z_{n+1}I_{s-n-1}}  = 0 \, ,
\end{split}
\ee
where here and in the rest of this appendix repeated covariant or contravariant indices denote a symmetrisation.
To analyse these equations it is convenient to begin from the divergence constraint they imply, 
\be \label{div-exp-fermi}
\left( z\,\pr_z - d + \frac{3}{2} \right) \psi_{z\m_{s-1}} + z\, \prd \psi_{\m_{s-1}} = 0 \, ,
\ee
which entails $\psi_{z_n I_{s-n}} \sim z^{\D + n}$. Even if the equations are of first order, two values of $\Delta$ are admissible due to the dependence on $\hat{\g}^z$ in \eqref{eom-exp-fermi}. Asymptotically one can indeed split each component of the field as
\be
\psi_{z_n I_{s-n}} = \psi^{+}_{z_n I_{s-n}} + \psi^{-}_{z_n I_{s-n}} \, ,
\ee
where the $\psi^\pm$ are eigenvectors of $\hat{\g}^z$, i.e.
\be \label{eigenvectors}
\hat{\g}^z \psi^{\pm}_{z_n I_{s-n}} = \mp\, \psi^{\pm}_{z_n I_{s-n}} \, .
\ee
Substituting this ansatz in \eqref{eom-exp-fermi}, the terms in the second line are subleading for $z \to 0$ and the first line vanishes provided that
\be \label{fall-off-fermi-field}
\psi^{\pm}_{z\cdots z\, I_1 \cdots I_{s-n}} \sim z^{\D_\pm+\,n} \quad \textrm{with} \ \
\left\{
\begin{array}{l}
\D_+ = d-\frac{5}{2} \\[5pt]
\D_- = \frac{3}{2}-2s
\end{array}
\right. .
\ee
This implies that asymptotically one has to force a projection as already noticed for $s = 3/2$ and $d= 4$ \cite{HT} (see also \cite{bnd_spin_5-2,bnd_spin_5-2_2} for the extension to arbitrary $d$ and \cite{AdS/CFT-spinors,boundary-dirac} for $s = 1/2$). A comparison with the fall-off conditions for Bose fields recalled in (C.9) of \cite{Campoleoni:2016uwr} shows that
\be
\D^{Fermi}_{\pm} = \D^{Bose}_{\pm} \pm \frac{1}{2} \, ,
\ee
while for $s = 0$ one recovers the asymptotic behaviour of a Dirac fermion of mass \mbox{$m^2 = - 2(d-3)$}.

\subsection{Residual gauge symmetry}\label{sec:gauge-fermi}

The fall-off conditions of the parameters of the residual gauge symmetry are fixed by eqs.~\eqref{fierz-gauge-fermi}. The divergence and trace constraints give
\begin{align}
\left( z\,\pr_z - d + \frac{3}{2} \right) \e_{z\m_{s-2}} + z\, \prd \e_{\m_{s-2}} & = 0 \, , \\
\hat{\g}^z\, \e_{z\m_{s-2}} + \hat{\g}\cdot \e_{\m_{s-2}} & = 0 \, ,
\end{align}
and the first condition implies $\e_{z_n j_{s-n-1}} \sim z^{\Th + n}$. By using these identities in \eqref{gauge-eom-fermi} one obtains
\be \label{gauge-exp-fermi}
\begin{split}
& \hat{\g}^z \left( z\,\pr_z - \frac{d-2(s-n)+1}{2} \right) \e_{z_n I_{s-n-1}} + \frac{d+2(s-1)}{2}\, \e_{z_n I_{s-n-1}}  \\
& + z\, \hat{\g}^J \pr_J \e_{z_n I_{s-n-1}} - (s-n-1)\, \hat{\g}_I \e_{z_{n+1}I_{s-n-2}}  = 0 \, .
\end{split}
\ee
This equation has the same form as \eqref{eom-exp-fermi}, apart from the shift $s \to s-1$ and a modification in the mass terms. As a result, by decomposing the gauge parameters as $\e = \e^+ + \e^-$ with 
\be \label{eigenvectors2}
\g^z \e^{\pm}{}_{z_n I_{s-n-1}} = \mp\, \e^{\pm}{}_{z_n I_{s-n-1}}
\ee
and following the same steps as above one obtains
\be \label{x-cov-fermi}
\e^\pm_{z\cdots z\, I_1 \cdots I_{s-n-1}} \sim z^{\Theta_\pm+\,n} \quad \textrm{with} \ \
\left\{
\begin{array}{l}
\Theta_+ = d-\frac{1}{2} \\[5pt]
\Theta_- = \frac{3}{2}-2s
\end{array}
\right. .
\ee
A comparison with the fall-off conditions in (C.12) of \cite{Campoleoni:2016uwr} shows that
\be
\Th^{Fermi}_{\pm} = \Th^{Bose}_{\pm} \pm \frac{1}{2}
\ee
also for gauge parameters.

One can compare these results with the conditions satisfied by a gauge transformation preserving the AdS background, for which
\be \label{killing-fermi}
\d \psi = s \left( D \e + \frac{1}{2}\, \g\, \e \right) = 0 \, ,
\qquad \slashed{\e} = 0 \, .
\ee
These constraints also imply $D\cdot\e =0$. Expanding \eqref{killing-fermi} one obtains
\be \label{killing-vs-boundary}
\left( z\,\pr_z + (2s-n-2) + \frac{1}{2}\, \hat{\g}^z \right) \e_{z_nI_{s-n-1}} = \cO(z^{\Th + n + 1}) \, .
\ee
This equation is analysed more in detail in section \textbf{\ref{sec:bnd}}; here it is worth noting that the solutions in the $\Theta_-$ branch of \eqref{x-cov-fermi} also solve \eqref{killing-vs-boundary}.

\subsection{Initial data at the boundary}\label{sec:leading-fermi}

In this subsection we display the constraints on the initial data at the boundary imposed by the equations of motion and the $\g$-trace constraint, and how the number of independent components is further reduced by the residual gauge symmetry. First of all, note that the solutions of \eqref{fierz-fermi} are generically of the form
\be
\psi_{z_mI_{s-m}} = \sum_{n=0}^\infty z^{\D_+ +m+n } q^{(m,n)}_{I_{s-m}}(x^k) 
\quad \textrm{or} \quad
\psi_{z_mI_{s-m}} = \sum_{n=0}^\infty z^{\D_- +m+n } \rho^{(m,n)}_{I_{s-m}}(x^k) \, ,
\ee 
where all spinor-tensors in the series have a definite (alternating) chirality:\footnote{The components of e.g.\ the $q^{(m,n)}_{I_{s-n}}$ can be considered as spinors defined on the $(d-1)$-dimensional boundary of AdS. When $d$ is odd, \eqref{proj-q-r} is a chirality projection. When $d$ is even, a priori the boundary values of $\psi_{IJ}$ would be collected in a couple of Dirac spinors and \eqref{proj-q-r} selects one of them.}
\be \label{proj-q-r}
\hat{\g}^z q^{(m,n)} = (-1)^{n+1} q^{(m,n)} \, , \qquad 
\hat{\g}^z \r^{(m,n)} = (-1)^n \r^{(m,n)} \, .
\ee
The $\g$-trace constraint \eqref{gamma-exp} then allows one to solve all components $\psi_{z_nI_{s-n}}$ with $n \geq 1$ in terms of the purely transverse one, $\psi_{I_s}$. The equation of motion \eqref{eom-exp-fermi} finally fixes the subleading components of $\psi_{I_s}$ in terms of the leading one.

Within the admissible fall-off conditions, the $\D_+$ branch is the one which is relevant for the analysis of surface charges. We denote its leading contributions in $\psi_{I_s}$ as 
\be \label{vev_fermi}
\psi_{I_s} = z^{\D_+} \cQ^-_{I_s}(x^K) + z^{\D_+ + 1} \cQ^+_{I_s}(x^K) + \cO(z^{\D_+ + 2}) \, , \qquad 
( 1 \mp \hat{\g}^z ) \cQ^\pm_{I_s} = 0 \, .
\ee
For completeness, we shall also analyse the constraints imposed on the leading contributions in the $\D_-$ branch, denoted as
\be \label{source_fermi}
\psi_{I_s} = z^{\D_-} \Psi^+_{I_s}(x^K) + z^{\D_- + 1} \Psi^-_{I_s}(x^J) + \cO(z^{\D_- + 2}) \, , \qquad 
( 1 \mp \hat{\g}^z ) \Psi^\pm_{I_s} = 0 \, .
\ee
The spinor-tensors $\cQ^-_{I_s}$ and $\Psi^+_{I_s}$ are boundary fields of opposite chirality (or, when the dimension of the boundary is odd, Dirac fields with different eigenvalues of $\hat{\g}^z$) of conformal dimensions, respectively, $\D_c = d+s-\frac{5}{2}$ and $\Delta_s = \frac{3}{2}-s$. They thus correspond to the fermionic \emph{conserved currents} and \emph{shadows fields} of \cite{shadows-fermi}.\footnote{If one performs a dilatation $x^\m \to \l x^\m$ then $\psi_{I'_s} = \l^{-s} \psi_{I_s}$, while on the right-hand side of \eqref{vev_fermi} or \eqref{source_fermi} one has $z'^{\D_\pm} = \l^{\D_\pm} z^{\D_\pm}$. As a result, both $\cQ$ and $\Psi$ must transform as $\cQ_{I'_s} = \l^{- ( \Delta_+ + s )} \cQ_{I_s}$, from where one reads the conformal dimensions. To compare them with eqs.~(5.8) and (6.6) of \cite{shadows-fermi}, consider that $d_{here} = (d+1)_{there}$.} 

Combining the e.o.m.\ (or, equivalently, the divergence constraint) and the $\g$-trace constraint gives
\be
\pr^J \cQ^-_{JI_{s-1}} = 0 \, , \qquad
\hat{\g}^J \cQ^-_{JI_{s-1}} = \h^{JK}\cQ^+_{JKI_{s-2}} = 0 \, .
\ee
Eq.~\eqref{eom-exp-fermi} also allows one to fix the $\g$-traceless component of $\cQ^+$ as
\be
\left(\cQ^+_{I_s}\right)^{Tr} \equiv \cQ^+_{I_s} - \frac{s}{d+2s-3}\, \hat{\g}^{\phantom{+}}_{I}\! \slashed{\cQ}^+_{I_{s-1}} = - \frac{1}{d+2s-3}\, \slashed{\pr} \cQ^-_{I_s} \, . 
\ee
The $\g$-trace of $\cQ^+$ remains free as the divergenceless part of $\cQ^-$. In the $\D_-$ branch the $\g$-trace constraint similarly imposes
\be
 \hat{\g}^J \Psi^+_{JI_{s-1}} = \h^{JK}\Psi^-_{JKI_{s-2}} =0 \, .
\ee
The full $\g$-traceless $\Psi^+$ remains instead unconstrained, while all $\Psi^-$ is now fixed as
\be
\Psi^-_{I_s} = - \frac{1}{d+2s-5}\left( \slashed{\pr} \Psi^+_{I_s} - \frac{s}{d+2s-4}\, \hat{\g}^{\phantom{+}}_{I}\! \prd \Psi^+_{I_{s-1}} \right) .
\ee
Note that, in analogy with Bose field \cite{Campoleoni:2016uwr}, the total number of independent data that asymptotically can be chosen arbitrarily\footnote{Of course, regularity  in the bulk should  fix the vev in terms of the source.  We do not discuss this issue here as we focus on the asymptotic behaviour of the theory. Integration in the bulk necessitates the full theory beyond the linear terms.} is the same in both branches, even if they are distributed in different ways in \eqref{vev_fermi} and \eqref{source_fermi}.

The number of independent initial data is further reduced by the residual gauge symmetry. The components of the field vary as
\be
\begin{split}
z\,\d \psi_{z_n I_{s-n}} & = n \left( z\,\pr_z + 2s - n - 1 + \frac{\hat{\g}^z}{2} \right) \e_{z_{n-1}I_{s-n}} + \frac{s-n}{2}\, \hat{\g}_I\! \left( 1 - \hat{\g}^z \right) \e_{z_{n}I_{s-n-1}} \\
& + (s-n) \left( z\,\pr_I \e_{z_nI_{s-n-1}} - (s-n-1) \h_{II} \e_{z_{n+1}I_{s-n-2}} \right) .
\end{split}
\ee
Gauge transformations generated by
\be
\e_{z_n I_{s-n-1}} = z^{\Th_+ + n}\, \xi^{(n)}_{I_{s-n-1}}(x^K) + \cO(z^{\Th_+ + n + 1}) \, , \qquad
(1+\hat{\g}^z)\, \xi^{(n)}_{I_{s-n-1}} = 0 \, ,  
\ee
naturally act on the $\D_+$ branch of solutions of the e.o.m.: they allow to set to zero the $\psi_{z_nI_{s-n}}$ components with $n \geq 1$ (and therefore also $\slashed{\cQ}^+$), while they leave $\cQ^-$ invariant. On the other hand, gauge transformations generated by
\be
\e_{z_n I_{s-n-1}} = z^{\Th_- + n}\, \varepsilon^{(n)}_{I_{s-n-1}}(x^K) + \cO(z^{\Th_- + n + 1}) \, , \qquad
(1-\hat{\g}^z)\, \varepsilon^{(n)}_{I_{s-n-1}} = 0 \, , 
\ee
naturally act on the $\D_-$ branch and they affect the leading contribution as
\be
\d \Psi^+_{I_s} = s\, \pr_I \ve_{I_{s-1}} - \frac{s}{d+2s-3} \left( \hat{\g}_I \slashed{\pr} \ve_{I_{s-1}} + (s-1) \h_{II} \prd \ve_{I_{s-2}} \right) ,
\ee
where we defined $\ve \equiv \ve^{(0)}$ and the variation is $\g$-traceless as it should.
This gauge freedom reduces the number of independent components $\Psi^+$ such that it becomes identical to that of the conserved current $\cQ^-$. It also leaves the coupling $i\! \left( \bar{\Psi}^+ \cQ^- + \bar{\cQ}^- \Psi^+ \right)$ invariant.

\newpage
\section{Conformal Killing spinor-tensors}\label{app:conformal-killing}

In this appendix we present the general solution of the conformal Killing equations \eqref{conformal-killing} for spinor-vectors in $d > 3$ (for $d=3$ see section \textbf{\ref{sec:3d}}). We then estimate the number of independent solutions of the conformal Killing equations \eqref{generic-conf-kill} for spinor-tensors of arbitrary rank. We conclude by proving the identities that we used in section \textbf{\ref{sec:example-symm}} to show that asymptotic symmetries are generated by parameters satisfying only the conditions \eqref{final-gauge} and \eqref{conformal-killing}. We also display recursion relations that, for arbitrary values of the rank, allow one to express all subleading components of the parameters generating asymptotic symmetries in terms of the leading ones.

\subsection{Conformal Killing spinor-vectors}\label{app:spinor-vectors}

Ignoring the spinorial index, eq.~\eqref{conf2-} has the same form as the conformal Killing vector equation in Minkowski space. It is therefore natural to consider the ansatz
\be \label{gen-sol-kill}
\z^+_I = v^+_I - \left( x_J \hat{\g}^J \right) v^-_I \, , \qquad
\z^-_I = v^-_I \, ,
\ee
where the $v^\pm_I$ have the same dependence on $x^I$ as conformal Killing vectors:
\be \label{v_i}
v^\pm_I \equiv a^\pm_I + \o^\pm_{IJ}\, x^J + b^\pm\,x_I + c^\pm_J \left( 2 x_I x^J - x^2 \d_I{}^J \right) , 
\qquad
(1 \mp \hat{\g}^r) v^\pm_I = 0 \, .
\ee
Eqs.~\eqref{gen-sol-kill} generalise the general solution of the conformal Killing spinor equations \cite{AdS-killing-spinors} and, indeed, they solve \eqref{conf2+} and \eqref{conf2-} for constant spinor-tensors $a^\pm_I$, $\o^\pm_{IJ}$, $b^\pm$ and $c^\pm_I$ only subjected to the chirality projections inherited from $\z^\pm_I$. The $\g$-trace conditions \eqref{constr2+} and \eqref{constr2-} impose relations between these spinor-tensors, that one can conveniently analyse by decomposing them in $\g$-traceless components. For instance, the $\g$-traceless projections of $a^\pm_I$ and of the antisymmetric $\o^\pm_{IJ}$ are defined as
\begin{subequations}
\begin{align}
\hat{a}^\pm_I & \equiv a^\pm_I - \frac{1}{d-1}\, \hat{\g}^{\phantom{\pm}}_I \!\slashed{a}^\pm \, , \\
\hat{\o}^\pm_{IJ} & \equiv \o^\pm_{IJ} + \frac{2}{d-3}\, \hat{\g}^{\phantom{\pm}}_{[I} \hat{\g}^K \o^\pm_{J]K} - \frac{1}{(d-2)(d-3)}\, \hat{\g}_{IJ} \hat{\g}^{KL} \o_{KL} \, .
\end{align}
\end{subequations}
The constraints relate the $\g$-traces of $a^\pm_I$, $\o^\pm_{IJ}$ and $c^\pm_I$ to other spinor-tensors in \eqref{v_i}, such that the general solution of the full system of equations \eqref{conformal-killing} is given by
\be
\begin{split}
\z^+_I & = \hat{a}^+_I + x^J\! \left\{ \hat{\o}^+_{IJ} - \hat{\g}^{\phantom{+}}_{(I} \hat{a}^-_{J)} + \frac{d+1}{d-3}\, \hat{\g}^{\phantom{+}}_{[I} \hat{a}^-_{J]} + \frac{d}{d-1}\, \h_{IJ} b^+ - \frac{d}{(d-1)(d-2)}\, \hat{\g}_{IJ} b^+ \right\} \\
& + x^Jx^K \left\{ \frac{2(d-2)}{d-3}\, \h^{\phantom{+}}_{I(J} \hat{c}^+_{K)} - \frac{d-1}{d-3}\, \h^{\phantom{+}}_{JK} \hat{c}^+_{I} - \frac{2}{d-3}\, \hat{\g}^{\phantom{+}}_{I(J} \hat{c}^+_{K)} + \hat{\g}^{\phantom{-}}_{(J} \hat{\o}^-_{K)I} \right. \\
& \left. - \frac{d}{d-2} \left( \h_{I(J} \hat{\g}_{K)} b^- - \frac{1}{d-1}\, \h_{JK} \hat{\g}_I b^- \right) \right\} + x^Jx^Kx^L \left\{ \h^{\phantom{-}}_{(JK} \hat{\g}^{\phantom{-}}_{L)} \hat{c}^-_I - 2\, \h^{\phantom{-}}_{I(J} \hat{\g}^{\phantom{-}}_K \hat{c}^-_{L)} \right\}
\end{split}
\ee
and
\be
\begin{split}
\z^-_I & = \hat{a}^-_I - \frac{1}{d-1}\, \hat{\g}_I b^+ + x^J \left\{ \hat{\o}^-_{IJ} - \frac{4}{d-3}\, \hat{\g}^{\phantom{+}}_{[I} \hat{c}^+_{J]} + \h_{IJ} b^- + \frac{2}{(d-1)(d-2)}\, \hat{\g}_{IJ} b^- \right\} \\
& + x^Jx^K \left\{ 2\,\h^{\phantom{-}}_{I(J} \hat{c}^-_{K)} - \h^{\phantom{-}}_{JK} \hat{c}^-_{I} \right\} .
\end{split}
\ee

\subsection{Comments on arbitrary rank}\label{app:conformal-general}

In the previous subsection we have seen that, in the rank-1 case, the general solution of the conformal Killing equations \eqref{generic-conf-kill} in $d-1$ dimensions depends on the integrations constants collected in the $\g$-traceless spinor-tensors $\hat{a}^\pm_I$, $\hat{\o}^\pm_{IJ}$, $b^\pm$ and $\hat{c}^\pm_I$. Hence, in analogy with what happens for ``bosonic'' conformal Killing tensors \cite{algebra}, there are as many integration constants as independent $\g$-traceless Killing spinor-tensors of a Minkowski space of dimensions $d$. The equations
\be
\d_{(flat)} \psi_{\m\n} = 2\,\pr_{(\m} \e_{\n)} = 0 \, , \qquad
\g^\m \e_\m = 0
\ee
are indeed solved by
\be \label{sol-flat-killing}
\e_\m = A_\m + B_{\m\n} x^\n \, , \qquad 
\g^\m A_\m = \g^\m B_{\m\n} = B_{(\m\n)} = 0 \, ,
\ee
and the number of independent components of the constants $A_\m$ and $B_{\m\n}$ in $d$ dimensions equates that of $\hat{a}^\pm_I$, $\hat{\o}^\pm_{IJ}$, $b^\pm$ and $\hat{c}^\pm_I$ in $d-1$ dimensions. This indicates that the number of independent $\g$-traceless Killing spinor-tensors on AdS and Minkowski backgrounds is the same (at least up to rank 1).
 
The pattern of independent spinor-tensors entering \eqref{v_i} can be understood from the branching rules for representations of the orthogonal group (see e.g.\ \S~8.8.A of \cite{Barut}). Denoting a Young tableau with $s$ boxes in the first row and $k$ boxes in the second by $\{s,k\}$, a $\g$-traceless $\{s,k\}$-projected spinor-tensor in $d+1$ dimensions decomposes in a sum of two-row projected spinor-tensors in $d$ dimensions as\footnote{This rule can be also checked by considering that the number of components of a $\{s,k\}$-projected $\g$-traceless Dirac spinor-tensor in $d$ dimensions is (see e.g.~(A.39) of \cite{review-mixed})
\[
\textrm{dim}_{O(d)}\{s,k\} = 2^{\left[\frac{d}{2}\right]}\frac{(s-k+1)}{(s+1)!k!}\, \frac{(d+s-3)!(d+k-4)!}{(d-2)!(d-4)!}\, (d+s+k-2) \, .
\]}
\be \label{branching}
\{s,k\}_{d} = \sum_{r=k}^s \sum_{l=0}^k\, n(d) \{r,l\}_{d-1} \, ,
\ee
where the multiplicity factor $n(d)$ is equal to 1 when $d$ is odd and to $2$ when $d$ is even. Applying this rule to $A_\m$ and $B_{\m\n}$, one recovers the spinor-tensors entering \eqref{v_i}. When $d$ is odd, the $\pm$ doubling in \eqref{v_i} allows to reproduce the components of a Dirac $\e_\m$ from two sets of Weyl spinor-tensors. When $d$ is even, the doubling accounts for the factor $n(d)$ in \eqref{branching}.

A full derivation of the solutions of the conformal Killing equations \eqref{generic-conf-kill} will be given elsewhere (see also \cite{superconformal} for related work based on superspace techniques). Here we assume that the pattern emerged in the rank $0$ and $1$ examples extends to arbitrary values of the rank. Accordingly, we assume that, for $d > 3$, the number of independent $\g$-traceless Killing spinor-tensors on AdS and Minkowski backgrounds is the same. In the limit $L \to \infty$ the solutions of the Killing equations \eqref{killing-ads-gen} are given by
\be \label{killing-flat}
\e_{\m_1 \cdots \m_s} = \sum_{k=0}^s \, A_{\m_1 \cdots \m_s | \n_1 \cdots \n_k} x^{\n_1} \cdots x^{\n_k} \, , 
\quad
\g^\r A_{\r\,\m_2 \cdots \m_{s} | \n_1 \cdots \n_k} = A_{(\m_1 \cdots \m_s | \m_{s+1}) \n_1 \cdots \n_{k-1}} = 0 \, .
\ee
Their number is therefore equal to the number of components of a $\g$-traceless (Weyl) spinor-tensor in $d+1$ dimensions with the symmetries of a rectangular $\{s,s\}$ Young tableau, that is to
\be
\textrm{dim}_{O(d+1)}\{s,s\} = 2^{\left[\frac{d}{2}\right]} \frac{(d+s-2)!(d+s-3)!(d+2s-1)}{s!(s+1)!(d-1)!(d-3)!}\,  \, .
\ee

\subsection{Independent conditions on asymptotic symmetries}\label{app:independent}

\subsubsection*{Identities involving conformal Killing spinor-vectors}

In order to verify that the conditions \eqref{final-gauge} and \eqref{conformal-killing} on the gauge parameter $\e^\m$ fully characterise asymptotic symmetries for spin-$5/2$ fields, one has to prove that \eqref{diff-constr} holds and that the second line in \eqref{dpsi-ij-exp} vanishes. This requires
\begin{align}
\frac{d-1}{d}\, \slashed{\pr} \prd \z^+ + (d+1)\, \prd\z^- & = 0 \, , \label{extra1} \\
\slashed{\pr} \prd\z^- & = 0 \, , \label{extra2} \\
\frac{d-1}{d}\, \pr_I \pr_J \prd\z^+ + 2\, \hat{\g}_{(I} \pr_{J)} \prd \z^- & = 0 \, , \label{extra3} \\
\pr_I \pr_J \prd \z^- & = 0 \, . \label{extra4}
\end{align}

We wish to prove that the identities above follow from \eqref{conformal-killing}. One can obtain scalars from these equations only by computing a double divergence or a divergence and a $\g$-trace (since the Killing equations are traceless). Eliminating $\slashed{\z}^\pm$ via \eqref{constr2+} and \eqref{constr2-}, the double divergences of \eqref{conf2+} and \eqref{conf2-} become, respectively,
\begin{align}
\frac{d-1}{d}\, \Box \prd\z^+ + \slashed{\pr} \prd \z^- & = 0 \, , \label{dd+} \\
\frac{d-2}{d-1} \, \Box \prd \z^- & = 0 \, . \label{dd-}
\end{align}
Computing a divergence and a $\g$-trace one obtains instead
\begin{align}
\frac{d-1}{d}\, \slashed{\pr} \prd\z^+ + (d+1) \prd \z^- & = 0 \, , \label{dg+} \\
\frac{d-1}{d}\, \Box \prd \z^+ - (d-3) \slashed{\pr} \prd \z^- & = 0 \, . \label{dg-}
\end{align}
Eq.~\eqref{dg+} directly shows that \eqref{extra1} is not independent from \eqref{conformal-killing}. Moreover, combining \eqref{dd+} and \eqref{dg-}, for $d > 2$ one obtains
\be \label{boxes}
\Box \prd\z^+ = \Box \prd \z^- = \slashed{\pr} \prd \z^- = 0 \, ,
\ee
so that \eqref{extra2} is not independent as well. All in all, this implies that the $\g$-trace constraint \eqref{diff-constr} is satisfied when the conformal Killing equations \eqref{conformal-killing} hold.

One can prove \eqref{conf2-} $\Rightarrow$ \eqref{extra4} by acting with a gradient on \eqref{conf2-} and manipulating the result as in Appendix~D of \cite{Campoleoni:2016uwr}:
\be
\begin{split}
0 & = 2\,\pr_K\! \left( \pr_{(I} \z_{J)}{}^{\!\!-} - \frac{1}{d-1}\,\h_{IJ}\, \prd \z^- \right) \\
& = 3\! \left( \pr_{(I} \pr_J \z_{K)}{}^{\!\!-} - \frac{2}{d-1}\,\h_{(IJ}\pr_{K)} \prd \z^- \right) - \pr_I \pr_J \z_{K}{}^{\!\!-} + \frac{4}{d-1}\, \h_{K(I} \pr_{J)} \prd \z^- \label{trick1} \\
& = 3\! \left( \pr_{(I} \pr_J \z_{K)}{}^{\!\!-} - \frac{1}{d-1}\,\h_{(IJ}\pr_{K)} \prd \z^- \right) - \pr_I \pr_J \z_{K}{}^{\!\!-} + \frac{2}{d-1}\, \h_{K(I} \pr_{J)} \prd \z^- \\
& \phantom{=}\, - \frac{1}{d-1}\, \h_{IJ} \pr_K \prd \z^- \, .
\end{split}
\ee
The terms between parentheses in the last step vanish independently because they are the symmetrisation of the first line. Contracting the other three terms with $\pr^K$ one then obtains (for $d > 1$)
\be
(d-3) \pr_I \pr_J \prd \z^- + \h_{IJ} \Box \prd \z^- = 0 \, .
\ee
The last addendum vanishes thanks to \eqref{dd-} (double divergence of \eqref{conf2-}), so that \eqref{extra4} is satisfied when $d > 3$. In $d =3$ the missing cancellation originates the variation of surface charges discussed at the end of section \textbf{\ref{sec:example-symm}}.

One can prove that \eqref{extra2} is not independent in a similar way. First of all, let us manipulate \eqref{conf2+} (here combined with \eqref{constr2-}) as in \eqref{trick1}:
\be \label{trick2}
\begin{split}
0 & = 2\,\pr_K \!\left( \pr_{(I} \z_{J)}{}^{\!\!+} - \frac{1}{d}\,\h_{IJ}\, \prd \z^+ + \hat{\g}_{(I} \z_{J)}{}^{\!\!-}  \right) \\
& = 3\! \left( \pr_{(I} \pr_J \z_{K)}{}^{\!\!+} - \frac{2}{d}\,\h_{(IJ}\pr_{K)} \prd \z^+ + 2\, \hat{\g}_{(I} \pr_J \z_{K)}{}^{\!\!-} \right) - \pr_I \pr_J \z_{K}{}^{\!\!+} + \frac{4}{d}\, \h_{K(I} \pr_{J)} \prd \z^+ \\
& \phantom{=}\, - 2\, \hat{\g}_{(I} \pr_{J)} \z_{K}{}^{\!\!-} - 2\, \hat{\g}_{K} \pr_{(I} \z_{J)}{}^{\!\!-} \\
& = 3\! \left( \pr_{(I} \pr_J \z_{K)}{}^{\!\!+} - \frac{1}{d}\,\h_{(IJ}\pr_{K)} \prd \z^+ + \hat{\g}_{(I} \pr_J \z_{K)}{}^{\!\!-} \right) - \pr_I \pr_J \z_{K}{}^{\!\!+} + \frac{2}{d}\, \h_{K(I} \pr_{J)} \prd \z^+ \\
& \phantom{=}\, - \frac{1}{d}\, \h_{IJ} \pr_{K} \prd \z^+ 
+ \hat{\g}_{(I|} \pr_{K} \z_{|J)}{}^{\!\!-} - \hat{\g}_{(I} \pr_{J)} \z_{K}{}^{\!\!-} - \hat{\g}_{K} \pr_{(I} \z_{J)}{}^{\!\!-} \, .
\end{split}
\ee
The terms between parentheses in the last step vanish because they are the symmetrisation of the first line. The remaining contributions thus yield another vanishing combination, whose contraction with $\pr^K$ gives
\be \label{trick3}
\frac{d-2}{d}\, \pr_I \pr_J \prd \z^+ + \hat{\g}_{(I} 
\pr_{J)} \prd \z^- - \Box \hat{\g}_{(I} \z_{J)}{}^{\!\!-} + \slashed{\pr} \pr_{(I} \z_{J)}{}^{\!\!-} = 0 \, .
\ee
One can show that $\slashed{\pr} \pr_{(I} \z_{J)}{}^{\!\!-}$ vanishes using first \eqref{conf2-} and then \eqref{boxes}. The term $\Box \hat{\g}_{(I} \z_{J)}{}^{\!\!-}$ is instead proportional to the second contribution, since the divergence of \eqref{conf2-} implies 
\be
\Box \z_{I}{}^{\!\!-} = - \frac{d-3}{d-1}\, \pr_I \prd\z^- \, .
\ee
All in all, \eqref{trick3} becomes
\be
\frac{d-2}{d-1} \left( \frac{d-1}{d}\, \pr_I\pr_J \prd\z^+ + 2\, \hat{\g}_{(I} \pr_{J)} \prd \z^- \right) = 0 \, ,
\ee
thus completing the proof that \eqref{extra3} is not independent from \eqref{conformal-killing}. Note that --~in contrast with the proof of \eqref{extra4}~-- this is true also in $d = 3$, as it is necessary to obtain a variation of $\psi_{IJ}$ satisfying the boundary conditions of section \textbf{\ref{sec:example-bnd}}.

\subsubsection*{Asymptotic Killing spinor-tensors}

In order to show that asymptotic symmetries are generated by gauge parameters of the form \eqref{gauge0-gen} that are fully characterised by $\z^\pm$, one should express the spinor-tensors $\a_k$ and $\b_k$ in terms of the former. This can be done by imposing the cancellation of the variations \eqref{varpsi-gen}. The first variation vanishes provided that 
\begin{subequations} \label{firstset}
\begin{align}
& \pr \a_k + \g\, \b_k + (s-1)\, \h \, \slashed{\b}{}_k = 0 \, , \label{eq1} \\[5pt]
& \pr \b_{k-1} - (s-1)\, \h \, \slashed{\a}{}_k = 0 \, , \label{eq2}
\end{align}
\end{subequations}
while $\d \psi_1$ vanishes provided that
\begin{subequations}
\begin{align}
& 2k\,\b_k + (s-1) \left( \pr \slashed{\a}{}_k + (s-2)\, \h\, \b^{\,\pe}_k \right) = 0 \, , \label{eq3} \\[5pt]
& 2k\,\a_k - (s-1) \left( \pr \slashed{\b}{}_{k-1} - \g\, \slashed{\a}{}_k - (s-2)\, \h\, \a'_k \right) = 0 \label{eq4} \, .
\end{align}
\end{subequations}
The last two equations allow one to fix recursively all $\a_k$ and $\b_k$ in terms of $\a_0 = \z^+$ and $\b_0 = \z^-$ (considering also that the $\g$-trace constraint \eqref{gamma-trace-symm} implies $\slashed{\a}{}_k^{[k]} = \b_k^{[k+1]} = 0$). Consistency with \eqref{firstset} is not manifest; yet, in analogy with what we proved above for $s=2$, it must follow from \eqref{generic-conf-kill}. Assuming compatibility of the full system of equations, one can first solve \eqref{eq2} by taking successive traces so as to obtain
\begin{align}
\slashed{\a}_k^{[n]} & = \!\sum_{j=0}^{k-n-1}\! C(n,j)\, \h^j \! \left\{ 2(n+j+1)\, \prd \b_{k-1}^{[n+j]} +  (s-2(n+j+1))\, \pr \b_{k-1}^{[n+j+1]} \right\} , \\
C(n,j) & = \frac{(-1)^j\, n! (s-2n-2)! (d+2(s-n)-2j-7)!!}{2^{j+1} (n+j+1)! (s-2n-2j-2)! (d+2(s-n)-5)!!} \, .
\end{align}
Combining this result with \eqref{eq4} one gets
\be \label{alpha}
\begin{split}
\a_k & = \frac{s-1}{2k} \left\{ \pr \slashed{\b}{}_{k-1} - \frac{1}{2(d+2s-5)}\, \g \left( 2\,\prd \b_{k-1} + (s-2)\, \pr \b^{\,\pe}_{k-1} \right) \right\} \\
& + \sum_{j=1}^{k-1} A(j,k) \left\{ \h^j \left[ 2j\, \prd\slashed{\b}{}_{k-1}^{[j-1]} + \slashed{\pr} \b_{k-1}^{[j]} + (s-2j-1)\, \pr \slashed{\b}{}_{k-1}^{[j]} \right] \right. \\
& \left. - \frac{(2j+1)(s-2j-1)}{2(j+1)(d+2s-2j-5)}\, \h^j \g \left[ 2(j+1)\, \prd \b_{k-1}^{[j]} + (s-2j-2)\, \pr \b^{[j+1]}_{k-1} \right] \right\}
\end{split}
\ee
with
\be
A(j,k) = \frac{(-1)^j (s-1)! (d+2s-2j-5)!!}{k\,2^{j+1} j! (s-2j-1)!(d+2s-5)!!} \, .
\ee
In a similar fashion, \eqref{eq3} gives directly
\begin{align} \label{sigma}
\b_k & = \sum_{j=0}^{k} B(j,k)\,  \h^j \left\{ 2j\, \prd \slashed{\a}{}_k^{[j-1]} + (s-2j-1)\, \pr \slashed{\a}{}_k^{[j]} \right\} \, , \\
B(j,k) & = \frac{(-1)^{j+1}(s-1)!}{2^{j+1}(s-2j-1)! \prod_{l=0}^j \left[ l(d+2s-2l-5) + k \right]} \, .
\end{align}

\end{subappendices}

\chapter*{Conclusion}

\addcontentsline{toc}{chapter}{Conclusion}

We have covered, in this work, a series of topics revolving around the Hamiltonin analysis of free higher spin gauge fields, gathering a few results along the way. Let us go through them one last time:

\begin{itemize}
\item We have obtained a complete set of (linearized) conformal invariants for three-dimensional bosonic fields of arbitrary spin, in the form of the Cotton tensor, whose properties we proved: it is a symmetric, traceless and divergenceless tensor, and it fully captures conformal curvature, as a necessary and sufficient condition for a higher spin field to be conformally flat is for its Cotton tensor to vanish; equivalently, any conformally invariant local function of a higher spin field is a function of its Cotton tensor and derivatives thereof. Finally, we showed that any tensor sharing the algebraic and differential properties of the Cotton tensor is the Cotton tensor of some higher spin field. This study should be extended to higher dimensions and to fields of mixed symmetry, and also to fermionic fields.

\item This construction allowed us to perform the Hamiltonian analysis of any bosonic higher spin gauge field over a flat background space-time in four dimensions. Beginning from Fronsdal's action, momenta, Hamiltonian and constraints were identified. The constraints were then solved through the introduction of prepotentials, which happen to exhibit conformal gauge invariance. Incidentally, this gauge freedom, together with explicit $SO(2)$ electric-magnetic duality invariance, completely fixes the Hamiltonian action written in terms of the prepotentials. Here too, an extension of this analysis to fields of mixed symmetry in arbitrary dimension and to fermions would be quite interesting.

\item This Hamiltonian action can actually also be seen as the one directly associated to a rewriting of higher spins equations of motion as twisted self-duality conditions. These conditions first appear as $s$th order covariant equations, but a complete first order subset of them can be extracted, equating the electric field of the particle to the magnetic field of its dual, and reciprocally (up to a sign). This kind of formalism could also be largely generalized.

\item We have modestly begun to expand these investigations to fermions, with a Hamiltonian study of four-dimensional (free) hypergravity, which selects as the superpartner of the graviton a spin $5/2$ field. The Hamiltonian analysis of this first fermionic higher spin was realized and its constraints solved, leading to a prepotential also enjoying conformal gauge invariance. The prepotential formalism turned out, as expected, to provide the appropriate framework to study how supersymmetry and electric-magnetic duality (a symmetry of the graviton's action) combined. The study of fermionic fields of arbitrary spin would be natural next step.

\item We have also begun to go beyond the realm of fully symmetric bosonic fields, by studying a chiral $(2,2)$-form in six dimensions, which appears to be an essential part of the maximally extended supersymmetric  $(4,0)$ theory. We rewrote the equations of motion of this mixed-symmetry field as twisted self-duality conditions, first in a manifestly covariant form and then in a first order form, which led us to the prepotential Hamiltonian action of this field, again exhibiting conformal gauge invariance. The rest of the fields intervening in the $(4,0)$ theory should be submitted to the same treatment, to improve our grasp on this fascinating theory.

\item Finally, we turned our attention to AdS in a computation of the surface charges of higher spin gauge fields over a constantly curved space-time of arbitrary dimension, through the Hamiltonian analysis of these theories. Explicit expressions were written down for these charges, for both bosonic and fermionic fields, in terms of the asymptotic values of the fields and of the parameters of improper gauge transformations. These charges could prove to be useful in the study of the thermodynamic properties of some possibly black hole-like exact solutions of Vasiliev's equations.

\end{itemize}

As this final overview hopefully suggests, higher spins remain, more than ever, a fascinating and open field of research in mathematical physics.

\bibliography{mybib}
\bibliographystyle{plain}

\end{document}